\documentclass[11pt]{article}
\usepackage[margin=1in]{geometry}
\usepackage{amsmath}
\usepackage{graphicx}
\usepackage{enumerate}
\usepackage{enumitem}
\usepackage{natbib}
\usepackage{url} % not crucial - just used below for the URL 

\usepackage{algorithmic} %For computer algorithm code in LATEX
\usepackage{algorithm} %For algorithm box around the algorithmic
\usepackage{setspace} %For setting spacing (e.g., double spacing, use \doublespacing)
\usepackage{rotating} %For rotating tables (provides sidewaysfigure and sidewaystable) 

\usepackage{amsfonts, amssymb} %Either provides Real number and other font'related stuff.
\usepackage{amsthm}
\usepackage[T1]{fontenc}
\usepackage{lmodern}

\usepackage{epstopdf} %for adding eps in pdfLatex	
\usepackage{hyperref} %For adding hyperlinks in documents
\usepackage{bm} %Bolds everything when uses \bm as a command
\usepackage{bbm}
\usepackage{multirow} %Have multiple rows in a table
\usepackage{latexsym} %Latex-based math symbols. Generally load this in conjunction with amssymb
\usepackage{rotating} %arbitrary rotate any object
\usepackage{titlesec} %centering section headings
\usepackage{caption} %captioning for tables
\usepackage{mathtools} %ceiling/floor commands
\usepackage{color} %Provides color for text.
\usepackage{diagbox}
\usepackage{pict2e}
\usepackage{etoolbox}
\usepackage[english]{babel}
\usepackage{empheq}

\usepackage{cases}
\usepackage{tikz}
\usetikzlibrary{arrows.meta,shapes,shapes.arrows,
shapes.geometric,shapes.multipart,decorations.pathmorphing,
positioning,shapes.swigs,}

\usepackage{lato}
\usepackage{pifont}
\usepackage{stmaryrd}

\newtheorem{theorem}{Theorem}[section]
\newtheorem{lemma}[theorem]{Lemma}
\newtheorem{proposition}[theorem]{Proposition}

\theoremstyle{definition}
\newtheorem{assumption}{Assumption}

\definecolor{mycolor}{RGB}{0,200,200} 

\newcommand{\indep}{\!\perp\!\!\!\perp\!}
\DeclareMathOperator*{\argmax}{arg\,max}
\DeclareMathOperator*{\argmin}{arg\,min}
\DeclareMathOperator*{\median}{median}
\newcommand{\nindep}{\not \hspace{-0.1cm} \rotatebox[origin=c]{90}{$\models$}\,}
\newcommand{\EXP}{E}
\newcommand{\VAR}{var}
\newcommand{\AVER}{\mathbbm{P}}
\newcommand{\EMP}{\mathbbm{G}}
\newcommand{\cond}{\, \big| \,}

\newcommand{\bcond}{\, \big| \,}
\newcommand{\con}{ ; }
\newcommand{\R}{\mathbbm{R}}
\newcommand{\ind}{\mathbbm{1}}
\newcommand{\T}{^\intercal}
\renewcommand{\Pr}{\text{pr}}

\newcommand{\bX}{{X}}
\newcommand{\bx}{{x}}
\newcommand{\bO}{{O}}
\newcommand{\bo}{{o}}
\newcommand{\Yo}{Y_1}
\newcommand{\Yz}{Y_0}
\newcommand{\pYo}{\potY{0}{1}}
\newcommand{\pYz}{\potY{0}{0}}
\newcommand{\InfFt}{\text{IF}}
\newcommand{\uncInfFt}{\phi}
\newcommand{\MM}{\Omega}
\newcommand{\btheta}{{\theta}} 
\newcommand{\InfFtAug}{\mathcal{U}}
\newcommand{\logit}{\text{logit}}
\newcommand{\expit}{\text{expit}}

\newcommand{\SPz}{\mathcal{S}_0(0)}
\newcommand{\SPo}{\mathcal{S}_1(0)}
\newcommand{\SPzt}{\mathcal{S}_0(1)}
\newcommand{\SPot}{\mathcal{S}_1(1)}

\newcommand{\SPg}{\mathcal{S}}

\newcommand{\SPoreceta}{\mathcal{S}_\alpha(\eta)}
\newcommand{\SPorecetastar}{\mathcal{S}_\alpha(\eta^*)}

\newcommand{\potY}[2]{Y_{#2}^{(#1)}}
\newcommand{\potf}[1]{f_{#1}}
\newcommand{\hpotf}[1]{\widehat{f}_{#1}}
\newcommand{\potg}[1]{f_{#1}}

\newcommand{\pote}[1]{e_{#1}}
\newcommand{\hpote}[1]{\widehat{e}_{#1}}
\newcommand{\LSS}{^{(-k)}}

\newcommand{\HL}[1]{\hyperlink{(#1)}{(#1)}}
\newcommand{\HT}[1]{\hypertarget{(#1)}{(#1)}}
\newcommand{\nHL}[1]{\hyperlink{#1}{#1}}
\newcommand{\nHT}[1]{\hypertarget{#1}{#1}}

\newcommand{\PI}{\text{PI}}
\newcommand{\PT}{\text{PT}}
\newcommand{\OREC}{\text{OREC}}
\newcommand{\EFF}{\text{Eff}}
\newcommand{\SUPP}{Supplementary Material}

\graphicspath{ {plot/} }% plot path

\definecolor{red1}{RGB}{255,64,64}
\definecolor{blue1}{RGB}{128,255,255}
\definecolor{green1}{RGB}{0,205,0}
\definecolor{myblue}{RGB}{34,151,230}

\numberwithin{table}{section}
\numberwithin{figure}{section}

\usepackage{rotating}
\usepackage{caption}
 
\usepackage{hyperref} %For adding hyperlinks in documents

\hypersetup{
colorlinks=true,
linkcolor=blue,
filecolor=blue,
urlcolor=blue,
citecolor=blue
}

\usepackage{enumitem}
\setlist[itemize]{leftmargin=0.5cm, labelsep=0.5em, itemsep=0.0cm, topsep=0.1cm, partopsep=0cm, parsep=0cm}  
\setlist[enumerate]{leftmargin=0.5cm, labelsep=0.5em, itemsep=0cm, topsep=0cm, partopsep=0cm, parsep=0cm}  

\usepackage{titlesec}

\allowdisplaybreaks

\begin{document}

\title{\vspace*{-1cm} A Nonparametric Framework for Universal Difference-in-Differences}
\author{
Chan Park$^{a}$, Eric J. Tchetgen Tchetgen$^{b}$\\[0.25cm]
\makebox[1cm][c]{{\footnotesize $^{a}$Department of Statistics, University of Illinois Urbana-Champaign, Champaign, IL 61820, U.S.A.}}\\
\makebox[1cm][c]{{\footnotesize $^{b}$Department of Statistics and Data Science, University of Pennsylvania, Philadelphia, PA 19104, U.S.A.}}
}
 \date{}
  \maketitle
\begin{abstract}
Difference-in-differences (DiD) is a popular approach to evaluate treatment effects in settings where both pre- and post-treatment measurements of the outcome are available. Despite its popularity, existing methods face important limitations. Specifically, they either:  (i) only apply to continuous outcomes and the average treatment effect on the treated; (ii) are sensitive to the transformation of the outcome; (iii) rely on a no unmeasured confounding assumption given pre-treatment covariates and outcome; (iv) lack semiparametric efficiency theory. In this paper, we introduce a novel framework for causal identification and inference in DiD settings that overcomes limitations (i)-(iv), making it the only existing framework that simultaneously satisfies these properties. Key to our framework is an \emph{odds ratio equi-confounding} assumption, which states that the generalized odds ratio function relating treatment and treatment-free potential outcome is stable across time periods, a form of distributional parallel trends assumption. Under this assumption, we establish nonparametric identification of virtually any standard treatment effect on the treated, including quantile treatment effects on the treated. We also develop corresponding consistent, asymptotically linear, and semiparametric efficient estimators that leverage modern statistical learning theory. We illustrate our framework through simulation studies and two real-world applications using Zika virus outbreak data and traffic safety data.
\end{abstract}
\noindent%
{\it Keywords:}  Average treatment effect on the treated,
Generalized odds ratio,
Mixed-bias property,
Quantile treatment effect on the treated

\newpage

\section{Introduction}			\label{sec:Intro}

Difference-in-differences (DiD) is one of the most widely used methods for assessing the causal impact of hypothetical policy interventions or external shocks. Standard DiD considers a setting with two time periods and pre- and post-treatment outcome measurements. In the canonical DiD setting, observed data can be divided into four groups based on time and treatment status (i.e., treated/control groups at pre/post-treatment time periods), and outcomes in these four groups are used in a specific manner to identify the additive average treatment effect on the treated group (ATT) in the follow-up period under certain conditions. Specifically, the so-called parallel trends (\nHL{PT}) assumption is a key justification for DiD identification of the ATT; \nHL{PT} states that, on average, the change in the treatment-free potential outcome in the treated group over time is equal to that in the control group. Under the \nHL{PT} assumption, the ATT is identified simply by comparing the average change in the outcomes in the treated group over time to that in the control group. In many cases, the treated and untreated groups may, in fact, have different pre-treatment covariate values that could lead to differences in potential outcome changes over time. Such confounding of the treatment effect on outcome trends poses an important threat to the plausibility of the \nHL{PT} assumption in practice, as it may invalidate the latter. In fact, in light of this concern, recent works (e.g., \citet{Heckman1997, Abadie2005, SantAnna2020}) have considered a \nHL{PT} assumption conditional on observed covariates, referred to as conditional \nHL{PT}; see Section 3 of \citet{DiD_Review2011} and Section 4 of \citet{DiD_Review2022} for comprehensive reviews. In what follows, we do not distinguish between marginal and conditional \nHL{PT} assumptions unless otherwise stated.

On the other hand, there has been a fast-growing literature on identifying and estimating treatment effects in DiD settings via an alternative formulation of the \nHL{PT} assumption. Notably, \citet{CiC2006} considered an alternative identifying assumption whereby the treatment-free potential outcomes are possibly nonlinear, monotone transformations of an unobserved confounder. \citet{Puhani2012} and \citet{Wooldridge2022} considered the nonlinear \nHL{PT} assumption, wherein the \nHL{PT} assumption is satisfied upon imposing a user-specified transformation relating the treatment and outcome variables; see Section \ref{sec:review} and {\SUPP} \ref{sec:supp:review} for a detailed review of these and other existing DiD approaches.

Despite a rich literature developed under the \nHL{PT} assumptions and other identifying assumptions, none of the aforementioned approaches is a DiD panacea as either (i) they only apply to continuous outcomes and the additive ATT; (ii) they fail to be transformation-invariant, i.e., their identifying assumptions do depend on how the outcome is transformed; (iii) they assume the absence of unmeasured confounders; or (iv) they lack semiparametric efficiency theory; see Section \ref{sec:KeyUDID} for details.  As a result, in order to potentially obtain more robust and efficient causal inference,  developing methods which overcome these limitations remains a research priority across many disciplines.

This paper develops a novel framework for identifying and estimating treatment effects in canonical DiD settings. Our approach centers on representing confounding bias as an association between the treatment and the treatment-free potential outcome, using a scale that is universally applicable across outcome types. This representation enables unbiased estimation of any causal effect on the treated, regardless of the nature of the outcome. Specifically, we show that confounding bias can be encoded through the association between treatment and the treatment-free potential outcome via a generalized odds ratio function \citep{Chen2007, TTRR2010}. This formulation accommodates discrete, continuous, and mixed-type outcomes, allowing for broad applicability in de-biasing arbitrary causal effects on the treated under the assumption we term \textit{odds ratio equi-confounding} (\nHL{OREC}). \nHL{OREC} states that the confounding bias, measured on a generalized odds ratio scale, can be identified by the generalized odds ratio association between the treatment and the pre-treatment outcome measure. In this sense, \nHL{OREC} provides a natural generalization of the \nHL{PT} assumption to the generalized odds ratio scale; see Assumption \ref{assumption:OREC} and related discussion. Importantly, \nHL{OREC} is neither strictly stronger nor weaker than other identifying assumptions in the DiD literature, including \nHL{PT}, because neither assumption appears to imply the other. Rather, \nHL{OREC} should be viewed as an \textit{alternative} identification condition. As such, identification and estimation strategies under \nHL{OREC} must be developed and studied independently from existing DiD approaches.

To the best of our knowledge, our proposed approach has special merit in that it is the only existing method that satisfies the following key properties: (i) it applies to the causal effects on the treated on any standard effect measure scale of potential interest (e.g., ATT, quantile effects on the treated); (ii) it applies to any outcome type including continuous outcomes, binary outcomes, count outcomes, or a mixture of these; (iii) the \nHL{OREC} assumption is transformation-invariant, i.e., it does not depend on how the outcome is transformed; (iv) it allows for the presence of an unmeasured confounder of the association between treatment and treatment-free potential outcome, and (v) we provide a complete semiparametric efficiency theory for estimation and inference. Throughout the paper, we refer to an estimation method for causal inference in DiD settings satisfying properties (i)-(v) as \emph{Universal DiD} (UDiD), adopting the terminology of \citet{UDID2024_Epi}. As our approach achieves all of these criteria, it can be viewed as a universal framework for estimating treatment effects in DiD settings.

    We note that \citet{UDID2024_Epi} introduced the UDiD framework and the \nHL{OREC} assumption, proposing estimators for the ATT based on parametric nuisance models. This paper complements and extends their work in several ways: (i) we derive the semiparametric efficiency bound for the ATT under \nHL{OREC} within a fully nonparametric model; (ii) we construct estimators that achieve this efficiency bound, with nuisance functions estimated nonparametrically using modern machine learning techniques; (iii) we present several practical approaches for assessing the identifying assumptions, including sensitivity analyses; and (iv) we extend the methodology to a broader class of standard counterfactual estimands, such as the quantile treatment effect on the treated (QTT), as well as settings involving outcomes that are missing not at random in longitudinal studies. In contrast to our approach, the doubly robust estimator proposed in \citet{UDID2024_Epi} relies on consistent estimation of the odds ratio function under a specified parametric model. Even if one were to instead estimate all nuisance functions nonparametrically, the estimator may still fail to be root-$N$ consistent and asymptotically linear, as it can exhibit first-order bias arising from estimation error in the odds ratio function; see {\SUPP} \ref{sec:supp-Other Estimators} for details. By contrast, our estimator is constructed from the efficient influence function, which ensures that the bias is second order in the nuisance estimation errors. This property enables the use of flexible nonparametric or machine learning methods for nuisance estimation while still yielding a root-$N$ consistent and asymptotically linear estimator, and achieving the semiparametric efficiency bound.
    
    In addition, all proposed methods are implemented in the \texttt{UDID} R package \citep{UDIDpackage}, which is publicly available at \url{https://github.com/qkrcks0218/UDID}, along with replication code for the simulation studies and data analysis.

\section{Preliminary}  \label{sec:setup review}

Let $N$ denote the number of observed units, indexed by subscript $i \in \{1,\ldots,N\}$, which we suppress in the notation unless necessary. For each unit, we observe independent and identically distributed (i.i.d.) variables $\bO = (Y_{0}, Y_{1}, A, \bX)$ where $\Yz$ and $\Yo$ are the outcomes at time 0 and 1, respectively, $A \in \{0,1\}$ is the indicator of whether a unit is treated between time 0 and 1, and $\bX \in \mathcal{X} \subseteq \R^d$ are observed $d$-dimensional pre-treatment covariates. Let $\potY{a}{t}$ be the potential outcome, which one would have observed had, possibly contrary to the fact, the treatment been set to $A=a$ at time 0 and 1. In the main text of the paper, we primarily focus on the ATT $\tau^* =
\EXP \big\{ \potY{1}{1} - \pYo \cond A=1 \big\} = \tau_1^* - \tau_0^*$ where  $
\tau_a^* 
=
\EXP \big\{ \potY{a}{1} \cond A = 1 \big\}$, to simplify the exposition. However, in {\SUPP} \ref{sec:General}, we extend the methods to establish inference for a general class of standard counterfactual estimands, including the quantile treatment effect on the treated (QTT). 

To describe our UDiD approach for $\tau^*$, we require notation used throughout. First, let $\mu^*(\bx) = \EXP \big\{ \pYo \cond A=1 , \bX = \bx \big\}$, which results in $\tau_0^* = \EXP \big\{ \mu^*(\bX) \cond A=1 \big\}$. Next, we let $\potf{ t}^*( y \cond a, \bx)$, $\potf{ t}^*( y, \bx \cond a)$, and $\potf{ t}^*( y , a, \bx)$ be the (conditional) density functions of $\potY{0}{t} \cond (A,\bX)$, $(\potY{0}{t},  \bX) \cond A$, and $(\potY{0}{t}, A, \bX)$, respectively. Let $\pote{ t}^*( a \cond y, \bx)$ be the conditional density functions of $A \cond (\potY{0}{t},\bX)$; hereafter, $\pote{ t}^*( a \cond y, \bx)$ is referred to as the \textit{extended propensity score}.  For our UDiD approach, we impose the following support condition:
\begin{assumption}[Support]  \label{assumption:support}
    The density $f_{t}^*(y,a,\bx)$ has the same support over $t \in \{0,1\}$ and $a \in \{0,1\}$, which is denoted by $\mathcal{S} = \big\{ (y,\bx) \cond \potf{t}^*(y,a,\bx) \in (0,\infty) \big\}$. 
\end{assumption}
\noindent Although this common support condition can in principle be relaxed, we introduce it to avoid unnecessary complications in the exposition; see {\SUPP} \ref{sec:supp-OREC} for details on how Assumption \ref{assumption:support} can be relaxed.

Let $y_R$ be a reference value for the outcome satisfying $(y_R,\bx) \in \mathcal{S}$. For $t \in \{0,1\}$, let $\beta_{t}^* (\bx) $ denote the baseline odds function of $A$ given $(\potY{0}{t}=y_R,\bX=\bx)$, and let $\alpha_{t}^* (y, \bx)$ denote the generalized odds ratio function \citep{Chen2007, TTRR2010} relating $\potY{0}{t}$ and $A$ given $\bX=\bx$, i.e.,  
\begin{align}				\label{eq-def-OR}
& \beta_{t}^* (\bx) 
=
\frac{ \pote{t}^* (1 \cond y_R , \bx) }{ \pote{t}^* (0 \cond y_R , \bx) }
%	=
%	\frac{ \potf{t}^* (y_R \cond 1, \bx) }{ \potf{t}^* (y_R \cond 0, \bx) }
%	\frac{\Pr(A=1 \cond \bX=\bx)}{\Pr(A=0 \cond \bX=\bx)}
\ , \
&&
\alpha_{t}^* (y,\bx)
=
%	\frac{ \potg{t}^* (y,1 \cond \bx) }{ \potg{t}^* (y,0 \cond \bx) }
%	\frac{ \potg{t}^* (y_R ,0 \cond \bx) }{ \potg{t}^* (y_R,1 \cond \bx) }
%	=
\frac{ \potf{t}^* (y \cond 1, \bx) }{ \potf{t}^* (y \cond 0, \bx) }
\frac{ \potf{t}^* (y_R \cond 0, \bx) }{ \potf{t}^* (y_R \cond 1, \bx) }
=
\frac{ \pote{t}^* (1 \cond y , \bx) }{ \pote{t}^* (0 \cond y , \bx) }
\frac{ \pote{t}^* (0 \cond y_R , \bx) }{ \pote{t}^* (1 \cond y_R , \bx) } 
\ . 
\end{align}
By definition, $\alpha_t^*(y,\bx)>0$  for $(y,\bx) \in \mathcal{S}$ and $\alpha_t^*(y_R,\bx)=1$ for all $\bx$. We define $\alpha_t^*(y,\bx)=0$ for $(y,\bx) \notin \mathcal{S}$ to avoid indeterminate $0/0$ cases. Of note, we have $\alpha_t^*(y,\bx)=1$ for all $(y,\bx) \in \mathcal{S}$ under exchangeability, i.e., $\potY{0}{t} \indep A \cond X$. In addition, while $y_R$ is involved in the definitions of both $\alpha_t^*$ and $\beta_t^*$, their product $\beta_t^*(\bx)\alpha_t^*(y,\bx) = \pote{t}^*(1 \cond y, \bx) / \pote{t}^*(0 \cond y, \bx)$ is invariant to the choice of $y_R$. Consequently, the specific value of $y_R$ is immaterial because the identification and estimation results depend on these terms only through this product; see, e.g.,  Theorem \ref{thm-EIF}. For implementation, we set $y_R$ to the empirical median of $Y \mid (A=0)$ for continuous $Y$ and $y_R = 0$ for binary $Y$.

Lastly, let $\logit(v) = \log\{ v/ (1-v) \}$ and $\expit(v) = 1/\{ 1+ \exp(-v) \}$. 
Let $\AVER_{\mathcal{I}} (V) = | \mathcal{I} |^{-1} \sum_{i \in \mathcal{I}} V_i $ be the empirical mean of $V$ over a set $\mathcal{I} \subseteq \{1,\ldots,N \}$. We denote $\AVER=\AVER_{\mathcal{I}}$ when $\mathcal{I} = \{1,\ldots,N\}$, i.e., the entire sample. For a sequence of random variables $\{ V_N \}$, let $V_N=O_P(r_N)$ indicate that $V_N/r_N$ is stochastically bounded, and let $V_N=o_P(r_N)$ indicate that $V_N/r_N$ converges to zero in probability as $N \rightarrow \infty$. Let $V_N \stackrel{D}{\rightarrow} W$ mean that $V_N$ weakly converges to a random variable $W$ as $N \rightarrow \infty$. Lastly, let $V \cond Z \stackrel{D}{=}W \cond Z$ mean that $V$ and $W$ are identically distributed conditional on $Z$.  

We make the following assumptions, which are commonly made in the DiD
literature:
\begin{assumption}[Consistency] \label{assumption:consistency}
For $t \in \{0,1\}$, $Y_t = \potY{A}{t}$  almost surely.
\end{assumption}
\begin{assumption}[No Anticipation] \label{assumption:no anticipation}
$\pYz = \potY{1}{0}$ almost surely.
\end{assumption}
\noindent Assumption \ref{assumption:consistency} states that the observed outcome matches the potential outcome corresponding to the observed treatment. Assumption \ref{assumption:no anticipation} states that the treatment does not causally impact the outcome before it is implemented.\footnote{In standard graphical causal models, such as the finest fully randomized causally interpreted structured tree graph \citep{Robins1986} and the nonparametric structural equation model with independent errors \citep{Pearl2009}, future interventions cannot affect past outcomes by construction, so that Assumption \ref{assumption:no anticipation} holds automatically. Nonetheless, it is often explicitly stated in the DiD literature, partly due to the practical concern that units’ outcomes may be affected by the ``planned'' treatment implementation; see \citet{PTTS2025} for a detailed discussion. We follow the common convention of stating it as an explicit assumption.} Note that, under Assumptions \ref{assumption:consistency} and \ref{assumption:no anticipation}, we have $\Yz = \pYz$ almost surely for all units regardless of their treatment status. 
%Interestingly, Assumption \ref{assumption:no anticipation} essentially follows from the definition of a counterfactual outcome in the most common graphical causal models \citep{PTTS2025}.

Under Assumptions \ref{assumption:support}-\ref{assumption:no anticipation}, some model parameters are identified. Specifically, the joint density $f_0^*(y,a,\bx)$ of $(Y_0,A,\bX)$ is identified, which in turn implies identification of $\beta_0^*(\bx)$ and $\alpha_0^*(y, \bx)$, and, likewise, the joint density $f_{1}^*(y,0,\bx)$ corresponding to $(Y_1,A=0,\bX)$ is also identified. Moreover, the first term of the ATT is identified as $\tau_1^* = \EXP ( A \Yo ) / \Pr(A=1)$. However, the joint density $f_{1}^*(y,1,\bx)$ corresponding to $(\potY{0}{1}=y,A=1,\bX=\bx)$, is not identified, and consequently $\tau_{0}^*$ remains unidentified. Therefore, identifying the ATT requires establishing identification of $\tau_0^*$, which we detail in the subsequent section.

\section{A Universal Difference-in-Differences Approach}

\subsection{Odds Ratio Equi-confounding} \label{sec-OREC}

We begin by introducing the key assumption of our UDiD approach:
\begin{assumption}[\hypertarget{OREC}{Odds Ratio Equi-confounding}] \label{assumption:OREC}
    $ \alpha_{0}^* (y,\bx) = \alpha_{1}^* (y, \bx)$ for all $(y,\bx) \in \SPg$, i.e., the generalized odds ratio function relating $A$ and $\potY{0}{t}$ is the same across time periods. 
\end{assumption}  
\noindent Assumption \ref{assumption:OREC} is the key identifying assumption, as it links the counterfactual quantity to the observed data. Specifically, under Assumptions \ref{assumption:support}-\ref{assumption:OREC}, $\alpha_1^*$ is identified through $\alpha_0^*$, since the latter is already identified under Assumptions \ref{assumption:support}-\ref{assumption:no anticipation}. In turn, identification of the ATT $\tau^*$ follows; see Section \ref{sec:Identification} for details.

The generalized odds ratio can be viewed as a distributional scale for measuring the degree of confounding bias for the association between $A$ and $\potY{0}{t}$. In particular, if the generalized odds ratio is equal to 1 for all $(y,\bx)$, it implies that there is no association between $A$ and $\potY{0}{t}$ conditional on $\bX=\bx$, i.e., no confounding bias given $\bX=\bx$. Accordingly, condition $\alpha_0^* = \alpha_1^*$ implies that the confounding bias in a generalized odds ratio scale is stable over time $t \in \{0,1\}$. Therefore, we aptly refer to the condition as the \textit{odds ratio equi-confounding} (\nHL{OREC}) assumption. It is important to notice that it is readily expressed in terms of potential outcomes, without explicit reference to the latent factor that confounds the association between $A$ and $\potY{0}{1}$.

Taking the logarithm on both sides of \nHL{OREC}, we obtain its alternative representation:
\begin{align}
\text{logit} 
\big\{ e_1^*(1 \cond y, \bx) \big\}
-
\text{logit} 
\big\{ e_1^*(1 \cond y_R, \bx) \big\}
=
\text{logit} 
\big\{ e_0^*(1 \cond y, \bx) \big\}
-
\text{logit} 
\big\{ e_0^*(1 \cond y_R, \bx) \big\} \ , \quad \forall (y,\bx) \in \SPg \ . 
\label{eq-EPS PT}
\end{align}
In words, the change in the log odds associated with the extended propensity score over time is the same across all $(y,\bx) \in \SPg$, i.e., a parallel relationship in the log odds of the extended propensity score over time. To better appreciate the condition, suppose that the conditional exposure model given $(\potY{0}{t},\bX)$ for $t \in \{0,1\}$ is given as $ \Pr(A = 1 \cond \potY{0}{t}, \bX) = \text{expit} \big\{ \gamma_{t0} + {\gamma}_{tX} \T \bX + \gamma_{tY} \potY{0}{t} \big\}$. Then, the \nHL{OREC} assumption is equivalent to $\gamma_{0Y} = \gamma_{1Y}$, indicating that the impact of the outcome on the treatment model is time-invariant on the logit scale. Therefore, \nHL{OREC} can be understood as a parallel trend (\nHL{PT}) condition of the extended propensity score on the logit scale. To the best of our knowledge, a \nHL{PT}-type condition on the treatment mechanism is new in the DiD literature. 

Equation \eqref{eq-EPS PT} is also useful as it indicates the transformation-invariance property of \nHL{OREC}. Specifically, consider a one-to-one transformation of $Y$, potentially nonlinear, denoted by $h(Y)$; then we have $e_t^*(a \cond y,\bx) = e_t^*(a \cond h(y),\bx) $, meaning that  \eqref{eq-EPS PT} still holds with the transformed outcome. Therefore, if \nHL{OREC} holds for $Y$, it continues to hold for any strictly monotone transformation $h(Y)$, thereby avoiding the need to specify the ``correct'' outcome transformation. More generally, the \nHL{OREC} condition can accommodate discrete, continuous, or even mixed-type outcomes, and it does not require that the outcome distribution belong to a certain class, say the exponential family. Therefore, it can be used broadly to de-bias arbitrary causal effects on the treated. 

Like many identifying assumptions in the DiD setting, \nHL{OREC} is generally not empirically testable. However, it can be refuted if certain support conditions are violated in the observed data; see Assumption \ref{assumption:support} and its relaxation in {\SUPP} \ref{sec:supp-OREC}. In addition, in {\SUPP} \ref{sec:supp-GM}, we provide several structural models in which the \nHL{OREC} assumption is satisfied, offering guidance for practitioners in evaluating its plausibility and interpreting it in specific applications.

\subsection{Identification}                    \label{sec:Identification}

In this section, we establish nonparametric identification of the ATT $\tau^*$. We begin with an identification result for $\beta_{1}^*(\bx)$  in \eqref{eq-def-OR} and $\mu^*(\bx)$, which are essential for identifying $\tau^*$. Recall that $\beta_{1}^* $ and $\mu^*$ involve the unobservable density $\potf{1}^*(y \cond 1, \bx)$, and consequently, these functions are not identifiable without an additional assumption. The \nHL{OREC} assumption is sufficient for identifying these functions in terms of nuisance functions identified from the observed data; Lemma \ref{lemma-beta_and_mu} formally states the result.
\begin{lemma}					\label{lemma-beta_and_mu}
Under Assumptions \ref{assumption:support}-\ref{assumption:no anticipation}, $\beta_{1}^*$ and $\mu^*$ are represented as
\begin{align}					\label{eq-beta1mu}
& 
\beta_{1}^*(\bX)
= 
\frac{ \Pr(A=1 \cond \bX) /  \Pr(A=0 \cond \bX)  }{\EXP \big\{ \alpha_1^* (\Yo, \bX) \cond A = 0 , \bX \big\}}
, \quad 
\mu^*(\bX) 
=
\frac{
\EXP \big\{ \Yo \alpha_1^*( \Yo, \bX) \cond A=0, \bX \big\}
}{
\EXP \big\{ \alpha_1^*( \Yo, \bX) \cond A=0, \bX \big\}
}
.
\end{align}
Therefore, under Assumptions \ref{assumption:support}-\ref{assumption:OREC}, $\beta_{1}^*$ and $\mu^*$ are identified by replacing $\alpha_1^*$ with $\alpha_0^*$.
\end{lemma}
\noindent From Lemma \ref{lemma-beta_and_mu}, we can directly represent $\tau^*$ as $\tau^* = \EXP \big[ A \big\{ \Yo - \mu^*(\bX) \big\} \big] / \Pr(A=1)$. Interestingly, by extending the approaches in \citet{Liu2020} to our setting, we can obtain other representations of the ATT under Assumptions \ref{assumption:support}-\ref{assumption:no anticipation}:
\begin{align} 
\tau^*
&
=
\EXP \big[ \big\{
A - 
(1-A) \beta_1^*(\bX) \alpha_1^*(\Yo , \bX)
\big\} \Yo \big] / \Pr(A=1)
 \label{eq-rep-IPW}
\\
& =
\EXP \big[
A
\big\{ \Yo - 
\mu^*(\bX) \big\}
\big] / \Pr(A=1)
 \label{eq-rep-OR}
\\
& =
\EXP 
\big[ \big\{ A - (1-A)  \beta_1^*(\bX) \alpha_1^*(\Yo,\bX) \big\} 
\big\{ \Yo -  \mu^*(\bX) \big\}
 \big] / \Pr(A=1)
.
\label{eq-rep-AIPW}
\end{align}
We refer to the three representations as inverse probability-weighting (IPW), outcome regression-based, and augmented inverse probability-weighting (AIPW) representations because \eqref{eq-rep-IPW} only uses the treatment odds at $(\Yo,\bX)$ (i.e., $\beta_1^*(\bX) \alpha_1^*(\Yo,\bX)$) as a weighting term,  \eqref{eq-rep-OR} only uses the outcome regression $\mu^*$, and \eqref{eq-rep-AIPW} uses both the treatment odds and outcome regression; see {\SUPP} \ref{sec:supp-id-PT} for details on these representations. We also note that the estimators in \citet{UDID2024_Epi} are developed based on these representations.

These representations align with those in \citet{Liu2020} who leveraged an instrumental variable to identify the extended propensity score; however, there are some notable differences between our setting and theirs. First, instrumental variables play a key identification role in their work, while our framework does not require them. Second, our setting considers a longitudinal setting with two time periods where the odds ratio at time 1 is identified under Assumptions \ref{assumption:support}-\ref{assumption:OREC}. On the other hand, they consider cross-sectional settings where the odds ratio is identified by leveraging key instrumental variable properties.

\section{Connections to Existing Approaches} \label{sec:review}

In this section, we compare our UDiD framework to the other existing approaches for DiD settings, especially in terms of the key identifying assumption. To this end, we first review the existing approaches. 

\subsection{Comparison to (Nonlinear) Parallel Trends} \label{sec:NPT}

First, the most well-known identifying assumption for the DiD setting is the (conditional) parallel trends (\nHL{PT}) assumption \citep{Heckman1997, Abadie2005, SantAnna2020, Callaway2021}. Specifically, upon accommodating covariates $\bX$, \nHL{PT} is expressed as: 
    \begin{align*}		\text{(\hypertarget{PT}{PT}):}
    \quad 
    &
\EXP \big\{ \pYo - \pYz \cond A = 1 , \bX \big\}
=
\EXP \big\{ \pYo - \pYz \cond A = 0 , \bX \big\}
\text{  almost surely}
\ .  
\end{align*}
The \nHL{PT} condition states that the time trends of the treatment-free potential outcomes (i.e., $\pYo-\pYz$) are, on average, identical in both treated and untreated groups conditional on observed covariates. Under Assumptions \ref{assumption:consistency}, \ref{assumption:no anticipation}, and \nHL{PT}, it is straightforward to show that $	\tau^* = \EXP
\{
\EXP ( \Yo | A = 1 , \bX )
- \EXP ( \Yo | A = 0 , \bX )
+ \EXP ( \Yz | A = 0 , \bX )
- \EXP ( \Yz | A = 1, \bX )
| A=1 \}$, justifying DiD.

It is well-known that the \nHL{PT} assumption can be understood as a condition related to the degree of confounding bias for the additive association between $A$ and $\potY{0}{1}$. To see this, we rearrange \nHL{PT} as $
\EXP \{ \pYz | A=1, \bX \}
- \EXP \{ \pYz | A=0, \bX \}
=
\EXP \{ \pYo | A=1, \bX \}
- \EXP \{ \pYo | A=0, \bX \}$. 
The right hand side would be zero if there were no confounding bias given $\bX$; therefore, non-null values of the latter reflect the magnitude of confounding bias on the additive scale, which cannot directly be observed. The equality states that the post-treatment additive confounding bias can be identified by the pre-treatment additive confounding bias. That is, the \nHL{PT} assumption is equivalent to the so-called bias stability condition \citep{Heckman1997, DiD_Review2011}, which is also referred to as the additive equi-confounding assumption \citep{Sofer2016} whereby the degree of confounding is assessed on the additive scale, i.e., the difference of counterfactual conditional means across observed treatment values.  In this sense, the \nHL{PT} assumption is analogous to \nHL{OREC} in spirit: both conditions posit that the degree of confounding bias is stable across time periods. However, they differ in the scale on which confounding is measured, specifically \nHL{PT} and  \nHL{OREC} characterize confounding on the additive scale and on the generalized odds ratio scale, respectively.

Despite its simplicity, the \nHL{PT} assumption may be incompatible with natural constraints of the outcome, say binary outcome; see an example in {\SUPP} \ref{sec:supp:review}. To account for the natural constraints of the outcome, \citet{Puhani2012} and \citet{Wooldridge2022} considered the so-called nonlinear parallel trends (\nHL{NPT}) assumption. The \nHL{NPT} assumption states that the transformed conditional expectations of potential outcomes satisfy \nHL{PT} where the transformation is defined via a monotone link function $\mathcal{L}$, i.e., 
\begin{align*}			
\text{(\nHT{NPT}):}\quad \quad 
&
\mathcal{L} \big(  \EXP \big\{ \pYo \cond A = 1 , \bX \big\} \big)
-
\mathcal{L} \big( \EXP \big\{ \pYz \cond A = 1 , \bX \big\} \big)		 
\\
&
=
\mathcal{L} \big( \EXP \big\{ \pYo \cond A = 0 , \bX \big\} \big)
-
\mathcal{L} \big( \EXP \big\{ \pYz \cond A = 0 , \bX \big\} \big)
\text{  almost surely}
\ .
\nonumber
\end{align*}
The \nHL{PT} assumption is a special case of \nHL{NPT} where the link function is the identity function. The link function is chosen to be compatible with the nature of the outcome. For example, when the outcome is binary, a common choice for the link function is the logit or probit function; for a count outcome, the log function is a standard choice. Using \nHL{NPT} as an identifying assumption, one can infer the treatment effect by adopting the methodologies developed under the \nHL{PT} assumption. This approach has gained popularity, especially for binary and count outcomes across a variety of fields such as statistics \citep{Taddeo2022}, epidemiology \citep{Supporting_Epi}, 
accounting \citep{Supporting_Accounting}, 
medicine \citep{Supporting_JAMA}, health policy \citep{Supporting_HealthPolicy}, and economics \citep{Puhani2012, Supporting_Economics, Wooldridge2022}.  

Despite this relaxation, \nHL{NPT} (including \nHL{PT}) is not transformation-invariant. Even if the assumption holds for the outcome $Y$, it does not necessarily hold for a transformed outcome $h(Y)$. In addition, \nHL{NPT} does not naturally extend to nonlinear treatment effects, such as the QTT detailed in {\SUPP} \ref{sec:General}.

The relationship between \nHL{OREC} and \nHL{NPT} (which includes \nHL{PT} as a special case) is worth clarifying. When the conditional density of the outcome belongs to an exponential family, \nHL{OREC} and \nHL{NPT} under the corresponding canonical link function are equivalent. For instance, if the outcome is binary and the link function is the logit function,  \nHL{OREC} and \nHL{NPT} are identical; see {\SUPP} \ref{sec:supp-Comparison} for a formal statement. In this sense, \nHL{OREC} can be interpreted as automatically selecting the appropriate scale of \nHL{NPT} based on the underlying outcome distribution, rather than requiring the analyst to specify a link function a priori. However, \nHL{OREC} and \nHL{NPT} are not nested in general. Specifically, there exist outcome distributions for which one assumption holds but not the other; see {\SUPP} \ref{sec:supp-Comparison} for data generating processes illustrating this non-nested relationship. Therefore, \nHL{OREC} and \nHL{NPT} should be viewed as related but distinct identifying conditions, and neither assumption implies the other in general.
 
\subsection{Comparison to Other Approaches} \label{sec:Other DiDs}

Alternative models for identifying treatment effects have been explored in nonlinear DiD settings. In particular, \citet{CiC2006} introduced the changes-in-changes (CiC) model for a continuous outcome, which posits that $\potY{0}{t} = h_t(U_t)$ and $U_{1} \cond A \stackrel{D}{=} U_{0} \cond A$; here, $U_{t}$ is a continuously distributed unobserved variable, and $h_t$ is a transformation at time $t\in \{0,1\}$. The transformation $h_t$ is assumed to be strictly monotone. Then, the counterfactual distribution of $\potY{0}{1} \cond (A=1)$ is nonparametrically identified from the observed data. Also see \citet{Sofer2016} and \citet{Sun2025} for related recent developments. It is important to note that the {CiC} model requires a structural model for the potential outcome, and point identification of causal effects is only established for continuous outcomes.\footnote{For discrete outcomes, CiC requires additional assumptions to achieve point identification of causal effects; see Section 4.2 of \citet{CiC2006} and {\SUPP} \ref{sec:supp:review}.} 

In {\SUPP} \ref{sec:supp:review}, we also review several other approaches for DiD settings, based on alternative identifying assumptions, including a PT assumption on the log characteristic function \citep{BonhommeSauder2011}, distributional DiD assumptions \citep{FanYu2012, CallawayLiOka2018, CallawayLi2019}, and the sequential ignorability assumption \citep{Ding2019}. Many of these alternative methods are only applicable to specific types of outcome distributions. Specifically, the former four assume that outcomes are continuous, so they are not applicable to other types of outcomes. In addition, their identifying assumptions are not transformation-invariant, as they are stated in terms of specific distributional features (e.g., characteristic functions, copulas, or additive differences) that change under nonlinear transformations of the outcome. While \citet{Ding2019} is free from this issue, it does not allow for unmeasured confounding between treatment and treatment-free potential outcome. 

In comparison to the {CiC} model and the other models in this subsection, \nHL{OREC} applies to both continuous and discrete outcomes, as the \nHL{OREC} assumption does not require specifying a generative model for the potential outcome nor a specific scale of the outcome. It is also crucial to note that \nHL{OREC} and the other identifying assumptions are non-nested; see {\SUPP} \ref{sec:supp-Comparison} for concrete examples.

\subsection{Key Properties of Universal Difference-in-Differences} \label{sec:KeyUDID}

We conclude the section by summarizing the key properties of our UDiD approach. As discussed in Sections \ref{sec:NPT} and \ref{sec:Other DiDs}, the \nHL{OREC} assumption is (i) compatible with any outcome type including continuous, discrete, and mixed-type outcomes; (ii) transformation-invariant (see Section \ref{sec-OREC}); and (iii) non-nested with the other identifying assumptions. Moreover, like most approaches in DiD settings, our approach accommodates unmeasured confounders of the association between $A$ and $\pYo$. From a theoretical perspective, our approach is fully nonparametric in that we do not make any parametric assumptions for identification and estimation purposes; we establish the semiparametric efficiency bound and provide sufficient conditions for our proposed estimator to attain it. Table \ref{tab:comparison} summarizes these points.

\begin{table}[!htp] 
\vspace*{-0.4cm}
\renewcommand{\arraystretch}{1.00} \centering
\setlength{\tabcolsep}{3pt}
\scalebox{0.65}{
\begin{tabular}{|c|cc|cc|cc|c|c|}
\hline
\multirow{2}{*}{Assumption} & \multicolumn{2}{c|}{\begin{tabular}[c]{@{}c@{}}Range of\\ Outcome\end{tabular}} & \multicolumn{2}{c|}{Estimand} & \multicolumn{2}{c|}{Semiparametric Efficiency} & \multirow{2}{*}{\begin{tabular}[c]{@{}c@{}}Transformation\\ Invariance\end{tabular}} & \multirow{2}{*}{\begin{tabular}[c]{@{}c@{}}Unmeasured\\ Confounder\end{tabular}} \\ \cline{2-7}
 & \multicolumn{1}{c|}{\makebox[0.7cm][c]{$\R$}} & \makebox[0.7cm][c]{$\{0,1\}$} & \multicolumn{1}{c|}{\makebox[0.7cm][c]{ATT}} & \makebox[0.7cm][c]{QTT} & \multicolumn{1}{c|}{ATT} & QTT & & \\ \hline 
\multirow{2}{*}{\nHL{PT}} & \multicolumn{1}{c|}{\multirow{2}{*}{$\checkmark$}} & \multirow{2}{*}{$\checkmark$} & \multicolumn{1}{c|}{\multirow{2}{*}{$\checkmark$}} & \multirow{2}{*}{\ding{55}} & \multicolumn{1}{c|}{$\checkmark$} & \multirow{2}{*}{\ding{55}} & \multirow{2}{*}{\ding{55}} & \multirow{2}{*}{$\checkmark$} \\
 & \multicolumn{1}{c|}{} & & \multicolumn{1}{c|}{} & & \multicolumn{1}{c|}{$\text{\citep{SantAnna2020}}$} & & & \\ \hline
\nHL{NPT} & \multicolumn{1}{c|}{\multirow{2}{*}{$\checkmark$}} & \multirow{2}{*}{$\checkmark$} & \multicolumn{1}{c|}{\multirow{2}{*}{$\checkmark$}} & \multirow{2}{*}{\ding{55}} & \multicolumn{1}{c|}{\multirow{2}{*}{\ding{55}}} & \multirow{2}{*}{\ding{55}} & \multirow{2}{*}{\ding{55}} & \multirow{2}{*}{$\checkmark$} \\
$\text{\citep{Puhani2012, Wooldridge2022}}$ & \multicolumn{1}{c|}{} & & \multicolumn{1}{c|}{} & & \multicolumn{1}{c|}{} & & & \\ \hline
{CiC} & \multicolumn{1}{c|}{\multirow{2}{*}{$\checkmark$}} & \multirow{2}{*}{\ding{55}} & \multicolumn{1}{c|}{\multirow{2}{*}{$\checkmark$}} & \multirow{2}{*}{$\checkmark$} & \multicolumn{1}{c|}{$\checkmark$} & $\checkmark$ & \multirow{2}{*}{$\checkmark$} & \multirow{2}{*}{$\checkmark$} \\
$\text{\citep{CiC2006}}$ & \multicolumn{1}{c|}{} & & \multicolumn{1}{c|}{} & & \multicolumn{2}{c|}{$\text{\citep{Sun2025}}$} & & \\ \hline
PT in $\log(\text{characteristic ft})$ & \multicolumn{1}{c|}{\multirow{2}{*}{$\checkmark$}} & \multirow{2}{*}{\ding{55}} & \multicolumn{1}{c|}{\multirow{2}{*}{$\checkmark$}} & \multirow{2}{*}{$\checkmark$} & \multicolumn{1}{c|}{\multirow{2}{*}{\ding{55}}} & \multirow{2}{*}{\ding{55}} & \multirow{2}{*}{\ding{55}} & \multirow{2}{*}{$\checkmark$} \\
$\text{\citep{BonhommeSauder2011}}$ & \multicolumn{1}{c|}{} & & \multicolumn{1}{c|}{} & & \multicolumn{1}{c|}{} & & & \\ \hline
Copula invariance & \multicolumn{1}{c|}{\multirow{3}{*}{$\checkmark$}} & \multirow{3}{*}{\ding{55}} & \multicolumn{1}{c|}{\multirow{3}{*}{$\checkmark$}} & \multirow{3}{*}{$\checkmark$} & \multicolumn{1}{c|}{\multirow{3}{*}{\ding{55}}} & \multirow{3}{*}{\ding{55}} & \multirow{3}{*}{\ding{55}} & \multirow{3}{*}{$\checkmark$} \\
$\text{\citep{CallawayLiOka2018}}$ & \multicolumn{1}{c|}{} & & \multicolumn{1}{c|}{} & & \multicolumn{1}{c|}{} & & & \\
$\text{\citep{CallawayLi2019}}$ & \multicolumn{1}{c|}{} & & \multicolumn{1}{c|}{} & & \multicolumn{1}{c|}{} & & & \\ \hline
Sequential ignorability & \multicolumn{1}{c|}{\multirow{2}{*}{$\checkmark$}} & \multirow{2}{*}{$\checkmark$} & \multicolumn{1}{c|}{\multirow{2}{*}{$\checkmark$}} & \multirow{2}{*}{$\checkmark$} & \multicolumn{1}{c|}{$\checkmark$} & $\checkmark$ & \multirow{2}{*}{$\checkmark$} & \multirow{2}{*}{\ding{55}} \\
$\text{\citep{Ding2019}}$ & \multicolumn{1}{c|}{} & & \multicolumn{1}{c|}{} & & \multicolumn{1}{c|}{$\text{\citep{Hahn1998}}$} & $\text{\citep{Firpo2007}}$ & & \\ \hline
\nHL{OREC} & \multicolumn{1}{c|}{\multirow{2}{*}{$\checkmark$}} & \multirow{2}{*}{$\checkmark$} & \multicolumn{1}{c|}{\multirow{2}{*}{$\checkmark$}} & \multirow{2}{*}{$\checkmark$} & \multicolumn{1}{c|}{\multirow{2}{*}{$\checkmark$}} & \multirow{2}{*}{$\checkmark$} & \multirow{2}{*}{$\checkmark$} & \multirow{2}{*}{$\checkmark$} \\
(Assumption \ref{assumption:OREC}) & \multicolumn{1}{c|}{} & & \multicolumn{1}{c|}{} & & \multicolumn{1}{c|}{} & & & \\ \hline
\end{tabular}}%
\caption{Comparison of Approaches for Difference-in-Differences Settings. 
The check mark ($\checkmark$) indicates that a criterion is achieved under the identifying assumption and additional conditions required by the prior works. The cross mark (\ding{55}) indicates that a criterion is not achieved.}
\label{tab:comparison}
\vspace*{-0.6cm}
\end{table}

As our approach is the only one that achieves all of these criteria, it can be viewed as a ``universal'' framework for estimating treatment effects in DiD settings. We emphasize that ``universal'' here refers exclusively to the simultaneous satisfaction of properties (i)-(v); it does not imply that \nHL{OREC} is preferable to \nHL{PT} or other assumptions in every application. Because \nHL{OREC} is non-nested with the other identifying assumptions (see {\SUPP} \ref{sec:supp:review}), it complements rather than replaces them.

\section{A Semiparametric Efficient Estimator}  \label{sec:Estimator}

\subsection{Semiparametric Efficiency Bound} \label{sec:SEB}

We now consider estimation of the ATT under the UDiD framework. One can begin by leveraging representations \eqref{eq-rep-IPW}-\eqref{eq-rep-AIPW}, as in \cite{UDID2024_Epi}, to construct an estimator where nuisance components are nonparametrically estimated. However, such estimators incur a bias dominated by the first-order bias of $\widehat{\alpha}_1$; see {\SUPP} \ref{sec:supp-Other Estimators} for details. Consequently, if $\alpha_1^*$ depends on continuous variables without a specified parametric form, the bias may not diminish sufficiently fast, preventing the estimator from achieving $N^{1/2}$-consistency.  To rectify this, we develop an influence function-based approach that eliminates first-order bias relative to the estimated nuisance components \citep{Robins2008_HOIF}. This framework ensures $N^{1/2}$-consistency even when nuisance parameters converge at rates significantly slower than the parametric rate.

To this end, we first derive the efficient influence function (EIF) for $\tau^*$ under $\mathcal{M}_{\text{OREC}}$, the collection of observed data laws satisfying Assumptions \ref{assumption:support}-\ref{assumption:OREC}.
\begin{theorem}						\label{thm-EIF}
The efficient influence function for $\tau^*$ in model $\mathcal{M}_{\OREC}$ is
\begin{align*}
&
\InfFt^*(\bO) = \frac{ A\Yo -  \uncInfFt_{0}^*(\bO) - A \tau^* }{\Pr(A=1)} \ , 
\quad  
\uncInfFt_{0}^*(\bO)
=
\left[
\begin{array}{l}
(1-A) 
\beta_1^*(\bX) \alpha_1^*(\Yo,\bX)
\big\{  \Yo - \mu ^*(\bX) \big\}
+
A \mu ^* (\bX) 
\\
+ (2A-1) R^*(\Yz,A,\bX) \big\{  \Yz - \mu ^*(\bX) \big\} 
\end{array}
\right] \ .
\end{align*}
Here, $ R^* (y,a,\bx)$ is the density ratio relating $(\pYo,A=1,\bX)$ to $(\Yz, A=a,\bX)$ with the form 
% $R^*(y,a,\bx)=0$ for $(y,\bx) \in \SPg^c$ and 
\begin{align}								\label{eq-densityratio} 
R^* (y,a,\bx)
= \frac{\potg{1}^* (y, 1 , \bx)}{ \potg{0}^* (y,a , \bx)}
& =
\beta_1^*(\bx)
\alpha_1^*(y,\bx)
\bigg\{
\frac{a}{\beta_{0}^*(\bx)
 \alpha_0^*(y,\bx)} + (1-a)
\bigg\}
\frac{ \potf{1}^* (y, 0 , \bx) }{ \potf{0}^* (y , 0, \bx) }
\ind \{ (y,x) \in \SPg \}
\nonumber
\\
& =
\beta_1^*(\bx)
\bigg\{
\frac{a}{\beta_{0}^*(\bx)} + (1-a)\alpha_1^*(y,\bx)
\bigg\}
\frac{ \potf{1}^* (y, 0 , \bx) }{ \potf{0}^* (y , 0, \bx) }
\ind \{ (y,x) \in \SPg \}
\ .
\end{align}
Equation \eqref{eq-densityratio}  follows from the \nHL{OREC} assumption. 
Consequently, the corresponding semiparametric efficiency bound for $\tau^*$ is $\VAR \big\{\InfFt^*(\bO)\big\}$.
\end{theorem}
\noindent We find that $\uncInfFt_{0}^*$ consists of three terms; the first two can be interpreted as the uncentered EIF for the functional $\EXP\{ A \pYo \}$ in a semiparametric model in which the generalized odds ratio function $ \alpha_1^*$  is completely known a priori \citep{Robins2000_Sensitivity}, while the third term formally reflects the uncertainty associated with the estimation of $\alpha_1^*$. To the best of our knowledge, this result is entirely novel in the DiD literature. Interestingly, the density ratio $R^*$, which appears in the third term of $\uncInfFt_{0}^*$, is effectively a Radon-Nikodym derivative, corresponding to a change of counterfactual probability measure from the conditional law of $(\Yz, A=a, \bX)$ to that of $(\pYo , A=1, \bX)$, and consequently, this augmentation term can be viewed as a projection of the counterfactual post-treatment sample space of $(\pYo,A=1,\bX)$ onto the sample space of the pre-treatment data $(\Yz,A,\bX)$. Technically speaking, this term is obtained by accounting for the fact that $\alpha_1^*$ is unknown and must be estimated using outcome data at time 0 upon leveraging the \nHL{OREC} assumption. Indeed, $Y_0$ only enters the EIF through this last term; see {\SUPP} \ref{sec:supp-EIF} for further details on the characterization of $\alpha_1^*$ in terms of the observed data distribution under \nHL{OREC}. The third term of $\uncInfFt_{0}^*$ is crucial to ensuring that the EIF-based estimator of $\tau^*$ constructed in Section \ref{sec:Conti-Improved} admits a bias at most of second-order.

As a simple approach to construct an estimator of $\tau^*$, one could in principle proceed in two stages, first by computing $\hpotf{0}(y \cond 0, \bx)$, $\hpotf{0}(y \cond 1, \bx)$, $\hpotf{1} (y \cond 0, \bx)$, and $\widehat{\Pr} (A = 1 \cond \bX = \bx)$ using nonparametric estimators of the corresponding unknown functions. %densities and the conditional probability, respectively. 
These density estimators could be obtained by standard nonparametric kernel  (conditional) density estimation techniques \citep{HallRacineLi2004, np2008,  LiRacine2008}, or via nonparametric kernel (conditional) density operators \citep{Song2013, Schuster2020}. 
Then, using relationships given by equations \eqref{eq-def-OR}, \eqref{eq-beta1mu}, and \eqref{eq-densityratio}, one could obtain plug-in estimators $\widehat{\beta}_0^{\text{PI}}$, $\widehat{\alpha}_1^{\PI}$, $\widehat{\beta}_1^{\PI}$, $\widehat{\mu}^{\PI}$, and $\widehat{R}^{\PI}$. Using the estimated nuisance functions, one might then estimate $\tau^*$ as 
$\widehat{\tau}^{\PI}
=
\big\{ \AVER (A) \big\}^{-1} 
\AVER \{ AY_1 - \widehat{\phi}_0 (\bO) \}
$ where $\widehat{\phi}_0$ is defined by 
\begin{align*}
\widehat{\phi}_0(\bO)
\!=\!
\widehat{\beta}_1^{\PI} (\bX) \widehat{\alpha}_1^{\PI}(\Yo,\bX)
(1-A) \big\{  \Yo - \widehat{\mu}^{\PI} (\bX) \big\}
\!+\!
A
\widehat{\mu}^{\PI} (\bX) 
\!+\!
(2A-1) \widehat{R}^{\PI} (\Yz,A,\bX) \big\{  \Yz - \widehat{\mu}^{\PI} (\bX) \big\} .
\end{align*}
Unfortunately, the simple substitution estimator $\widehat{\tau}^{\PI}$ is unlikely to perform well in finite samples mainly because, except for $\widehat{\mu}^{\PI}$, all nuisance functions used in $\widehat{\tau}^{\PI}$ involve density ratios. As a result, the simple substitution estimator may be overly sensitive to density estimators appearing in denominators, potentially leading to instability in the weights and in the corresponding estimator of the functional of interest. Therefore, in the following section, we propose an estimator that performs better in finite samples. 

\subsection{Proposed Estimator}			\label{sec:Conti-Improved}

In short, the proposed estimator is derived from the EIF and adopts the cross-fitting approach of \citet{Schick1986}, recently popularized by \citet{Victor2018}.  We implement cross-fitting in this paper as follows. We randomly split the observed data into non-overlapping folds, denoted by $\big\{ \mathcal{I}_1,\ldots, \mathcal{I}_K \big\}$. For each $k \in \{1,\ldots,K\}$, we use all folds other than $\mathcal{I}_k$, i.e., $\mathcal{I}_k^c$, and referred to as estimation fold, to estimate the nuisance functions and evaluate the estimator of the target estimand over $\mathcal{I}_k$, referred to as evaluation fold, using the estimated nuisance functions based on the estimation fold $\mathcal{I}_k^c$. To fully use the data, we aggregate $K$ estimators of the target estimand by simple averaging. We consider the following three steps to estimate the nuisance functions in the estimation fold $\mathcal{I}_k^c$. In the first step, we obtain estimators of $f_0^*(y \cond 0,\bx)$ and $\Pr(A =a \cond \bX=\bx)$ using nonparametric estimators. In the second step, we directly estimate the density ratio, instead of estimating the densities in the denominator and numerator, to reduce the risk of unstable density ratio estimators. Using the estimated density ratios, we obtain an estimator of $\alpha_1^*$ which is considerably more stable than the plug-in method described in the previous section. Lastly, we evaluate these estimated nuisance functions over the evaluation fold and obtain our estimator of ${\tau}^*$ based on the estimated EIF. The rest of the section provides details on the estimation procedure.

\textit{Step 1: Estimation of $f_{0}^*(y \cond 0,\bx)$ and $\Pr(A=a \cond \bX=\bx)$}: We denote the estimation fold by $\mathcal{I}_{k}^c$, and its subset corresponding to treatment group $A = a$ by $\mathcal{I}_{ka}^c = \mathcal{I}_k^c \cap \{ i | A_i = a \}$ for $a \in \{0,1\}$. To estimate the conditional density $f_{0}^*(y | 0, \bx)$, we employ a nonparametric kernel conditional density estimation method implemented in the \texttt{np} R package \citep{np2008}. Specifically, we treat $Y_0$ as the outcome and $\bX$ as the covariates, and fit the estimator using the observed data in $\mathcal{I}_{k0}^c$, i.e., units with $A = 0$ in the estimation fold. To estimate the propensity score $\Pr(A = 1 \cond \bX = \bx)$, we model $A$ as the response and $\bX$ as the predictors, using probabilistic machine learning methods and their ensemble via superlearner \citep{SL2007}; see {\SUPP} \ref{sec:supp-BinaryEIF} for details on the machine learning algorithms included in the \texttt{UDID} R package \citep{UDIDpackage}. The resulting estimators are denoted by $\widehat{f}_{0}\LSS(y \mid 0, \bx)$ and $\widehat{\Pr}\LSS(A = 1 \mid \bX = \bx)$.

\textit{Step 2: Estimation of $\alpha_1^*$}: Let $r_{0}^* (y,\bx) = \potf{0}^*(y,\bx \cond 1)/  \potf{0}^*(y,\bx \cond 0)$ for $(y,\bx) \in \SPg$ be the density ratio that we need to estimate; we remind the reader that $\potf{t}^*(y,\bx \cond a)$ is the density of $(\potY{0}{t} , \bX) \cond (A=a)$. 
Various methods exist for directly estimating such density ratios efficiently, with minimal computational cost. In this work, we adopt the Kullback-Leibler (KL) importance estimation approach \citep{DensityRatio2007_1,DensityRatio2007_2, DensityRatio2010_1}, as implemented in the \texttt{densratio} R package \citep{densratiopackage}. The idea of KL importance estimation approach is to view $r_0^*$ as the minimizer of the KL divergence from the normalized numerator density $ \potf{0}^*(y,\bx \cond 1)$ and that induced by the denominator and the density ratio, i.e., $ \potf{0}^* (y,\bx \cond 0) \cdot r_{0}^*(y,\bx)$. Therefore, an estimator of $r_0^*$, say $\widehat{r}_0$, can be obtained by solving the following constrained optimization problem:
\begin{align}				\label{eq-KL-DR}
\widehat{r}_0
=
& 
\argmax_{r_0 \in \mathcal{H}_{0 X}} 
\EXP \big[  \log \{ r_0(\Yz,\bX) \} \cond A=1 \big]
\text{ subject to }
\EXP \big\{ r_0 (\Yz,\bX) \cond A=0 \big\} = 1 \ .
\end{align}
where $\mathcal{H}_{0 X}$ is a function space over $(\Yz,\bX)$ that is sufficiently rich to approximate any possible ground-truth for $r_0^*$; we choose to work with a Reproducing Kernel Hilbert Space (RKHS) associated with a Gaussian kernel $\mathcal{K}$ as $\mathcal{H}_{0 X}$. Then, an estimator  $\widehat{r}_{0,\text{KL}} \LSS$ is obtained from an empirical analogue of \eqref{eq-KL-DR} based on the estimation fold $\mathcal{I}_k^c$, which has a form of $\widehat{r}_{0,\text{KL}} \LSS (y,\bx) =	\sum_{j \in \mathcal{I}_{k1}^c } \widehat{\gamma}_{j} \LSS \cdot \mathcal{K} \big( (y,\bx), (y_j,\bx_j) \big)$; here, the non-negative coefficients $\widehat{\gamma}\LSS = \big\{ \widehat{\gamma}_j \LSS \big\}_{j \in \mathcal{I}_{k1}^c} \in \R^{|\mathcal{I}_{k1}^{c}|}$ are obtained from
\begin{align}			\label{eq-KL-DR2}
\widehat{\gamma} \LSS
=
\argmax_{\gamma} \AVER_{\mathcal{I}_{k1}^c} \big[ \log \big\{ K_{11} \LSS \gamma \big\} \big] \text{ subject to } 
\AVER_{\mathcal{I}_{k0}^c} \big\{ K_{01} \LSS \gamma \big\} = 1 \ , \
\gamma \geq 0 \ ,
\end{align} 
where $K_{aa'}\LSS$ is the gram matrix of which $(i,j)$th entry is $\mathcal{K} \big( (y_{i},\bx_{i}), (y_{j},\bx_{j} ) \big)$ for $i \in \mathcal{I}_{ka}^c$ and $j \in \mathcal{I}_{ka'}^c$ for $(a,a') \in \{0,1\}^{\otimes 2}$. Using the estimated density ratio $\widehat{r}_{0,\text{KL}} \LSS$, we obtain the estimated baseline odds of $A$ at time 0 and the estimated odds ratio as $\widehat{\beta}_0\LSS(\bx) =  \widehat{r}_{0,\text{KL}}\LSS(y_R,\bx) \AVER_{\mathcal{I}_k^c}(A)/\AVER_{\mathcal{I}_k^c}(1-A)$ and 
$\widehat{\alpha}_1\LSS(y,\bx) = \widehat{r}_{0,\text{KL}}\LSS(y,\bx)/\widehat{r}_{0,\text{KL}}\LSS(y_R,\bx)$, 
respectively.  Similarly, we obtain an estimator of $r_{1}^* (y,\bx) = \potf{1}^* (y,\bx \cond 0) / \potf{0}^* (y,\bx \cond 0)$, denoted by $\widehat{r}_{1,\text{KL}}\LSS$. We provide alternative estimation strategies in {\SUPP} \ref{sec:supp-DR-Estimation}.

Given the estimated nuisance components, we construct estimators of $\beta_1^*$, $\mu^*$, and $R^*$ based on the relationships in \eqref{eq-beta1mu} and \eqref{eq-densityratio}:
\begin{align*}
&
\widehat{\beta}_1\LSS(\bx)
=
\frac{ \widehat{\Pr}\LSS(A=1 \cond \bX=\bx) }{\widehat{\Pr}\LSS(A=0 \cond \bX=\bx) }
\frac{1}{ \widehat{\EXP} \LSS \big\{ \widehat{\alpha}_1\LSS(Y_1,\bX) \cond A=0, \bX=\bx \big\}}
\ , \\
&
\widehat{\mu}\LSS(\bx)
=
\frac{ \widehat{\EXP} \LSS \big\{ \Yo \widehat{\alpha}_1\LSS(Y_1,\bX) \cond A=0, \bX=\bx \big\}}{ \widehat{\EXP} \LSS \big\{ \widehat{\alpha}_1\LSS(Y_1,\bX) \cond A=0, \bX=\bx \big\}} \ , \\
& 
\widehat{R}\LSS (y,a,\bx)
= \widehat{\beta}_1 \LSS(\bx) 
\big[
a \big\{ \widehat{\beta}_0\LSS(\bx)
\big\}^{-1}
+ (1-a)
\widehat{\alpha}_1\LSS(y,\bx)
\big] \widehat{r}_{1,\text{KL}}\LSS(y,\bx) \ ;
\end{align*}
here, the expectation operator $\widehat{\EXP}\LSS$ is taken with respect to the estimated conditional density of $Y_1 \mid (A = 0, \bX = \bx)$, which is $\widehat{f}_{1}\LSS (y \cond 0,\bx) = \widehat{f}_{0}\LSS (y \cond 0,\bx) \widehat{r}_{1,\text{KL}}\LSS(y,\bx)$.

\textit{Step 3: Estimation of $\tau^*$}: In the previous steps, we described our proposed approach for computing  $\widehat{\alpha}_1\LSS$, $\widehat{\beta}_1\LSS$, $\widehat{\mu}\LSS$, and $\widehat{R}\LSS$. For each $k \in \{1,\ldots,K\}$, we use the evaluation fold $\mathcal{I}_k$ to estimate $\tau^*$, which we average to obtain the final estimator $\widehat{\tau}$ as follows:
\begin{align}\label{eq-Estimator}
    &
    \widehat{\tau}
    =
    \frac{1}{N}
    \sum_{k=1}^{K}
    \sum_{i \in \mathcal{I}_k}
    \frac{
    A_i Y_{1i} - \widehat{\phi}_0\LSS(O_i)
    }{\AVER(A)}
    \ , \\
    &
    \widehat{\phi}_0\LSS(O)
    =
    \left[
\begin{array}{l} 
\widehat{\beta}_1\LSS (\bX) \widehat{\alpha}_1 \LSS(\Yo,\bX)
(1-A) \big\{ \Yo - \widehat{\mu}\LSS (\bX) \big\} 
+
A
\widehat{\mu}\LSS (\bX) 
\\
+ (2A-1) \widehat{R}\LSS(\Yz,A,\bX) \big\{  \Yz - \widehat{\mu}\LSS(\bX) \big\} 
\end{array}	
\right] .
\nonumber
\end{align}

For binary and polytomous outcomes, the discrete nature of the data leads to several simplifications in the estimation procedure; detailed derivations for these cases are provided in {\SUPP} \ref{sec:supp-BinaryEIF}. Furthermore, implementing Steps 1 and 2 requires selecting several hyperparameters, including the bandwidth parameters for the kernel density estimator and the RKHS kernels used in density ratio estimation. Although these parameters can be tuned via cross-validation, doing so is often computationally intensive, particularly for large datasets. To reduce this burden, investigators may instead rely on data-driven heuristics to select these parameters efficiently; see {\SUPP} \ref{sec:supp-Median Heuristic} for a comprehensive discussion of these practical alternatives. We adopt such heuristics in the simulation and real-world applications in Sections \ref{sec:simulation} and \ref{sec:Application}.

\subsection{Statistical Properties of the Estimator}                                    \label{sec:property}
To better describe statistical properties of our proposed estimator, let $\{ \widehat{\alpha}_1  \LSS (y,\bx)$, $\widehat{\beta}_0 \LSS (\bx)$, $\hpotf{0} \LSS (y \cond 0, \bx)$, $\widehat{\beta}_1 \LSS (\bx)$, $\hpotf{1}\LSS(y \cond 0, \bx) \}$ denote estimators of $\{ \alpha_1^*(y,\bx)$, $\beta_0^*(\bx)$, $\potf{0}^*(y \cond 0, \bx)$, ${\beta}_1^*(\bx)$, $\potf{1}^*(y \cond 0, \bx) \}$, respectively. For a function $\xi(y,\bx)$, let $\big\| \xi (\Yo,\bX) \big\|_{P,2} $ denote the $L_2( P )$-norm with respect to the conditional distribution $(\Yo , \bX) \cond (A=0)$. Let $r_{\alpha,N}\LSS  = \big\| \widehat{\alpha}_1\LSS(\Yo,\bX)  - \alpha_1^*(\Yo,\bX) \big\|_{P ,2}$, $r_{\beta_t,N}\LSS  = \big\| \widehat{\beta}_t\LSS(\bX)  - \beta_t^*(\bX) \big\|_{P,2}$, and $r_{\potf{t},N}\LSS  = \big\| \hpotf{t}\LSS(\Yo \cond 0,\bX) - \potf{t}^*(\Yo \cond 0, \bX) \big\|_{P,2}$ for $t \in \{0,1\}$ and $k \in \{1,\ldots,K\}$. Suppose the following conditions hold for the true densities and the estimated nuisance functions for all $k \in \{1,\ldots,K\}$. 

\begin{assumption}[Strong Overlap] \label{assumption:SO}
    The support $\SPg$ is a bounded, compact subset of $\R \otimes \mathcal{X}$. Additionally, there exists a constant $c \in (0,\infty)$ satisfying  $\potg{t}^*(y,a,\bx) \in [c^{-1},c]$ for all $(y,\bx) \in \SPg$ and $(a,t) \in \{0,1\}^{\otimes 2}$.
\end{assumption}

\begin{assumption}[Boundedness] \label{assumption:Boundedness}
    The support of $\hpotf{1}\LSS(y \cond 0, \bx)$ has the same support as $\potf{1}^*(y \cond 0, \bx)$. Additionally, there exists a constant $c \in (0,\infty)$ satisfying $\hpotf{t} \LSS (y \cond 0, \bx) \in [c^{-1},c]$, $ \widehat{\beta}_t \LSS (\bx) \in [c^{-1},c]$, and $\widehat{\alpha}_1 \LSS(y,\bx) \in [c^{-1},c]$ for all $(y,\bx) \in \SPg$ and $t \in \{0,1\}$.
\end{assumption}

\begin{assumption}[Consistent Estimation] \label{assumption:ConsistentEstimation} $r_{\alpha,N} \LSS$, $r_{\beta_0,N} \LSS$, $r_{f_0,N} \LSS$, $r_{\beta_1,N} \LSS$, and $r_{f_1,N} \LSS$ are $o_P(1)$. 
\end{assumption}

\begin{assumption}[Pre-treatment Cross-product Rate] \label{assumption:PreCross} $r_{\alpha,N} \LSS \cdot r_{\beta_0,N} \LSS$, $r_{\alpha,N} \LSS  \cdot 
r_{\potf{0},N} \LSS$, and $r_{\beta_0,N} \LSS \cdot r_{\potf{0},N} \LSS$ are $o_P(N^{-1/2})$.
\end{assumption}

\begin{assumption}[Post-treatment Cross-product Rate] \label{assumption:PostCross} $r_{\alpha,N} \LSS \cdot r_{\beta_1,N} \LSS$, $r_{\alpha,N} \LSS \cdot 
r_{\potf{1},N} \LSS$, and $r_{\beta_1,N} \LSS \cdot r_{\potf{1},N} \LSS$ are $o_P(N^{-1/2})$.
\end{assumption}

\noindent Assumption \ref{assumption:SO} implies a strong overlap condition, i.e., the density $\potf{t}^*(y,a,\bx)$ is bounded below by $c^{-1}$ and bounded above by  $c$ over its support $\SPg$, thus their ratios are well-behaved. This condition is sufficient for achieving regular, asymptotically linear estimators. Assumption \ref{assumption:Boundedness} states that the estimated density of $Y_1 \cond (A=0,\bX)$ has its support over that of the true density, and the estimated nuisance functions are uniformly bounded. Assumptions \ref{assumption:SO} and \ref{assumption:Boundedness} ensure that the density ratios are uniformly bounded away from zero and infinity. The boundedness of density ratios is essential in the proof of Theorem \ref{thm-AsympNormal}, as it allows for establishing finite upper bounds and employing various inequalities, such as the H\"older's inequality, to handle quantities involving the density ratios.

Assumption \ref{assumption:ConsistentEstimation} states that the estimated nuisance functions converge to their true functions as the sample size increases, which is plausible under nonparametric estimation strategies. Assumption \ref{assumption:PreCross} states that the cross-product rates of pre-treatment nuisance function estimators are $o_P(N^{-1/2})$. Importantly, if two out of the three nuisance functions are estimated at sufficiently fast rates (e.g., small $r_{\alpha,N}\LSS$ and $r_{\beta_0,N}\LSS$), the other nuisance function is allowed to converge at a substantially slower rate (e.g., large $r_{f_0,N}\LSS$) provided that the cross-products remain $o_P(N^{-1/2})$. This is an instance of the mixed-bias property described by \citet{Rotnitzky2020}. The condition can be interpreted as requirements on the smoothness of the nuisance components.
For simplicity, we consider a toy example in which the outcome is binary, the $d$-dimensional covariate $\bX$ has its support as $[0,1]^d$, and the nuisance functions $f_0^*(y \cond 0,\bx)$, $\beta_0^*(\bx)$, and $\alpha_1^*(y,\bx)$ belong to the H\"older spaces with the smoothness exponent $\delta_{f_0}$, $\delta_{\beta_0}$, and $\delta_{\alpha}$, respectively. The estimators of these nuisance functions based on the kernel density estimator or other nonparametric estimators, such as series estimators, yield a convergence rate of $O_P(N^{-2\delta/(2\delta+d)})$ in terms of mean squared error, which is minimax optimal; see \citet{Stone1980} and Chapter 1 of \citet{Tsybakov2009} for details. Consequently, Assumption \ref{assumption:PreCross} reduces to the following condition when the aforementioned estimators are used:
\begin{align}
\frac{ 2 \delta_{f_0} }{ 2 \delta_{f_0} + d }
+
\frac{ 2 \delta_{\beta_0} }{ 2 \delta_{\beta_0} + d } 
> \frac{1}{2}
\ , \quad
\frac{ 2 \delta_{f_0} }{ 2 \delta_{f_0} + d }
+
\frac{ 2 \delta_{\alpha} }{ 2 \delta_{\alpha} + d } 
> \frac{1}{2}
\ , \quad
\frac{ 2 \delta_{\beta_0} }{ 2 \delta_{\beta_0} + d } 
+
\frac{ 2 \delta_{\alpha} }{ 2 \delta_{\alpha} + d } 
> \frac{1}{2} \ . 
\label{eq-smooth}
\end{align}
Therefore, in terms of smoothness, if any two of the three nuisance functions are smooth enough (e.g., large $\delta_{\alpha}$ and $\delta_{\beta_0}$), the other nuisance function is allowed to be less smooth (e.g., small $\delta_{f_0}$) as long as condition \eqref{eq-smooth} is satisfied.  Assumption \ref{assumption:PostCross} imposes similar conditions on post-treatment nuisance function estimators.  Assumptions \ref{assumption:ConsistentEstimation}-\ref{assumption:PostCross} are satisfied if all nuisance functions are estimated at $o_P(N^{-1/4})$ rates, which may be attainable under certain conditions. For example, in the aforementioned toy example, condition \eqref{eq-smooth} is satisfied if all nuisance parameters are smooth enough in that their H\"older smoothness exponents are greater than $d/2$, i.e., half of the covariate dimension.

Of note, $r_{\alpha,N}\LSS$ appears in both Assumptions \ref{assumption:PreCross} and \ref{assumption:PostCross} because the odds ratio function $\alpha_1^*$ serves as a nuisance function in both the pre- and post-treatment periods under the \nHL{OREC} assumption. Furthermore, $\alpha_1^*$ is coupled with all remaining nuisance functions, making its estimation the primary determinant for the plausibility of Assumptions \ref{assumption:PreCross} and \ref{assumption:PostCross}. If $\widehat{\alpha}_1\LSS$ converges at a slower rate---perhaps due to the nonsmoothness of $\alpha_1^*$ or a suboptimal estimation method---these assumptions can only be satisfied if the other nuisance functions across both periods are estimated at sufficiently fast rates. This is precisely why the density ratio estimation methods in Section \ref{sec:Conti-Improved} are employed to ensure good convergence of $\widehat{\alpha}_1\LSS$, which attains the minimax optimal rate \citep{DensityRatio2010_1}.

Theorem \ref{thm-AsympNormal} establishes the asymptotic property of the proposed estimator of $\tau^*$.
\begin{theorem}						\label{thm-AsympNormal}
Suppose that Assumptions \ref{assumption:support}-\ref{assumption:ConsistentEstimation} hold. Then, the estimator in \eqref{eq-Estimator} is consistent for $\tau^*$. If Assumptions \ref{assumption:PreCross} and \ref{assumption:PostCross} additionally hold, the estimator is asymptotically normal as $N^{1/2} \big( \widehat{\tau} - \tau^* \big) 
\stackrel{D}{\rightarrow} 
N \big( 0, \sigma^2 \big)$, where the variance $\sigma^2$ is equal to the semiparametric efficiency bound for $\tau^*$ under model $\mathcal{M}_{\OREC}$ defined in Theorem \ref{thm-EIF}, i.e., $ \sigma^2 = \VAR \big\{ \InfFt^*(\bO) \big\}$. Additionally, under Assumptions \ref{assumption:support}-\ref{assumption:PostCross}, a consistent estimator of $\sigma^2$ is given by
\begin{align*}
&	\widehat{\sigma}^2 
=
\frac{1}{N} \sum_{k = 1}^{K} \sum_{i \in \mathcal{I}_k}
\bigg[
\bigg\{
\frac{ A_{i} Y_{1i} - \widehat{\uncInfFt}_{0} \LSS (\bO_i) - A_i \widehat{\tau}}{ \AVER \big( A \big)}
\bigg\}^{2}
\bigg] \ .
\end{align*} 	 
\end{theorem}
\noindent 
The proposed estimator is asymptotically normal, provided that at least two of the three nuisance components for each time period are estimated at sufficiently fast rates, although potentially considerably slower than the parametric rate. The result, therefore, implies that our proposed estimator has the mixed-bias property discussed above.  Using the variance estimator $\widehat{\sigma}^2$, a valid $100(1-q)$\% confidence interval for $\tau^*$ is given by $ \big( \widehat{\tau} + z_{q/2} \widehat{\sigma}/  N^{1/2} ,  \widehat{\tau} + z_{1-q/2} \widehat{\sigma}/  N^{1/2} \big)$, where $z_q$ is the $100q$-th percentile of the standard normal distribution. 
%Alternatively, one can estimate standard errors and construct confidence intervals using the multiplier bootstrap; see {\SUPP} \ref{sec:supp-Variance} for details. 

The cross-fitting estimator depends on a specific sample split and may produce outlying estimates if some split samples do not represent the entire data. To resolve the issue, one may consider the median adjustment  from multiple cross-fitting estimates;  see {\SUPP} \ref{sec:supp-Median Heuristic} for details. 

When \nHL{OREC} is violated, the asymptotic normality result stated in Theorem \ref{thm-AsympNormal} no longer holds. To account for this, we develop a sensitivity analysis to assess the robustness of causal conclusions to potential violations of \nHL{OREC}. Specifically, rather than assuming $\alpha_1^*/\alpha_0^*=1$, we allow it to vary within a specified range, for example, $\alpha_1^*/\alpha_0^* \in [\Gamma^{-1}, \Gamma]$ with $\Gamma \geq 1$, and then construct corresponding sharp bounds for the ATT. Details of this sensitivity analysis are provided in {\SUPP} \ref{sec:supp-SA}.

When multiple pre-treatment periods are available, these periods can be used to assess the plausibility of \nHL{OREC}. First, one can perform a placebo test. Second, the pre-treatment data can be used to calibrate the sensitivity parameter $\Gamma$, which helps interpret the results of the sensitivity analysis. Examples illustrating the implementation and interpretation of these approaches are given in Section \ref{sec:Application} and {\SUPP} \ref{sec:supp-data}.

Lastly, in {\SUPP} \ref{sec:General} and \ref{sec:Missing}, we consider extensions of our approach. First, we consider the estimation of general causal estimands that include both the ATT discussed above and nonlinear causal estimands such as the QTT. Second, we consider settings where one aims to identify and make inferences about functionals of an outcome that is missing not at random in longitudinal settings.

\section{Simulation} \label{sec:simulation}

We conducted simulation studies to evaluate the finite-sample performance of the proposed ATT estimators for both continuous and binary outcomes. Pre-treatment covariates $\bX \in \R^2$, binary treatment $A \in \{0,1\}$, and potential outcomes $Y_t^{(a)} \in \R$ or $Y_t^{(a)} \in \{0,1\}$ for $t \in \{0,1\}$ and $a \in \{0,1\}$ were generated according to a specified data-generating process; details of the data-generating process are provided in {\SUPP} \ref{sec:supp:simulation}. The data-generating process satisfies the \nHL{OREC} condition, making it compatible with the UDiD framework, but it does not satisfy the \nHL{PT} condition.

We explored moderate to large sample sizes $N$, taking values in $\{500,1000,2000,4000\}$. Using the simulated data, we computed the proposed ATT estimator based on the procedure outlined in Section \ref{sec:Conti-Improved}, denoted by $\widehat{\tau}_{\text{UDID}}$. To provide a benchmark, we also computed $\widehat{\tau}_{\text{UDID}}^*$, an ATT estimator based on the EIF using the true nuisance functions;  therefore, it serves as an oracle efficient estimator. As a competing estimator, we considered the estimator developed in \citet{SantAnna2020} and \citet{Callaway2021} under the \nHL{PT} assumption, which is implemented in the \texttt{did} R package \citep{didpackage}, which is denoted by $\widehat{\tau}_{\text{DID}}$. For a fair comparison, $\widehat{\tau}_{\text{DID}}$ was implemented using machine learning methods and their ensemble via superlearner \citep{SL2007}. For $\widehat{\tau}_{\text{UDID}}$ and $\widehat{\tau}_{\text{DID}}$, we implemented the median adjustment by repeating cross-fitting 3 times. We evaluated the performance of each estimator based on 1000 repetitions for each value of $N$.  

\begin{figure}[!htb]
\centering
\includegraphics[width=1\textwidth]{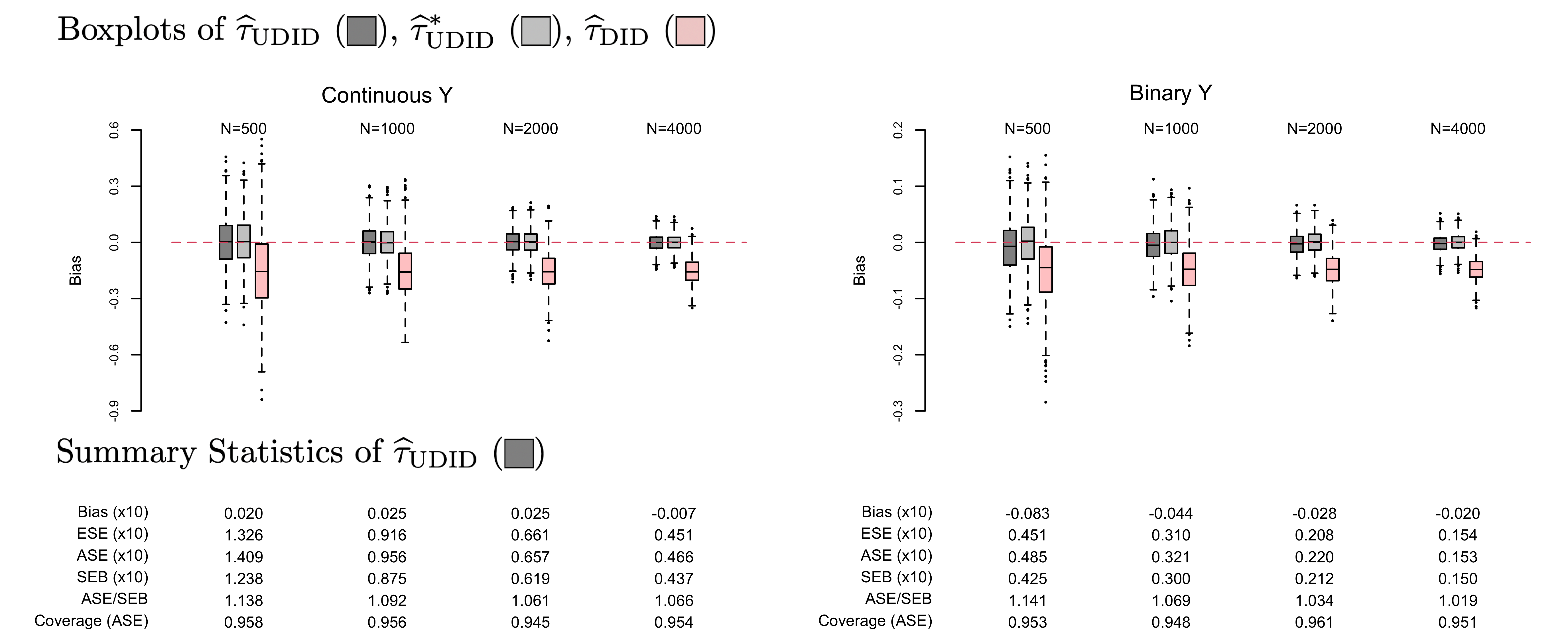}
\vspace*{-0.4cm}
\caption{Graphical summary of the simulation results. The left and right panels show results for continuous and binomial outcomes, respectively. In the top panel, each column gives boxplots of biases of the proposed estimator $\widehat{\tau}_{\text{UDID}}$, the oracle efficient estimator $\widehat{\tau}_{\text{UDID}}^*$ (which attains the semiparametric efficiency bound in Theorem \ref{thm-EIF}), and the competing estimator $\widehat{\tau}_{\text{DID}}$ for  $N \in \{500,1000,2000,4000\}$, respectively. The bottom panel provides numerical performance metrics for $\widehat{\tau}_{\text{UDID}}$, including: empirical bias; asymptotic standard error (ASE); empirical standard error (ESE); the semiparametric efficiency bound (SEB), computed as the ASE of $\widehat{\tau}_{\text{UDID}}^*$; the ASE/SEB ratio; and the empirical coverage of 95\% confidence intervals based on the ASE. The bias and standard errors are scaled by a factor of 10.}
\label{Fig-ContiSim1} 
\vspace*{-0.75cm}
\end{figure}

The top panel of Figure \ref{Fig-ContiSim1} summarizes the bias results. For both outcome types, the proposed estimator $\widehat{\tau}_{\text{UDID}}$ yields negligible bias across all sample sizes. Notably, $\widehat{\tau}_{\text{UDID}}$ performs competitively even when compared to the oracle efficient estimator $\widehat{\tau}_{\text{UDID}}^*$. In contrast, the estimator developed under the \nHL{PT} assumption $\widehat{\tau}_{\text{DID}}$ is biased, which is expected due to the simulation setup. The bottom panel of Figure \ref{Fig-ContiSim1}  provides numerical summaries of our estimator $\widehat{\tau}_{\text{UDID}}$. As $N$ increases, all standard errors of $\widehat{\tau}_{\text{UDID}}$ decrease, and their values are similar to each other. Furthermore, the asymptotic standard error (ASE) of $\widehat{\tau}_{\text{UDID}}$ is remarkably similar to the ASE of $\widehat{\tau}_{\text{UDID}}^*$. This similarity corroborates the consistency of our variance estimator established in Theorem \ref{thm-AsympNormal}. Lastly, we find that the empirical coverage from 95\% confidence intervals based on the asymptotic standard error is close to the nominal coverage. Based on these simulation results, the performance of the proposed estimator was found to align with the asymptotic properties established in Section \ref{sec:property}. 

In {\SUPP} \ref{sec:supp:simulation}, we also compare the computation times of $\widehat{\tau}_{\text{UDID}}$ and $\widehat{\tau}_{\text{DID}}$. We find that $\widehat{\tau}_{\text{UDID}}$ requires a similar computation time as $\widehat{\tau}_{\text{DID}}$, sometimes even faster, demonstrating that our estimator is fully competitive with existing approaches from a computational perspective.
 
\section{Application: Zika Virus Outbreak in Brazil} 			\label{sec:Application}

We illustrate our methodology with two real-world applications: one involving a continuous outcome related to a Zika virus outbreak in Brazil, and another with a binary outcome concerning traffic safety. We provide details about the Zika virus application below, while the traffic safety analysis is detailed in {\SUPP} \ref{sec:Application:BinomY}.

Infection with the Zika virus during pregnancy can be transmitted from a pregnant woman to her fetus and may severely impact fetal brain development, leading to conditions such as microcephaly, i.e., an abnormally small head \citep{Zika2016}. Brazil was among the countries most severely affected by the virus. In particular, the 2015 outbreak led to over 200,000 reported cases in Brazil by 2016 \citep{Zika2018}. In response, numerous studies have investigated whether the Zika outbreak led to a decline in birth rates (e.g., \citealp{Zika2018_3, Taddeo2022, UDID2024_Epi, SPC2024}). We revisit this question using our proposed methods. 

%We estimated the effect of the Zika virus outbreak on birth rate in Brazil. 
The data we analyzed includes sociodemographic summary statistics for 1,823 municipalities in 11 states in the northern, northeastern, and southern regions of Brazil between 2013 and 2016. According to a report from the Brazilian Ministry of Health \citep{Zika2017_Report}, the epidemic was more severe in the northeastern region of Brazil compared with the other regions. Specifically, over 20\% of municipalities in the five northeastern states in the dataset have confirmed cases of the Zika virus, while the six states in the northern and southern regions report confirmed cases in no more than 2\% of their municipalities; see Figure \ref{Fig-Brazil} in {\SUPP} \ref{sec:supp-data} for a graphical summary. Based on this information, we defined 752 municipalities in the northeastern states as the treated group and 1,071 municipalities in the northern and southern states as the control group. 
For each municipality, the pre-treatment covariates included: log-transformed population size and population density, the proportion of females, and an indicator for whether the municipality’s gross domestic product (GDP) exceeded Brazil's national GDP; these variables were measured in 2013. The outcome variable was the birth rate, defined as the number of live births per 1,000 individuals.  %While our dataset is similar to those used in 
Notably, we included a larger number of municipalities and an additional covariate (namely the GDP-related indicator) compared to  \citet{Taddeo2022}, \citet{UDID2024_Epi}, and \citet{SPC2024}.

%\begin{figure}[!htb]
%\centering  
%\vspace*{-0.4cm}
%\includegraphics[width=0.7\textwidth]{Brazil.png}
%\vspace*{-0.2cm}
%\caption{Graphical summary of Zika virus data. Each state label indicates the state name, the number of municipalities, and the proportion of municipalities with confirmed Zika cases.}
%\label{Fig-Brazil}
%\vspace*{-0.5cm}
%\end{figure}	

Similar to the simulation studies, we compare two estimators $\widehat{\tau}_{\text{UDID}}$ and $\widehat{\tau}_{\text{DID}}$. For both estimators, we use the median adjustment by repeating cross-fitting 100 times. We empirically confirmed the support condition (both Assumption \ref{assumption:support} and its relaxation in {\SUPP} \ref{sec:supp-OREC}), which appears to be satisfied; see {\SUPP} \ref{sec:supp-data} for details. 

For the main analysis, we defined 2014 and 2016 as time periods 0 and 1, respectively, consistent with prior studies \citep{Taddeo2022, UDID2024_Epi, SPC2024}. We also conducted a placebo analysis restricted to the pre-treatment period, during which the causal effect was expected to be null. Specifically, we estimated the ATT under the \nHL{OREC} and \nHL{PT} assumptions by treating 2013 and 2014 as time periods 0 and 1, respectively. This placebo test is analogous to the \nHL{PT} test commonly used in the DiD literature \citep{Roth2022}. It is often regarded in practice that failing to reject the null effect in a pre-treatment period, such as 2014 in our case, can be interpreted as supportive of the identifying assumptions. However, we emphasize that such pre-treatment tests are neither necessary nor sufficient for these assumptions to hold in the post-treatment period. Consequently, we interpret the results solely as circumstantial evidence in support of the effect estimates obtained under each identifying assumption.

Table \ref{Table-Conty-Data} reports the ATT estimates based on the two estimators, $\widehat{\tau}_{\text{UDID}}$ and $\widehat{\tau}_{\text{DID}}$. In the placebo analysis, where 2013 and 2014 are treated as time periods 0 and 1, respectively, the ATT estimate from $\widehat{\tau}_{\text{UDID}}$ is not statistically significant at 5\% level. Consistent with the preceding discussion, this result does not necessarily validate the \nHL{OREC} assumption but at least suggests no strong empirical evidence against it. In contrast, $\widehat{\tau}_{\text{DID}}$ is statistically significant at 5\% level, raising concerns about validity of the \nHL{PT} assumption. In the main analysis, where 2014 and 2016 are treated as time periods 0 and 1, respectively, two ATT effect estimates yield comparable results in that the Zika virus outbreak reduced birth rate in the five northeastern states of Brazil, and the estimates are significant at 5\% level. The findings are consistent with findings in the literature \citep{Zika2018_3,Taddeo2022,UDID2024_Epi,SPC2024}.

\begin{table}[!htp]
\renewcommand{\arraystretch}{1.0} 
\centering 
\setlength{\tabcolsep}{4pt}
\scalebox{0.8}{
\begin{tabular}{|c|cc|cc|}
\hline
Estimator   & \multicolumn{2}{c|}{$\widehat{\tau}_{\text{UDID}}$}                      & \multicolumn{2}{c|}{$\widehat{\tau}_{\text{DID}}$}                        \\ \hline
Years ($t=0, t=1$) & \multicolumn{1}{c|}{(2013,\,2014)} & (2014,\,2016) & \multicolumn{1}{c|}{(2013,\,2014)} & (2014,\,2016) \\ \hline
 Estimate & \multicolumn{1}{c|}{0.006} & -0.915 & \multicolumn{1}{c|}{-0.305} & -0.906 \\ \hline
 Standard Error & \multicolumn{1}{c|}{0.131} & 0.118 & \multicolumn{1}{c|}{0.154} & 0.139 \\ \hline
 95\% CI & \multicolumn{1}{c|}{(-0.251,\,0.264)} & (-1.146,\,-0.684) & \multicolumn{1}{c|}{(-0.607,\,-0.003)} & (-1.178,\,-0.633) \\ \hline
\end{tabular}}%
\caption{Analysis results of the Zika virus outbreak in Brazil. The reported standard errors and 95\% confidence intervals of $\widehat{\tau}_{\text{UDID}}$ are obtained from the consistent variance estimator in Theorem \ref{thm-AsympNormal}.}
\label{Table-Conty-Data} 
\vspace*{-0.5cm}
\end{table} 

We further evaluated the conclusions using the sensitivity analysis described in {\SUPP} \ref{sec:supp-SA}. Consistent with the placebo analysis, we leveraged the multiple pre-treatment periods to calibrate the sensitivity parameter, which aids in interpreting the sensitivity analysis. Overall, the causal findings based on $\widehat{\tau}_{\text{UDID}}$ in this application appear robust, even when the \nHL{OREC} assumption is severely violated relative to the natural variation in the odds ratio across pre-treatment periods. In contrast, the causal findings based on $\widehat{\tau}_{\text{DID}}$ appear sensitive to plausible violations of the \nHL{PT} assumption. See {\SUPP} \ref{sec:supp-data} for additional details.

\section{Concluding Remarks}        \label{sec:conclusion}

In this paper, we have proposed a framework for identifying and estimating the ATT under the \nHL{OREC} assumption in standard DiD settings. We derived the EIF and the semiparametric efficiency bound for the ATT in a nonparametric model for the observed data. We constructed a corresponding estimator of the ATT, which has a desirable mixed-bias structure in the sense that the estimator is $N^{1/2}$-consistent and asymptotically normal if two out of three, but not necessarily all, nuisance components at each time point are estimated at sufficiently fast rates. We verified all theoretical properties via simulation studies for both continuous and binary outcomes and two important empirical applications.

While the UDiD framework exhibits desirable statistical properties, as summarized in Section \ref{sec:KeyUDID}, it is only one of several DiD approaches, each relying on specific assumptions. It is important to reiterate that \nHL{OREC} is not nested within any other assumptions; thus, the UDiD framework should be understood as an alternative to existing DiD methods, complementing rather than subsuming them. Moreover, while \nHL{OREC} is untestable based solely on observed data, it can still be falsified, just like other assumptions. In particular, \nHL{OREC} cannot hold if the support condition (Assumption \ref{assumption:support} and its relaxation in {\SUPP} \ref{sec:supp-OREC}) is violated. Consequently, we recommend that researchers carefully assess the support condition, as the validity of UDiD depends on it; see Section \ref{sec:Application} and {\SUPP} \ref{sec:supp-data} for examples from the Zika virus application.

We end the paper by suggesting some future directions worth investigating. First, the proposed methodology could be extended to settings with multiple time periods and a general class of treatment patterns, including both staggered and non-staggered adoption, by formulating a group-time \nHL{OREC} condition. To illustrate the key idea, let $t \in \{0,1,\ldots,T\}$ be the calendar time index and let $G \in \{1,\ldots,T,\infty\}$ denote the cohort defined by the time period at which units first receive treatment, where $G=\infty$ denotes the never-treated group. Then, $\alpha_{g,t}^*(y,\bx) = \{ \Pr(G=g \cond \potY{0}{t}=y,\bX=\bx) \Pr(G=\infty \cond \potY{0}{t}=y_R,\bX=\bx) \} / \{ \Pr(G=g \cond \potY{0}{t}=y_R,\bX=\bx) \Pr(G=\infty \cond \potY{0}{t}=y,\bX=\bx) \}$ can be viewed as a group-time odds ratio function relating cohort membership $G$ to the treatment-free potential outcome $\potY{0}{t}$ given $\bX$. The staggered \nHL{OREC} condition would require that $\alpha_{g,t}^*(y,\bx)$ is stable across time $t$ for cohort $g$. Under this condition, identification of causal effects for the treated can be established using arguments analogous to those developed in this paper. Second, it might be possible to relax the \nHL{OREC} assumption substantially, by assuming the two odds ratio functions are connected through a more general relationship, say $\alpha_1^* = \phi(\alpha_0^*)$ where the link function $\phi$ may not be the identity function. We postulate that genuine instrumental variables \citep{Hernan2006}, bespoke instrumental variables  \citep{Dukes2022_BespokeIV},  or proxy variables \citep{TT2020_Proximal} may be useful to identify $\phi$  in such settings. Finally, the key assumptions required for the asymptotic normality of our estimator (Assumptions \ref{assumption:PreCross} and \ref{assumption:PostCross}) may fail in the presence of high-dimensional covariates. In such cases, one may need to impose structured parametric models or adopt nonparametric approaches that adaptively exploit sparsity or other low-dimensional structure. Alternatively, as is often done in practice with DiD, the UDiD approach can be implemented without covariates, with the resulting conclusions subsequently evaluated through sensitivity analyses in Section \ref{sec:supp-SA}. We leave these potential extensions of \nHL{OREC}-based methods to future work.

\newpage 
\newpage

\appendix

\renewcommand{\theequation}{S.\arabic{equation}}
\setcounter{equation}{0}

\section*{Supplementary Material}

This document contains the supplementary materials for ``A Nonparametric Framework for Universal Difference-in-Differences.'' Section \ref{sec:supp-Additional} provides additional details of the main paper. Section \ref{sec:supp:prmain} presents the proofs of the lemmas and theorems of the main paper, and Section \ref{sec:supp-proof-supp} presents the proofs of the lemmas and theorems introduced in the supplementary material.

\section{Additional Details of the Main Paper}									\label{sec:supp-Additional}

\subsection{Details on Assumptions}					\label{sec:supp-OREC}

We first introduce a relaxed condition of Assumption \ref{assumption:support}. In order to do so, let $\mathcal{S}_t(a)$ be the support of the law of $ (\potY{0}{t},A=a,\bX)$, i.e., $\mathcal{S}_t(a) = \big\{ (y,\bx) \cond \potf{t}^*(y,a,\bx) \in (0,\infty) \big\}$.
\begin{assumption}[Post-treatment Overlap] \label{assumption:support-1}
    $\SPot \subseteq \SPo$, i.e., the support of $(\potY{0}{1}, A=1, \bX)$ is included in that of $(\potY{0}{1}, A=0, \bX)$.
\end{assumption}
\begin{assumption}[Cross-time Overlap] \label{assumption:support-2}
For $a=0,1$, $\mathcal{S}_{1}(a) \subseteq \mathcal{S}_{0}(a)$, i.e., the support of $(\pYo , A=a, \bX)$ is included in that of $(\Yz, A=a, \bX)$.
\end{assumption}

Assumptions \ref{assumption:support-1} and \ref{assumption:support-2} are depicted in Figure \ref{Fig-Support}. It is trivial that Assumption \ref{assumption:support} in the main paper implies Assumptions \ref{assumption:support-1} and \ref{assumption:support-2}. Therefore, the latter two constitute a weaker condition than the former.

\begin{figure}[!htp]
\centering
\scalebox{0.65}{\begin{tikzpicture}[scale=0.8]
\begin{scope}[fill opacity = 1,text opacity=1]
\draw[color=red] (1,0) ellipse (3.0cm and 2cm);
\draw[color=red] (-1,0) ellipse (2.75cm and 2cm);
\draw (0.5,0) ellipse (2.25cm and 1.75cm);
\draw (0,0) circle[radius=1.25];
% (0.8, 0) circle[radius=1.2];
\node[fill = white] at (-3.75, 0) (A) {{\color{red}$\SPzt$}};   
\node[fill = white] at (4.25, 0) (B) {{\color{red}$\SPz$}};   
\node[fill = white] at (2.3, 1.15) (C) {$\SPo$};   
\node[fill = white] at (0, 0) (D) {$\SPot$};   

\draw[color=red] (1+10,0) ellipse (3.0cm and 2cm);
\draw[color=red] (-1+10,0) ellipse (2.75cm and 2cm);
\draw (0.5+10,0) ellipse (2.25cm and 1.75cm);
\draw (0+10,0) circle[radius=1.25];
% (0.8, 0) circle[radius=1.2];
\node[fill = white] at (-3.75+10, 0) (AA) {{\color{red}$\SPzt$}};   
\node[fill = white] at (4.25+10, 0) (BB) {{\color{red}$\SPz$}};   
\node[fill = white] at (2.3+10, 1.15) (CC) {$\SPo$};   
\clip (-1+10,0) ellipse (2.75cm and 2cm);
\clip (0.5+10,0) ellipse (2.25cm and 1.75cm);
\fill[black!20!white] (5+10,5) rectangle (-5+10,-5);
\node[fill = none] at (0+10, 0) (DD) {$\SPot$};  

\end{scope}
\end{tikzpicture}}
\caption{Graphical Illustrations of the Four Supports Under Assumptions \ref{assumption:support-1}, \ref{assumption:support-2} (left panel), and Under Assumptions \ref{assumption:support-1}, \ref{assumption:support-2}, \ref{assumption:OREC} (right panel). The black ellipses depict supports at time 1, and the red ellipses depict supports at time 0.}
\label{Fig-Support}
\end{figure}

Assumption \ref{assumption:support-1} means that the probability law of $(\potY{0}{1}, A=1, \bX) $ is dominated by that of $(\potY{0}{1}, A=0, \bX)$, and the condition is required to ensure that the odds ratio at time 1 does not involve division by zero. Under the condition, any realized value of $\pYo$ under $A=1$ is also realizable under $A=0$, and it is a necessary condition for identifying the distribution of the counterfactual outcomes using the observed data. We remark that the condition is not testable because $\mathcal{S}_1(1)$ is counterfactual.  

Assumption \ref{assumption:support-2} for $a=0$ implies that the denominator of $\alpha_0^*(y,\bx)$ is positive if that of $\alpha_1^*(y,\bx)$ is positive, and it guarantees that $\alpha_0^* (y,\bx)$ is well-defined (i.e., avoid division by zero) if $\alpha_1^* (y,\bx)$ is well-defined. Similarly, the cross-time overlap for $a=1$ implies that the numerator of $\alpha_0^*(y,\bx)$ is positive if that of $\alpha_1^*(y,\bx)$ is positive, and it is required to avoid a case where $\alpha_1^* (y,\bx)$ is positive while $\alpha_0^*(y,\bx)$ is zero. We note that $\SPo \subseteq \SPz$ is testable from the observed data whereas $\SPot \subseteq \SPzt$ is untestable because $\SPot$ is counterfactual. However, combining Assumptions \ref{assumption:support-1} and \ref{assumption:support-2}, $\SPzt \cap \SPo$ must be non-empty as it contains $\SPot$. Therefore, one can empirically verify whether $\SPzt \cap \SPo$ is non-empty, ensuring that Assumptions \ref{assumption:support-1} and \ref{assumption:support-2} are not falsified.

We compare our support conditions (Assumptions \ref{assumption:support-1} and \ref{assumption:support-2}) to those in \citet{CiC2006}. In their work, the treatment-free potential outcomes $\potY{0}{t}$ are determined as $\potY{0}{t}=h(U,t)$ where $U$ is a latent variable of a study unit and $h(u,t)$ is a nonlinear function that is strictly increasing in $u$. In their framework, the support of $U \cond (A=1,t)$ is included in that of $U \cond (A=0,t)$, i.e., $\text{supp}( U \cond A=1,t ) \subseteq \text{supp}( U \cond A=0,t )$ for $t=0,1$. As a consequence, the supports of the potential outcomes satisfy $\text{supp}( \potY{0}{0} \cond A=1 ) \subseteq \text{supp}( \potY{0}{0} \cond A=0 )$ and $\text{supp}( \potY{0}{1} \cond A=1 ) \subseteq \text{supp}( \potY{0}{1} \cond A=0 )$. The former condition corresponds to Assumption \ref{assumption:support-1} when covariates are not considered. Our framework does not require the latter, but \citet{CiC2006} does not require Assumption \ref{assumption:support-2}, either. Therefore, there is no deterministic relationship between our and their support conditions. However, their work requires an additional condition that the support of $U \cond (A,t)$ is not allowed to change over time, i.e.,
\begin{align}       \label{eq-CiC_support}
\text{supp}(U \cond A=a,\color{red}t=0) = \text{supp}(U \cond A=a,t=1)
\quad \text{ for } a=0,1 \ .
\end{align}
Condition \eqref{eq-CiC_support} can be stronger than Assumption \ref{assumption:support-2} according to the form of $h(u,t)$. For instance, if $h(\cdot , t)$ is the identity function for both $t=0,1$, \eqref{eq-CiC_support} implies $\SPo=\SPz$ and $\SPot=\SPzt$, which is stronger than Assumption \ref{assumption:support-2}.

Next, we show that the \nHL{OREC} assumption does not impose an additional restriction on the observed data if Assumptions \ref{assumption:support}-\ref{assumption:no anticipation} (or \ref{assumption:support-1}, \ref{assumption:support-2}, \ref{assumption:consistency}, \ref{assumption:no anticipation}) are satisfied. We can rewrite the \nHL{OREC} assumption as
\begin{align}   \label{eq-OREC-2}
\underbrace{
\frac{ \potf{0}^*(y \cond 1, \bx) }{ \potf{0}^*(0 \cond 1, \bx) }
\frac{ \potf{0}^*(0 \cond 0, \bx) }{ \potf{0}^*(y \cond 0, \bx) }
\frac{ \potf{1}^*(y \cond 0, \bx) }{ \potf{1}^*(0 \cond 0, \bx) }
}_{=: \mathcal{L}(y,\bx), \text{ observed data}}
=
\underbrace{
\frac{ \potf{1}^*(y \cond 1, \bx) }{ \potf{1}^*(0 \cond 1, \bx) }
}_{=: \mathcal{R}(y,\bx),  \text{ counterfactual data}} 
\text{ for } (y,\bx) \in \SPo
\end{align}
First, suppose Assumptions \ref{assumption:support}-\ref{assumption:no anticipation} are satisfied. Then, the left hand side is well-defined over $(y,\bx) \in \SPz$ (to rule out division by zero cases), and is non-zero over $(y,\bx) \in \SPzt \cap \SPo$. The right hand side is non-zero over $(y,\bx) \in \SPot \subseteq \SPzt \cap \SPo$, and it must satisfy the following two restrictions for any counterfactual density $\potf{1}^*(y \cond 1,\bx)$: (R1) $\mathcal{R}(0,\bx) = 1$ and (R2) $\mathcal{R}(y,\bx) = 0$ for $(y,\bx) \in \big\{ \SPzt \cap \SPo \big\}^c$. Restriction (R1) is from the definition of the odds function, and restriction (R2) is from $\SPot \subseteq \{ \SPzt \cap \SPo \}$. Restrictions other than (R1) and (R2) are dependent on the form of the counterfactual density $\potf{1}^*(y \cond 1,\bx)$ and thus cannot be empirically verified.

If the \nHL{OREC} assumption (Assumption \ref{assumption:OREC}) is also assumed on top of Assumptions \ref{assumption:support}-\ref{assumption:no anticipation}, the left hand side now must satisfy the following restrictions: (i) the supports of the left and right hand sides are equal, i.e., $\SPzt \cap \SPo = \SPot$ and (ii) the left hand side must satisfy the two restrictions (R1) and (R2) that are satisfied by the right hand side. For (i), note that we have $\SPot = \{ \SPzt \cap \SPo \} \subseteq \SPz$ and $\SPot \subseteq \{ \SPzt \cap \SPo \} \subseteq \SPz$ with and without \nHL{OREC}, respectively. Therefore, regardless of \nHL{OREC}, the three supports of the observed data (i.e., $\SPz$, $\SPzt$, and $\SPo$) are not further restricted by assuming \nHL{OREC}. For (ii), the left hand side already satisfies restrictions (R1) and (R2) without the \nHL{OREC} assumption because (R1): $\mathcal{L}(0,\bx) = 1$ due to its form and (R2): $\mathcal{L}(y,\bx)=0$ over $(y,\bx) \in \SPzt^c$ or $(y,\bx) \in \SPo^c$, implying $\mathcal{L}(y,\bx)=0$ for  $\big\{ \SPzt \cap \SPo\big\}^c$. Therefore, even though we further invoke \nHL{OREC} on top of Assumptions \ref{assumption:support}-\ref{assumption:no anticipation}, there is no additional restriction on the left hand side, implying that the \nHL{OREC} assumption does not restrict the observed data.

\subsection{Review of Approaches in Difference-in-Differences Settings}
\label{sec:supp:review}			

\begin{itemize}
    \item[1.] (\textit{Standard DiD under the \textit{\nHL{PT} Assumption}}) 
    
    Suppressing covariates, consider $\potY{0}{t}$ is generated from the following model for $t \in \{0,1\}$:
\begin{align}
    \text{(\nHT{DiD} model):}
    \quad \quad 
    &
    \potY{0}{t} = h_t(U_t) \ , 
    &&
    h_{t}(u) = u + b_T\, t \ , \quad  
    \quad U_{t} = b_0 + b_{A}\, A + \epsilon_{t} \ ,
    \label{eq-DiD1}
    \\
    &
    \epsilon_{t} \text{ satisfies either }
    &&
    \text{time invariance:}
    \quad 
     \epsilon_{1} \cond A \stackrel{D}{=} \epsilon_{0} \cond A \quad \text{or, }     
    \label{eq-DiD2}
     \\
     &
    &&
    \text{treatment invariance:}
    \quad 
     \epsilon_{t} \cond (A=0) \stackrel{D}{=} \epsilon_{t} \cond (A=1)   \ .     
    \label{eq-DiD3}
\end{align}
Here, $\epsilon_t$ is an unobserved error at time $t$ that is independent
of time or treatment. Therefore, $U_{t}$ is also unobserved. Note that $\potY{0}{t}$ is a deterministic linear function of $U_{t}$, but the exposure mechanism $A$ given $U_{t}$ is unrestricted. In addition, the \nHL{DiD} model implies rank preservation, which rules out any additive interaction between $A$ and $U_{t}$ in causing $\potY{0}{t}$. Note that the \nHL{DiD} model satisfies the \nHL{PT} assumption, which reads as follows upon reintroducing covariates:
    \begin{align*}		\text{(\nHL{PT}):}
    \quad 
    &
\EXP \big\{ \pYo - \pYz \cond A = 1 , \bX \big\}
=
\EXP \big\{ \pYo - \pYz \cond A = 0 , \bX \big\}
\text{  almost surely}
\ .  
\end{align*}

    The \nHL{PT} assumption may be incompatible with natural constraints of the outcome. To illustrate, ignoring covariates, consider a binary outcome setting where the conditional distributions of $\potY{0}{t}$ given $A$ are $\pYz \cond A \sim \text{Ber}(0.2 + 0.6A)$ and $\pYo \cond A \sim \text{Ber}(0.6+0.36A)$, respectively. Under the \nHL{PT} assumption, the conditional mean $\EXP \big\{ \pYo \cond A=1 \big\}$ is evaluated as 1.2. Of course, not only does it differ from the true value of 0.96, but it also falls beyond its natural unit interval range. Therefore, the \nHL{PT} assumption is violated in this context, illustrating that the \nHL{PT} assumption may not be plausible for certain DiD problems, especially when the outcome has restrictions on its range. Another major limitation of the \nHL{PT} assumption is that it does not naturally extend to nonlinear treatment effects, such as the QTT.

\item[2.] (\textit{Changes-in-changes}) 

We restate the changes-in-changes (\nHL{CiC}) model for continuous outcomes here for exposition:
\begin{align}
    \text{(\nHT{CiC} model):}
    \quad \quad 
    &
    \potY{0}{t} = h_t(U_t) \ ,   
    \label{eq-CiC1}
    \\
    &
    U_{1} \cond A \stackrel{D}{=} U_{0} \cond A \ , \label{eq-CiC2}
    \\
    &
    \text{supp}(U_t \cond A=1) \subseteq \text{supp}(U_t  \cond A=0) \ .
    \label{eq-CiC3}
\end{align}

For point identification of the counterfactual distribution with a discrete outcome, the CiC framework requires additional conditional independence assumptions. Specifically, it assumes either $U_t \indep A \mid (Y_t, t)$ for $t = 0, 1$, or a structural model of the form $\potY{0}{t} = h_t(U_t, \bX)$, where $\bX$ denotes continuous covariates satisfying $U_t \indep \bX \mid A$.

\item[3.] (\textit{Parallel Trends in the Log Characteristic Function}) 

\citet{BonhommeSauder2011} considered a case where the outcome is continuous and is generated from an additive model and the log characteristic functions of $\potY{0}{t}  \cond A$ satisfy the PT condition, i.e.,
\begin{align}	\label{eq-BS2011}
&
\log \Psi_{\potY{0}{1} | A =1} (s)
-
\log \Psi_{\potY{0}{0} | A =1} (s)
=
\log \Psi_{\potY{0}{1} | A =0} (s)
-
\log \Psi_{\potY{0}{0} | A =0} (s)  \ ,
\end{align}
where $\Psi_{\potY{0}{t}|A} (s)$ is a characteristic function of $\potY{0}{t}$ given $A$, i.e., $\EXP \big\{ \exp \big( i s \potY{0}{t} \big) \cond A \big\}$ with $i=\sqrt{-1}$. Therefore, the characteristic function of $\potY{0}{t} \cond (A=1)$ is identified based on the PT condition on the log scale, and therefore the distribution of $\potY{0}{t} \cond (A=1)$ is identified leveraging the one-to-one relationship between a characteristic function and a distribution.

\item[4.] (\textit{Copula Invariance}) 

Leveraging the continuous nature of the outcome, \citet{FanYu2012} focused on the change in the treatment-free potential outcomes over time and assumed the change is identically distributed in both treated and untreated groups; this assumption is referred to as a distributional difference-in-differences assumption by others \citep{CallawayLiOka2018, CallawayLi2019}. Specifically, for $\Delta_t^{(0)}  = \potY{0}{t} - \potY{0}{t-1}$, we have
\begin{align*}
&
(\text{Distributional DiD}):
\qquad 
\Delta_t^{(0)} \indep A \ .
\end{align*}
Unfortunately, the distributional difference-in-differences assumption is insufficient for identifying the counterfactual distribution of $\potY{0}{t} \cond (A=1)$. To identify the counterfactual distribution of $\potY{0}{t} \cond (A=1)$, \citet{CallawayLiOka2018} and \citet{CallawayLi2019} further introduced a copula stability assumption, which is expressed as 
\begin{align*}
    (\text{Copula Stability}):
\qquad 
C_{\Delta_1^{(0)} , \potY{0}{0} | A = 1} (u,v)
=
C_{\Delta_1^{(0)} , \potY{0}{0} | A = 0} (u,v) \ ;
\end{align*}
here, $C_{V,W|A=a}(u,v)$ is the conditional copula function of $(V,W)$ given $A=a$ at $(u,v) \in [0,1]^2$. Therefore, the copula stability assumption implies the dependence structure of the pre-treatment outcome and the change in the treatment-free potential outcomes over time is the same for both treated and untreated groups.

\item[5.] (\textit{Sequential Ignorability}) 

\citet{Ding2019} used the sequential ignorability condition (see \citet{HR2020} for a textbook definition) to the canonical DiD settings as an identifying assumption which states that there is no unmeasured confounder of the association between the post-treatment treatment-free outcome and the treatment other than pre-treatment outcome and covariates, i.e., $\potY{0}{t} \indep A \cond ( \Yz, \bX)$. 
\end{itemize}

 \subsection{Examples of Generative Models Satisfying the Odds Ratio Equi-Confounding Assumption} 	\label{sec:supp-GM}
 
In this section, we present several structural models that satisfy the \nHL{OREC} assumption. We begin with three examples that accommodate the most common outcome types encountered in practice:

\begin{itemize}
\item[1.] (\textit{Continuous})

Suppose the outcomes are generated from the following model:
\begin{align*}
&
\potY{0}{0} = b_0 (\bX) + U_0
\ , 
&&
\potY{0}{1} = b_1 (\bX) + U_1 \ .
\end{align*}
Here, $(U_0,U_1)$ are unobserved random variables following $U_t \cond (A,\bX) \sim N( \nu(A,\bX) ,\sigma_U^2(\bX) )$, and $(U_0,U_1)$ are allowed to have an arbitrary correlation structure. The function $b_t(\bX)$ models the unit's time-specific mean level. Therefore, we find		
\begin{align*}
&
\potY{0}{0} \cond (A,\bX) \sim N \big( b_0 (\bX) + \nu(A,\bX) , \sigma_U^2(\bX) \big)
\\
&
\potY{0}{1} \cond (A,\bX) \sim N \big( b_1(\bX) + \nu(A,\bX) , \sigma_U^2(\bX) \big) \ ,
\end{align*}
which satisfies the \nHL{OREC} and \nHL{PT} assumptions as well.
%	Under the choice of $L(s) = s$, we find \nHL{NPT} is satisfied. 

\item[2.] (\textit{Binary})

Suppose the outcomes are generated from the following latent variable model:
\begin{align*}
&
\potY{0}{0} = \ind \big\{  b_0(\bX)  + U _0 \geq 0 \big\} 
\ , 
&&
\potY{0}{1} = \ind \big\{  b_1(\bX) + U_1 \geq 0 \big\} \ ,
\end{align*}
Here, $(U_0,U_1)$ are unobserved random variables following $U_t \cond (A,\bX) \sim \text{Logistic}( \nu(A,\bX) ,\sigma_U(\bX) )$, and $(U_0,U_1)$ are allowed to have an arbitrary correlation structure. The function $b_t(\bX)$ models the unit's time-specific base level. The treatment-free potential outcomes are discretized values in the indicator functions. Then, after some algebra, we find
\begin{align*}
&
\potY{0}{0} \cond ( A, \bX )
\sim \text{Ber} \Bigg(  \text{expit} \bigg( 
\frac{  b_0 (\bX) + \nu(A, \bX) }{\sigma_U(\bX)}  \bigg)
\Bigg)
\ , \\
&
\potY{0}{1} \cond (A, \bX)
\sim
\text{Ber} \Bigg(
 \text{expit} \bigg( 
\frac{  b_1 (\bX) + \nu(A, \bX) }{\sigma_U(\bX)}  \bigg) 
\Bigg)
\ .
\end{align*}
which satisfies the \nHL{OREC} assumption.

\item[3.] (\textit{Count})

Suppose the outcomes are generated from the following binomial model:
\begin{align*}
&
\potY{0}{0} \cond ( U_0, U_1, A, \bX ) \sim \text{Binomial} \big( U_0 , b_0(\bX) \big)
\ ,
&&
\potY{0}{1} \cond ( U_0, U_1, A, \bX ) \sim \text{Binomial} \big( U_1 , b_1(\bX) \big) \ .
\end{align*}
Here, $(U_0,U_1)$ are unobserved random variables following $U_t \cond (A,\bX) \sim \text{Poisson}( \nu(A,\bX) )$. Additionally, $(U_0,U_1)$ are allowed to have an arbitrary correlation structure, and so too are the two outcomes $(\potY{0}{0},\potY{0}{1})$. The function $b_t(\bX)$ models the unit's time-specific success probability for each trial. Then, after some algebra, we find
\begin{align*}
&
\potY{0}{0} \cond ( A, \bX )
\sim \text{Poisson} \big( \nu(A,\bX) b_0(\bX) \big)
\ , 
&&
\potY{0}{1} \cond (A, \bX)
\sim \text{Poisson} \big( \nu(A,\bX) b_1(\bX) \big)
\ .
\end{align*}
which satisfies the \nHL{OREC} assumption.

\item[4.] (\textit{The UDiD Model; \citealp{UDID2024_Epi}}) 

Next, we consider a structural model similar in spirit to the \nHL{DiD} and \nHL{CiC} models. Again, suppressing covariates, consider the following model first introduced in  \citet{UDID2024_Epi}:
\begin{align}
\text{(\nHT{UDiD} model):}
\qquad 
    &
    \potY{0}{t} \indep A \cond \bX, U_t \ , \ t=0,1  \ ,
    \label{eq-UDiD1}
    \\
    &
    A \cond (U_1= u, \bX) \stackrel{D}{=} A \cond (U_0=u, \bX) \ , \ \forall u
    \ , 
    \label{eq-UDiD2}
    \\
    &
    U_1 \cond (A=0,Y_1=y, \bX) \stackrel{D}{=} U_0 \cond (A=0,Y_0=y, \bX)
    \ , \ \forall y
    \label{eq-UDiD3}
     \ .
\end{align}
Compared to the \nHL{DiD} and \nHL{CiC} models, the \nHL{UDiD} model differs notably in its required assumptions. First, condition \eqref{eq-UDiD1} is a form of latent ignorability condition, which states that the ignorability condition is satisfied conditional on a latent variable $U_t$. Unlike \eqref{eq-DiD1} and \eqref{eq-CiC2}, which state that $\potY{0}{t}$ is a fixed function of $U_t$, \eqref{eq-UDiD1} does not impose any restriction on the relationship between $\potY{0}{t}$ and $U_{t}$. As a result, \eqref{eq-UDiD1} is a significant relaxation of the latter conditions. Next, \eqref{eq-UDiD2} states that the treatment mechanism $A \cond U_{t}$ is invariant over time, a condition not assumed in the \nHL{DiD} and \nHL{CiC} models.  Lastly, condition \eqref{eq-UDiD3} assumes that the conditional distribution of $U_t$ given $(A=0,Y_t)$ is stable over time but otherwise unrestricted. This is similar to condition \eqref{eq-DiD2} of the \nHL{DiD} model and condition \eqref{eq-CiC2} of the \nHL{CiC} model. However, \eqref{eq-UDiD3} presents notable differences from \eqref{eq-DiD2} and \eqref{eq-CiC2}. First, \eqref{eq-UDiD3} is only related to untreated units, whereas \eqref{eq-DiD2} and \eqref{eq-CiC2} are related to both treated and untreated units. Second, unlike \eqref{eq-DiD2} and \eqref{eq-CiC2}, \eqref{eq-UDiD3} requires the observed outcome in the conditioning argument. Therefore, \eqref{eq-DiD2} and \eqref{eq-CiC2} can be considered marginal counterparts of \eqref{eq-UDiD3} that incorporate both treated and control units. Of note, the former two conditions and \eqref{eq-UDiD3} are not nested, similar to the non-nested relationship between marginal and conditional \nHL{PT} conditions. 

From a modeling perspective, the \nHL{UDiD} model, like the \nHL{DiD} and \nHL{CiC} models, allows for selection on unobservables by permitting the distribution of $U$ to differ between treated and control units. However, unlike the \nHL{DiD} model, both the \nHL{UDiD} and \nHL{CiC} models are transformation-invariant in that any monotone transformation of an outcome satisfying the model remains within the model. Moreover, the \nHL{UDiD} model does not impose restrictions on additive interactions between treatment $A$ and the latent variable $U$ for the outcome model, unlike the \nHL{DiD} model. Additionally, the \nHL{UDiD} model is compatible with outcomes of any type, whereas the \nHL{DiD} and \nHL{CiC} models are limited to continuous outcomes due to their respective assumptions \eqref{eq-DiD1} and \eqref{eq-CiC1}. Finally, the \nHL{UDiD} model differs from the \nHL{DiD} and \nHL{CiC} models by assuming that the treatment mechanism $A \cond U_t$ is time-invariant, a restriction not imposed by the other two models.
 
It remains to establish that the \nHL{UDiD model} is compatible with \nHL{OREC}, which we outline below. For time $t=0,1$, the odds of treatment at $\potY{0}{t}=y, \bX=\bx$ is
\begin{align}
&
\frac{ 
\Pr(A=1 \cond \potY{0}{t} = y, \bX=\bx)
}{ 
\Pr(A=0 \cond \potY{0}{t} = y, \bX=\bx)
}	
\nonumber
\\
&
=
\int 
\frac{
P(\potY{0}{t} = y, A=1, U_t=u, \bX=\bx)
}{ 
P (\potY{0}{t}=y , A=0, \bX=\bx)
} 
\, du
\nonumber
\\
&
=
\int 
\frac{
P(\potY{0}{t} = y, A=1, U_t=u, \bX=\bx)
}{ 
P(\potY{0}{t} = y, A=0, U_t=u, \bX=\bx)
}
\frac{ 
P (\potY{0}{t}=y , A=0, U_t=u, \bX=\bx)
}{
P (\potY{0}{t}=y , A=0, \bX=\bx)
} 
\, du 
\nonumber
\\  
&
=
\int 
\frac{
\Pr(A=1 \cond \potY{0}{t} = y, U_t=u, \bX=\bx)
}{ 
\Pr(A=0 \cond \potY{0}{t} = y, U_t=u, \bX=\bx)
}
P (U_t = u \cond \potY{0}{t}=y, A=0, \bX=\bx )
\, du
\nonumber
\\  
&
=
\int 
\frac{
\Pr(A=1 \cond  U_t=u, \bX=\bx)
}{ 
\Pr(A=0 \cond U_t=u, \bX=\bx)
}
P (U_t = u \cond \potY{0}{t}=y, A=0, \bX=\bx)
\, du \ .
\label{eq-model OREC cond1}
\end{align}
The first four lines are trivial. 
The fifth line is from \eqref{eq-UDiD1}. We also find that \eqref{eq-UDiD3} implies
\begin{align} \label{eq-model OREC cond3} 
    & P (U_1 = u \cond \potY{0}{1}=y, A=0, \bX=\bx)
    \nonumber
    \\
    &
    = P (U_1 = u \cond Y_1=y, A=0, \bX=\bx)
    \nonumber
    \\
    &
    = P (U_0 = u \cond Y_0=y, A=0, \bX=\bx)
    \nonumber
    \\
    &
    = P (U_0 = u \cond \potY{0}{0}=y, A=0, \bX=\bx)
     \ .
\end{align}
Therefore, we establish that
\begin{align*}
&
\frac{ 
\Pr(A=1 \cond \potY{0}{0} = y, \bX=\bx)
}{ 
\Pr(A=0 \cond \potY{0}{0} = y, \bX=\bx)
}
\\
&
=
\int 
\frac{
\Pr(A=1 \cond U_0=u, \bX=\bx)
}{ 
\Pr(A=0 \cond U_0=u, \bX=\bx)
}
P (U_0 = u \cond \potY{0}{0}=y, A=0 , \bX=\bx)
\, du
\\
&
=
\int 
\frac{
\Pr(A=1 \cond U_1=u, \bX=\bx)
}{ 
\Pr(A=0 \cond U_1=u, \bX=\bx)
}
P (U_1 = u \cond \potY{0}{1}=y, A=0 , \bX=\bx)
\, du
\\
&
=
\frac{ 
\Pr(A=1 \cond \potY{0}{1} = y, \bX=\bx)
}{ 
\Pr(A=0 \cond \potY{0}{1} = y, \bX=\bx)
}	\ .
\end{align*}
The first and third identities are from \eqref{eq-model OREC cond1}. The second identity is from \eqref{eq-model OREC cond3} and \eqref{eq-UDiD2}. Therefore, this implies that the odds ratio is the same over time:
\begin{align*} 
\alpha_0^*(y,\bx)
& =
\log \bigg\{
\frac{ 
\Pr(A=1 \cond \potY{0}{0} = y, \bX=\bx)
}{ 
\Pr(A=0 \cond \potY{0}{0} = y, \bX=\bx)
}	
\frac{ 
\Pr(A=0 \cond \potY{0}{0} = y_R, \bX=\bx)
}{ 
\Pr(A=1 \cond \potY{0}{0} = y_R, \bX=\bx)
}	
\bigg\}
\\
&
=
\log \bigg\{
\frac{ 
\Pr(A=1 \cond \potY{0}{1} = y, \bX=\bx)
}{ 
\Pr(A=0 \cond \potY{0}{1} = y, \bX=\bx)
}	
\frac{ 
\Pr(A=0 \cond \potY{0}{1} = y_R, \bX=\bx)
}{ 
\Pr(A=1 \cond \potY{0}{1} = y_R, \bX=\bx)
} 
\bigg\}
=\alpha_1^*(y,\bx)
  \ .
\end{align*} 

\item[5.] (\textit{Discrete Choice Models; \citealp{McFadden1973, Train2009}}) \\
The \nHL{OREC} condition is also compatible with widely used models in practice. For example, consider the binary logit discrete choice model, a standard framework in econometrics that was part of the work for which Daniel L. McFadden received the 2000 Nobel Prize in Economics \citep{McFadden1973, Train2009}. Let $V_{tj}$ be the utility of alternative $j \in \{0,1\}$ at time $t \in \{0,1\}$, which is modeled as $V_{tj} = b_{tj}(\bX) + U_{tj}$. Here, $b_{tj}(\bX)$ is the observed component of alternative $j$'s utility at time $t$ that depends on covariates, and $U_{tj}$ is the unobserved component of alternative $j$'s utility at time $t$. Suppose that $U_{tj}$ is generated as $U_{tj} \sim \text{EV}( \nu_j (A,\bX), 1)$ where $\text{EV}(\mu,\sigma)$ is the type I extreme value distribution with the location parameter $\mu$ and the scale parameter $\sigma$. Furthermore, we assume that $U_{t0} \indep U_{t1}$, i.e., the unobserved components of the two alternatives' utility are independent; however, $U_{0j}$ and $U_{1j}$ can be arbitrarily correlated. Let $\potY{0}{t}$ be the indicator of choosing alternative 1, i.e., $\potY{0}{t} = \ind \big( V_{t1} > V_{t0} \big)$. Then, we have that 
\begin{align*}
\potY{0}{t} \cond ( A, \bX ) \sim \text{Ber} \big( \text{expit} \big\{ b_{t1} (\bX) - b_{t0} (\bX) + \nu_{1}(A, \bX) - \nu_{0}(A, \bX) \big\} \big) \ ,
\end{align*}
thus satisfying the \nHL{OREC} condition. Note that the distributions of $U_{tj}$ are allowed to differ across the treated groups, demonstrating that the \nHL{OREC} condition can easily incorporate so-called ``selection on unobservables.''

\end{itemize}

\subsection{Comparison between Identifying Assumptions in the Difference-in-Differences Setting and the Odds Ratio Equi-confounding Assumption} 	\label{sec:supp-Comparison}

\begin{itemize}
\item[1.] \textit{(Comparison to \nHL{PT})} 

We begin by comparing the \nHL{PT} and \nHL{OREC} assumptions. At a higher level, the \nHL{PT} and \nHL{OREC} assumptions play a common purpose, which is to establish a relationship between the unobserved treatment-free potential outcome at time 1 (i.e., $\pYo \cond (A=1)$) and the observed treatment-free potential outcomes (i.e., $\pYz$ and $\pYo \cond (A=0)$). Additionally, neither \nHL{PT} nor \nHL{OREC} assumptions is generally empirically testable with the key exception being when the outcome is a priori known to satisfy certain support conditions that may conflict with the \nHL{PT} or \nHL{OREC} assumption in the observed sample, therefore refuting the assumption.  Specifically, if the counterfactual mean $\EXP \big\{ \potY{0}{1} \cond A=1, \bX \big\}$ implied by the \nHL{PT} assumption is guaranteed to fall within the support of the outcomes for all laws in the specified model, the \nHL{PT} assumption cannot be falsified. Likewise, if $\SPo \subseteq \SPz$ and $\SPzt \cap \SPo$ is not empty, as implied by Assumptions \ref{assumption:support-1} and \ref{assumption:support-2}, for all laws in the specified model, the \nHL{OREC} assumption cannot be falsified.

The two assumptions also have notable differences. First, the \nHL{OREC} assumption is sufficient to characterize the counterfactual distribution of $\potY{0}{1} \cond (A=1)$. Therefore, under \nHL{OREC}, investigators can infer any causal effects on the treated, such as the counterfactual mean of a transformed outcome (i.e.,  $\EXP \big\{ \mathcal{G}(\potY{0}{t}) \cond A=1 \big\}$ where $\mathcal{G}$ is an integrable function) and the counterfactual median (i.e., $\text{median} \{ \potY{0}{t} \cond A=1 \big\}$). On the other hand, the \nHL{PT} assumption is insufficient to characterize the counterfactual distribution of $\potY{0}{1} \cond (A=1)$ because it is only related to the counterfactual mean of the original outcome. Therefore, \nHL{PT} cannot identify general causal estimands, including the transformed outcome's counterfactual mean and the counterfactual median.  Second, the \nHL{OREC} assumption has the transformation-invariance property. More concretely, let $\widetilde{Y}_t^{(0)}$ be a monotone transformation of the original outcome $\potY{0}{t}$. Then, \nHL{OREC} with respect to $\widetilde{Y}_t^{(0)}$ implies \nHL{OREC} with respect to $\potY{0}{t}$, and vice versa. However, the \nHL{PT} assumption is not transformation-invariant in that \nHL{PT} with respect to $\widetilde{Y}_t^{(0)}$ does not imply \nHL{PT} with respect to $\potY{0}{t}$, and vice versa, unless the transformation is linear.  

It is interesting to study the relationship between the \nHL{PT} and \nHL{OREC} assumptions. In general, \nHL{PT} and \nHL{OREC} assumptions do not imply each other, i.e., they are not nested. Therefore, the \nHL{OREC} assumption can be satisfied even though the \nHL{PT} assumption is violated, and vice versa. For instance, let us consider a case where potential outcomes are normally distributed as $\potY{0}{t} \cond (A,\bX) \sim N \big( \mu_t(A,\bX), \sigma_t^2(\bX) \big)$. From straightforward algebra, we establish that \nHL{PT} and \nHL{OREC} are equivalent to the following conditions, respectively:
\begin{align*}
& \text{PT} && \Leftrightarrow &&
\mu_{1}(1,\bx) - \mu_{1}(0,\bx) = \mu_{0}(1,\bx) - \mu_{0}(0,\bx) \ , \  
\\
& \text{OREC} && \Leftrightarrow &&
\sigma_1^{-2}(\bx) \big\{ \mu_{1}(1,\bx) - \mu_{1}(0,\bx) \big\} = \sigma_0^{-2}(\bx) \big\{ \mu_{0}(1,\bx) - \mu_{0}(0,\bx) \big\}   \ .
\end{align*}
In this example, \nHL{PT} and \nHL{OREC} imply that unweighted and weighted difference-in-means of treated and untreated groups are identical across times. Therefore, \nHL{PT} and \nHL{OREC} do not imply each other unless $\sigma_0^2(\bx)=\sigma_1^2(\bx)$ for all $\bx$. 
Notably, the main difference between \nHL{PT} and \nHL{OREC} is that the latter accounts for potential differences of scale by standardizing with the inverse variance $\sigma_t^{-2}(\bx)$ so that the assumptions are rendered equivalent if $\sigma_t^{2}(\bx)$ is constant over time for given $\bx$. 

It is worth mentioning the binary outcome case as well. Suppose the potential outcomes are distributed as $\potY{0}{t} \cond (A,\bX) \sim \text{Ber}\big( p_t(A,\bX) \big)$. Then, \nHL{PT} and \nHL{OREC} reduce to
\begin{align*}
& \text{PT} &&  \Leftrightarrow &&
p_{1}(1,\bx) - p_{1}(0,\bx) = p_{0}(1,\bx) - p_{0}(0,\bx) \ , \  
\\
& \text{OREC} &&  \Leftrightarrow &&
\text{logit} \big\{ p_1(1,\bx) \big\} 
-
\text{logit}  \big\{ p_1(0,\bx) \big\} 
=
\text{logit}  \big\{ p_0(1,\bx) \big\} 
-
\text{logit}  \big\{ p_0(0,\bx) \big\} 
  \ .
\end{align*} 
Similarly, suppose that the potential outcomes are count data, and are distributed as $\potY{0}{t} \cond (A,\bX) \sim \text{Poisson}\big( \lambda_t(A,\bX) \big)$. Then, \nHL{PT} and \nHL{OREC} reduce to
\begin{align*}
& \text{PT} && \Leftrightarrow &&
\lambda_{1}(1,\bx) - \lambda_{1}(0,\bx) 
= \lambda_{0}(1,\bx) - \lambda_{0}(0,\bx) \ , \  
\\
& \text{OREC} && \Leftrightarrow &&
\log\{  \lambda_1(1,\bx) \} - \log \{ \lambda_1(0,\bx) \}
=
\log\{   \lambda_0(1,\bx) \} -  \log\{   \lambda_0(0,\bx) \} 
\ .    
\end{align*}
Therefore, \nHL{OREC} is a generalization of \nHL{PT} with respect to logit and log link functions in these examples; these results are induced from the nonlinear \nHL{PT} formulation in \nHL{NPT}. An appealing property of \nHL{OREC} in the binary and count outcome cases is that the counterfactual parameters $p_1(1,\bx)$ and $\lambda_1(1,\bx)$ always belong to its natural range (i.e., $(0,1)$ and $(0,\infty)$, respectively) under \nHL{OREC} whereas they may go beyond the range under \nHL{PT}. Based on these two examples, \nHL{OREC} may be interpreted as a transformation-adapted generalization of \nHL{PT}. In fact, this transformation-adaptive property appears to apply quite broadly to distributions in the exponential family including geometric, negative binomial, Exponential, and Gamma distributions; see the examples below:

\begin{itemize}
\item[(i)] (\textit{Gaussian}) $\potY{0}{t} \cond (A,\bX) \sim N \big( \mu_t(A,\bX), \sigma_t^2(A) \big)$
\begin{align}
& \text{PT} && \Leftrightarrow &&
\mu_{1}(1,\bX) - \mu_{1}(0,\bX) = \mu_{0}(1,\bX) - \mu_{0}(0,\bX) \ ,
\nonumber
\\
& \text{OREC} && \Leftrightarrow &&
\sigma_1^{-2}(\bX) \big\{ \mu_{1}(1,\bX) - \mu_{1}(0,\bX) \big\} = \sigma_0^{-2}(\bX) \big\{ \mu_{0}(1,\bX) - \mu_{0}(0,\bX) \big\} \ .
\label{eq-Normal PT OREC}
\end{align}
\item[(ii)] (\textit{Binomial}) $\potY{0}{t} \cond (A,\bX) \sim \text{Bin}\big( M(\bX) , p_t(A,\bX) \big)$
\begin{align*}
& \text{PT} && \Leftrightarrow &&
p_{1}(1,\bX) - p_{1}(0,\bX) = p_{0}(1,\bX) - p_{0}(0,\bX) \ ,
\\
& \text{OREC} && \Leftrightarrow &&
\frac{ p_1(1,\bX) \{ 1- p_1(0,\bX) \} }{ \{1-p_1(1,\bX)\} p_1(0,\bX) } 
=
\frac{ p_0(1,\bX) \{ 1- p_0(0,\bX) \} }{ \{1-p_0(1,\bX)\} p_0(0,\bX) }
\ .    
\end{align*}
We remark that the Bernoulli distribution is a special case when $M(\bX)=1$.

\item[(iii)] (\textit{Negative Binomial}) $\potY{0}{t} \cond (A,\bX) \sim \text{NegBin}\big( M(\bX) , p_t(A,\bX) \big)$
\begin{align*}
& \text{PT} && \Leftrightarrow &&
\frac{p_{1}(1,\bX)}{1-p_{1}(1,\bX)} - \frac{p_{1}(0,\bX)}{1-p_{1}(0,\bX)} = \frac{p_{0}(1,\bX)}{1-p_{0}(1,\bX)} - \frac{p_{0}(0,\bX)}{1-p_{0}(0,\bX)}  \ ,
\\
& \text{OREC} && \Leftrightarrow &&
\frac{1-p_1(1,\bX)}{1-p_1(0,\bX)} 
=
\frac{1-p_0(1,\bX)}{1-p_0(0,\bX)} 
\ .    
\end{align*}
We remark that the geometric distribution is a special case when $M(\bX)=1$.

\item[(iv)] (\textit{Poisson}) $\potY{0}{t} \cond (A,\bX) \sim \text{Poisson}\big( \mu_t(A,\bX) \big)$
\begin{align*}
& \text{PT} && \Leftrightarrow &&
\mu_{1}(1,\bX) - \mu_{1}(0,\bX) = \mu_{0}(1,\bX) - \mu_{0}(0,\bX) \ ,
\\
& \text{OREC} && \Leftrightarrow &&
\mu_1(1,\bX) / \mu_1(0,\bX)
=
\mu_0(1,\bX) / \mu_0(0,\bX)
\ .    
\end{align*}

\item[(v)] (\textit{Gamma}) $\potY{0}{t} \cond (A,\bX) \sim \text{Gamma}\big( \kappa_t(\bX) , \lambda_t(A,\bX) \big)$
\begin{align*}
& \text{PT} && \Leftrightarrow &&
\kappa_1(\bX) \big\{ \lambda_1(1,\bX) - \lambda_1(0,\bX)  \big\}
=
\kappa_0(\bX) \big\{ \lambda_0(1,\bX) - \lambda_0(0,\bX)  \big\}
\\
& \text{OREC} && \Leftrightarrow &&
\lambda_1^{-1}(1,\bX) - \lambda_1^{-1}(0,\bX) 
=
\lambda_0^{-1}(1,\bX) - \lambda_0^{-1}(0,\bX) 
\ .    
\end{align*}

We remark that the exponential distribution is a special case when $\kappa_t(\bX)=1$.

\end{itemize}
Note that these relationships serve as examples demonstrating that the \nHL{PT} and \nHL{OREC} assumptions are not nested.

More generally, \nHL{OREC} can be understood as a \nHL{PT} condition of the extended propensity score in the logit scale. Specifically, taking the logarithm on both sides in \nHL{OREC}, we obtain the following conditions for all $(y,\bx) \in \SPg$:
\begin{align*}
\text{logit} 
\big\{ e_1^*(1 \cond y, \bx) \big\}
-
\text{logit} 
\big\{ e_1^*(1 \cond y_R, \bx) \big\}
=
\text{logit} 
\big\{ e_0^*(1 \cond y, \bx) \big\}
-
\text{logit} 
\big\{ e_0^*(1 \cond y_R, \bx) \big\} \ .
\end{align*}
In words, the change in the log odds associated with the extended propensity score over time is the same across all $(y,\bx) \in \SPg$, i.e., parallel relationship in the log odds of the extended propensity score over time. To better appreciate the condition, suppose that the conditional exposure model given $(\potY{0}{t},\bX)$ for $t=0,1$ is given as follows:
\begin{align*}
A \cond (\potY{0}{t}, \bX) \sim \text{Ber} \big( \text{expit} \big\{ \gamma_{t0} + {\gamma}_{tX} \T \bX + \gamma_{tY} \potY{0}{t} \big\} \big) \ .
\end{align*}
Then, the \nHL{OREC} assumption is equivalent to $\gamma_{0Y} = \gamma_{1Y}$, indicating that, upon conditioning on $\bX$, the impact of $\potY{0}{t}$ on $A$ in the logit scale is the same over time. To the best of our knowledge, a \nHL{PT}-type condition on the treatment mechanism is new in the DiD literature. 

% We illustrate the relationship between \nHL{PT} and \nHL{OREC} assumptions under standard Gaussian models. Suppose the potential outcomes are normally distributed as $\potY{0}{t} \cond A,\bX \sim N \big( \mu_t (A,\bX), \sigma^2(\bX) \big)$. 
% The \nHL{PT} assumption states that $ \mu_1 (1,\bX) - \mu_1 (0,\bX) = \mu_0 (1,\bX) - \mu_0 (0,\bX)$. Moreover, the odds ratio at time $t$ is given by
% \begin{align*}
% \alpha_t (y,\bX)
% & =
% \exp \bigg[
% - \frac{ 
% \big\{ y- \mu_t(1,\bX) \big\}^2 + \big\{ 0 - \mu_t(0,\bX) \big\}^2
% -
% \big\{ 0- \mu_t(1,\bX) \big\}^2 - \big\{ y - \mu_t(0,\bX) \big\}^2
% }{2\sigma^2(\bX)}
% \bigg]
% \\
% & =
% \exp \bigg[
% \frac{ y \big\{ \mu_t(1,\bX) - \mu_t(0,\bX) \big\}  }{\sigma^2(\bX)}
% \bigg] \ .
% \end{align*}
% Therefore, the \nHL{OREC} assumption is equivalent to $\mu_1 (1,\bX) - \mu_1 (0,\bX) = \mu_0 (1,\bX) - \mu_0 (0,\bX)$, which is identical to the \nHL{PT} assumption. 

% Next, suppose that the outcome is heteroskedastic across time, i.e., $\potY{0}{t} \cond A,\bX \sim N \big( \mu_t (A,\bX), \sigma_t^2(\bX) \big)$. Note that the \nHL{PT} assumption remains the same as $\mu_1 (1,\bX) - \mu_1 (0,\bX) = \mu_0 (1,\bX) - \mu_0 (0,\bX)$. However, the odds ratio at time $t$ is $\alpha_t (y,\bX) = 
% \exp \big[  y \sigma_t^{-2}(\bX) \big\{ \mu_t(1,\bX) - \mu_t(0,\bX) \big\} \big]$. Therefore, the \nHL{OREC} assumption is equivalent to  $\sigma_1^{-2}(\bX)\{\mu_1(1,\bX)-\mu_1(0,\bX)\} 
% = 
% \sigma_0^{-2}(\bX)\{\mu_0(1,\bX)-\mu_0(0,\bX)\}$ which in this case differs from the \nHL{PT} assumption.

Under additional conditions, we can establish an interesting relationship between the \nHL{PT} and \nHL{OREC} assumptions. To ensure that both odds ratio functions are well-defined,  suppose that Assumption \ref{assumption:support} is satisfied, i.e., $\mathcal{S}_t(a)$ are identical for $(a,t) \in \{0,1\}^{\otimes 2}$ throughout this Section.  Following \citet{Chen2007} and \citet{TTRR2010}, one may parametrize the conditional distribution of $\potY{0}{t} \cond (A,\bX)$ in terms of the odds ratio $\alpha_{t}(y,\bx)$ and the outcome's baseline density $\potf{t0}(y \cond \bx) := \potf{t}(y \cond 0,\bx)$ for  $t=0,1$. This is because $\potf{t}(y \cond 1,\bx)$, the conditional density of $\potY{0}{t}$ given $(A=1,\bX=\bx)$, admits the following representation in terms of $\alpha_t$ and $\potf{t0}$ as $\potf{t} ( y \cond 1, \bx) = \{\alpha_{t}(y,\bx) \potf{t0}(y \cond \bx) \}/\{ \int \alpha_{t}(z,\bx) \potf{t0}(z \cond \bx) \, dz\}$ given that $\int \alpha_{t}(z,\bx) \potf{t0}(y \cond \bx) \, dz < \infty$. A key property of this parametrization is that  $\alpha_t$ and $\potf{t0}$ are variationally independent, meaning that the functional form of one parameter does not restrict the functional form of the other. Therefore, the specification of one nuisance component does not restrict one's ability to specify the other. An important property of the \nHL{OREC} assumption is that it solely restricts the relationship between $\alpha_0$ and $\alpha_1$, and consequently, $\alpha_t$ and $\potf{t0}$ are guaranteed to remain variationally independent for each $t$ and so are $\potf{00}$ and $\potf{10}$. In contrast, in order to ensure variational independence between $\alpha_t$ and $\potf{t0}$ under \nHL{PT}, not only must $\alpha_0$ and $\alpha_1$ be related to each other, but so must $\potf{00}$ and $\potf{10}$. Specifically, there must be a deterministic relationship between $\alpha_0$ and $\alpha_1$ that does not depend on the baseline densities $(\potf{00},\potf{10})$, and likewise, there must be a relationship linking $(\potf{00},\potf{10})$. Figure \ref{Fig-OR_PT_OREC} visually describes this result. The three models are submodels of $\mathcal{M}_{\text{VI}}
=
\big\{ P(\bO) \cond 
\text{$\alpha_t$ and $\potf{t0}$ are variationally independent for $t=0,1$}
\big\}$ with the following forms:  $\mathcal{M}_{\text{PT}} = \big\{ P(\bO) \in \mathcal{M}_{\text{VI}}  \cond
\text{The \nHL{PT} assumption holds}  \big\} $, $\mathcal{M}_{\text{OREC}} = \big\{ P(\bO) \in \mathcal{M}_{\text{VI}}  \cond \text{The \nHL{OREC} assumption holds} \big\}$, $\mathcal{M}_{\text{OR}} = \big\{ P(\bO) \in \mathcal{M}_{\text{VI}}  \cond \alpha_1 \text{ and } \alpha_0$  have a deterministic relationship that does not depend on $(\potf{00},\potf{10}) \big\}$.
\begin{figure}[!htp]
\centering
\scalebox{0.8}{\begin{tikzpicture}
\begin{scope}[fill opacity = 1,text opacity=1]
\draw[draw = black, even odd rule] (-3.5, -1.75) rectangle (3.5, 1.75) 
node[below left=1ex] {$\mathcal{M}_{\text{OR}}$};
\draw[draw = black,even odd rule] (-0.8, 0) circle[radius=1.2]
(0.8, 0) circle[radius=1.2];

\node[fill = white] at (-2, 0) (D) {$\mathcal{M}_{\OREC}$};
\node[fill = white] at (2.2, 0) (C) {$\mathcal{M}_{\PT}$};   
\end{scope}
\end{tikzpicture}}
\caption{Visual Illustration of Lemma \protect{\ref{lemma-PT_OREC}}.}
\label{Fig-OR_PT_OREC}
\end{figure}

To formally state the result, we introduce additional notation. Let $\mathcal{F}_{t X}$ and $\mathcal{F}_{X}$ be collections of functions defined over the supports of $(\potY{0}{t},\bX)$ and $\bX$, respectively. For each $\bx$, let $\mathcal{T}_{\bx}: \mathcal{F}_{t X}^{\otimes 2} \rightarrow \mathcal{F}_X$ be an operator of the form 
\begin{align*}
\mathcal{T}_{\bx} \big( \alpha_t,\potf{t0} \big)
=
\frac{ \int y \alpha_{t}(y,\bx) \potf{t0}(y \cond \bx) \, dy }{ \int \alpha_{t}(y,\bx) \potf{t0}(y \cond \bx) \, dy} -  \int y \potf{t0}(y \cond \bx) \, dy \ .
\end{align*}
The operator $\mathcal{T}_{\bx}(\alpha_t,\potf{t0})$ measures confounding bias on the additive scale at time $t$ corresponding to given $(\alpha_t,\potf{t0})$ functions, at a given value of $\bx$ i.e., $\mathcal{T}_{\bx} (\alpha_t,\potf{t0}) = \EXP_{\alpha_t,\potf{t0}} \big\{ \potY{0}{t} \cond A=1,\bX=\bx \big\} - \EXP_{\alpha_t,\potf{t0}} \big\{ \potY{0}{t} \cond A=0,\bX=\bx \big\}$ where $\EXP_{\alpha_t,\potf{t0}}$ is the expectation operator evaluated with respect to the conditional density of $\potY{0}{t} \cond (A=a,\bX=\bx)$ parametrized by $(\alpha_t,\potf{t0})$; see the last paragraph of this Section for details on the operator $\mathcal{T}_{\bx}$. Lemma \ref{lemma-PT_OREC} states the formal result. 
\begin{lemma}				\label{lemma-PT_OREC}

Suppose $\mathcal{S}_t(1) \subseteq \mathcal{S}_t(0)$ $(t=0,1)$ and the following \textit{injectivity} condition hold: (\textit{Injectivity}) for each $(y,\potf{t0})$, there exists an injective mapping $\varphi_{y,\potf{t0}}: \mathcal{F}_X \rightarrow \mathcal{F}_{t X}$ satisfying $\varphi_{y,\potf{t0}} (\mathcal{T}_{\bx} (\alpha_t,\potf{t0})) = \alpha_t(y,\bx)$. 	If the \nHL{PT} assumption holds, then $\alpha_t$ can be variationally independent with $\potf{t0}$ for $t=0,1$ if and only if there exists a one-to-one function $\phi:\mathcal{F}_{0X} \rightarrow \mathcal{F}_{1X}$ satisfying $\alpha_1(y,\bx) = \phi(\alpha_0(y,\bx))$ where $\phi$ does not depend on $(\potf{00},\potf{10})$.  Additionally, the baseline densities $\potf{00}$ and $\potf{10}$ are variationally dependent under the \nHL{PT} assumption through the restriction $\mathcal{T}_{\bx}(\phi(\alpha),\potf{10}) - \mathcal{T}_{\bx}(\alpha,\potf{00}) = 0$ for any odds ratio function $\alpha$.
\end{lemma}

\noindent  We remark that the injectivity condition is satisfied for a wide collection of standard outcome distributions including the normal and binomial distributions as well as the exponential likelihood family satisfying certain regularity conditions; see the next paragraph. Lemma \ref{lemma-PT_OREC} implies that, under the \nHL{PT} assumption and injectivity condition, the relationship in Figure \ref{Fig-OR_PT_OREC} holds, i.e., the odds ratios at times 0 and 1 must necessarily be related to each other for variational independence to hold between the odds ratio $\alpha_t$ and the outcome's baseline density $\potf{t0}$ for both time periods. In other words, for the odds ratio parameter and the outcome's baseline density to be freely specified under \nHL{PT}, it is necessary to restrict the relationship between the odds ratio functions across time periods. The result essentially states that identifying the ATT will necessarily involve restricting the relationship between the two odds ratio functions even under the common \nHL{PT} assumption. Furthermore, unlike \nHL{OREC}, \nHL{PT} further induces a relationship linking potential outcomes' baseline densities. We plan to explore the scope for identification when $\phi$ is unspecified in future research.

Next, we consider the following three examples that satisfy the injectivity condition in Lemma \ref{lemma-PT_OREC}.

\begin{itemize}
\item[(i)] (\textit{Example 1: Gaussian}) Suppose that $\potY{0}{t} \cond A, \bX \sim N( \mu_t(A,\bX), \sigma_t^2(\bX) )$. Then, %we find the odds ratio and the operator $\mathcal{T}_{\bx}$ have the following forms:
\begin{align*}
&
\alpha_t(y,\bx) = \exp \bigg[
\frac{ y \big\{ \mu_t(1,\bx) - \mu_t(0,\bx) \big\}  }{\sigma_t^2(\bx)}
\bigg] 
\ , 
\
&&
\mathcal{T}_{\bx}( \alpha_t,\potf{t0} )
=
\mu_t(1,\bx) - \mu_t(0,\bx) \ .
\end{align*}
Therefore, %$\varphi_{y,\potf{t0}}$ and $\varphi_{y,\potf{t0}}^{-1}$ are
\begin{align*}
&
\varphi_{y,\potf{t0}} \big( \mathcal{T}_{\bx}(\alpha_t,\potf{t0}) \big) 
=  \exp \bigg\{ 	\frac{ y \cdot \mathcal{T}_{\bx}(\alpha_t,\potf{t0}) }{ \sigma_t^2(\bx) }  \bigg\}
\ , 
&&
\varphi_{y,\potf{t0}}^{-1} \big( \alpha_t(y,\bx) \big) 
=
\frac{ \log \big\{ \alpha_t(y,\bx) \big\} \sigma_t^2(\bx) }{ y } \ .
\end{align*}
Note that these mappings are injective when $(y, \potf{t0})$ are fixed. Therefore, the relationship between $\alpha_1$ and $\alpha_0$ is
\begin{align*}
\alpha_1(y,\bx) 
=		
\varphi_{y,\potf{10}} \big( \varphi_{y,\potf{00}}^{-1} \big( \alpha_0(y,\bx) \big) \big) 
=
\big\{ \alpha_0(y,\bx) \big\}^{ \sigma_0^2(\bx)/\sigma_1^2(\bx) } \ .
\end{align*}
As a conclusion, $\alpha_t$ and $\potf{t0}$ are variationally independent if $\sigma_0^2(\bx)/\sigma_1^2(\bx)$ is equal to a function $c(\bx)$ that does not depend on $(\potf{10}, \potf{00})$.

\item[(ii)] (\textit{Example 2: Binomial}) Suppose that $\potY{0}{t} \cond A, \bX \sim \text{Ber}( p_t(A,\bX) )$. Then, we find the odds ratio and the operator $\mathcal{T}_{\bx}$ have the following forms:
\begin{align*}
&
\alpha_t(y,\bx) 
=  \frac{ p_t(1,\bx) \{ 1- p_t(0,\bx) \} }{ \{ 1-  p_t(1,\bx) \} p_t(0,\bx) } y + (1-y) 
\ , 
\
&&
\mathcal{T}_{\bx}(\alpha_t,\potf{t0})
=
p_t(1,\bx) - p_t(0,\bx) \ .
\end{align*}
Therefore, $\varphi_{y,\potf{t0}}$ and $\varphi_{y,\potf{t0}}^{-1}$ are
\begin{align*}
&
\varphi_{y,\potf{t0}} \big( \mathcal{T}_{\bx}(\alpha_t,\potf{t0}) \big) 
=
\frac{ \mathcal{T}_{\bx}(\alpha_t,\potf{t0}) + p_t(0,\bx) }{1 - \mathcal{T}_{\bx}(\alpha_t,\potf{t0}) - p_t(0,\bx) } \frac{1-p_t(0,\bx)}{p_t(0,\bx)} y + (1-y)
\ , 
\\
&
\varphi_{y,\potf{t0}}^{-1} \big( \alpha_t(y,\bx) \big) 
=
\frac{ p_t(0,\bx) \{ 1- p_t(0,\bx)  \}  \{ \alpha_t(y,\bx) -1  \}  }{ \alpha_t(y,\bx) p_t(0,\bx) - p_t(0,\bx) + 1 } 
\ .
\end{align*}
Note that these mappings are injective when $(y, \potf{t0})$ are fixed. Therefore, the relationship between $\alpha_1$ and $\alpha_0$ is
\begin{align*}
\alpha_1(y,\bx) 
& =		
\varphi_{y,\potf{10}} \big( \varphi_{y,\potf{00}}^{-1} \big( \alpha_0(y,\bx) \big) \big) 
\\
&
=
\frac{  \displaystyle{ \frac{ p_0(0,\bx) \{ 1- p_0(0,\bx)  \}  \{ \alpha_0(y,\bx) -1  \}  }{ \alpha_0(y,\bx) p_0(0,\bx) - p_0(0,\bx) + 1 }  + p_1(0,\bx) } }{ \displaystyle{ 1 -   \frac{ p_0(0,\bx) \{ 1- p_0(0,\bx)  \}  \{ \alpha_0(y,\bx) -1  \}  }{ \alpha_0(y,\bx) p_0(0,\bx) - p_0(0,\bx) + 1 }   - p_1(0,\bx) } } \frac{1-p_1(0,\bx)}{p_1(0,\bx)} y + (1-y)
\ .
\end{align*}
Therefore, $\alpha_t$ and $\potf{t0}$ are variationally independent if $p_1(\bx) = p_0(\bx)$; in this case, $\alpha_1=\alpha_0$.

\item[(iii)] (\textit{Example 3: Linear Exponential Family}) Suppose the conditional density of the outcome belongs to an exponential family of the following form with appropriately chosen $\eta_t$, $\xi_t$, $g_t$, and $\mathcal{A}_t$:
\begin{align*}
\potf{t} (y \cond a,\bx) \propto \exp \big\{ \eta_t (a,\bx) y + \xi_t (\bx) g_t (y) - \mathcal{A}_t \big( \eta_t (a,\bx), \xi_t (\bx) \big) \big\} \ ,
\end{align*}
where the derivative of the log partition function $\mathcal{A}_t$ with respect to the first argument is an injective mapping of the first argument, i.e., $\mathcal{B}_t(\eta,\xi) := \big\{\partial \mathcal{A}_t(w,\xi) / \partial w \big\} \big|_{w=\eta} $ satisfies that $\eta \neq \eta'$ implies $\mathcal{B}_t (\eta, \xi) \neq \mathcal{B}_t (\eta', \xi)$. We find the odds ratio and the operator $\mathcal{T}_{\bx}$ are given as
\begin{align*}
& 
\alpha_t(y,\bx)  = \exp \big[ y \big\{  \eta_t (1,\bx) -\eta_t (0,\bx) \big\} \big] \ ,
&&
\mathcal{T}_{\bx}(\alpha_t,\potf{t0}) = \mathcal{B}_t \big( \eta_t (1,\bx) , \xi_t (\bx) \big) - \mathcal{B}_t \big( \eta_t (0,\bx) , \xi_t (\bx) \big) \ .
\end{align*}
Let $\mathcal{B}_{t}^{-1}  $ be the inverse map of $\mathcal{B}_t( \cdot, \xi_t (\bx) )$ that satisfies $\mathcal{B}_{t} \big( \mathcal{B}_{t}^{-1} \big( b(y,\bx) \big)  , \xi_t (\bx) \big) = b(y,\bx)$ for any function $b$. Note that $\mathcal{B}_{t} \big( \eta_t (0,\bx) , \xi_t (\bx) \big)$ depends on $\potf{t0}$ through $\eta_t (0,\bx)$ and $\xi_t (\bx)$, and $\mathcal{B}_{t}^{-1}$ depends on the baseline density $\potf{t0}$ through $\xi_t (\bx)$. We find 
\begin{align*}
\eta_t (1,\bx) = \mathcal{B}_{t}^{-1} \big( \mathcal{T}_{\bx}(\alpha_t,\potf{t0}) + \mathcal{B}_{t} \big( \eta_t (0,\bx), \xi_t (\bx) \big) \big) \ .
\end{align*}
Consequently, $\varphi_{y,\potf{t0}}$ and $\varphi_{y,\potf{t0}}^{-1}$ are
\begin{align*}
\varphi_{y,\potf{t0}} \big( \mathcal{T}_{\bx}(\alpha_t,\potf{t0}) \big) 
& =  \exp \big[ y \big\{  \eta_t (1,\bx) -\eta_t (0,\bx) \big\} \big]
\\
& = \exp \big[
y \big\{  \mathcal{B}_{t}^{-1} \big( \mathcal{T}_{\bx}(\alpha_t,\potf{t0}) + \mathcal{B}_{t} \big( \eta_t (0,\bx), \xi_t (\bx) \big) \big)  - \eta_t (0,\bx) \big\}
\big] \ , \\
\varphi_{y,\potf{t0}}^{-1} \big( \alpha_t(y,\bx) \big) 
& = \mathcal{B}_{t} \big( \{\log \alpha_t(y,\bx) \}/ y + \eta_t (0,\bx) , \xi_t (\bx) \big) -  \mathcal{B}_{t} \big( \eta_t (0,\bx) , \xi_t (\bx) \big)  \ .
\end{align*}
Therefore, we find $\alpha_1(y,\bx) = \varphi_{y,\potf{10}} \big( \varphi_{y,\potf{00}} ^{-1} \big( \alpha_0(y,\bx) \big) \big)$ where
\begin{align*}
\alpha_1(y,\bx) 
& = \varphi_{y,\potf{10}} \big( \varphi_{y,\potf{00}} ^{-1} \big( \alpha_0(y,\bx) \big) \big)
\\
& =
\exp \Bigg[
y \Bigg\{  \mathcal{B}_{1}^{-1} \bigg( 
\begin{array}{l}
 \mathcal{B}_{0} \big( \{\log \alpha_0(y,\bx) \}/ y + \eta_0(0,\bx) , \xi_0(\bx) \big) 
 \\
 -  \mathcal{B}_{0} \big( \eta_0(0,\bx) , \xi_0(\bx) \big) 
 + \mathcal{B}_{1} \big( \eta_1(0,\bx), \xi_1(\bx) \big)
\end{array} \bigg)  - \eta_1(0,\bx) \Bigg\}	\Bigg] \ .
\end{align*}
To make $\alpha_t$ and $\potf{t0}$  variationally independent, $ \varphi_{y,\potf{10}} \circ \varphi_{y,\potf{00}} ^{-1}$ should not depend on $\mathcal{B}_{t}(\cdot, \xi_t(\bx))$, $\eta_{t}(0,\bx)$, and $\xi_{t}(\bx)$. Note that (\textit{Example 1: Gaussian}) and (\textit{Example 2: Binomial}) are special cases of (\textit{Example 3: Linear Exponential Family}) with
\begin{itemize}
\item[] (\textit{Example 1: Gaussian})
\begin{align*}
&
\eta_t(a,\bx) = \frac{\mu_t(a,\bx)}{\sigma_t^2(\bx)} 
\ , \
\xi_t(\bx) = - \frac{1}{2 \sigma_t^2(\bx)}
\ , \ 
g_t(y) = y^2
\ , \
\mathcal{A}_t(\eta_t,\xi_t)
=
-\frac{\eta_t^2}{4\xi_t} - 0.5 \log (-2\xi_t) \ .
\end{align*}

\item[] (\textit{Example 2: Binomial})
\begin{align*}
&
\eta_t(a,\bx) = \log \frac{p_t(a,\bx)}{1-p_t(a,\bx)}
\ , \
\xi_t(\bx) = 0
\ , \ 
g_t(y) = 0
\ , \
\mathcal{A}_t(\eta_t,\xi_t)
=
\log \{ 1 + \exp (\eta_t) \} \ .
\end{align*}
\end{itemize}
\end{itemize}
 
\item[2.] \textit{(Comparison to \nHL{NPT})}  

Suppose the conditional density of the outcome belongs to an exponential family of the following form with appropriately chosen $\eta_t$, $\xi_t$, $g_t$, and $\mathcal{A}_t$:
\begin{align*}
\potf{t} (y \cond a,\bx) \propto \exp \big\{ \eta_t (a,\bx) y + \xi_t (\bx) g_t (y) - \mathcal{A}_t \big( \eta_t (a,\bx), \xi_t (\bx) \big) \big\} \ .
\end{align*}
Let $\mathcal{L}$ be the canonical link satisfying $\mathcal{L}\{ \EXP(Y | A=a,\bX=\bx) \} = \eta_{t}(a,\bx)$, i.e., $\eta_{t}(a,\bx) = \mathcal{L}^{-1} \big(  \EXP(Y | A=a,\bX=\bx)  \big)$. Suppose \nHL{NPT} holds under this canonical link, which yields the condition $\eta_{1}(1,\bx) - \eta_{0}(1,\bx)
    =
    \eta_{1}(0,\bx) - \eta_{0}(0,\bx) $. Meanwhile, the odds ratio in this model is given by $\alpha_t(y,\bx)  = \exp \big[ y \big\{  \eta_t (1,\bx) -\eta_t (0,\bx) \big\} \big]$.
Thus, \nHL{OREC} implies the condition $
     \eta_1 (1,\bx) -\eta_1 (0,\bx)
     =
      \eta_0 (1,\bx) -\eta_0 (0,\bx)$, which coincides with \nHL{NPT} in this setting.

However, these two assumptions are not generally nested. To illustrate this, consider the following two data-generating processes:
\begin{align*}
    &
    \textbf{DGP 1}:
    &&
    \potY{0}{t} \cond A \sim N \big( A + t, 1 \big)
    \\
    &
    \textbf{DGP 2}:
    &&
    \potY{0}{t} \cond (A,\bX) \sim N \big( A+t+3At, 3t+1 \big)
\end{align*} 
From \eqref{eq-Normal PT OREC}, we observe that \nHL{NPT} holds for \textbf{DGP 1} with the identity link, but not for \textbf{DGP 2}. Conversely, we find that \nHL{OREC} is satisfied for \textbf{DGP 2}, but not for \textbf{DGP 1}.

\item[3.] \textit{(Comparison to Changes-in-Changes)}  

We consider the following two data generating processes:
\begin{align*}
&
\textbf{DGP 3}:
&&
\potY{0}{0}  = U_0 
\ , \
&&
\potY{0}{1}  = U_1  
\ , \ 
&&
U_0 \cond A \sim N(A,1)
\ , \ 
&&
U_1 \cond A \sim N(2A,2)
\\
&
\textbf{DGP 4}:
&&
\potY{0}{0}  = U_3
\ , \
&&
\potY{0}{1}  = 2 U_3
\ , \ 
&&
U_3 \cond A \sim N(A,1)
\ .
\end{align*}
We first focus on \textbf{DGP 3} in which $\potY{0}{0} \, | \, A \sim N(A,1)$ and $\potY{0}{1} \, | \, A \sim N(2A,2)$. From \eqref{eq-Normal PT OREC}, we find the \nHL{OREC} assumption is satisfied. However, the \nHL{CiC} model is violated because the latent variables at time 0 and 1 (which are $U_0$ and $U_1$, respectively) do not have the same distribution conditioning on $A$, i.e., $U \, | \, (A,t=0) = U_0 \, | \, A \stackrel{D}{\neq} U_1 \, | \, A = U \, | \, (A,t=1)$. Next, we consider \textbf{DGP 4}  in which $\potY{0}{0} \, | \, A \sim N(A,1)$ and $\potY{0}{1} \, | \, A \sim N(2A,4)$. Again, from \eqref{eq-Normal PT OREC}, we find the \nHL{OREC} assumption is violated. On the other hand, it satisfies all conditions of \citet{CiC2006}, implying that it is a valid \nHL{CiC} model. Therefore, these two data generating processes imply that the \nHL{OREC} condition and the \nHL{CiC} model are not nested.
 
\item[4.] \textit{(Comparison to Parallel Trends in the Log Characteristic Function)}  

Consider $\potY{0}{t} \cond A \sim N(\mu_t(A),\sigma_t^2)$ of which characteristic function is $\Psi_{\potY{0}{t} | A } (s) = \exp \big\{ i s \mu_t(A) - 0.5 \sigma_t^2 s^2 \big\}$. Therefore, parallel trends in the log characteristic function reduces to $
i s \big\{ \mu_1(1) - \mu_1(0) \big\} 
=
i s \big\{ \mu_0(1) - \mu_0(0) \big\}$, which reduces to the \nHL{PT} assumption. From \eqref{eq-Normal PT OREC}, this implies that the \nHL{OREC} assumption and the \nHL{PT} condition in the log characteristic function are not nested because the \nHL{OREC} assumption is equivalent to $\{ \mu_1(1) -\mu_1(0) \big\}/\sigma_1^2 = \{ \mu_0(1) -\mu_0(0) \big\}/\sigma_0^2$.

\item[5.] \textit{(Comparison to Copula Invariance)}  

Consider the following data generating processes:
\begin{align*}
& 
\textbf{DGP 5}: 
&&
\potY{0}{0} = U_1 U_2 + \epsilon_0 ,
&&
\potY{0}{1} = U_1 U_2 + \epsilon_1 ,
&&
A = U_1 \sim \text{Ber}(0.5) ,
\\
&
&& 
(U_2,\epsilon_0,\epsilon_1)
\stackrel{i.i.d.}{\sim} N(0,1) ,
&&
(U_2,\epsilon_0,\epsilon_1) \indep U_1
\\ 
& 
\textbf{DGP 6}: 
&&
\potY{0}{0} = U_3 + \epsilon_0
,
&&
\potY{0}{1} = Y_0 + \epsilon_1
, 
&&
A = U_3 \sim \text{Ber}(0.5)
,
\\
&
&&
(\epsilon_0,\epsilon_1) \stackrel{i.i.d.}{\sim}N(0,1)
,
&&
(\epsilon_0,\epsilon_1) \indep U_3 \ .
\end{align*}
Under \textbf{DGP 5}, we find
\begin{align*}
&
\begin{pmatrix}
\potY{0}{0} \\ \potY{0}{1}
\end{pmatrix}
\bigg| (A=0) 
\sim \text{MVN}_2 
\left(
\begin{pmatrix}
0 \\ 0
\end{pmatrix}
,
\begin{pmatrix}
1 & 0 \\ 0 & 1
\end{pmatrix}
\right)
\ ,
&&
\begin{pmatrix}
\potY{0}{0} \\ \potY{0}{1}
\end{pmatrix}
\bigg| (A=1) 
\sim \text{MVN}_2 
\left(
\begin{pmatrix}
0 \\ 0
\end{pmatrix}
,
\begin{pmatrix}
2 & 1 \\ 1 & 2
\end{pmatrix}
\right)
\\
&
\begin{pmatrix}
\potY{0}{0} \\ \Delta_{1}^{(0)}
\end{pmatrix}
\bigg| (A=0) 
\sim \text{MVN}_2 
\left(
\begin{pmatrix}
0 \\ 0
\end{pmatrix}
,
\begin{pmatrix}
1 & -1 \\ -1 & 2
\end{pmatrix}
\right)
\, 
&&
\begin{pmatrix}
\potY{0}{0} \\ \Delta_{1}^{(0)}
\end{pmatrix}
\bigg| (A=1) 
\sim \text{MVN}_2 
\left(
\begin{pmatrix}
0 \\ 0
\end{pmatrix}
,
\begin{pmatrix}
2 & -1 \\ -1 & 2
\end{pmatrix}
\right)
\end{align*}
Therefore, the \nHL{OREC} assumption is satisfied with $\alpha_1^*(y) = \alpha_0^*(y) = 1$. However, the copula stability assumption is violated because the copula of the untreated group is $C_{\Delta_1^{(0)} , \potY{0}{0} | A = 0} (u,v) = \Phi_2 \big( \Phi^{-1}(u), \Phi^{-1}(v) \con -1/\sqrt{5} \big)$ whereas the copula of the treated group is $C_{\Delta_1^{(0)} , \potY{0}{0} | A = 1} (u,v) = \Phi_2 \big( \Phi^{-1}(u), \Phi^{-1}(v) \con -1/2 \big)$; here, $\Phi$ is the cumulative distribution function of $N(0,1)$ and $\Phi_2 (\cdot ,\cdot  \con \rho)$ is the cumulative distribution function of $\text{MVN}_2 \left( \begin{pmatrix}
0 \\ 0
\end{pmatrix}, \begin{pmatrix}
1 & \rho \\ \rho & 1
\end{pmatrix} \right)$.

On the other hand, under \textbf{DGP 6}, we find
\begin{align*}
&
\begin{pmatrix}
\potY{0}{0} \\ \potY{0}{1}
\end{pmatrix}
\bigg| (A=0) 
\sim \text{MVN}_2 
\left(
\begin{pmatrix}
0 \\ 0
\end{pmatrix}
,
\begin{pmatrix}
1 & 1 \\ 1 & 2
\end{pmatrix}
\right)
\ ,
&&
\begin{pmatrix}
\potY{0}{0} \\ \potY{0}{1}
\end{pmatrix}
\bigg| (A=1) 
\sim \text{MVN}_2 
\left(
\begin{pmatrix}
1 \\ 1
\end{pmatrix}
,
\begin{pmatrix}
1 & 1 \\ 1 & 2
\end{pmatrix}
\right)
\\
&
\begin{pmatrix}
\potY{0}{0} \\ \Delta_{1}^{(0)}
\end{pmatrix}
\bigg| (A=0) 
\sim \text{MVN}_2 
\left(
\begin{pmatrix}
0 \\ 0
\end{pmatrix}
,
\begin{pmatrix}
1 & 0 \\ 0 & 1
\end{pmatrix}
\right)
\ , 
&&
\begin{pmatrix}
\potY{0}{0} \\ \Delta_{1}^{(0)}
\end{pmatrix}
\bigg| (A=1) 
\sim \text{MVN}_2 
\left(
\begin{pmatrix}
1 \\ 0
\end{pmatrix}
,
\begin{pmatrix}
1 & 0 \\ 0 & 1
\end{pmatrix}
\right) \ .
\end{align*}
Therefore, the \nHL{OREC} assumption is violated with $\alpha_0^*(y) = \exp(y) \neq \exp(y/2) = \alpha_1^*(y)$. However, the distributional DiD and copula stability assumptions are satisfied with the copulas of the treated and untreated groups as $C_{\Delta_1^{(0)} , \potY{0}{0} | A = 1} (u,v) = C_{\Delta_1^{(0)} , \potY{0}{0} | A = 0} (u,v) = uv$. These two data generating processes show that \nHL{OREC} and the identifying assumptions in \citet{CallawayLiOka2018} and \citet{CallawayLi2019} are not nested.
 
\item[6.] \textit{(Comparison to Sequential Ignorability)}  

Consider the following two data generating processes:
\begin{align*}
&
\textbf{DGP 7}: 
\quad
Y_0 = U + \epsilon_0
\ , \quad
\potY{0}{1} = U + Y_0 + \epsilon_1
\\
&
\textbf{DGP 8}: 
\quad
Y_0 = U + \epsilon_0
\ , \quad
\potY{0}{1} = Y_0 + \epsilon_1
\end{align*}
where $
A = U \sim \text{Ber}(0.5)$, $(\epsilon_0,\epsilon_1) \stackrel{i.i.d.}{\sim}N(0,1)$, $	(\epsilon_0,\epsilon_1) \indep U$. 

For \textbf{DGP 7}, we find 
\begin{align*}
&
\potY{0}{1} \cond (A = 1, Y_0)
\stackrel{D}{=}
\potY{0}{1} \cond (U = 1, Y_0)
\sim N(Y_0 + 1 , 1)  
\\
&
\potY{0}{1} \cond (A = 0, Y_0)
\stackrel{D}{=}
\potY{0}{1} \cond (U = 0, Y_0)
\sim N(Y_0  , 1)
\ ,
\end{align*}
indicating that the sequential ignorability is not satisfied. On the other hand, the \nHL{OREC} assumption is satisfied from \eqref{eq-Normal PT OREC}.

On the other hand, for \textbf{DGP 8}, we have
\begin{align*}
&
\potY{0}{1} \cond (A = 1, Y_0)
\stackrel{D}{=}
\potY{0}{1} \cond (U = 1, Y_0)
\sim N(Y_0 , 1)
\\
&
\potY{0}{1} \cond (A = 0, Y_0)
\stackrel{D}{=}
\potY{0}{1} \cond (U = 0, Y_0)
\sim N(Y_0  , 1)
\ ,
\end{align*}
indicating that the sequential ignorability condition is satisfied. On the other hand, the \nHL{OREC} assumption is violated with $\alpha_0^*(y) = \exp(y) \neq \exp(y/2) = \alpha_1^*(y)$, which is a direct consequence of \eqref{eq-Normal PT OREC}.  These two data generating processes show that the \nHL{OREC} condition and the sequential ignorability condition are not nested.

\end{itemize}

\subsection{Identification of the ATT under the Parallel Trends and Odds Ratio Equi-confounding Assumptions}							\label{sec:supp-id-PT}

We consider a broader class of causal effects of the form $\EXP \big\{ \mathcal{G}(Y_1^{(1)}) - \mathcal{G}(\pYo) \cond A=1 \big\}$ where $\mathcal{G}(\cdot)$ is a fixed, square-integrable function. The first term is identifiable via $\EXP \big\{ \mathcal{G}(Y_1^{(1)}) \cond A=1 \big\} = \EXP \big\{ A \mathcal{G}(Y_1^{(1)})  \big\}/ \Pr(A=1)$, so our focus is on the identifying formula of the counterfactual mean $\EXP \big\{ \mathcal{G}(\pYo) \cond A=1 \big\}$. 

For convenience, we revisit the inverse probability weighting (IPW), outcome regression, and augmented inverse probability weighting (AIPW) representations of this counterfactual mean  $\EXP \big\{ \mathcal{G}(\pYo) \cond A=1 \big\}$:
\begin{align}
&
\EXP \big\{ \mathcal{G}(\pYo) \cond A=1 \big\}
\nonumber
\\
& = 
\EXP \big\{
(1-A) \beta_1^*(\bX) \alpha_1^*(\Yo , \bX)
\mathcal{G}(\Yo) 
\big\} / \Pr(A=1)
&&
\label{supp:eq-rep-IPW}
\\
& =
\EXP \big\{
A
\mu^*(\bX)
\big\} / \Pr(A=1)
&&
\label{supp:eq-rep-OR} 
\\
& =
\EXP \big[
(1-A) 
\beta_1^*(\bX) \alpha_1^*(\Yo,\bX)
\big\{  \mathcal{G}(\Yo) - \mu^*(\bX) \big\}
+
A
\mu^*(\bX)
\big] / \Pr(A=1)
\ .
&&
\label{supp:eq-rep-AIPW} 
\end{align}

We first establish why these three representations are valid. We first show the IPW representation result:
\begin{align*} 
\EXP \big\{
(1-A) 
\beta_1^*(\bX) \alpha_1^*(\Yo , \bX)
\mathcal{G}(\Yo) 
\big\}
& = 
\EXP \big[
\Pr(A=0 \cond \bX)
\beta_1^*(\bX) 
\EXP \big\{
\alpha_1^*(\Yo , \bX) \mathcal{G}(\Yo) 
\cond A=0 , \bX
\big\}
\big]
\\
& = 
\EXP \big[\Pr(A=1 \cond \bX)\EXP \big\{
\mathcal{G}(\pYo) 
\cond A=1 , \bX
\big\}
\big]
\\
& = 
\Pr(A=1)
\EXP \big \{ \mathcal{G}(\pYo) \cond A=1 \big\} \ .
\end{align*}
The third line is from \eqref{eq-IPW-basis}, which we establish later. Similarly, we find the outcome regression-based representation:
\begin{align*}
\EXP \big\{ A \mu^*(\bX)
\big\}
& = 
\Pr(A=1)
\EXP \big\{ \mu^*(\bX) \cond A=1 \big\}
\\
& = 
\Pr(A=1)
\EXP \big[ \EXP \big\{ \mathcal{G} (\pYo) \cond A=1 , \bX \big\} \cond A=1 \big]
\\
& = 
\Pr(A=1) \EXP \big\{ \mathcal{G} (\pYo) \cond A=1   \big\}  \ .
\end{align*}
Lastly, we show the AIPW representation:
\begin{align*}
& \EXP \big[
\beta_1^*(\bX) \alpha_1^*(\Yo , \bX)
(1-A) \big\{ \mathcal{G}(\Yo) - \mu^*(\bX) \big\}
+ A \mu^*(\bX)
\big]
\\
& = 
\Pr(A=1)
\EXP \big \{ \mathcal{G}(\pYo) - \mu^*(\bX) \cond A=1 \big\} +  \Pr(A=1) \EXP \big\{ \mathcal{G} (\pYo) \cond A=1   \big\}
\\
& =
\Pr(A=1) \EXP \big\{ \mathcal{G} (\pYo) \cond A=1   \big\}  \ .
\end{align*}
where the second line is based on the IPW and OR representations.

The first representation is in the form of a weighted average with weights applied to $\mathcal{G}(\Yo)$ given by $\beta_1^*(\bX) \alpha_1^*(y,\bX)  = \pote{1}^*(1 \cond y,\bX) /\pote{1}^*(0 \cond y,\bX)$. Here, $\pote{1}^*(a \cond y,\bx) = \Pr(A=a \cond \pYo=y,\bX=\bx)$ can be viewed as an extended propensity score relating $A$ with $(\pYo, \bX)$. From this representation, we can find the relationship between \eqref{supp:eq-rep-IPW} and the representation in \citet{Abadie2005} where $\EXP \big\{ \mathcal{G}(\pYo) \cond A=1 \big\}$ is identified under the \nHL{PT} assumption. In particular, if the \nHL{PT} assumption holds for $\mathcal{G}(\pYo)$ and $\mathcal{G}(\pYz)$, then we get
\begin{align*}
&
\EXP \bigg[ 
\frac{ \Pr(A=1 \cond \bX) (1-A) \mathcal{G}(\Yo) }{ \Pr(A=0 \cond \bX) }
+
\frac{ \big\{ A - \Pr(A=1 \cond \bX) \big\} \mathcal{G}(\Yz) }{ \Pr(A=0 \cond \bX) }
\, \bigg| \, \bX
\bigg]
\\
& =		
\frac{ \Pr(A=1 \cond \bX) }{ \Pr(A=0 \cond \bX) }
\EXP \big\{ \mathcal{G}(\Yo) \cond A=0 , \bX \big\} \Pr(A=0 \cond \bX) 
\\
&
\hspace*{1cm}
-
\frac{ \Pr(A=1 \cond \bX) }{ \Pr(A=0 \cond \bX) }
\EXP \big\{  \mathcal{G}(\Yz) \cond A=0, \bX \big\}
\Pr(A=0 \cond \bX) 
+
\EXP \big\{  \mathcal{G}(\Yz) \cond A=1, \bX \big\}
\Pr(A=1 \cond \bX) 
\\
& =		
\Pr(A=1 \cond \bX)
\Big[
\EXP \big\{ \mathcal{G}(\Yo) \cond A=0 , \bX \big\}
+
\EXP \big\{ \mathcal{G}(\Yz) \cond A=1 , \bX \big\}
-
\EXP \big\{ \mathcal{G}(\Yz) \cond A=0 , \bX \big\} 
\Big]
\\
& =		
\Pr(A=1 \cond \bX) \EXP \big\{ \mathcal{G}(\pYo) \cond A=1 , \bX \big\}
\end{align*}
where the last line is straightforward from the \nHL{PT} assumption on $\mathcal{G}(\pYo)$ and $\mathcal{G}(\pYz)$. Therefore, we find
\begin{align}                       \label{supp:PT-IPW}
&
\frac{1}{\Pr(A=1)}
\EXP \bigg[ 
\frac{ \Pr(A=1 \cond \bX) (1-A) \mathcal{G}(\Yo) }{ \Pr(A=0 \cond \bX) }
+
\frac{ \big\{ A - \Pr(A=1 \cond \bX) \big\} \mathcal{G}(\Yz) }{ \Pr(A=0 \cond \bX) }
\, \bigg| \, \bX
\bigg]
\\
& = 
\frac{1}{\Pr(A=1)}
\EXP \Bigg[
\EXP \bigg[ 
\frac{ \Pr(A=1 \cond \bX) (1-A) \mathcal{G}(\Yo) }{ \Pr(A=0 \cond \bX) }
+
\frac{ \big\{ A - \Pr(A=1 \cond \bX) \big\} \mathcal{G}(\Yz) }{ \Pr(A=0 \cond \bX) }
\, \bigg| \, \bX
\bigg] \Bigg]
\nonumber
\\
& = 
\frac{ \EXP \big[
\Pr(A=1 \cond \bX) \EXP \big\{ \mathcal{G}(\pYo) \cond A=1 , \bX \big\}
\big] }{\Pr(A=1)}		
\nonumber
= \frac{\EXP \big[ \EXP \big\{ A \mathcal{G}(\pYo) \cond \bX \big\}	\big]}{\Pr(A=1)}
= \EXP \big\{ \mathcal{G}(\pYo) \cond A=1 \big\} \ .
\nonumber
\end{align}

Comparing \eqref{supp:PT-IPW} and \eqref{supp:eq-rep-IPW}, we can immediately conclude that they are similar in that $(1-A)\mathcal{G}(\Yo)$ are weighted by the ratio of propensity scores. However, \eqref{supp:eq-rep-IPW} involves the extended propensity score $e^*$ whereas the standard propensity score $\Pr(A=1 \cond \bX)$ are used in \eqref{supp:PT-IPW}. Additionally, \eqref{supp:eq-rep-IPW} does not have an additional term related to $\Yz$ once the odds ratio $\alpha_1^*$ is provided.

The second representation has the outcome regression-based (OR) form, which is comparable to the standard representation from the \nHL{PT} assumption:
\begin{align} \label{supp:PT-OR}
& \EXP \big\{ \mathcal{G}(\pYo) \cond A=1 \big\}
\nonumber
\\
&
=
\EXP
\big[
\EXP \big\{ \mathcal{G}(\Yo) \cond A = 0 , \bX \big\}
- \EXP \big\{ \mathcal{G}(\Yz)  \cond A = 0 , \bX \big\}
+ \EXP \big\{ \mathcal{G}(\Yz) \cond A=1, \bX \big\}
\cond A=1
\big] \ . 
\end{align}
Similar to the IPW case, the two OR style representations differ in how $\Yz$ is used in the representation. Specifically, \eqref{supp:eq-rep-OR} only uses the outcome regression $\mu^* (\bX)$, which is indirectly related to $\Yz$ via the odds ratio function $\alpha_1^*$. On the other hand, $\Yz$ is directly employed in \eqref{supp:PT-OR}.

We referred to the last representation as AIPW representation because equation \eqref{supp:eq-rep-AIPW} involves both odds of $A$ (i.e., $\beta_1^* \alpha_1^*$) and outcome regression (i.e., $\mu^*$). We also remark that \eqref{supp:eq-rep-AIPW} recovers the target parameter so long as (i) $\alpha_1^*$ is correctly specified and (ii) either $\beta_{1}^* $ or $\potf{1}^*(y \cond 0,\bx)$ is correctly specified; Lemma \ref{lemma-DR} formally states the result.
\begin{lemma}				\label{lemma-DR}

Suppose that Assumptions \ref{assumption:support}-\ref{assumption:no anticipation} hold. Let $\InfFt$ be an influence function for the functional $\EXP \big\{ \mathcal{G}(\pYo) \cond A=1 \big\}$ in a semiparametric model proposed in \citet{Robins2000_Sensitivity} where the odds ratio function at time 1, $\alpha_1^*$, is a priori known and observed data distribution is unrestricted, i.e.,  
\begin{align}						\label{eq-DR-Moment}
&
\InfFt ( \bO_{1} \con \beta_1, \potf{10} )
\\
&
=
\frac{\beta_1(\bX) \alpha_1^*(\Yo,\bX)
(1-A) \big\{  \mathcal{G}(\Yo) - \mu(\bX \con \potf{10}) \big\}
+
A
\big[
\mu(\bX \con \potf{10}) - \EXP \big\{ \mathcal{G}(\pYo) \cond A=1 \big\} \big] }{\Pr(A=1)}
\nonumber
\end{align}
where $\beta_1(\bx)$ and $\potf{10}(y \cond  \bx)$ are working models of $\beta_1^*$ and $\potf{1}^*(y \cond 0, \bx)$, respectively, and \\
$ \mu(\bx \con \potf{10} )	 = 	\big\{ \int \mathcal{G}(y) \alpha_1^*(y,\bx) \potf{10} (y \cond \bx) \, dy  \big\} / \big\{ \int \alpha_1^*(y,\bx) \potf{10}(y \cond \bx) \, dy  \big\}$.  Then, $\EXP \big\{ \InfFt(\bO_{1} \con \beta_1, 
\potf{10} ) \big\} = 0$ if $\beta_1(\bx) =\beta_1^*(\bx) $ or $ \potf{10} (y \cond  \bx) = \potf{1}^*(y \cond 0, \bx)$, but not necessarily both.
\end{lemma}

The proof of Lemma \ref{lemma-DR} is in Section \ref{sec:supp-proof-supp}. We remark that a similar robustness property is discussed in \citet{Liu2020}.

The AIPW representation \eqref{supp:eq-rep-AIPW} parallels the AIPW representations discovered under the \nHL{PT} assumption presented in Section 2.2 of \citet{SantAnna2020}. Specifically, the identification strategy of the ATT under the \nHL{PT} assumption is given as follows:
\begin{align*}
&
\EXP \big\{ \mathcal{G}(\potY{1}{1}) - \mathcal{G}(\pYo) \cond A=1 \big\}
\\
& =
\frac{1}{\Pr(A=1)}
\EXP \bigg[
\bigg[
A
-
\bigg[\underbrace{ \EXP \bigg\{ \frac{ \pi(\bX) (1-A) }{1-\pi(\bX)} \bigg\} } _{=:C_\pi} \bigg]^{-1}
\bigg\{ \frac{ \pi(\bX) (1-A) }{1-\pi(\bX)} \bigg\}
\bigg]
\Big[
\big\{ \mathcal{G}( \Yo ) -  \mathcal{G}( \Yz ) \big\}
-
\Delta(\bX)
\Big]
\bigg]  \  , \
\end{align*}
where $\pi(\bX)$ and $\Delta(\bX)$ are working models of $\Pr(A=1 \cond \bX)$ and $\EXP \big\{ \mathcal{G}( \Yo ) -  \mathcal{G}( \Yz ) \cond A=0, \bX \big\}$, respectively. The identity holds if $\pi$ or $\Delta$, but not necessarily both, is correctly specified. As a result, we have the AIPW identification of $\EXP \big\{ \mathcal{G}(\pYo) \cond A=1 \big\}$ as follows:
\begin{align}       \label{supp:PT-DR}
& \EXP \big\{ \mathcal{G}(\pYo) \cond A=1 \big\}
\\
& = 
\frac{1}{\Pr(A=1)}
\EXP \bigg[
\bigg[
A
-
C_{\pi}^{-1}
\bigg\{ \frac{ \pi(\bX) (1-A) }{1-\pi(\bX)} \bigg\}
\bigg]
\Big[
\big\{ \mathcal{G}( \Yo ) -  \mathcal{G}( \Yz ) \big\}
-
\Delta(\bX)
\Big]
-
A \mathcal{G}(\Yo)
\bigg] \ .
\nonumber
\end{align}
Again, in \eqref{supp:PT-DR}, the pre-treatment outcome $\Yz$ is directly employed in the representation, whereas the AIPW representation \eqref{supp:eq-rep-AIPW} (which corresponds to the influence function \eqref{eq-DR-Moment}) indirectly uses the pre-treatment outcome $\Yz$ via the pre-treatment odds ratio $\alpha_0^*$, which replaces $\alpha_1^*$ under the \nHL{OREC} assumption.

\subsection{Details on the Characterization of the Odds Ratio Function at Time 0}							\label{sec:supp-EIF}

We characterize the odds ratio at time 0, $\alpha_0^*$, as the solution to a moment equation. Under Assumptions \ref{assumption:support}-\ref{assumption:OREC} (or Assumptions \ref{assumption:support-1}, \ref{assumption:support-2}, \ref{assumption:consistency}-\ref{assumption:OREC} with weaker support conditions), the odds ratio at time 1, $\alpha_1^*$, is also characterized as the solution to the same moment equation. For any set $\SPg \subseteq \SPzt$, let $\overline{\EXP}_{\SPg}$ be the expectation operator only over $\SPg \cap \SPz$ where the odds ratio $\alpha_0^*$ is well-defined and positive, i.e., for a function $\mathfrak{m}$, we define
\begin{align*}
\overline{\EXP}_{\SPg} \big\{ \mathfrak{m}(\Yz,A,\bX) \}
=
\iint_{\SPg \cap \SPz} \mathfrak{m}(y,a,\bx) \, P(\Yz=y,A=a,\bX=\bx) \, d(y,\bx) \ .
\end{align*}
With the new notation, the following Lemma provides a moment equation characterizing the restriction over $\mathcal{S}$ along with its properties.
\begin{lemma}					\label{lemma-OR-DR}
Suppose Assumptions \ref{assumption:consistency}-\ref{assumption:OREC} hold. For a set $\SPg \subseteq \SPzt$, let $\Psi_{\SPg}(\bO_{0} \con \alpha,\potf{00},\pote{00},{\mathfrak{m}})$ be the following function for any integrable function ${\mathfrak{m}}(\Yz,\bX)$:
\begin{align}					\label{eq-OR-DR}
&
\Psi_{\SPg} ( \bO_{0} \con \alpha, \potf{00}, \pote{00}, {\mathfrak{m}})
\\
&
=
\big[ {\mathfrak{m}}(\Yz,\bX) - \overline{\EXP}_{\SPg,\potf{00}} \big\{ {\mathfrak{m}}(\Yz, \bX) \cond A=0, \bX \big\} \big] 
\big\{ \alpha (\Yz,\bX) \big\}^{-A} 
\big\{ A- \pote{00}(A \cond \bX) \big\} \ ,
\nonumber
\end{align}
where $\alpha(y,\bx)$, $\potf{00}(y \cond \bx)$, and $\pote{00}(a \cond \bx)$ are working models of $\alpha_0^*(y,\bx)$, $\potf{0}^*(y \cond 0, \bx)$, and $\pote{0}^*(a \cond 0, \bx)$, respectively, for $(y,\bx) \in \SPg \cap \SPz$, and $\overline{\EXP}_{\SPg,\potf{00}} \big\{ {\mathfrak{m}}(\Yz, \bX) \cond A=0, \bX \big\} =\int_{\SPg \cap \SPz} {\mathfrak{m}}(y,\bX) \potf{00} (y \cond \bX) \, dy$. Then, we have the following results:
\begin{itemize}
\item[(i)] $\overline{\EXP}_{\SPg} \big\{ \Psi_{\SPg} ( \bO_{0} \con \alpha_0^*, \potf{0}^*( \cdot \cond 0, \cdot) , \pote{0}^*( \cdot \cond 0,\cdot) , {\mathfrak{m}}) \big\} = 0$ for any ${\mathfrak{m}}$, i.e., the odds ratio at time $0$ is the solution to the moment equation $\overline{\EXP}_{\SPg} \big\{ \Psi_{\SPg} ( \bO_{0} \con \alpha, \potf{0}^*( \cdot \cond 0, \cdot) , \pote{0}^*( \cdot \cond 0,\cdot) , {\mathfrak{m}}) \big\} = 0$;
\item[(ii)] Suppose that $\alpha^\dagger$ satisfies (a) $\overline{\EXP}_{\SPg} \big\{ \Psi_{\SPg} ( \bO_{0} \con \alpha^\dagger, \potf{0}^*( \cdot \cond 0, \cdot) , \pote{0}^*( \cdot \cond 0,\cdot)  , {\mathfrak{m}}) \big\} = 0$ for any ${\mathfrak{m}}$ and (b) $\alpha^\dagger(0,\bx) = 1$ for all $\bx$. Then, we have $\alpha^\dagger (y,\bx) = \alpha_0^* (y,\bx)$ almost surely for $(y,\bx) \in \SPg \cap \SPz$;
\item[(iii)] $\overline{\EXP}_{\SPg} \big\{ \Psi_{\SPg} ( \bO_{0} \con \alpha_0^*,  \potf{00}, \pote{00} , {\mathfrak{m}}) \big\} = 0$ for any ${\mathfrak{m}}$ if $\potf{00}(y \cond \bx) =\potf{0}^*( y \cond 0, \bx) $ or $\pote{00}(a \cond \bx) = \pote{0}^*( a \cond 0,\bx)$, but not necessarily both,  for $(y,\bx) \in \SPg \cap \SPz$.
\end{itemize}		
\end{lemma}
\noindent The proof of Lemma \ref{lemma-OR-DR} is in Section \ref{sec:supp-proof-supp}. Result (i) means that the moment equation \eqref{eq-OR-DR} provides an alternative characterization of the odds ratio function besides its definition when the two baseline densities $\potf{0}^*( \cdot \cond 0,\cdot)$ and $\pote{0}^*( \cdot \cond 0,\cdot)$ are correctly specified. Result (ii) implies that $\alpha_0^*$ is the unique solution to the moment equation among the collection of functions that satisfies the boundary condition of the odds ratio (i.e., $\alpha^\dagger(0,\bx)=1$ for all $\bx$). These three results indicate that the odds ratio at time 0 is restricted by the moment restriction $\overline{\EXP}_{\SPg} \big\{ \Psi_{\SPg} ( \bO \con \alpha_0^*, \potf{0}^*( \cdot \cond 0,\cdot), \pote{0}^*( \cdot \cond 0,\cdot), {\mathfrak{m}}) \big\} = 0$, and is uniquely defined if both baseline densities are correctly specified. Result (iii) shows the moment equation is AIPW against misspecification of  the baseline densities and the moment restriction is still valid for $\alpha_0^*$. 

Under the \nHL{OREC} assumption, $\alpha_1^*(y,\bx)$ is equivalent to $\alpha_0^*(y,\bx)$ over $\SPzt \cap \SPo$, and consequently, it can be characterized by using the moment equation. If we take $\SPg=\SPot$ (which is equal to $\SPzt \cap \SPo$ under Assumptions \ref{assumption:support-1}, \ref{assumption:support-2}, \ref{assumption:consistency}-\ref{assumption:OREC}), $\alpha_1^*(y,\bx)$ is characterized as the solution to the moment equation over $(y,\bx) \in \SPot$ and is equal to zero over $(y,\bx) \in \R \cap \SPot^c$, i.e.,
\begin{align}          \label{eqref-alpha1-constraint}
& \text{For } (y,\bx) \in \SPot, &&
\alpha_1^* \text{ solves }
\overline{\EXP}_{\SPot} \big\{ \Psi_{\SPot} ( \bO_{0} \con \alpha, \potf{0}^*( \cdot \cond 0,\cdot) , \pote{0}^*( \cdot \cond 0,\cdot) , \mathfrak{m}) \big\} = 0 \ ,
\nonumber
\\
& \text{For } (y,\bx) \in \R \cap \SPot^c, &&
\alpha_1^*(y,\bx) = 0 \ .
\end{align}

\subsection{Details on Density Ratio Estimation} 							\label{sec:supp-DR-Estimation}

In Section \ref{sec:Conti-Improved}, we use the KL divergence as the distance measure between the numerator $\potf{0}^*(y,\bx \cond 0)$ and the denominator scaled by the density ratio $\potf{0}^*(y,\bx \cond 0) r_0^*(y, \bx)$. Instead, we can use other distance measures to construct an estimator of $r_0^*$. For instance, one can consider the least-squares importance fitting \citep{DensityRatio2008} by minimizing the squared loss:
\begin{align*}
r_0^* =
\argmin_{r \in \mathcal{H}_{0 X} }
\iint_{\SPz \cap \SPzt} \Big\{ r(y,\bx) - r^*(y,\bx) \Big\}^2 \potf{0}^*(y,\bx \cond 0) \, d(y,\bx)  \ .
\end{align*}
The empirical counterpart of the solution to the equation is $\widehat{r}_{0,\text{MSE}} \LSS (y,\bx)  = \sum_{j \in \mathcal{I}_{k1}^c } \widehat{\gamma}_{j}\LSS \cdot \mathcal{K} \big( (y,\bx), (y_j,\bx_j) \big)$ where the coefficients $\widehat{\gamma}_i$s are obtained by solving the following optimization problem including  $L_2$-regularization term:
\begin{align*}
&	
\widehat{{\gamma}}\LSS 	:=
\argmin_{{\gamma}}
\bigg[
\AVER_{\mathcal{I}_{k0}^c} \Big\{ \Big( K_{1 \cdot}\LSS {\gamma} \Big)^2	\Big\} 
- 2 \AVER_{\mathcal{I}_{k1}^c} \Big\{ \Big( K_{0 \cdot}\LSS {\gamma} \Big)^2	\Big\}
+
\lambda
{\gamma}\T K_{\cdot\cdot}\LSS {\gamma}
\bigg]
\ , \\
&
K_{a \cdot} \LSS
=
\Big[ \mathcal{K} \big( (y_{i},\bx_{i}), (y_{j},\bx_{j} ) \big) \Big]_{ \substack{i \in \mathcal{I}_{ka}^c \\ j \in \mathcal{I}_{k}^c} } \ , 
K_{\cdot \cdot} \LSS
=
\Big[ \mathcal{K} \big( (y_{i},\bx_{i}), (y_{j},\bx_{j} ) \big) \Big]_{ i, j \in \mathcal{I}_{k}^c } \ ,
\end{align*}
where $\lambda$ is the regularization parameter that can be chosen from cross-validation. The solution can be efficiently obtained via re-framing it as quadratic programming, and $\widehat{r}_{\text{MSE}}\LSS$ converges to the optimal function in $\mathcal{H}_{0 X}$ with $o_P(N^{-1/4})$ rate under additional conditions; see \citet{DensityRatio2012_1} for details.

Instead of targeting the conditional density of $\Yz$ given $(A,\bX)$, we may see the problem from another by focusing on the conditional probability of $A$ given $(\Yz,\bX)$. Specifically, based on the Bayes formula and the definition of $\pote{0}^*(a \cond y,\bx) =  \Pr(A=a \cond \Yz=y, \bX=\bx)$, we find
\begin{align*}
r_0^*(y,\bx)
= 
\frac{ \Pr(A=0 \cond \bX = \bx) }{ \Pr(A=1 \cond \bX = \bx) }
\frac{ \pote{0}^*(1 \cond y,\bx) }{ \pote{0}^*(0 \cond y,\bx) } 
\end{align*}
Consequently, we can obtain a density ratio estimator based on any probabilistic classification machine learning (ML) methods, i.e., 
\begin{align*}
\widehat{r}_{0,\text{ML}}\LSS (y , \bx) 
=
\frac{ \widehat{\Pr}\LSS(A=0 \cond \bX) }{ \widehat{\Pr}\LSS(A=1 \cond \bX) }
\frac{ \hpote{0}\LSS(1 \cond y,\bx) }{ \hpote{0}\LSS(0 \cond y,\bx) }  \ . 
\end{align*}
Under mild conditions, the ML-based estimator $\widehat{r}_{0,\text{ML}}\LSS$ achieves $o_P(N^{-1/4})$ when the conditional probabilities are estimated by Lasso \citep{Belloni2011, Belloni2013}, random forests \citep{Wager2016, Syrgkanis2020}, neural networks \citep{ChenWhite1999, Farrell2021}, and boosting \citep{Luo2016}. We may use ensemble learners of many ML methods based on the superlearner algorithm \citep{SL2007}; see \citet{ESL} for details on various ML methods.

We can also use an ensemble of density ratio estimators for improved estimation. In particular, we consider the following weighted geometric mean: 
\begin{align*}
\widehat{r}_0\LSS(y,\bx \con {w})
=
\prod_{j=1}^{J}
\Big\{ \widehat{r}_{0,j}\LSS (y,\bx) \Big\}^{w_j} \ , \ \sum_{j=1}^{J} w_j = 1 \ , \ {w} = (w_1,\ldots,w_J)\T \geq 0_{J} \ .
\end{align*}
where $\widehat{r}_{0,j} \LSS$ is the $j$th density ratio estimator such as $\widehat{r}_{0,\text{KL}}\LSS$, $\widehat{r}_{0,\text{MSE}} \LSS$, and $\widehat{r}_{0,\text{ML}}\LSS$. We can choose ${w}$ by focusing on the alternative representations of $\potf{0}^* (y \cond 1,\bx)$. Let $\hpotf{0} \LSS (y,\bx \cond 0)$ and $\hpotf{0} \LSS (y,\bx \cond 1)$ be the density estimates obtained from nonparametric density estimation methods. By the definition of $r_0^*$, we may choose $\widehat{{w}}$ so that the two density estimators  $\widehat{r}_0\LSS  (y,\bx \con \widehat{{w}}) \cdot \hpotf{0} \LSS (y,\bx \cond 0)$ and $\hpotf{0} \LSS (y,\bx \cond 1)$, are similar to each other in the $L_2$ distance or the KL divergence. Using these density ratio estimators, we can obtain estimators of the baseline odds of $A$ at time 0 and the corresponding odds ratio. We also obtain an estimator of $r_1^*$ from a similar estimation procedure.

We conclude the section by discussing how to select reference outcome values to improve estimation performance. Occasionally, the conditional density of $( \Yz, A=0 ) \cond \bX $ at the reference outcome value $y_R$, i.e.,  $\potg{0}^*(y_R, 0 \cond \bx)$, can be extremely small. This issue arises when some covariates are strongly predictive of the outcome, and make the conditional support of $\Yz \cond (A=0,\bX=\bx)$ much narrower than the marginal support of $\Yz \cond A=0$. Then, the reference outcome value $y_R$ may not belong to the conditional support of $\Yz \cond (A=0,\bX=\bx)$ for some covariates $\bx$ even though $y_R$ belongs to the marginal support of $\Yz \cond A=0$. In this case, nuisance components having $\potg{0}^* (y_R,0 \cond \bx)$ as the denominator might be ill-posed. To resolve this issue,  we tune the reference outcome value for each $\bx$, say $y_R(\bx)$, so that $\potg{0}^* (y_R(\bx),0 \cond \bx)$ is sufficiently large. For example, we may select $y_R(\bx)$ as the median of the empirical conditional distribution $\Yz \cond (A=0,\bX=\bx)$. This choice yields a considerably more stable estimator of the density ratios and the odds ratio functions. 

\subsection{Simplified Estimation for Binary Outcomes}							\label{sec:supp-BinaryEIF}

The procedure in Section \ref{sec:Conti-Improved} is valid for binary outcomes, but some steps can be simplified due to the discrete nature of the outcome. First, the odds ratio can be estimated based on the probabilistic ML methods. Specifically, using the probabilistic ML methods and their ensemble via superlearner \citep{SL2007}, we obtain the estimates for $p_0^* (y,a  \cond  \bX) = \Pr \big( \Yz=y,A=a \cond \bX \big)$ using the estimation fold $\mathcal{I}_k^c$, denoted by $\widehat{p}_0 \LSS (y,a \cond \bX)$. The baseline odds function of $A$ at time 0 and odds ratio estimators are given as 
\begin{align*}
\widehat{\beta}_0 \LSS (\bx)
=
\frac{ \widehat{p}_0\LSS (0,1 \cond \bx) }{ \widehat{p}_0\LSS (0,0 \cond \bx) }
\quad , \quad
\widehat{\alpha}_1 \LSS (y,\bx)
=
\frac{ \widehat{p}_0\LSS (1,1 \cond \bx) }{ \widehat{p}_0\LSS (1,0 \cond \bx) }
\frac{ \widehat{p}_0\LSS (0,0 \cond \bx) }{ \widehat{p}_0\LSS (0,1 \cond \bx) } \ .
\end{align*}
Similarly, we can obtain the estimates for $p_1^* (y,0  \cond  \bX) = \Pr(\Yo=y,A=0 \cond \bX)$, denoted by $\widehat{p}_1\LSS (y,0 \cond \bX) $. From relationships \eqref{eq-beta1mu}, one can obtain estimators of $\beta_1^*$ and $\mu^*$. 

For the binary outcome simulation and application, we include the following machine learning methods in the superlearner library: linear regression via \texttt{glm}, lasso/elastic net via \texttt{glmnet} \citep{glmnet}, spline via \texttt{earth} \citep{earth} and \texttt{polspline} \citep{polspline}, generalized additive model via \texttt{gam} \citep{gam}, boosting  via \texttt{xgboost} \citep{xgboost} and \texttt{gbm} \citep{gbm}, random forest via \texttt{ranger} \citep{ranger}, and neural net via \texttt{RSNNS} \citep{RSNNS}.

Also, an alternative form of the EIF for the ATT is available when the outcome is binary. Note that $\potg{1}^*(\Yz, 1 \cond \bX) \big\{ \Yz - \mu^*(\bX) \big\}$ is given as
\begin{align*}
\potg{1}^*(\Yz, 1 \cond \bX) \big\{ \Yz - \mu^*(\bX) \big\}
& =	
\Bigg\{
\begin{array}{ll}
\Pr(A=1 \cond \bX) \mu^*(\bX) \big\{ 1 - \mu^*(\bX) \big\} 
&
\text{if $\Yz=1$}
\\
- \Pr(A=1 \cond \bX) \big\{ 1- \mu^*(\bX) \big\}  \mu^*(\bX) 
&
\text{if $\Yz=0$}
\end{array}
\\
& = (2\Yz-1) \big\{ p_0^*(0,1 \cond \bX) + p_0^*(1,1  \cond  \bX) \big\} \mu^*(\bX) \big\{ 1 - \mu^*(\bX) \big\} \ . 
\end{align*}
Therefore, we find the augmentation term is equivalent to
\begin{align*}
(2A -1 )
R^*(\Yz, A, \bX) \big\{ \Yz - \mu^*(\bX) \big\}
% &
% =
% (2A -1 )
% \frac{ \potg{1}^*(\Yz, 1 \cond \bX) \big\{ \Yz - \mu^*(\bX) \big\} }{ \potg{0}^*(\Yz, A \cond \bX) } 
% \\
& =
\frac{ 
(2A -1 )
(2\Yz-1) \big\{ p_0^*(0,1,\bX) + p_0^*(1,1,\bX) \big\} \mu^*(\bX) \big\{ 1 - \mu^*(\bX) \big\}
}{ p_0^*(\Yz,A \cond \bX) } \ .
\end{align*}
Consequently, the EIF for the ATT under a binary outcome is represented as 
\begin{align*}
& \InfFt^*(\bO)
=
\frac{ 1}{\Pr(A=1)}
\Bigg[
\begin{array}{l}
\big\{ A 
-
\beta_1^*(\bX) \alpha_1^*(\Yo,\bX)
(1-A) \big\} \big\{  \Yo - \mu ^*(\bX) \big\}
- A \tau^* 
\\
- (2A-1)  (2\Yz-1) \frac{ 
\{ p_0^*(0,1,\bX) + p_0^*(1,1,\bX)  \} \mu^*(\bX)  \{ 1 - \mu^*(\bX)  \}
}{ p_0^*(\Yz,A \, | \, \bX) }
\end{array}	
\Bigg]
\ .
\end{align*}
The corresponding estimator of the ATT can be obtained based on this alternative form of the EIF.

\subsection{Bias Structure of the Cross-fitting Estimators based on Representations \eqref{eq-rep-IPW}-\eqref{eq-rep-AIPW}}							\label{sec:supp-Other Estimators}

In this section, we provide details on the leading biases of the cross-fitting estimators for $\tau^*$ based on the three representations \eqref{eq-rep-IPW}, \eqref{eq-rep-OR}, and \eqref{eq-rep-AIPW}. These estimators have the following forms:
\begin{align*}
&
\widehat{\tau}_{m}
=
\frac{1}{K} \sum_{k=1}^{K} \widehat{\tau}_{m}^{(k)}
\  , \ m \in \{ \text{IPW}, \text{OR}, \text{AIPW} \}
\\
&
\widehat{\tau}_{\text{IPW}}^{(k)}
=
\Big\{ \AVER(A) \Big\}^{-1} 
\AVER_{\mathcal{I}_k}
\left[
\begin{array}{l} 
\big\{ A  - 
(1-A) \widehat{\beta}_1\LSS (\bX) \widehat{\alpha}_1 \LSS(\Yo,\bX)
\big\}
\Yo
\end{array}	
\right]
\\
&
\widehat{\tau}_{\text{OR}}^{(k)}
=
\Big\{ \AVER(A) \Big\}^{-1} 
\AVER_{\mathcal{I}_k}
\left[
\begin{array}{l} 
A \big\{ \Yo - 
\widehat{\mu}\LSS (\bX)
\big\}
\end{array}	
\right]
\\
&
\widehat{\tau}_{\text{AIPW}}^{(k)}
=
\Big\{ \AVER(A) \Big\}^{-1} 
\AVER_{\mathcal{I}_k}
\left[
\begin{array}{l} 
\big\{ A - 
\widehat{\beta}_1\LSS (\bX) \widehat{\alpha}_1 \LSS(\Yo,\bX)
(1-A) \big\} \big\{ \Yo - \widehat{\mu}\LSS (\bX) \big\}
\end{array}	
\right]
\end{align*}
Following the calculation in Section \ref{sec:supp:thm-AsmptoticNormal}, we find
\begin{align*}
\big\| \widehat{\tau}_{\text{IPW}}^{(k)} - \tau^* \big\|
&
=
O_P
\Big(
N^{1/2} \cdot \big\{
\big\| \widehat{\beta}_1 \LSS - \beta_1^* \big\|_{P,2}
+
\big\| \widehat{\alpha}_1 \LSS - \alpha_1^* \big\|_{P,2}
\big\}
\Big)
\\
\big\| \widehat{\tau}_{\text{OR}}^{(k)} - \tau^* \big\|
&
=
O_P
\Big(
N^{1/2} \cdot \big\{
\big\| \widehat{f}_1 \LSS - f_1^* \big\|_{P,2}
+
\big\| \widehat{\alpha}_1 \LSS - \alpha_1^* \big\|_{P,2}
\big\}
\Big)
\\
\big\| \widehat{\tau}_{\text{AIPW}}^{(k)} - \tau^* \big\|
&
=
O_P
\left(
N^{1/2} \cdot \left\{
\begin{array}{l}
        \big\| \widehat{\beta}_1 \LSS - \beta_1^* \big\|_{P,2}
\big\| \widehat{\alpha}_1 \LSS - \alpha_1^* \big\|_{P,2}
+
\big\| \widehat{f}_1 \LSS - f_1^* \big\|_{P,2}
\big\| \widehat{\alpha}_1 \LSS - \alpha_1^* \big\|_{P,2}
\\
+
\big\| \widehat{\beta}_1 \LSS - \beta_1^* \big\|_{P,2}
\big\| \widehat{f}_1 \LSS - f_1^* \big\|_{P,2}
+
\big\| \widehat{\alpha}_1 \LSS - \alpha_1^* \big\|_{P,2}
\end{array}
\right\}
\right)
\end{align*}
where $f_1^*$ and $\widehat{f}_1\LSS$ are shorthand for $f_1^*(y_1 \cond A=0,\bX)$ and $\widehat{f}_1\LSS(y_1 \cond A=0,\bX)$, respectively. Therefore, $\widehat{\tau}_{\text{IPW}}$ is $N^{1/2}$-consistent for $\tau^*$ if the convergence rates of $ \beta_1^* $ and $\alpha_1^*$ are $o_P(N^{-1/2})$. Similarly, $\widehat{\tau}_{\text{OR}}$ is $N^{1/2}$-consistent for $\tau^*$ if the convergence rates of $f_1^* $ and $\alpha_1^*$ are $o_P(N^{-1/2})$. Lastly, $\widehat{\tau}_{\text{AIPW}}$ is $N^{1/2}$-consistent for $\tau^*$ if the cross-product convergence rates of the post-treatment nuisance functions are $o_P(N^{-1/2})$ (which is the same as Assumption \ref{assumption:PostCross}) and the convergence rate of $\alpha_1^*$ is $o_P(N^{-1/2})$. However, it is well-known that this rate is not feasible (e.g., \citet{Stone1980}). Therefore, these cross-fitting estimators cannot be $N^{1/2}$-consistent for $\tau^*$. We remark that $\widehat{\tau}_{\text{AIPW}}$ can be $N^{1/2}$-consistent if $\alpha_1^*$ is known (i.e., $\big\| \widehat{\alpha}_1 \LSS - \alpha_1^* \big\|_{P,2}=0$) and the cross-product convergence rate $\big\| \widehat{\beta}_1 \LSS - \beta_1^* \big\|_{P,2}
\big\| \widehat{f}_1 \LSS - f_1^* \big\|_{P,2}$ is $o_P(N^{-1/2})$. This coincides with the robustness property in Lemma \ref{lemma-DR}. 

\subsection{A Practical Guideline for Hyperparameter Tuning} 							\label{sec:supp-Median Heuristic}

In practice, the bandwidth parameters for both the kernel density estimator and the RKHS-based density ratio estimator can be selected using standard heuristics. We consider two prominent examples:

\begin{itemize}
	\item For continuous variables, bandwidth selection for the kernel density estimator may be guided by Silverman’s rule-of-thumb \citep{Silverman1986}. As implemented in the \texttt{np} R package, the heuristic for the $j$th continuous variable, $\kappa_{\text{Silverman}, j}$, is given by:
\begin{align*}
\kappa_{\text{Silverman}, j} = 1.06 \sigma_j N^{-1/(2p+\ell)}
\end{align*}
where $\sigma_j$ is an adaptive measure of spread for the $j$th covariate, $N$ is the sample size used for estimation, $p$ is the kernel order, and $\ell$ is the dimension of the continuous variable vector.

\item Bandwidth selection for the RKHS in density ratio estimation can be performed via the median heuristic \citep{Garreau2017}. For instance, when tuning the bandwidth of a Gaussian kernel, $\exp \big\{ - \| \bX_i - \bX_j \|_2^2 / \kappa \big\}$, this heuristic suggests setting $\kappa$ to the median of the squared pairwise distances:
\begin{align*}
    \kappa_{\text{median}} = \text{median} \left\{ \| \bX_{i} - \bX_{j} \|_2^2 \mid i < j, \,\, i,j \in \text{Training Data} \right\} \ .
\end{align*}

\end{itemize}

\subsection{Median Adjustment of Cross-fitting Estimators}							\label{sec:supp-MedAdjustment}    

Because of its design, the cross-fitting estimator depends on a specific sample split and may produce outlying estimates if some split samples do not represent the entire data. To resolve the issue, \citet{Victor2018} proposes to use median adjustment from multiple cross-fitting estimates.  First, let $\widehat{\tau}_{s}$ $(s=1,\ldots,S)$ be the $s$th cross-fitting estimate with a variance estimate $\widehat{\sigma}_{s}^2$. Then, the median-adjusted cross-fitting estimate and its variance estimate are defined as follows: 
\begin{align*}
& \widehat{\tau}_{\median}
:=
\median_{s=1,\ldots,S} \widehat{\tau}_{s}
\ , \quad \widehat{\sigma}_{\median}^2
:=
\median_{s=1,\ldots,S} \big\{ \sigma_s^2 + (\widehat{\tau}_s - \widehat{\tau}_{\median} )^2 \big\} \ .
\end{align*}
These estimates are more robust to the particular realization of the sample partition.

\subsection{Sensitivity Analysis} \label{sec:supp-SA}

The UDiD framework relies on the \nHL{OREC} condition, which states that $\alpha_0^*(y,\bx) = \alpha_1^*(y,\bx)$ for all $(y,\bx)$. Since \nHL{OREC} is an untestable assumption, we develop a sensitivity analysis to assess the robustness of causal conclusions to potential violations of \nHL{OREC}. In spirit, our approach is similar to the sensitivity analyses of \citet{Yadlowsky2022} and \citet{Rambachan2023}, who considered departures from the no unmeasured confounding and \nHL{PT} assumptions, respectively.

Specifically, we replace the \nHL{OREC} condition with the following relaxed assumption:
\begin{align}
\frac{ 
	\alpha_1^*(y,\bx)
	}{ \alpha_0^*(y,\bx) }
	\in [ \Gamma^{-1}, \Gamma] \ , \quad \forall 
	(y,\bx) \in 
	\SPot
	\ .
	\label{eq-supp-sensitivity}
\end{align}
where $\Gamma \geq 1$ is a user-specified sensitivity parameter that characterizes the magnitude of departure from \nHL{OREC}. When $\Gamma=1$, condition \eqref{eq-supp-sensitivity} reduces to \nHL{OREC}. The goal of the sensitivity analysis is to derive bounds on the ATT as a function of $\Gamma$ under \eqref{eq-supp-sensitivity}.

To derive these bounds, we proceed as follows. First, we consider the continuous outcome. \citet{Chen2007} established that the density $\potg{1}^* (y \cond 1, \bx) = f^* (\potY{0}{1}=y \cond A=1, \bX=\bx)$ is proportional to
\begin{align*}
	\potg{1}^* (y \cond 1, \bx) \  \propto  \
	\alpha_1^*(y, \bx)
	\beta_1^*(\bx) f_1^*(y \cond 0,\bx) \ .
\end{align*}
Letting $w(y,\bx) = \alpha_1^*(y,\bx)/\alpha_0^*(y,\bx) \in [\Gamma^{-1},\Gamma]$, we obtain
\begin{align*}
	\potg{1}^* (y \cond 1, \bx \con w) \  \propto  \
	w(y,\bx)
	\alpha_0^*(y, \bx)
	\beta_1^*(\bx) f_1^*(y \cond 0,\bx) \ .
\end{align*}
The conditional counterfactual mean $\EXP \{ \potY{0}{1} \cond A=1, \bX=\bx \}$ is then represented as
\begin{align*}
	\mu^*(\bx \con w)
	=
	\frac{ 
		\int y w(y,\bx)
	\alpha_0^*(y, \bx) f_1^*(y \cond 0,\bx) \, dy
	}{
	\int  
	w(y,\bx)
	\alpha_0^*(y, \bx) f_1^*(y \cond 0,\bx) 
	\, dy 
	} \ .
\end{align*}

Let $w^{\max}(y,\bx)$ be the function that makes $\mu^*(\bx \con w)$ as large as possible, i.e.,
\begin{align*}
	w^{\max}(y,\bx) \in \argmax_{w \in [\Gamma^{-1},\Gamma]} \mu^*(\bx \con w) \ .
\end{align*}
Then, we find
\begin{align*}
	\mu^*(\bx \con w^{\max})
	& =
	\frac{ 
		\int y w^{\max}(y,\bx)
	\alpha_0^*(y, \bx)  f_1^*(y \cond 0,\bx) \, dy
	}{
	\int  
	w^{\max}(y,\bx)
	\alpha_0^*(y, \bx)  f_1^*(y \cond 0,\bx) 
	\, dy 
	}
	\\
	&
	\geq  
	\frac{ 
		\int y w(y,\bx)
	\alpha_0^*(y, \bx) f_1^*(y \cond 0,\bx) \, dy
	}{
	\int  
	w(y,\bx)
	\alpha_0^*(y, \bx) f_1^*(y \cond 0,\bx) 
	\, dy 
	} 
	=	
	\mu^*(\bx \con w)
	  \ , \quad \forall w \in [\Gamma^{-1},\Gamma ]
\end{align*}
Rearranging the term, we find
\begin{align}
	0
	& 
	=
	\int w^{\max}(y,\bx) \big\{ y - \mu^*(\bx \con w^{\max}) \big\} 
	\alpha_0^*(y, \bx) f_1^*(y \cond 0,\bx) \, dy 
	\nonumber
	\\
	& 
	\geq 
	\int w(y,\bx) 
	\underbrace{ \big\{ y - \mu^*(\bx \con w^{\max}) \big\} 
	\alpha_0^*(y, \bx) f_1^*(y \cond 0,\bx) 
	}_{=:\mathcal{D}(y,\bx)}
	\, dy 
	  \ , \quad \forall w \in [\Gamma^{-1},\Gamma ] \ .
	\label{eq-supp-sensitivity 2}
\end{align} 
To maximize the functional $ \int w(y,\bx) \mathcal{D}(y,\bx) \, dy$ in \eqref{eq-supp-sensitivity 2}, $w(y,\bx)$ must take the largest admissible value whenever $\mathcal{D}(y,\bx) > 0$ and the smallest admissible value whenever $\mathcal{D}(y,\bx)<0$. This yields
\begin{align*}
	w^{\max}(y,\bx)
	=
	\left\{
		\begin{array}{ll}
		\Gamma & \text{if } y \geq m^{\max}(\bx) \quad \text{and} \quad y \neq y_{R}
		\\
		\Gamma^{-1} & \text{if } y < m^{\max}(\bx) \quad \text{and} \quad y \neq y_{R}
		\\
		1 & \text{if } y =y_{R}
		\end{array}
	\right. 
\end{align*}
where $m^{\max}(\bx) = \mu^*(\bx \con w^{\max})$. The condition $w^{\max}(y_R,\bx) = 1$ encodes the margin constraint $\alpha_{t}^*(y_R,\bx)=1$. The threshold $m^{\max}(\bx)$ is characterized as the unique solution to $\mathcal{E}^{\max}(m^{\max}(\bx)) = 0$, where
\begin{align*}
	\mathcal{E}^{\max}(m) 
	&
	=
	\Gamma 
	\int_{m(\bx)}^{\infty}
	y \, \alpha_0^*(y,\bx)  \, f_1^*(y \cond 0,\bx) \, dy
	+
	\Gamma^{-1}
	\int_{-\infty}^{m(\bx)}
	y \, \alpha_0^*(y,\bx) \, f_1^*(y \cond 0,\bx) \, dy
	\\
	&
	\qquad \qquad
	-
	m(\bx)
	\bigg\{
		\Gamma 
	\int_{m(\bx)}^{\infty}
	\alpha_0^* (y,\bx) \, f_1^*(y \cond 0,\bx) \, dy
	+
	\Gamma^{-1}
	\int_{-\infty}^{m(\bx)}
	\alpha_0^* (y,\bx) \, f_1^*(y \cond 0,\bx) \, dy
	\bigg\}
	\\
	&
	=
	\Gamma 
	\int_{m(\bx)}^{\infty}
	\big\{ y - m(\bx) \big\} \, \alpha_0^*(y,\bx)  \, f_1^*(y \cond 0,\bx) \, dy
	+
	\Gamma^{-1}
	\int_{-\infty}^{m(\bx)}
	\big\{ y - m(\bx) \big\} \, \alpha_0^*(y,\bx) \, f_1^*(y \cond 0,\bx) \, dy 
	 \ .
\end{align*}
The derivative of  $\mathcal{E}^{\max}$ can be derived by the Leibniz integral rule:
\begin{align*}
 \frac{\partial \mathcal{E}^{\max}(m) }{\partial m}
 =
- 
	\bigg\{
		\Gamma 
	\int_{m(\bx)}^{\infty}
	y  \, f_1^*(y \cond 0,\bx) \, dy
	+
	\Gamma^{-1}
	\int_{-\infty}^{m(\bx)}
	y \, f_1^*(y \cond 0,\bx) \, dy
	\bigg\} < 0 \ .
\end{align*}
In addition, $\mathcal{E}^{\max}(\infty) = -\infty$ and $\mathcal{E}^{\max}(-\infty) = \infty$. 
Therefore, $m^{\max}(\bx)$ is the unique solution and can be efficiently computed via the bisection method.

Likewise, let $w^{\min}(y,\bx)$ be the function that minimizes $\mu^*(\bx \con w)$, i.e.,
\begin{align*}
	w^{\min}(y,\bx) \in \argmin_{w \in [\Gamma^{-1},\Gamma]} \mu^*(\bx \con w) \ .
\end{align*}
By an analogous argument, $w^{\min}(y,\bx)$ is characterized as
\begin{align*}
	w^{\min}(y,\bx)
	=
	\left\{
		\begin{array}{ll}
		\Gamma^{-1} & \text{if } y \geq m^{\min}(\bx) \quad \text{and} \quad y \neq y_{R}
		\\
		\Gamma & \text{if } y < m^{\min}(\bx) \quad \text{and} \quad y \neq y_{R}
		\\
		1 & \text{if } y =y_{R}
		\end{array}
	\right. 
\end{align*}
where $m^{\min}(\bx)$ is characterized as the solution to $\mathcal{E}^{\min}(m^{\min}(\bx)) =  0$ where 
\begin{align}
	\mathcal{E}^{\min}(m) 
	& = 
	\Gamma^{-1}
	\int_{m(\bx)}^{\infty}
	y \, \alpha_0^*(y,\bx)  \, f_1^*(y \cond 0,\bx) \, dy
	+
	\Gamma
	\int_{-\infty}^{m(\bx)}
	y \, \alpha_0^*(y,\bx) \, f_1^*(y \cond 0,\bx) \, dy
	\nonumber
	\\
	&
	\qquad \qquad
	-
	m(\bx)
	\bigg\{
		\Gamma^{-1}
	\int_{m(\bx)}^{\infty}
	y  \, f_1^*(y \cond 0,\bx) \, dy
	+
	\Gamma
	\int_{-\infty}^{m(\bx)}
	y \, f_1^*(y \cond 0,\bx) \, dy
	\bigg\}
	\nonumber
	\\
	&
	=
	\Gamma^{-1} 
	\int_{m(\bx)}^{\infty}
	\big\{ y - m(\bx) \big\} \, \alpha_0^*(y,\bx)  \, f_1^*(y \cond 0,\bx) \, dy
	+
	\Gamma
	\int_{-\infty}^{m(\bx)}
	\big\{ y - m(\bx) \big\} \, \alpha_0^*(y,\bx) \, f_1^*(y \cond 0,\bx) \, dy 
	 \ .
	 \label{eq-def of m min}
\end{align}

Second, for the binary outcome, the reference value has a positive probability; thus, it suffices to tilt the non-reference value. Without loss of generality, let $y_R=0$. Then, $w^{\max}$ and $w^{\min}$ are characterized as
\begin{align*}
	&
	w^{\max}(y,x) = 1 \ind(y=y_R) + \Gamma \ind(y \neq y_R)
	\ , 
	\\
	&
	w^{\min}(y,x) = 1 \ind(y=y_R) + \Gamma^{-1} \ind(y \neq y_R) \ .
\end{align*}

Given $w^{\max}$ and $w^{\min}$, the lower and upper bounds of the ATT under \eqref{eq-supp-sensitivity}, denoted by $\tau^{\text{LB}}$ and $\tau^{\text{UB}}$, are attained by setting $\alpha_1^*$ to $
	\alpha_1^{\text{LB}}(y,\bx)
	=
	w^{\max}(y,\bx) \alpha_0^*(y,\bx)$
	and $
	\alpha_1^{\text{UB}}(y,\bx)
	=
	w^{\min}(y,\bx) \alpha_0^*(y,\bx)$, respectively, i.e.,
	\begin{align*}
	\tau^{\dagger}
	=
	\frac{ \EXP \big[ A \{ \Yo - \mu^\dagger(\bX) \} \big] }
	{ \Pr(A=1) }
	\ ,\qquad 
\mu^{\dagger}(\bX)
=
\frac{ 
\EXP \big\{ \Yo \alpha_1^{\dagger} (\Yo, \bX) \cond A = 0 , \bX \big\}
}{
\EXP \big\{ \alpha_1^{\dagger} (\Yo, \bX) \cond A = 0 , \bX \big\}
}
\ , \qquad 
\dagger \in \{\text{LB},\text{UB} \} \ .
	\end{align*}
By construction, $\mu^{\text{LB}}(\bX)$ and $\mu^{\text{UB}}(\bX)$ are the tightest possible smallest and largest values for $\EXP \{ \pYo \cond A=1,\bX \}$. Therefore, the interval $[\tau^{\text{LB}},\tau^{\text{UB}}]$ constitutes the sharp bound for the ATT under the sensitivity model \eqref{eq-supp-sensitivity}.

The corresponding plug-in version of the EIFs in Theorem \ref{thm-EIF} are given by
\begin{align}
&
\InfFt^{\dagger}(\bO) = \frac{ A\Yo -  \uncInfFt_{0}^{\dagger}(\bO) - A \tau^{\dagger} }{\Pr(A=1)} 
\ , 
\qquad \dagger \in \{ \text{LB},\text{UB} \}
\label{eq-supp-EIF bound}
\end{align}
where
\begin{align*} 
&
\uncInfFt_{0}^{\dagger} (\bO)
=
\left[ 
	\begin{array}{l}
(1-A) 
\beta_1^{\dagger} (\bX) \alpha_1^{\dagger} (\Yo,\bX)
\big\{  \Yo - \mu^{\dagger} (\bX) \big\}
+
A \mu^{\dagger} (\bX) 
\\
+ (2A-1) R^{\dagger} (\Yz,A,\bX) \big\{  \Yz - \mu^{\dagger}(\bX) \big\} 	
	\end{array}
\right]
\ , 
\\
&
\beta_1^{\dagger}(\bx)
= 
\frac{ \Pr(A=1 \cond \bX) /  \Pr(A=0 \cond \bX)  }
{ 
\EXP \big\{ \alpha_1^{\dagger} (\Yo, \bX) \cond A = 0 , \bX \big\}
}
\ , 
\\
&
R^{\dagger} (y,a,\bx)
=
\beta_1^{\dagger}(\bx)
\alpha_1^{\dagger}(y,\bx)
\bigg\{
\frac{a}{\beta_{0}^*(\bx)
 \alpha_0^*(y,\bx)} + (1-a)
\bigg\}
\frac{ \potf{1}^* (y, 0 , \bx) }{ \potf{0}^* (y , 0, \bx) }  \ .
\end{align*}
Note that $\InfFt^{\dagger}(\bO)$ in \eqref{eq-supp-EIF bound} is indeed the EIF for $\tau^{\dagger}$, as formalized in the following theorem.
\begin{theorem}						\label{thm-EIF SB}
The efficient influence function for $\tau^{\dagger}$ for $\dagger \in \{ \text{LB},\text{UB}\}$ in model $\mathcal{M}_{\OREC}$ is $\InfFt^{\dagger}(\bO)$ in \eqref{eq-supp-EIF bound}.
\end{theorem}

Based on $\InfFt^{\dagger}(\bO)$, we construct estimators for $\tau^{\text{LB}}$ and $\tau^{\text{UB}}$ along with their corresponding standard errors, following the same approach as in Section \ref{sec:Conti-Improved}. We denote these estimators by $\widehat{\tau}^{\dagger}$ and $\widehat{\sigma}^{\dagger}$ for $\dagger \in \{\text{LB},\text{UB}\}$. Accordingly, the estimated bounds for the ATT and the associated $100(1-q)\%$ confidence interval are given by
\begin{align*}
& \big[ \widehat{\tau}^{\text{LB}} , \widehat{\tau}^{\text{UB}} \big]
\qquad \text{ and } \qquad 
\bigg[ \widehat{\tau}^{\text{LB}} - z_{1-q/2}
\frac{ 
\widehat{\sigma}^{\text{LB}} } { N^{1/2} } ,
\widehat{\tau}^{\text{UB}} + z_{1-q/2}
\frac{ 
\widehat{\sigma}^{\text{UB}} } { N^{1/2} }
\bigg]  \ .
\end{align*}

One may also apply the median adjustment described in Section \ref{sec:supp-MedAdjustment}. For $\dagger \in \{\text{LB}, \text{UB} \}$, define
\begin{align*}
&
\widehat{\tau}_{\text{median}}^{\dagger} 
=
\median_{s=1,\ldots,S} \widehat{\tau}_{s}^{\dagger} 
\ ,
\qquad 
\widehat{\sigma}_{\text{median}}^{\dagger,2} 
=
\median_{s=1,\ldots,S} \big\{ \sigma_s^{\dagger,2} + (\widehat{\tau}_s^{\dagger} - \widehat{\tau}_{\median}^{\dagger} )^2 \big\}  \ .
\end{align*}
The median-adjusted bounds and the corresponding confidence interval are then given by
\begin{align}
& \big[ \widehat{\tau}_{\text{median}}^{\text{LB}} , \widehat{\tau}_{\text{median}}^{\text{UB}} \big]
\qquad \text{ and } \qquad 
\bigg[ \widehat{\tau}_{\text{median}}^{\text{LB}} - z_{1-q/2}
\frac{ 
\widehat{\sigma}_{\text{median}}^{\text{LB}} } { N^{1/2} } ,
\widehat{\tau}_{\text{median}}^{\text{UB}} + z_{1-q/2}
\frac{ 
\widehat{\sigma}_{\text{median}}^{\text{UB}} } { N^{1/2} }
\bigg]  \ .
\label{eq-supp-sensitivity 4}
\end{align}

If an additional pre-treatment time period, say $t=-1$, is available, the sensitivity parameter $\Gamma$ can be empirically calibrated as follows. In particular, one may consider a time-lapse version of \eqref{eq-supp-sensitivity} by comparing the two pre-treatment periods $t=0$ and $t=-1$:
\begin{align*}
\frac{ {\alpha}_0^* (y,\bx) }
{ {\alpha}_{-1}^* (y,\bx) }
\in [\Gamma^{-1},\Gamma]
\ , \quad 
\forall (y,\bx) \in \SPzt \ .
\end{align*}
This relationship suggests that a plausible range for $\Gamma$ can be informed by empirical estimates. Specifically, consider the ratios of the estimated odds ratios:
\begin{align}
\widehat{\rho}_{i}
:= 
\frac{ \widehat{\alpha}_0 (Y_{0,i},X_i) }
{ \widehat{\alpha}_{-1} (Y_{0,i},X_i) }
\quad \text{ where } \quad A_i = 1 \ .
\label{eq-supp-sensitivity 3}
\end{align}
The odds ratios $\widehat{\alpha}_{0}$  and $\widehat{\alpha}_{-1}$ can be estimated using density ratio estimation methods (e.g., KLIEP; see Section \ref{sec:Conti-Improved}). 

The empirical distribution of $ \mathcal{R} := \{ \widehat{\rho}_i \cond A_i =1 \}$ then provides a data-driven guide for selecting a plausible range of $\Gamma$. For example, a reference value of $\Gamma$ (denoted by $\Gamma_{R}$) can be taken as the maximum deviation of $\mathcal{R}$ from 1, say 
\begin{align*}
	\Gamma_{R}
	= 
	\max \big\{ 
	\exp \big( | \log \min \mathcal{R} | \big)
	,
	\exp \big( | \log \max \mathcal{R} | \big)
	\big\}  
	\ ,
\end{align*}
However, this $\Gamma_{R}$ can be sensitive to estimation error, as $\widehat{\rho}_i$ may take extreme values. In particular, density ratio estimates can be unstable in regions of the outcome-covariate space where observations are sparse, which can in turn produce extreme values of $\widehat{\rho}_i$.

To mitigate this issue, we recommend taking $\Gamma_{R}$ as interior quantiles rather than the full range of $\mathcal{R}$. For instance, one may remove the most extreme 2.5\% on each side and take the 
\begin{align} \label{eq-supp-sensitivity 5}
	\Gamma_{R}
	= 
	\max \big\{ 
	\exp \big( | \log \mathcal{R}^{(0.025)} | \big)
	,
	\exp \big( | \log \mathcal{R}^{(0.975)} | \big)
	\big\}  
	\ ,
\end{align}
where $\mathcal{R}^{(q)}$ is the $q$th quantile of $\mathcal{R}$. Such trimming mitigates the influence of estimation error and yields a more stable and robust calibration of the sensitivity parameter.

A similar approach can be constructed for the \nHL{PT}-based DiD analyses. Consider the following relaxed assumption:
\begin{align}
\EXP \{ Y_1^{(0)} - Y_0^{(0)} \cond A=1,\bX=\bx \}
-
\EXP \{ Y_1^{(0)} - Y_0^{(0)} \cond A=0,\bX=\bx \}
\in [-\Gamma,\Gamma] \ , \quad \forall \bx \ ,
\label{eq-supp-sensitivity did}
\end{align}
where $\Gamma \geq 0 $ is a user-specified sensitivity parameter representing the maximum allowable departure from the \nHL{PT} assumption. Then, the ATT is no longer point-identified but is instead bounded as follows:
\begin{align*}
\tau^* 
&
= 
\EXP \{ Y_1^{(1)} - Y_1^{(0)} \cond A=1 \}
\\
&
= 
\EXP (Y_1 - Y_0 \cond A=1) -
\EXP \big[ \EXP \big\{ Y_1^{(0)} - Y_0^{(0)} \cond A=1,\bX \} \cond A=1 \big]
\\
&
\in 
\Bigg[ 
\begin{array}{l}
\EXP (Y_1 - Y_0 \cond A=1) - 
\EXP \big\{ \EXP ( Y_1 - Y_0 \cond A=0,\bX ) \cond A=1 \big\}
-
\Gamma
, 
\\
\qquad \qquad 
\EXP (Y_1 - Y_0 \cond A=1) - 
\EXP \big\{ \EXP ( Y_1 - Y_0 \cond A=0,\bX ) \cond A=1 \big\}
+
\Gamma
\end{array}
\Bigg] \ .
\end{align*}
Therefore, we may obtain the EIF-based estimator for these lower and upper bounds of the ATT. 

If an additional pre-treatment time period, say $t=-1$, is available, the sensitivity parameter $\Gamma$ can be empirically calibrated. Specifically, a reference value $\Gamma_{R}$ can be set as interior quantiles of $\mathcal{D} = \{ \widehat{\nu}_i \cond i \in \{1,\ldots,N\} \}$ represents the empirical distribution of the estimated differences $\widehat{\nu}_i = \widehat{\EXP}(Y_{0} - Y_{-1} \cond A=1,\bX_i ) - \widehat{\EXP}(Y_{0} - Y_{-1} \cond A=0,\bX_i )$. Consistent with the approach in \eqref{eq-supp-sensitivity 5}, we discard the extreme quantiles to ensure that the calibrated parameter is not driven by outliers or estimation error.

\subsection{Details of the Simulation} \label{sec:supp:simulation}

\subsubsection{Data-generating Process}

We provide details of the simulation data-generating process. First, we generated two observed covariates $\bX = (X_1,X_2)$ where each component is independent standard normal. We then generated the treatment indicator $A$ from $\text{Ber} \big( \text{expit}\{ 0.1(X_1+X_2) \big\} \big)$. 

For the continuous outcome setting, the potential outcomes were generated from the following models:
\begin{align*}
&
\potY{0}{0} \cond (A,\bX) \sim 
N \big(
3 + 0.01 (5+2X_1+2X_2) A + 0.1(X_1+X_2) , 4
\big) \ , \ 
&& \potY{1}{0} = \potY{0}{0} \ , \
\\
& 
\potY{a}{1} \cond (A,\bX) \sim
N \big(
3.5 + 0.5 a + 0.01 (5+2X_1+2X_2) A + 0.1(X_1+X_2) , 1
\big) \ , \ 
&& a \in \{0,1\} \ .
\end{align*}
Note that the potential outcomes are not conditionally independent of the treatment given covariates, indicating that the conditional ignorability condition is violated. Additionally, the \nHL{OREC} assumption is satisfied with the odds ratio function $\alpha_1^*(y,\bx) = \exp\big\{ 0.01 y (5+2x_1+2x_2) \big\}$, whereas the \nHL{PT} assumption is violated. The true ATT is 0.5. 

For the binary outcome setting, we generated the potential outcomes from the following models:
\begin{align*}
&
\potY{0}{0} \cond (A,\bX) \sim 
\text{Ber}
\big(
\text{expit} \big\{ -0.5 + (1-0.2X_1-0.2X_2) A + 0.1 X_1 + 0.1X_2 \big\}
\big)
\ , \ 
&& \potY{1}{0} = \potY{0}{0} \ , \
\\
& 
\potY{a}{1} \cond (A,\bX) \sim
\text{Ber}
\big(
\text{expit} \big\{ 0.5 + (1-0.2X_1-0.2X_2) A + 0.1 X_1 + 0.1X_2  \big\}
\big) \ , \ 
&& a \in \{0,1\} \ .
\end{align*}
Again, the potential outcomes are not conditionally independent of the treatment given covariates, and the \nHL{OREC} assumption is satisfied with an odds ratio function $\alpha_1^*(y,\bx) = \exp \big\{y(1-0.2x_1-0.2x_2) \big\}$, whereas the \nHL{PT} assumption is violated. The true ATT is 0 because $\potY{0}{1}$ and $\potY{1}{1}$ have identical distributions.

\subsubsection{Comparison of Computation Time}

We compare the computation times of $\widehat{\tau}_{\text{UDID}}$ and $\widehat{\tau}_{\text{DID}}$ across 1000 simulation repetitions. Table \ref{supp:Table-CompTime} presents the mean and median computation times. We find that $\widehat{\tau}_{\text{UDID}}$ requires a similar computational effort as $\widehat{\tau}_{\text{DID}}$ and is sometimes even faster, demonstrating that our estimator is fully competitive from a computational perspective.

\begin{table}[!htp]
\small
\renewcommand{\arraystretch}{1.1} \centering
\setlength{\tabcolsep}{5pt}
\begin{tabular}{|c|c|cc|cc|}
\hline
\multirow{2}{*}{$N$}  & \multirow{2}{*}{Metric} & \multicolumn{2}{c|}{Continuous $Y$}                                                 & \multicolumn{2}{c|}{Binary $Y$}                                                     \\ \cline{3-6} 
                      &                         & \multicolumn{1}{c|}{$\widehat{\tau}_{\text{UDID}}$} & $\widehat{\tau}_{\text{DID}}$ & \multicolumn{1}{c|}{$\widehat{\tau}_{\text{UDID}}$} & $\widehat{\tau}_{\text{DID}}$ \\ \hline
\multirow{2}{*}{500}  & Mean                    & \multicolumn{1}{c|}{\phantom{0}74.83}                          & 123.80                         & \multicolumn{1}{c|}{188.80}                          & 179.75                        \\ \cline{2-6} 
                      & Median                  & \multicolumn{1}{c|}{\phantom{0}77.11}                          & 127.66                        & \multicolumn{1}{c|}{195.99}                         & 186.84                        \\ \hline
\multirow{2}{*}{1000} & Mean                    & \multicolumn{1}{c|}{119.14}                         & 174.58                        & \multicolumn{1}{c|}{273.37}                         & 249.85                        \\ \cline{2-6} 
                      & Median                  & \multicolumn{1}{c|}{124.80}                          & 182.28                        & \multicolumn{1}{c|}{283.86}                         & 258.95                        \\ \hline
\multirow{2}{*}{2000} & Mean                    & \multicolumn{1}{c|}{229.16}                         & 280.48                        & \multicolumn{1}{c|}{435.83}                         & 398.59                        \\ \cline{2-6} 
                      & Median                  & \multicolumn{1}{c|}{239.60}                          & 292.33                        & \multicolumn{1}{c|}{458.14}                         & 417.61                        \\ \hline
\multirow{2}{*}{4000} & Mean                    & \multicolumn{1}{c|}{541.23}                         & 524.70                         & \multicolumn{1}{c|}{820.70}                          & 739.05                        \\ \cline{2-6} 
                      & Median                  & \multicolumn{1}{c|}{557.66}                         & 542.98                        & \multicolumn{1}{c|}{838.42}                         & 758.06                        \\ \hline
\end{tabular}
\caption{Comparison of mean and median computation times (in seconds) for $\widehat{\tau}{\text{UDID}}$ and $\widehat{\tau}{\text{DID}}$ across different sample sizes and outcome types.}
\label{supp:Table-CompTime}
\end{table}

\subsection{Details on the Data Analysis}							\label{sec:supp-data}

We provide details on the Zika virus outbreak data. First, Table \ref{supp:Table-Conti-Data1} shows the list of pre-treatment covariates.
\begin{table}[!htp]
\small
\renewcommand{\arraystretch}{1.1} \centering
\setlength{\tabcolsep}{5pt}
\begin{tabular}{|c|c|c|c|}
\hline
Type                        & Characteristics & Details                                             & Notation  \\ \hline
Binary                      & GDP             & $\ind(\text{GDP} \geq \text{Brazil's GDP in 2013})$ & $X_{\text{gdp}}$ \\ \hline
\multirow{3}{*}{Continuous} & Population      & $\log(\text{Population})$                           & $X_{\text{pop}}$ \\ \cline{2-4} 
                            & Density         & $\log(\text{Population Density})$                   & $X_{\text{den}}$ \\ \cline{2-4} 
                            & Female          & Proportion of Female                                & $X_{\text{pf}}$  \\ \hline
\end{tabular}
\caption{Details of Pre-treatment Covariates in the Zika Virus Outbreak Data.}
\label{supp:Table-Conti-Data1}
\end{table}

\begin{figure}[!htb]
\centering  
\includegraphics[width=0.9\textwidth]{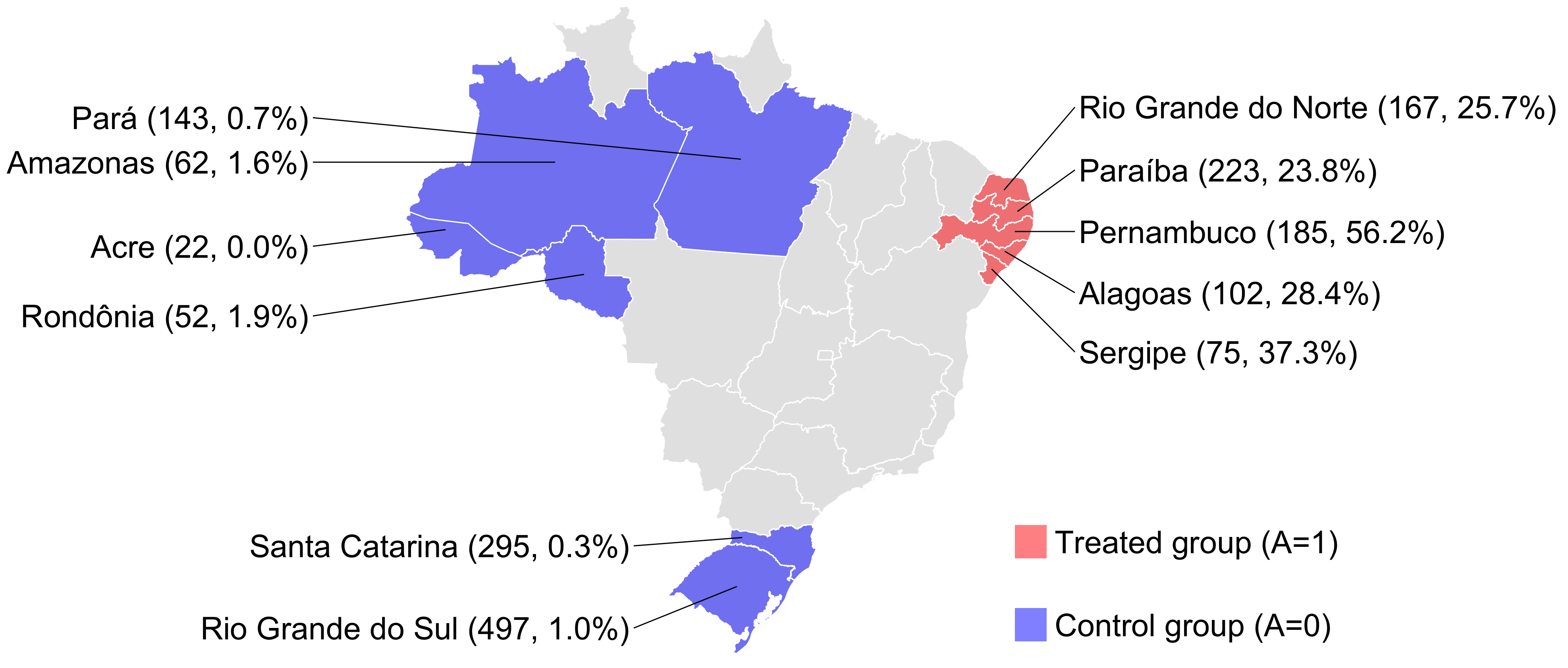}
\caption{Graphical summary of Zika virus data. Each state label indicates the state name, the number of municipalities, and the proportion of municipalities with confirmed Zika cases.}
\label{Fig-Brazil}
\end{figure}

Second, Figure \ref{Fig-Brazil} provides a graphical summary of the treated and control groups.

%Third, we provide details on how we use the pre-treatment covariates to obtain $\widehat{\tau}_{\text{DID}}$. We use the \texttt{att\_gt} function in the \texttt{did} R package to obtain the ATT estimator, and we can specify the formula of covariates in the \texttt{att\_gt} function through the \texttt{xformla} argument. We include the original covariate and the second-order interactions, i.e.,
%\begin{align*}
%\texttt{xformula}= \ \sim \ &
%(X_{\text{gdp}} + X_{\text{pop}} + 
%X_{\text{den}} + X_{\text{pf}})^2
%\end{align*}

\begin{figure}[!htb]
\centering
\includegraphics[width=\textwidth]{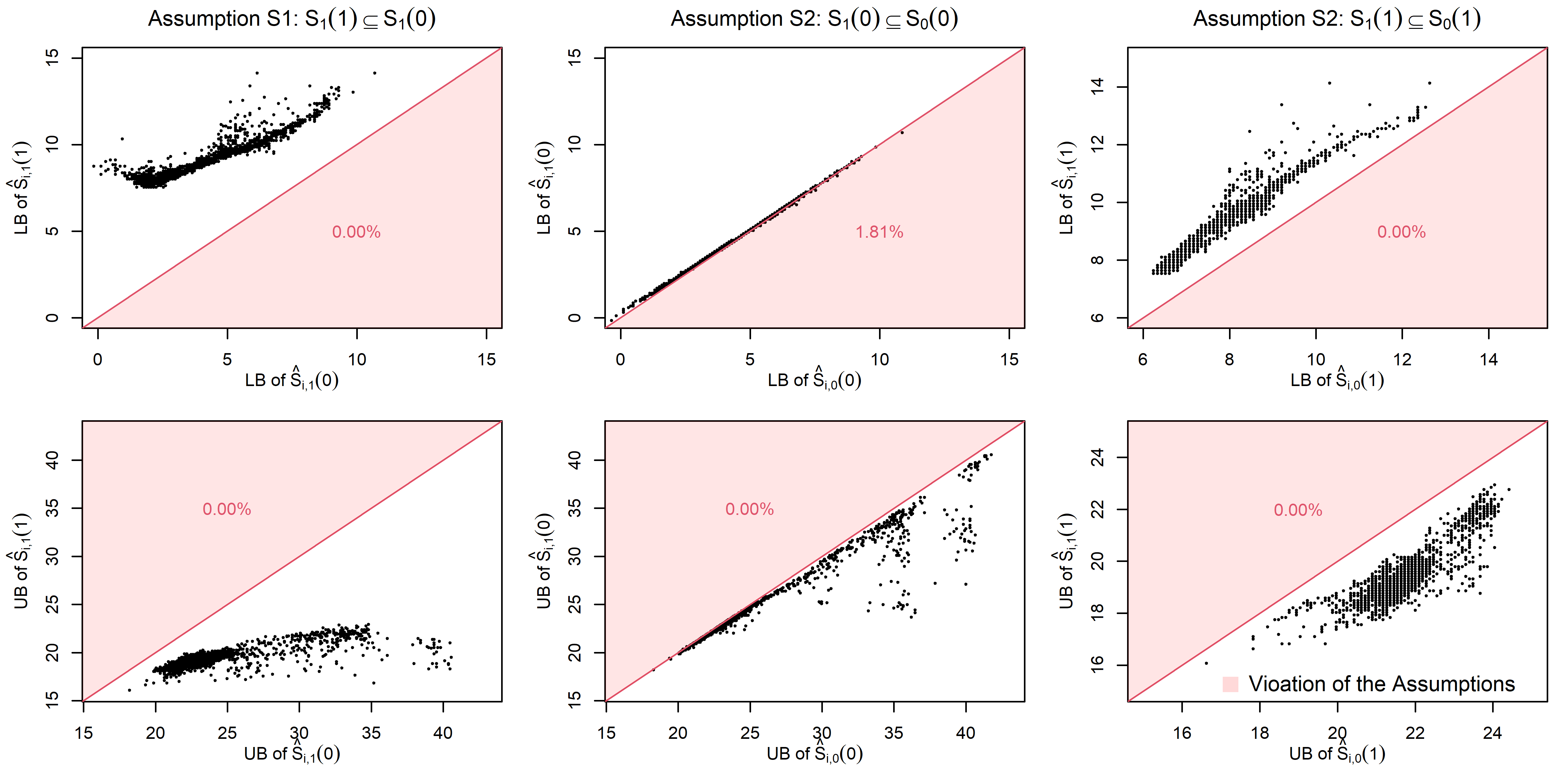}
\caption{Validation of Assumptions \ref{assumption:support-1} and \ref{assumption:support-2} in the Zika Virus Outbreak Data. The left column of plots evaluates Assumption \ref{assumption:support-1}, while the middle and right columns evaluate Assumption \ref{assumption:support-2}. Red-shaded areas indicate regions where the assumptions are violated. The numbers within these red areas represent the proportion of observations falling into those regions.}
\label{Fig-Conti-Overlap}
\end{figure}

Third, we provide visual evidence that the overlap assumption (Assumptions \ref{assumption:support-1} and \ref{assumption:support-2}) is plausible for the data. Using the observed data, we estimate the conditional densities of $\Yz \cond (A=0,\bX)$, $\Yz \cond (A=1,\bX)$, and $\pYo \cond (A=0,\bX)$, denoted by $\widehat{f}_0(y \cond 0,\bX)$, $\widehat{f}_0(y \cond 1,\bX)$, and $\widehat{f}_1(y \cond 0,\bX)$, respectively. Under the \nHL{OREC} assumption, we could obtain estimates of the conditional density of $\pYo \cond (A=1,\bX)$, denoted by $\widehat{f}_1(y \cond 1, \bX )$. For observed covariates $\bX_i$ $(i=1,\ldots,N)$, we define the support as the range of $y$ that makes the estimated density greater than $10^{-3}$, i.e., $\widehat{\mathcal{S}}_{i,t}(a) := \big\{ y \cond \widehat{f}_t(y \cond a,\bX_i) \geq 10^{-3} \big\}$ for $t=0,1$ and $a=0,1$. Figure \ref{Fig-Conti-Overlap} provides an empirical assessment of Assumptions \ref{assumption:support-1} and \ref{assumption:support-2}, both of which appear to be reasonably well satisfied.

Fourth, we provide the details on the assessment of the plausibility of the \nHL{UDiD model}. The procedure is largely similar to Appendix A.10 of \citet{UDID2024_Epi}, but we include the details here for completeness. Specifically,  the third condition of the \nHL{UDiD model}---condition \eqref{eq-UDiD3}---requires that the conditional distribution $U_t|(A=0, Y_t)$ remains stable over time. To assess this, one can examine whether the conditional distribution of measured covariates $X | (A=0, Y_t)$ is stable over time. Although these empirical checks are not formal tests of condition \eqref{eq-UDiD3}, the condition is more plausible if it holds for observed covariates. 

We have considered versions of these simple empirical checks in the context of the Zika virus application as follows. We denote the four covariates in Table \ref{supp:Table-Conti-Data1} by $X_{\text{gdp}}, X_{\text{pop}}, X_{\text{den}}, X_{\text{pf}}$, respectively. Likewise, we denote the birth rate
at time $t \in \{0,1\}$ of a municipality by $Y_t$. Recall that we only use untreated municipalities because it suffices to check the condition only among
control units.  

\begin{figure}[!htp]
    \centering
    \includegraphics[width=0.7\linewidth]{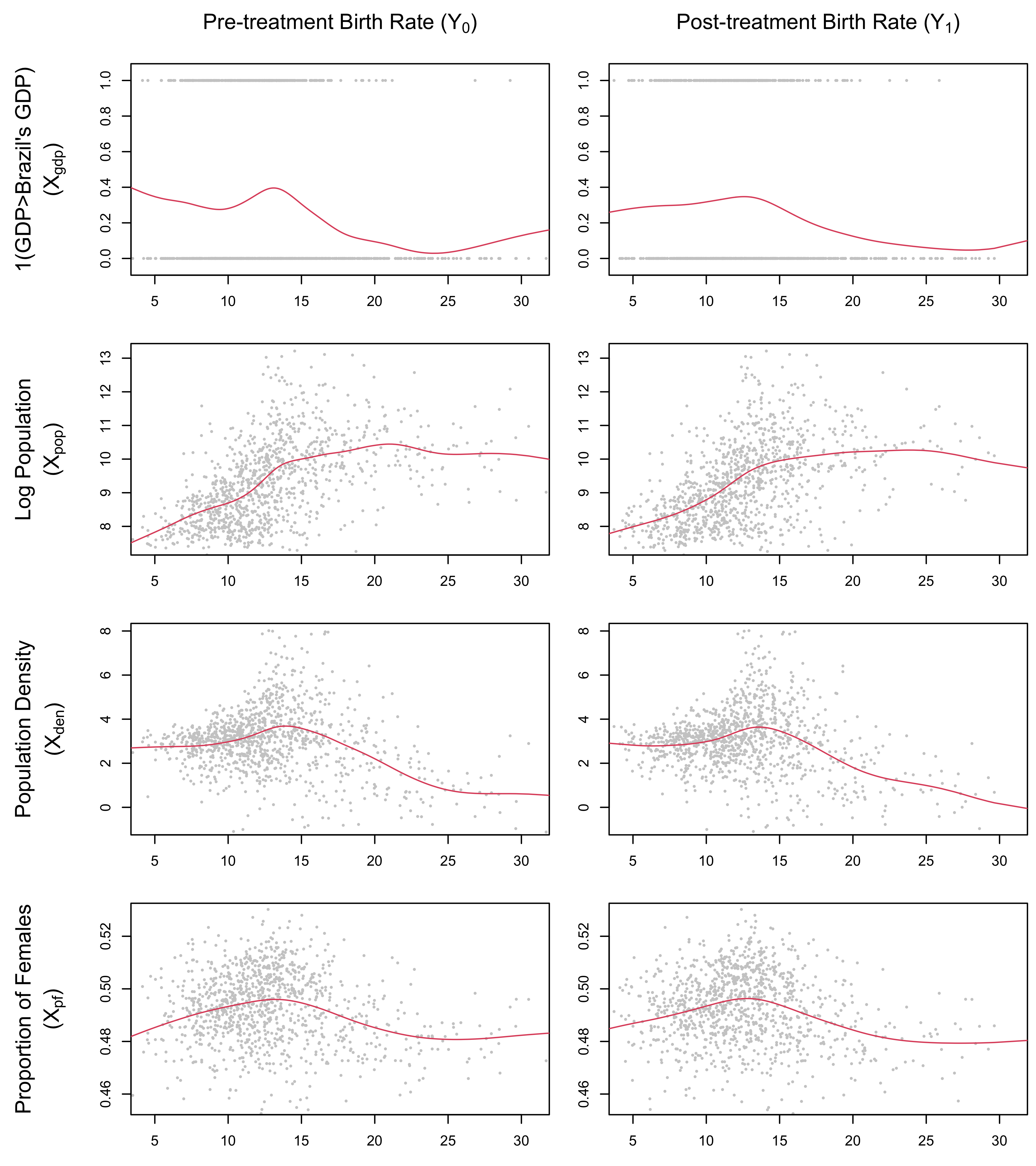}
    \caption{Graphical Summary of the Relationship between $Y_t$ and $X$. Each row corresponds to one of the four covariates in Table \ref{supp:Table-Conti-Data1}. The left and right columns display the pre- and post-exposure outcomes (birth rate), respectively. The $x$- and $y$-axes represent the outcomes and covariates, respectively. Plots are restricted to the middle 99\% of observed values. The red line depicts a cubic smoothing spline fit from regressing $X$ on $Y_t$.}
    \label{fig:UDID Check}
\end{figure}

First, we visually assess whether the relationship between $X$ and $Y_t$ does not dramatically change over time. Figure \ref{fig:UDID Check} graphically summarizes the empirical joint distribution of $X$ and $Y_t$. Upon visual examination, we observe that the associations between covariates and outcomes remain relatively stable across time periods. 

Next, we conduct statistical tests based on a parametric regression model. Specifically, we treat $X$ as the response variable and include the intercept, $Y_t$, $Y_t^2$, and $Y_t^3$ as explanatory variables, i.e.,
\begin{align*}
    \EXP (
    X_{j} 
    \cond Y_{t} )
    =
    \beta_{jt0} + \beta_{jt1} Y_{t} + \beta_{jt2} Y_{t}^2 + \beta_{jt3} Y_{t}^3 \ , 
    \quad 
    j \in \{ \text{gdp, pop, den, pf} \} \ , \quad t \in \{0,1\} \ .
\end{align*}
The quadratic and cubic terms are included to account for nonlinear relationships shown in Figure \ref{fig:UDID Check}. We then conduct the following statistical tests for $j \in \{ \text{gdp, pop, den, pf} \}$:
\begin{align}
    &
    H_{0,jk}: \beta_{j0k} = \beta_{j1k} 
    &&
    \text{versus}
    &&
    H_{a,jk}: \text{Not }H_{0,jk} 
    \ , 
    &&
    k \in \{0,1,2,3 \} \ , 
    \nonumber
    \\
    &
    H_{0,j4}: 
    \bigcap_{k=0}^{3} H_{0,jk}
    &&
    \text{versus}
    &&
    H_{a,j4}: \text{Not }H_{0,j4}    \ .
    \label{eq-Hypothesis}
\end{align}
At 5\% significance level, the null hypothesis $H_{0,jk}$ is rejected if the corresponding Wald statistic is greater than $\chi_{1,0.95}^{2} = 3.84$, and the null hypothesis $H_{0,j4}$ is rejected if the corresponding Wald statistic is greater than $\chi_{4,0.95}^{2} = 9.49$, where $\chi_{p,\alpha}^{2}$ denotes the $\alpha$th percentile of the chi-squared distribution with $p$ degrees of freedom. Table \ref{table:wald} reports the Wald statistics for the null hypotheses in \eqref{eq-Hypothesis}. For all four covariates, we find that there is no statistical evidence that the regression model varies over time at 5\% level. These empirical checks suggest that there is no significant evidence against the \nHL{UDiD model}.

\begin{table}[!htp]
\small
\renewcommand{\arraystretch}{1.1} \centering
\setlength{\tabcolsep}{5pt}
\begin{tabular}{|cc|cccc|}
\hline
\multicolumn{2}{|c|}{\multirow{2}{*}{Wald statistic for $H_{0,jk}$ in \eqref{eq-Hypothesis}}} & \multicolumn{4}{c|}{$j$}                                                                     \\ \cline{3-6} 
\multicolumn{2}{|c|}{}                                                                                              & \multicolumn{1}{c|}{gdp}   & \multicolumn{1}{c|}{pop}   & \multicolumn{1}{c|}{den}   & pf    \\ \hline
\multicolumn{1}{|c|}{\multirow{5}{*}{$k$}}                              & 0 (intercept)                             & \multicolumn{1}{c|}{0.069} & \multicolumn{1}{c|}{0.114} & \multicolumn{1}{c|}{2.327} & 0.647 \\ \cline{2-6} 
\multicolumn{1}{|c|}{}                                                  & 1 ($Y_t$)                                 & \multicolumn{1}{c|}{0.097} & \multicolumn{1}{c|}{0.129} & \multicolumn{1}{c|}{1.165} & 0.335 \\ \cline{2-6} 
\multicolumn{1}{|c|}{}                                                  & 2 ($Y_t^2$)                               & \multicolumn{1}{c|}{0.153} & \multicolumn{1}{c|}{0.579} & \multicolumn{1}{c|}{0.405} & 0.164 \\ \cline{2-6} 
\multicolumn{1}{|c|}{}                                                  & 3 ($Y_t^3$)                               & \multicolumn{1}{c|}{0.153} & \multicolumn{1}{c|}{0.579} & \multicolumn{1}{c|}{0.405} & 0.164 \\ \cline{2-6} 
\multicolumn{1}{|c|}{}                                                  & 4 (intercept, $Y_t$, $Y_t^2$, $Y_t^3$)                                    & \multicolumn{1}{c|}{0.063} & \multicolumn{1}{c|}{0.247} & \multicolumn{1}{c|}{2.606} & 0.722 \\ \hline
\end{tabular}
\caption{Wald Statistics of the Null Hypotheses in \eqref{eq-Hypothesis}. Rows correspond to the null hypotheses for the intercept, linear ($Y_t$), quadratic ($Y_t^2$), and cubic ($Y_t^3$) terms. Columns correspond to the null hypotheses for each of the four covariates in Table \ref{supp:Table-Conti-Data1}.}
\label{table:wald}
\end{table}

Lastly, we report sensitivity analysis results in order to assess the robustness of the conclusion in the main paper. Details of the sensitivity analysis can be found in Section \ref{sec:supp-SA}. We considered the sensitivity model \eqref{eq-supp-sensitivity} with $\Gamma \in [0,\exp(0.5)]$. For each $\Gamma$, we consider the median-adjusted bound estimates for the ATT and the corresponding 95\% confidence interval defined in \eqref{eq-supp-sensitivity 4} from 100 cross-fitting estimates. 

\begin{figure}[!htb]
 \centering
 \includegraphics[width=1\textwidth]{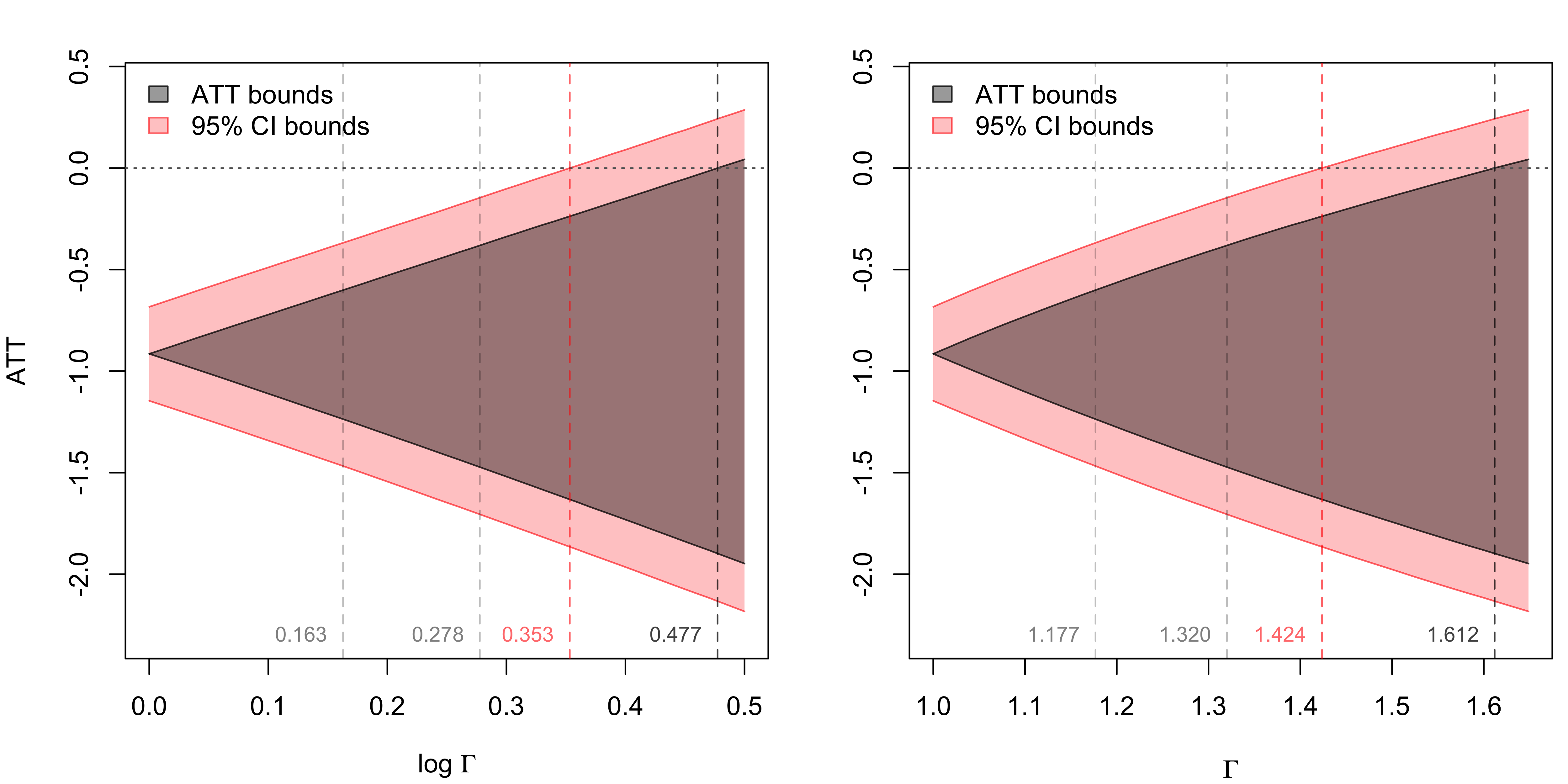}
 \caption{Graphical summary of the sensitivity analysis results for the \protect\nHL{OREC}-based UDiD estimator. The left and right panels use $\log \Gamma$ and $\Gamma$ as the x-axis scales, respectively. For each value of $\Gamma$, the black region denotes the estimated ATT bounds, while the red region shows the corresponding 95\% confidence intervals. The black and red dashed vertical lines, together with the labeled x-axis values, indicate the breakdown points at which the null hypothesis $H_0: \tau^* = 0$ is no longer rejected and the ATT bound includes zero. The gray dashed vertical lines and its labeled x-axis values indicate the reference sensitivity parameter values $\Gamma_{R,0.95}$ and  $\Gamma_{R,0.98}$, respectively.}
 \label{fig-sensitivity} 
 \end{figure}

Figure \ref{fig-sensitivity} visually reports the sensitivity analysis result. First, when $\Gamma$ is greater than 1.424, (equivalently, $\log \Gamma \geq 0.353$), the null hypothesis $H_0: \tau^* = 0$ is no longer significant at 5\% level, as the upper bound of the 95\% confidence interval in \eqref{eq-supp-sensitivity 4} contains zero. In words, even though the \nHL{OREC} assumption is violated, but as long as it is within the range of $\alpha_1^*/ \alpha_0^* \in [1/1.424,1.424]$, one can still conclude that the Zika virus outbreak resulted in a decrease in birth rate in the five northeastern states of Brazil at 5\% significance level. Second, when $\Gamma$ is greater than 1.612, (equivalently, $\log \Gamma \geq 0.477$), the ATT bound \eqref{eq-supp-sensitivity 4} contains zero. In words, even though the \nHL{OREC} assumption is violated, but as long as it is within the range of $\alpha_1^* / \alpha_0^* \in [1/1.612, 1.612]$, one can still conjecture that the Zika virus outbreak resulted in a decrease in birth rate in the five northeastern states of Brazil, as the resulting ATT bound remains negative.

To better interpret the sensitivity parameter values, we follow a data-driven reference value $\Gamma_{R}$ outlined in Section \ref{sec:supp-SA}. Using the $t=2013$ and $t=2014$ as two pre-treatment time periods, we calibrate the reference sensitivity parameter defined in \eqref{eq-supp-sensitivity 5} using the 95\% and 98\% interior quantiles, respectively. These values are
\begin{align*}
	&
	\Gamma_{R,0.95}
	= 
	\max \big\{ 
	\exp \big( | \log \mathcal{R}^{(0.025)} | \big)
	,
	\exp \big( | \log \mathcal{R}^{(0.975)} | \big)
	\big\}  
	=
	\max \big\{ 
		\exp(0.115), \exp(0.163)
	\big\}
	=
	1.177  \ ,
	\\
	&
	\Gamma_{R,0.98}
	= 
	\max \big\{ 
	\exp \big( | \log \mathcal{R}^{(0.01)} | \big)
	,
	\exp \big( | \log \mathcal{R}^{(0.99)} | \big)
	\big\}  
	=
	\max \big\{ 
		\exp(0.145), \exp(0.278)
	\big\}
	=
	1.320  \ .
\end{align*}
These reference values help contextualize the breakdown values $1.424$ and $1.612$. Taking the first reference value $\Gamma_{R,0.95} = 1.177$ and the first breakdown value $1.424$ as an example, the plausible range of $\alpha_{0}^*/\alpha_{-1}^*$ is $[0.850,1.177]$, whereas in order to nullify the causal conclusion in the post-treatment time period, we need $\alpha_{1}^*/\alpha_0^* > 1.424$ or $\alpha_{1}^*/\alpha_0^* < 1/1.424$. Therefore, in terms of log-scale, $\log(1.424) = 0.353$ is substantially larger than the 95\% pre-treatment fluctuation of $\log(1.177) = 0.163$. This indicates that for a structural shift to nullify the estimated causal effect, its magnitude would need to be more than twice as large ($0.353 / 0.163 = 2.166$) as the natural variations observed during the pre-treatment period. 

A similar conclusion holds when evaluating with the second breakdown value $1.612$. The log-scale breakdown value of $\log(1.612) = 0.477$ greatly exceeds the calibrated natural fluctuation of $\log(1.177) = 0.163$. Therefore, for a structural shift to make the estimated causal effect positive, its magnitude would need to be nearly three times as large ($0.477 / 0.163 = 2.926$) as the historical variations. The more conservative 98\% reference value ($\Gamma_{R,0.98} = 1.320$) yields similarly robust interpretations.

In short, the magnitude of hidden structural change in the post-treatment periods required to overturn the causal conclusion established under \nHL{OREC} would have to be substantially larger than the maximum imbalance observed during the pre-treatment periods. Therefore, the causal findings in the Zika virus application appear robust to plausible violations of the \nHL{OREC} assumption.

\begin{figure}[!htb]
 \centering
 \includegraphics[width=0.5\textwidth]{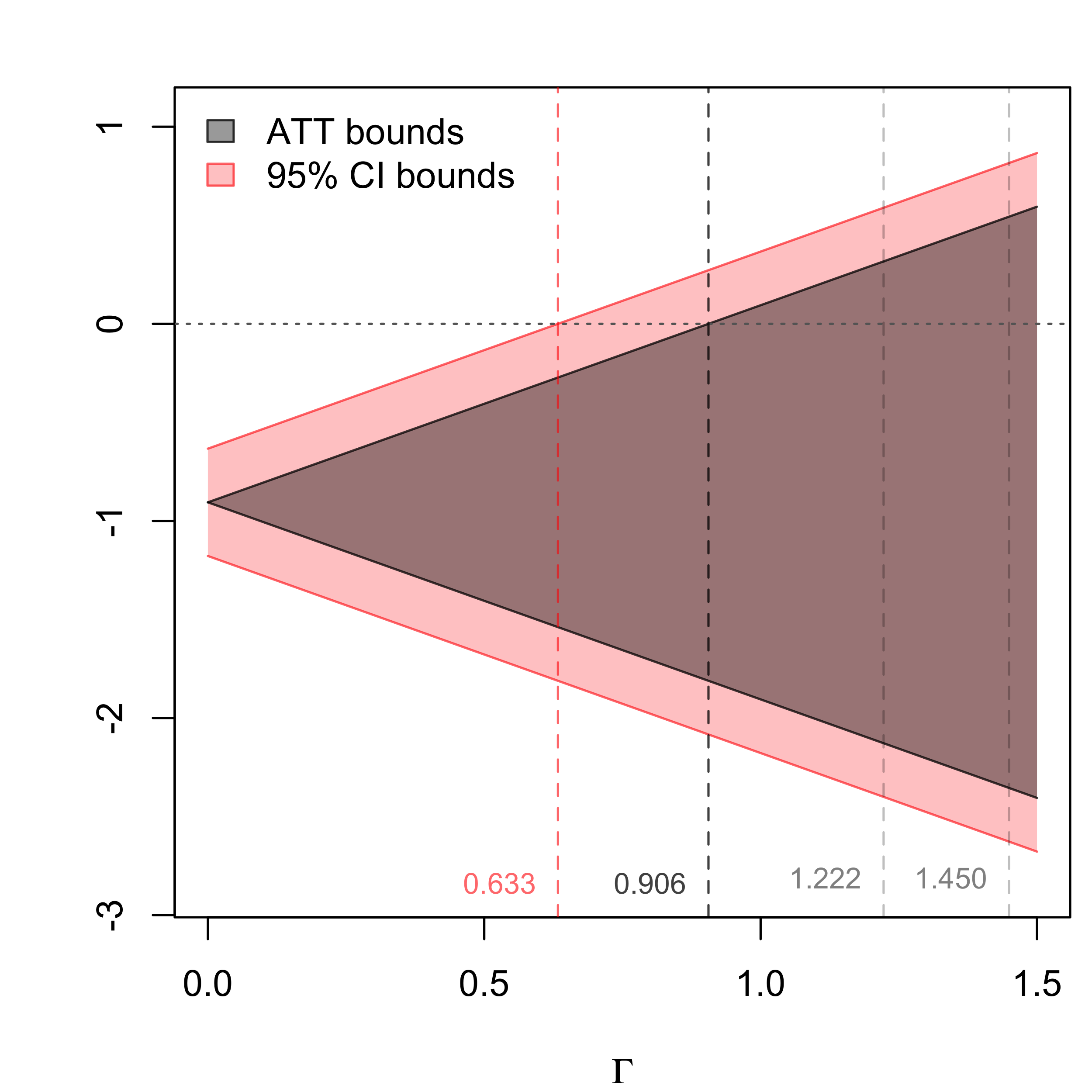}
 \caption{Graphical summary of the sensitivity analysis results for the \protect\nHL{PT}-based DiD estimator. For each value of $\Gamma$, the black region denotes the estimated ATT bounds, while the red region shows the corresponding 95\% confidence intervals. The black and red dashed vertical lines, together with the labeled x-axis values, indicate the breakdown points at which the null hypothesis $H_0: \tau^* = 0$ is no longer rejected and the ATT bound includes zero. The gray dashed vertical lines and its labeled x-axis values indicate the reference sensitivity parameter values $\Gamma_{R,0.95}$ and  $\Gamma_{R,0.98}$, respectively.}
 \label{fig-sensitivity-did} 
 \end{figure}

Similarly, we also report sensitivity analysis results for the \nHL{PT}-based DiD estimator under the sensitivity model \eqref{eq-supp-sensitivity did}. Figure \ref{fig-sensitivity-did} visually reports the sensitivity analysis result. First, when $\Gamma$ is greater than 0.633, the null hypothesis $H_0: \tau^* = 0$ is no longer significant at 5\% level, as the upper bound of the 95\% confidence interval in \eqref{eq-supp-sensitivity 4} contains zero. Second, when $\Gamma$ is greater than 0.906, the ATT bound \eqref{eq-supp-sensitivity 4} contains zero. To better interpret the sensitivity parameter values, we follow a data-driven reference value $\Gamma_{R}$. Following the description in Section \ref{sec:supp-SA}, we obtain $\Gamma_{R,0.95} = 1.222$ and $\Gamma_{R,0.98} = 1.450$. Notably, the breakdown values identified above are smaller than these empirically grounded benchmarks. This comparison suggests that a structural shift capable of overturning our causal conclusions would only need to be of a similar magnitude to the historical variations observed in the pre-treatment periods. Consequently, unlike the \nHL{OREC}-based UDiD approach, the causal findings in this Zika virus application appear sensitive to plausible violations of the \nHL{PT} assumption, as the magnitude of hidden structural change required to invalidate the results is within the range of observed historical imbalance.

\subsection{Additional Data Analysis with a Binary Outcome: Pennsylvania Traffic Data} \label{sec:Application:BinomY}

We analyze the Pennsylvania traffic data that is studied in \citet{LiLi2019} and \citet{Ding2019}. 
The data consist of 1986 traffic sites in Pennsylvania of which crash histories and site-specific characteristics are measured in 2008 and 2012. We define the years 2008 and 2012 as time $0$ and $1$, respectively. From 2009 to 2011, centerline and shoulder rumble strips were installed in 331 traffic sites, which we consider as the treated group,  whereas the other 1655 traffic sites did not receive rumble strips before 2012, which we consider the control group. Table \ref{supp:Table-Binary-Data1} shows the contingency tables across times. 
\begin{table}[!htp]
\small
\renewcommand{\arraystretch}{1.1} \centering
\setlength{\tabcolsep}{5pt}
\begin{tabular}{|cc|cc|cc|}
\hline
\multicolumn{2}{|c|}{\multirow{2}{*}{}}                  & \multicolumn{2}{c|}{Outcome at $t=0$} & \multicolumn{2}{c|}{Outcome at $t=1$} \\ \cline{3-6} 
\multicolumn{2}{|c|}{}                                   & \multicolumn{1}{c|}{$\Yz=0$} & $\Yz=1$ & \multicolumn{1}{c|}{$\Yo=0$} & $\Yo=1$ \\ \hline
\multicolumn{1}{|c|}{\multirow{2}{*}{Treatment}} & $A=1$ & \multicolumn{1}{c|}{1101}        &  554 & \multicolumn{1}{c|}{1106}        &   549  \\ \cline{2-6} 
\multicolumn{1}{|c|}{}                           & $A=0$ & \multicolumn{1}{c|}{232}        &   99   & \multicolumn{1}{c|}{241}        &  90  \\ \hline
\end{tabular}
\caption{Contingency Table of $(A,Y)$ Across Pre- and Post-treatment Periods in the Pennsylvania Traffic Data}
\label{supp:Table-Binary-Data1}
\end{table}

Using the data, we study the effect of installing shoulder rumble strips on crash incidence at a traffic site. Specifically, we define the outcome as the indicator of whether there has been at least one crash at a site. As pre-treatment covariates, we consider the following characteristics: speed limit, segment length in miles, pavement width, average shoulder width, number of driveways, the existence of intersections, the existence of curves, the average degree of curvature, and the average annual daily traffic volume. Table \ref{supp:Table-Binary-Data2} shows the list of pre-treatment covariates, and how those are used in the analysis.
\begin{table}[!htp]
\small
\renewcommand{\arraystretch}{1.1} \centering
\setlength{\tabcolsep}{5pt}
\begin{tabular}{|c|c|c|c|}
\hline
Type                        & Characteristics        & Details        & Notation                                        \\ \hline
\multirow{5}{*}{Binary}     & Speed limit            & $\ind(\text{Speed sign}\geq 45 \text{ miles})$  & $X_{\text{sl}}$       \\ \cline{2-4} 
                & \multirow{2}{*}{Width} & $\ind(\text{Widths}\in (20,23] \text{ feet})$    & $X_{\text{wd.1}}$     \\ \cline{3-4} 
                &                        & $\ind(\text{Widths} > 23 \text{ feet})$        & $X_{\text{wd.2}}$           \\ \cline{2-4} 
                & Intersection           & $\ind(\text{Number of intersections} \geq 1)$     & $X_{\text{int}}$          \\ \cline{2-4} 
                & Curve                  & $\ind(\text{Number of curves} \geq 1)$       & $X_{\text{cv}}$                      \\ \hline
\multirow{5}{*}{Continuous} & Length                 & Roadway segment length in mile     & $X_{\text{len}}$                      \\ \cline{2-4} 
                & Shoulder width         & $\log(1+\text{Average shoulder width in feet})$        & $X_{\text{sw}}$                          \\ \cline{2-4} 
                & Number of driveways    & $\log(1+\text{Number of driveways})$       & $X_{\text{nd}}$                           \\ \cline{2-4} 
                & Degree of curvature    & $\log(1+\text{Average degree of curvature})$        & $X_{\text{dc}}$                             \\ \cline{2-4} 
                & Traffic volume         & $\log(1+\text{Annual average daily traffic volume in 2008})$      & $X_{\text{tv}}$                       \\ \hline
\end{tabular}
\caption{Details of Pre-treatment Covariates in the Pennsylvania Traffic Data}
\label{supp:Table-Binary-Data2}
\end{table}

Since the outcome is binary, the overlap assumption (Assumptions \ref{assumption:support-1} and \ref{assumption:support-2}) is plausible, but we empirically assess these assumptions. Using the observed data, we estimate the conditional probabilities $\Pr(\Yz=1 \cond A=0,\bX)$, $\Pr(\Yz=1 \cond A=1,\bX)$, and $\Pr(\pYo=1 \cond A=0,\bX)$ from machine learning methods, denoted by $\widehat{f}_0(1 \cond 0,\bX)$, $\widehat{f}_0 (1 \cond 1,\bX)$, and $\widehat{f}_{1}(1 \cond 0, \bX)$. Under the \nHL{OREC} assumption, we could obtain estimates of the conditional probability $\Pr(\pYo=1 \cond A=1, \bX)$, denoted by $\widehat{f}_1(1 \cond 1,\bX)$. We study the range of the estimated conditional probabilities across observed covariates $\bX_i$ $(i=1,\ldots,N)$. Table \ref{supp:Table-Binary-Overlap} shows the summary statistics of these conditional probabilities. We find that Assumptions \ref{assumption:support-1} and \ref{assumption:support-2} appear plausible.

\begin{table}[!htp]
\renewcommand{\arraystretch}{1.1} \centering
\small
\setlength{\tabcolsep}{5pt}
\begin{tabular}{|c|c|c|c|c|}
\hline
& $\widehat{f}_0(1 \cond 0,\bX_i)$ & $\widehat{f}_0(1 \cond 1,\bX_i)$ & $\widehat{f}_1(1 \cond 0,\bX_i)$ & $\widehat{f}_0(1 \cond 1,\bX_i)$ (under \nHL{OREC}) \\ \hline
Minimum & 0.114                            & 0.103                            & 0.075                            & 0.076                                                                \\ \hline
$Q_1$   & 0.236                            & 0.241                            & 0.220                            & 0.224                                                                \\ \hline
$Q_2$   & 0.307                            & 0.310                            & 0.299                            & 0.303                                                                \\ \hline
Mean    & 0.327                            & 0.331                            & 0.326                            & 0.330                                                                \\ \hline
$Q_3$   & 0.400                            & 0.403                            & 0.416                            & 0.417                                                                \\ \hline
Maximum & 0.715                            & 0.721                            & 0.774                            & 0.779                                                                \\ \hline
\end{tabular}
\caption{Validation of the Overlap Assumption in the Pennsylvania Traffic Data. The summary measures are obtained from the observed covariates $\bX_i$ for $i=1,\ldots,N$.}
\label{supp:Table-Binary-Overlap}
\end{table}

Similar to the previous application, we compare the two estimators  $\widehat{\tau}_{\text{UDID}}$ and $\widehat{\tau}_{\text{DID}}$. For both estimators, we use machine learning methods in Section \ref{sec:supp-BinaryEIF} and the median adjustment in Section \ref{sec:supp-MedAdjustment} by repeating cross-fitting 100 times.

Table \ref{Table-Binary-Data} summarizes the results. The first two rows show the estimates of the ATT. We find that the two estimates are similar to each other. We find that $\widehat{\tau}_{\text{UDID}}$ has a slightly larger standard error prior to rounding. The ATT estimates from both methods are not significant at 5\% level, agreeing with the inconclusive findings in \citet{LiLi2019}.

\begin{table}[!htp]
\renewcommand{\arraystretch}{1.1} \centering
\small
\setlength{\tabcolsep}{4pt} 
\begin{tabular}{|c|c|c|}
\hline
Estimator & $\widehat{\tau}_{\text{UDID}}$ & $\widehat{\tau}_{\text{DID}}$ \\ \hline
Estimate  & -1.62                    & -2.60                  \\ \hline
ASE       & 4.24                     & 4.24                   \\ \hline
95\% CI     & (-9.93,6.69)            & (-10.91,5.70)           \\ \hline
\end{tabular}
\caption{Summary of the Analysis of the Pennsylvania Traffic Data. The reported standard errors and 95\% confidence intervals of $\widehat{\tau}_{\text{UDID}}$ are obtained from the consistent variance estimator in Theorem \ref{thm-AsympNormal}.}
\label{Table-Binary-Data}
\end{table}

\subsection{Extension: General Estimands} \label{sec:General}

As an extension, we consider general causal estimands that include both the ATT discussed above as well as nonlinear causal estimands such as quantile causal effects on the treated. To formalize the framework, let $\btheta^*$ denote the estimand of interest, defined as the solution to a counterfactual population moment equation $\EXP \big\{ \MM(\pYo,\bX \con \btheta) \cond A=1 \big\}=0$. Two concrete examples are:
\begin{itemize}
\item[(i)] (\textit{Example 1: Counterfactual Mean}) In this case, $\MM(\pYo,\bX \con \btheta) = \pYo - \theta$. The solution is $\btheta^* = \tau_0^*$, the counterfactual mean considered in the previous sections.
\item[(ii)] (\textit{Example 2: Counterfactual Quantile}) Suppose $\pYo$ is continuous. In this case, $\MM(\pYo,\bX \con \btheta) = \ind \big\{ \pYo \leq \theta \big\} - q$ for a user-specified value for $q \in (0,1)$ of interest. The corresponding solution defines the $q$th quantile for the treatment-free counterfactual distribution in the treated, $\pYo \cond A=1$, i.e., $\btheta^* = \tau_{0,q}^*$ satisfies $\Pr \big\{ \pYo \leq \tau_{0,q}^* \cond A=1 \big\} = q$. 
\end{itemize}
The following theorem offers a generalization of Theorem \ref{thm-EIF-General}, and provides the EIF for $\btheta^*$; %Theorem \ref{thm-EIF-General} states the result.
\begin{theorem}                                      \label{thm-EIF-General}
Suppose Assumptions \ref{assumption:support}-\ref{assumption:OREC} and regularity conditions in Section \ref{sec:supp-Regularity} hold. Then, the efficient influence function for $\btheta^*$ in $\mathcal{M}_\OREC$ is
$\InfFt^*(\bO \con \btheta^*)   =    - \big\{ V_\EFF (\btheta^*) \big\}^{-1} \MM_\EFF^* (\bO \con \btheta^*)$ where
\begin{align}               \label{eq-EffMoment}
&
\MM_\EFF^* (\bO \con \btheta)
=
\left[
\begin{array}{l}
(1-A) 
\beta_1^*(\bX) \alpha_1^*(\Yo,\bX)
\big\{  \MM(\Yo,\bX \con \btheta) - \mu_\MM^*(\bX \con \btheta) \big\}
+
A
\mu_\MM^*(\bX \con \btheta)
\\
+ (2A-1) R^*(\Yz,A,\bX) \big\{  \MM(\Yz,\bX \con \btheta) - \mu_\MM^*(\bX \con \btheta) \big\}
\end{array}	
\right], 
\\
&
\mu_\Omega^*(\bX \con \btheta) = \frac{ \EXP \big\{ \MM (\Yo , \bX \con \btheta) \alpha_1^*(\Yo, \bX) \cond A=0 , \bX \big\} }{  \EXP \big\{ \alpha_1^*(\Yo, \bX) \cond A=0 , \bX \big\} } \ , \ \nonumber
V_\EFF^* (\btheta)
=
\frac{\partial \EXP \big\{ \MM_\EFF^*(\bO \con \btheta) \big\} }{\partial \btheta \T} \ .
\end{align}
Consequently, the semiparametric efficiency bound for $\btheta^*$ is $\VAR \big\{ \InfFt^*(\bO \con \btheta^*) \big\}$.
\end{theorem}
In the previous two examples, applying Theorem \ref{thm-EIF-General} yields:
\begin{itemize}
\item[(i)] (\textit{Example 1: Counterfactual Mean}) Straightforward algebra confirms that $\mu_\MM^*(\bX \con \btheta) = \mu^*(\bX) - \theta$ and $V_\EFF^*(\btheta) = - \Pr(A=1)$. Therefore, we recover the EIF of Theorem \ref{thm-EIF}. 
\item[(ii)] (\textit{Example 2: Counterfactual Quantile}) From some algebra, we find $\mu_\MM^*(\bX \con \btheta) = F_{1}^* (\theta \cond 1 , \bX) - q$ and $V_\EFF^*(\btheta) = \Pr(A=1) \partial \EXP \big\{ \pYo \leq \theta \cond A=1 \big\} / \partial \theta = \Pr(A=1) \potf{1|A}^* (\theta \cond 1)$ where $F_{1}^*(y \cond 1, \bx)$ is the conditional cumulative function of $\pYo \cond A=1,\bX$ and  $\potf{1|A}^*(y \cond 1)$ is the conditional density of $\pYo \cond A=1$. Therefore, the EIF of $\tau_{0,q}^*$ is 
\end{itemize}%
\begin{align}           \label{eq-EIF_Quantile}
&
\InfFt^* (\bO \con \tau_{0,q}^*)
=
- \frac{ \displaystyle{\left[
\begin{array}{l}		
(1-A) \beta_1^*(\bX) \alpha_1^*(\Yo,\bX)
\big\{
\ind \big(  \Yo \leq \tau_{0,q}^* \big) - F_1^*( \tau_{0,q}^* \cond 1 , \bX)
\big\}
\\
+
A \big\{ F_1^*( \tau_{0,q}^* \cond 1 , \bX) - q \big\}
\\
+ (2A-1) R^*(\Yz,A,\bX)
\big\{
\ind \big(  \Yz \leq \tau_{0,q}^* \big) - F_1^*( \tau_{0,q}^* \cond 1 , \bX)
\big\}
\end{array}
\right]} }{\Pr(A=1) \potf{1|A}^*(\tau_{0,q}^* \cond 1)}        \ .
\end{align}

Next, we consider the estimation of $\btheta^*$ using the EIF as a moment equation, where all nuisance parameters are estimated. Let us consider the following cross-fitting estimator $\widehat{\btheta}^{(k)}$ that (asymptotically) solves the estimating equation $\AVER_{\mathcal{I}_k} \big\{ \widehat{\MM}_\EFF \LSS (\bO \con \widehat{\btheta}^{(k)} ) \big\} = r_{N} $ where $r_N$ is $o_P(N^{-1/2})$ and $\widehat{\MM}_\EFF \LSS $ is the efficient moment equation \eqref{eq-EffMoment} using estimated nuisance functions, i.e., 
\begin{align*}
&
\widehat{\MM}_\EFF\LSS (\bO \con \btheta)
=
\left[
\begin{array}{l}
(1-A) \
\widehat{\beta}_1\LSS (\bX) \widehat{\alpha}_1\LSS(\Yo,\bX)
\big\{  \MM(\Yo,\bX \con \btheta) - \widehat{\mu}_\MM\LSS( \bX \con \btheta) \big\}
+
A
\widehat{\mu}_\MM\LSS (\bX \con \btheta)
\\
+ (2A-1) \widehat{R}\LSS (\Yz,A,\bX) \big\{  \MM(\Yz,\bX \con \btheta) - \widehat{\mu}_\MM \LSS (\bX \con \btheta) \big\}
\end{array}	
\right]
\ , \\
&
\widehat{\mu}_\Omega \LSS (\bX \con \btheta) = \frac{ \EXP \big\{ \MM (\Yo , \bX \con \btheta) \widehat{\alpha}_1\LSS (\Yo, \bX) \cond A=0 , \bX \big\} }{  \EXP \big\{ \widehat{\alpha}_1\LSS(\Yo, \bX) \cond A=0 , \bX \big\} } \ .
\end{align*}
Then, the aggregated cross-fitting estimator across $K$-folds, i.e., $\widehat{\btheta} = K^{-1} \sum_{k=1}^{K} \widehat{\btheta}^{(k)}$ is a semiparametric efficient estimator for $\btheta^*$ under additional conditions; Theorem \ref{thm-AsympNormal-General} states the result.
\begin{theorem}                                                     \label{thm-AsympNormal-General}
Suppose Assumptions \ref{assumption:support}-\ref{assumption:PostCross} and regularity conditions in Section \ref{sec:supp-Regularity} hold. Then, the aggregated cross-fitting estimator $\widehat{\btheta}$ is asymptotically normal as $\sqrt{N} ( \widehat{\btheta } - \btheta^* ) \stackrel{D}{\rightarrow} N ( 0, \Sigma) $ where the variance matrix $\Sigma$ is equal to the semiparametric efficiency bound under model $\mathcal{M}_\OREC$, i.e., $\Sigma = \VAR \big\{ \InfFt^* (\bO \con \btheta^*) \big\} $. 
\end{theorem}
\noindent A consistent variance estimator $\widehat{\Sigma}$ is given in Section \ref{sec:supp-Regularity}.

It may be challenging to find a solution to the estimating equation based on $\MM_\EFF$ due to its complex form. For example, the EIFs of the counterfactual quantile in \eqref{eq-EIF_Quantile} are non-linear and non-smooth functions of the target parameter, indicating that the solution may not be easily obtained. In this case, we recommend a one-step estimator which updates a preliminary consistent estimator $\widetilde{\btheta}^{(k)}$, obtained by solving an inefficient but simpler moment equation. Such a simpler consistent estimator may be obtained by solving an inverse probability moment equation obtained from equation \eqref{eq-EIF_Quantile} by setting $F_{1}^*$  and $R^*$ to zero. Then, the one-step estimator $\widehat{\btheta}^{(k)}$ is given by $\widehat{\btheta}^{(k)} = \widetilde{\btheta}^{(k)} - \big\{ \widehat{V}_\EFF\LSS(\widetilde{\btheta}^{(k)}) \big\}^{-1} \AVER_{\mathcal{I}_k} \big\{ \MM_\EFF\LSS(\bO, \widetilde{\btheta}^{(k)}) \big\}$ where $\widehat{V}_\EFF\LSS(\widetilde{\btheta}\LSS)$ is an estimator of the Jacobian matrix of $\MM_\EFF$ evaluated at $\widetilde{\btheta}^{(k)}$; see Section \ref{sec:supp-Regularity}.

\subsubsection{Regularity Conditions for the Estimation of General Estimands}       \label{sec:supp-Regularity}

We first introduce the regularity conditions for the consistency of $\widehat{\btheta}$:

\begin{itemize}
    \item[(i)] \textit{Regularity Conditions for the Consistency of $\widehat{\btheta}$}
\end{itemize}

\begin{itemize}
\item[\HT{R1}] The parameter space $\Theta$ is a compact subset in $\R^p$, and true parameter $\btheta$ is in the interior of $\Theta$.

\item[\HT{R2}] $\MM(y,\bx \con \btheta)$ is uniformly bounded for $(y,\bx,\btheta) \in \big\{ \SPz \cup \SPzt \big\} \otimes \Theta$.

\item[\HT{R3}] $\inf_{\btheta: \| \btheta - \btheta^* \| \geq \epsilon } \big\| \EXP \big\{ \MM_\EFF^* (\bO \con \btheta) \big\} \big\| > 0 =   \big\| \EXP \big\{ \MM_\EFF^* (\bO \con \btheta^*) \big\} \big\| $ for every $\epsilon>0$.

\end{itemize}	

Note that Regularity conditions \HL{R1}-\HL{R3} are standard in M-estimation literature to guarantee the consistency of the estimator; see Chapter 5 of \citet{Vaart1998} and \citet{Stenfanski2002} for details. Next, we introduce regularity conditions needed for establishing asymptotic normality of $\widehat{\btheta}$ and consistency of the variance estimator:

\begin{itemize}
    \item[(ii)] \textit{Regularity Conditions for the Asymptotic Normality of $\widehat{\btheta}$ and the Consistency of the Proposed Variance Estimator}
\end{itemize}

Let ${\eta}$ be the nuisance components $(\alpha_1,\beta_0,\beta_1,\potf{0|AX},\potf{1|AX})$. Let the expectation operator of $\MM$ be 
\begin{align*}
&
\mu_\MM(\bx \con \btheta,{\eta})
=
\bigg[ \int_{\SPo} \alpha_1(y,\bx) \potf{1|AX}(y \cond 0,\bx)  \bigg]^{-1}
\bigg[ \int_{\SPo} \MM(y,\bx \con \btheta) \alpha_1(y,\bx) \potf{1|AX}(y \cond 0,\bx) \bigg] \ .
\end{align*}
Then, for a fixed ${\eta}$, we assume the following conditions:

\begin{itemize}
\item[\HT{R4}] $\mu_\MM(\bx \con \btheta, {\eta})$ is differentiable with respect to $\btheta \in \Theta$ with the Jacobian matrix $\mathcal{J}(\bx \con \btheta, {\eta}) := \nabla_\theta\T \mu_\MM(\bx \con \btheta, {\eta})$. The Jacobian matrix $\mathcal{J}(\bx \con \btheta, {\eta})$ is uniformly bounded over $(\bx, \btheta) \in \mathcal{X} \otimes \Theta$, and   $\EXP \big\{ \mathcal{J}(\bX \con \btheta, {\eta}^*) \big\}$ is invertible for $\btheta$ in the neighborhood of $\btheta^*$. 
\item[\HT{R5}]  $\big\{ \MM_\EFF(\bO \con \btheta, {\eta}) \cond \btheta \in \Theta \big\}$ is $P$-Donsker.

\item[\HT{R6}] There exists a function $\omega (\bx \con {\eta}) $ that is uniformly bounded over $\bx \in \mathcal{X}$ satisfying the following result for all $\btheta_1,\btheta_2 \in \Theta$ and $\bx \in \mathcal{X}$:
\begin{align*} 
&
\bigg\| \nabla_\theta \T \int_{\SPo} \big\{ \MM(y,\bx \con \btheta_1) -  \MM(y,\bx \con \btheta_2) \big\} \alpha_1(y,\bx) \potf{1|AX}(y | 0,\bx) \, dy	 \bigg\|_2 ^2\leq \omega (\bx \con  {\eta}) \cdot \big\| \btheta_1 - \btheta \big\|_2^2 \ 
\\
&
\bigg| \int_{\SPo} \big\| \MM(y,\bx \con \btheta_1) -  \MM(y,\bx \con \btheta_2) \big\|_2^2 \alpha_1(y,\bx) \potf{1|AX}(y | 0,\bx) \, dy		\bigg| 
\leq 
\omega (\bx \con  {\eta})
\cdot 
\big\| \btheta_1 - \btheta_2 \big\|_2^2
\ . 	\end{align*}

\end{itemize}

Regularity conditions \HL{R4}-\HL{R6} are non-standard compared to the usual conditions required for M-estimators because we allow our estimating equation $\MM$ to be non-smooth (e.g., the estimating equation for quantiles).  Regularity condition \HL{R4} means that, even though the original estimating equation $\MM(y,\bx \con \btheta)$ is non-smooth, its conditional expectation $\EXP_{{\eta}} \big\{ \MM(\pYo, \bX \con \btheta) \cond A=1, \bX \big\}$ is smooth with respect to $\btheta$. Regularity condition \HL{R5} means that the efficient estimating equation at given nuisance functions over $\theta \in \Theta$ is not overly complex. Regularity condition \HL{R6} means that the Jacobian $\mathcal{J}(\bX \con \btheta)$ in \HL{R4} is Lipschitz continuous, and the conditional expectation of the $L_2$ distance between original estimating equations at two parameters, $\EXP_{{\eta}}\big\{ \big\| \MM(\pYo,\bX \con \btheta_1) - \MM(\pYo,\bX \con \btheta_2) \big\|_2^2 \cond A=1, \bX \big\}$, also has the Lipschitz continuity property. A more interpretable condition can replace these assumptions if $\MM$ is continuously differentiable; for instance, it is sufficient to assume the following condition that is satisfied for many smooth estimating equations:
\begin{itemize}
\item[\HT{R7}] For any $\btheta_1$ and $\btheta_2$, we have a bounded function $\omega(\bo, {\eta})$ satisfying $\big\| \MM_\EFF(\bo \con \btheta_1 , {\eta} ) - \MM_\EFF(\bo \con \btheta_2 , {\eta} ) \big\|_2 \leq \omega(\bo,{\eta}) \big\| \btheta_1 - \btheta_2 \big\|_2 $. Additionally, $\MM_\EFF(\bo, \btheta, {\eta})$ is twice differentiable, and $\EXP \big\{ \nabla_\theta\T \MM_{\EFF}(\bO \con \btheta, {\eta}^*) \big\} \big|_{\btheta=\btheta^*}$ is invertible. 
\end{itemize}

We introduce the consistent variance estimator $\widehat{\Sigma} = \widehat{\Sigma}_B^{-1} \widehat{\Sigma}_M \widehat{\Sigma}_B^{-\intercal}$ :
\begin{align*}
& 
\widehat{\Sigma}_B
=
K^{-1} \sum_{k=1}^{K}  \widehat{\Sigma}_B \LSS
\ , \
&&
\widehat{\Sigma}_B^{(k)} 
=  
\AVER_{\mathcal{I}_k} \Big\{ A \widehat{\mathcal{J}} \LSS (\bX \con \widehat{\btheta}) \Big\}
\ , \\
&
\widehat{\Sigma}_M
=
K^{-1} \sum_{k=1}^{K} \widehat{\Sigma}_M^{(k)}
\ , \
&&
\widehat{\Sigma}_M^{(k)} = \AVER_{\mathcal{I}_k}
\Big\{ \widehat{\MM}_\EFF\LSS (\bO \con \widehat{\btheta} ) ^{\otimes 2} \Big\}
\end{align*}
Note that $\widehat{\Sigma}_B^{(k)}$ uses the estimated Jacobian matrix, and this is to incorporate non-smooth estimating equations. If the original estimating equation $\MM$ satisfies Regularity condition \HL{R7}, we can use $\widehat{\Sigma}_B^{(k)} 
=  
\AVER_{\mathcal{I}_k} \Big\{ \nabla_\theta\T \widehat{\MM}_\EFF\LSS (\bO \con \btheta)  \Big\} \Big|_{\btheta=\widehat{\btheta} }$. 
Additionally, for the one-step estimator, we choose $\widehat{V}_\EFF\LSS(\widetilde{\btheta}^{(k)}) = \AVER_{\mathcal{I}_k} \big\{ A \widehat{\mathcal{J}} \LSS (\bX \con \widetilde{\btheta}^{(k)}) \big\}$; if $\MM$ is differentiable, we can take  $\widehat{V}_\EFF\LSS(\widetilde{\btheta}^{(k)}) = \AVER_{\mathcal{I}_k} \big\{ \nabla_\theta\T \widehat{\MM}_\EFF\LSS (\bO \con \widetilde{\btheta}^{(k)})  \big\}$.

\subsection{Extension: Missing Data Setting}  \label{sec:Missing}

Next, we consider settings where study units' outcomes are measured multiple times, but some subjects drop out before the end study period. Let us consider a simple data structure $\{ \bX, \Yz, 1-A, (1-A) \Yo \}$ where $\bX$ is a collection of baseline covariates,  $Y_t$ for $t \in \{0,1\}$ is an outcome of interest measured at time $t$, and $A$ is an indicator of whether $\Yo$ is missing $(A=1)$ or measured $(A=0)$; note that we define $A$ to make the discussion below concordant to the results in the previous sections. Let $\EXP(\Yo)$ be the target estimand. The most popular identifying assumptions to address missing data is that they are missing completely at random (MCAR) or missing at random (MAR) in that $\Yo \indep A$ or $\Yo \indep A \cond (\Yz, \bX)$, respectively. Under these assumptions, identification of $\EXP(\Yo)$ is straightforward using the fact that $\EXP (\Yo \cond A=1) - \EXP (\Yo \cond A=0) = 0$ and $\EXP (\Yo \cond A=1, \Yz, \bX)=\EXP (\Yo \cond A=1, \Yz, \bX) = 0$. A more challenging case arises when the drop-out mechanism is missing not at random (MNAR) in that $\Yo \nindep A \cond (\Yz,\bX)$, and an additional assumption is needed to identify the target estimand. For example, an approach recently introduced in \citet{Oliver_MNAR2022} is motivated by the DiD strategy. Specifically, they assume the following \nHL{PT} assumption on the outcomes $(\Yz,\Yo)$ holds, i.e., $
\EXP \big( \Yo - \Yz \cond A=0, \bX \big) 
=
\EXP \big( \Yo - \Yz \cond A=1, \bX \big)$. Then, the estimand is identified by the usual DiD estimator as  $
\EXP \big( \Yo \big)=
\EXP \big[ \EXP \big( \Yo \cond A=0, \bX \big) 
-
\big\{
\EXP \big( \Yz \cond A=0, \bX \big) 
-
\EXP \big( \Yz \cond A=1, \bX \big) 
\big\} \Pr (A=1 \cond \bX)
\big]$. 

The \nHL{OREC} identification framework introduced in this paper, therefore, provides an alternative identification strategy to \nHL{PT} when missingness is not at random. Note that $\EXP(\Yo) = \EXP(\Yo \cond A=0) \Pr(A=0) + \EXP(\Yo \cond A=1) \Pr(A=1)$; therefore, to identify the mean of $Y_1$, it clearly suffices to identify $ \EXP(\Yo \cond A=1)$, the conditional mean of the outcome in the subset of the population with missing outcome. Suppose that the missing mechanism satisfies the \nHL{OREC} condition in that $\alpha_0^* (y,\bx)$ is equal to $\alpha_1^* (y,\bx)$ where $\alpha_t^* $ is the odds ratio relating $Y_t$ with $A$ given $\bX$. Under Assumptions \ref{assumption:support}-\ref{assumption:OREC} tailored to the missing data setting, we can identify $\EXP (\Yo \cond A=1)$ by using $\alpha_0^*$ instead of $\alpha_1^*$ in the three representations \eqref{eq-rep-IPW}, \eqref{eq-rep-OR}, and \eqref{eq-rep-AIPW}. Moreover, Theorem \ref{thm-EIF} provides the EIF of $\EXP (\Yo \cond A=1)$ under the model that is only restricted by the \nHL{OREC} assumption. Therefore, we can obtain the estimators for $\EXP(\Yo \cond A=1)$ and $\EXP(\Yo)$ by following the approaches in Section \ref{sec:Conti-Improved}, and these estimators are consistent and asymptotically normal under the stated conditions. Lastly, we can likewise identify and estimate other causal quantities (e.g., quantiles) of the outcome subject to missingness using the approaches described in the prior Section; details are omitted as the extension is somewhat straightforward.

\newpage

\section{Proof of the Main Paper} \label{sec:supp:prmain}

In this section, we use the following shorthand for the conditional distributions for $t=0,1$:
\begin{align*}
& \potg{t}^* (y,a \cond \bx) = \potf{tA|X}^*(y,a \cond \bx)  = P(\potY{0}{t}=y,A=a \cond \bX=\bx)
\ , \\
& \potf{t}^* (y \cond a,\bx) = \potf{t|AX}^* (y \cond a, \bx) = P(\potY{0}{t}=y \cond A=a,\bX=\bx)
\ , \\
& 
\pote{t}^* (a \cond y,\bx) = \potf{A|tX}^*(a \cond y,\bx) = \Pr(A=a \cond \pYz=y, \bX=\bx)
\ .
\end{align*}
That is, we unify the density notation by using $f$, and let the subscript indicate the conditional distribution. Similarly, we denote
\begin{align*}
& \potf{t|X}^* (y \cond \bx) = P(\potY{0}{t}=y \cond \bX=\bx)
\ , 
&& \potf{A|X}^* (a \cond \bx) = \Pr(A=a \cond \bX=\bx) \ .
\end{align*}

\subsection{Proof of Lemma \ref{lemma-beta_and_mu}}                            \label{sec:supp-beta_and_mu}
  
% Since $\tau_{1}^* = \EXP \{ \potY{1}{1} \cond A=1 \}$ is identified by $\tau_{1}^* = \EXP (A \Yo) / \Pr(A=1)$, it suffices to establish identification for $\tau_{0}^* = \EXP \{ \potY{0}{1} \cond A=1 \}$. In this proof, we establish identification of $\EXP \{ \mathcal{G}(\potY{0}{1}) \cond A=1 \}$ where $\mathcal{G}(\cdot)$ is a fixed, integrable function. 

We find
\begin{align*}
\beta_{1}^*(\bX) 
=
\frac{ \potf{1A|X}^*(0,1 \cond \bX) }{ \potf{1A|X}^*(0,0 \cond \bX) }
\ , \
\alpha_1^*(y, \bX)
=
\left\{
\begin{array}{ll}
\frac{ \potf{1A|X}^*(y,1 | \bX) }{ \potf{1A|X}^*(y,0 | \bX) }
\frac{ \potf{1A|X}^*(0,0 | \bX) }{ \potf{1A|X}^*(0,1 | \bX) } &
\text{for } y \in \SPot 
\\
0  & \text{for } y \in \R \cap \SPot^c
\end{array}
\right.
\ .
\end{align*}
Consequently, we obtain the following result for all $y \in \SPot$:
\begin{align*}
\beta_{1}^*(\bX) 
\alpha_1^*(y, \bX)
\potf{1|AX}^*(y \cond 0 , \bX) 
\potf{A|X}^*(0 \cond \bX)
& =
\beta_{1}^*(\bX) 
\alpha_1^*(y, \bX)
\potf{1|AX}^*(y , 0 \cond \bX) 	
\\
&
=
\potf{1A|X}^*(y, 1 \cond \bX) 
=
\potf{1|AX}^*(y \cond 1 , \bX)  \potf{A|X}^*(1 \cond \bX) \ ,
\end{align*}
and $\beta_{1}^*(\bX) 
\alpha_1^*(y, \bX)
\potf{1|AX}^*(y \cond 0 , \bX) 
\potf{A|X}^*(0 \cond \bX)=0$ for $y \in \R \cap \SPot^c$.

We integrate both sides with respect to $y$ over $\SPo$, and we get
\begin{align*}
\beta_{1}^*(\bX) 
\EXP \big\{ \alpha_1^*(\Yo, \bX) \cond A=0, \bX \big\}
\potf{A|X}^*(0 \cond \bX)
=
\potf{A|X}^*(1 \cond \bX) \ .
\end{align*}
This proves the result related to $\beta_{1}^*$.

Next, we find the following result for any integrable $\mathcal{G}$:
\begin{align}                       \label{eq-IPW-basis}
\beta_{1}^* (\bX) \EXP \big\{ \mathcal{G}(\Yo) \alpha_1^* (\Yo, \bX)  \cond A=0, \bX \big\} 
& =
\int_{\SPo} \mathcal{G}(y) \beta_{1}^*(\bX) \alpha_1^*(y, \bX) \potf{1|AX}^* (y \cond 0, \bX) \, dy
\nonumber
\\
& = 
\frac{ \potf{A|X}^*(1 \cond \bX) }{ \potf{A|X}^*(0 \cond \bX) }
\int_{\SPot} \mathcal{G}(y) \potf{1|AX}^* (y \cond 1, \bX) \, dy
\nonumber
\\
& = 
\frac{ \potf{A|X}^*(1 \cond \bX) }{ \potf{A|X}^*(0 \cond \bX) }
\EXP \big\{ \mathcal{G}(\pYo ) \cond A=1, \bX \big\} \ .
\end{align}
As a consequence, we get
\begin{align*}
&
\frac{\EXP \big\{ \mathcal{G}(\Yo) \alpha_1^* (\Yo, \bX)  \cond A=0, \bX \big\}}{\EXP \big\{ \alpha_1^* (\Yo, \bX)  \cond A=0, \bX \big\}}
\\
&
=
\bigg[ \frac{ \potf{A|X}^*(1 \cond \bX) }{ \potf{A|X}^*(0 \cond \bX) }
\EXP \big\{ \mathcal{G}(\pYo ) \cond A=1, \bX \big\} \bigg]
\bigg\{ \frac{ \potf{A|X}^*(1 \cond \bX) }{ \potf{A|X}^*(0 \cond \bX) }
\EXP \big( 1 \cond A=1, \bX \big)
\bigg\}^{-1}
\\
&
= \EXP \big\{ \mathcal{G}(\pYo ) \cond A=1, \bX \big\} \ ,
\end{align*}
where the first identity is from taking $\mathcal{G}$ in the denominator as the constant function 1, i.e., $\mathcal{G}(y) \equiv 1$. This concludes the proof.

\subsection{Proof of Theorem \ref{thm-EIF}}                                 \label{sec:supp-Proof-thm-EIF}

In the proof, we show a more general result by characterizing the EIF for $ \tau^*(\mathcal{G}) := \EXP \{ \mathcal{G}(\potY{1}{1}) - \mathcal{G}(\potY{0}{1}) \cond A=1 \}$, where $\mathcal{G}(\cdot)$ is a fixed, integrable function. With a slight abuse of notation, we denote  $\mu^*(\bX) = \EXP \big\{ \mathcal{G}(\potY{0}{1}) \cond A=1 , \bX \big\}$.

We find $\mathcal{M}_{\OREC}$ is characterized as a regular model of the observed data of the form:
\begin{align*}
\mathcal{M}_{\OREC}
=
\Big\{ P(\bO) \, \Big| \, \mathcal{S}_0(1) \cap \mathcal{S}_1(0) \neq \emptyset \text{ and } \mathcal{S}_1(0) \subseteq \mathcal{S}_0(0) \text{ where } \mathcal{S}_{t}(a) = \text{support}(Y_{t},\bX \cond A=a) \Big\} \ .
\end{align*}
where the support conditions are from Assumption \ref{assumption:support-1} and \ref{assumption:support-2}. We consider a parametric submodel of $\mathcal{M}_{\OREC}$ parametrized by a one-dimensional parameter $\eta$:
\begin{align}							\label{eq-Model-OREC-ParaSubmodel}
&
\mathcal{M}_{\OREC}
(\eta)
\\
&
=
\Big\{ P (\bO \con \eta) \, \Big| \, \mathcal{S}_0(1) \cap \mathcal{S}_1(0) \neq \emptyset \text{ and } \mathcal{S}_1(0) \subseteq \mathcal{S}_0(0) \text{ where } \mathcal{S}_{t}(a) = \text{support}(Y_{t},\bX \cond A=a \con \eta)  \Big\} \ .
\nonumber
\end{align}

Before going into more detail, we define a few more notations related to $\eta$. Let $\nabla_\eta h(\cdot \con \eta)$ denote the derivative of $h( \cdot \con \eta)$ with respect to $\eta$, and $\EXP^{(\eta)}\{ h(\bO) \big\}$ denote the expectation of function $h$ with respect to the distribution $P(\bO \con \eta)$. Let $f_O (\bO \con \eta)$ be the density of the parametric submodel $P(\bO \con \eta)$. We suppose that the true distribution of the observed data $P(\bO)$ is recovered at $\eta^*$, i.e., $P^*(\bO) = P(\bO \con \eta^*)$. 

Since the restrictions on the supports do not change the tangent space, we find the tangent space of the model $\mathcal{M}_{\OREC}$ is given as
\begin{align}                           \label{eq-tangentspace}
\mathcal{T}_\OREC 
=
\big\{ S(\bO) 
\cond
\EXP\big\{ S(\bO) \big\} = 0 , \ 
\EXP\big[ \big\{ S(\bO) \big\}^2 \big] < \infty
\big\}
\ ,
\end{align}
where the expectations in $\mathcal{T}_\OREC $ are evaluated at the true distribution $P^*(\bO)$ satisfying the support conditions. In other words, $\mathcal{T}_{\OREC}$ is the entire Hilbert space of mean-zero, square-integrable functions of $\bO$ with $\mathcal{S}_0(1) \cap \mathcal{S}_1(0) \neq \emptyset$ and  $\mathcal{S}_1(0) \subseteq \mathcal{S}_0(0)$.

Since the model is nonparametric, there is a unique influence function for $\tau^*(\mathcal{G})$, and it is the EIF in $\mathcal{M}_{\OREC}$. Therefore, to establish that $\InfFt^*(\bO) = \InfFt_{1}^*(\bO) - \InfFt_{0}^*(\bO)$ is the EIF for $\tau^*(\mathcal{G})$, it suffices to show that $\tau_{a}^*(\mathcal{G}) := \EXP \{ \mathcal{G}(\potY{a}{1}) \cond A=1 \}$ is a differentiable parameter \citep{Newey1990}, i.e., 
\begin{align}						\label{eq-diffpara}
\frac{ \partial }{\partial \eta}  
\EXP^{(\eta)} \big\{ \mathcal{G}(\potY{a}{1}) \cond A=1 \big\} \bigg|_{\eta=\eta^*}
=
\EXP \big\{
s_O(\bO \con \eta^*) \InfFt_{a}^*(\bO)		
\big\} \ ,  \quad 
\InfFt_{a}^*(\bO) \in \mathcal{T}_{\OREC} \ , 
\end{align}
where $s_O(\bO \con \eta) = \nabla_\eta f_O(\bO \con \eta)/f_O(\bO \con \eta)$. 

First, from straightforward algebra, one can find
\begin{align*}
    \InfFt_{1}^*(\bO) = \frac{A \mathcal{G}(\Yo) - A \tau_{1}^*(\mathcal{G}) }{\Pr(A=1)}
\end{align*}
satisfies \eqref{eq-diffpara}. Therefore, it suffices to find the EIF for the counterfactual mean $\tau_{0}^*(\mathcal{G})$. 

We first provide an alternative form of the right hand side. The (conjectured) EIF is written as
\begin{align*} 
\InfFt^*(\bO) 
&
= \widetilde{\uncInfFt}(\bO_{1} \con \eta^*) 
- \frac{A \EXP \big\{ \mathcal{G}(\pYo) \cond A=1 \big\} }{\Pr(A=1)}
+ \InfFtAug (\bO_{0} \con \eta^* )  \ , 
\end{align*}
where $\widetilde{\uncInfFt}(\bO_1 \con \eta)$ and $\InfFtAug (\bO_0 \con \eta)$ are
\begin{align*}
\widetilde{\uncInfFt}(\bO_{1} \con \eta)
&
=
\frac{ 1}{\Pr(A=1 \con \eta)}
\Big[
\beta_1(\bX \con \eta) \alpha_1(\Yo,\bX \con \eta)
(1-A) 		 \big\{  \mathcal{G}(\Yo) - \mu(\bX \con \eta) \big\}
+
A
\mu(\bX  \con \eta)
\Big] \ .
\\
\InfFtAug (\bO_{0} \con \eta  ) 
&
=
\frac{(2A-1) R(\Yz,A,\bX \con \eta) \big\{  \mathcal{G}(\Yz) - \mu(\bX \con \eta) \big\}}{\Pr(A=1 \con \eta)} 
\\
&
=
\frac{  
(2A-1) \beta_1(\bX \con \eta) \alpha_1(\Yz,\bX \con \eta) 
}{
\Pr(A=1 \con \eta) 
}		
\frac{ \potf{1 A|X}(\Yz,0 \cond \bX \con \eta ) }{ \potf{0 A|X}(\Yz, A \cond \bX \con \eta ) }
\big\{  \mathcal{G}(\Yz) - \mu (\bX \con \eta) \big\}
\ .
\end{align*}
Here, $\alpha_1(\Yo,\bX \con \eta)$ is the solution to the moment equation \eqref{eqref-alpha1-constraint} at $\eta$, i.e., for $\SPoreceta := \mathcal{S}_0(1) \cap \mathcal{S}_1(0)$, i.e., 
\begin{align}        \label{eqref-alpha1-constraint-eta}
& \text{For } (y,\bx) \in \SPoreceta, 
&&
\alpha_1^{(\eta)} \text{ solves }
\overline{\EXP}_{\SPoreceta}^{(\eta)} \big\{ \Psi_{\SPoreceta} ( \bO_{0} \con \alpha, \potf{0}( \cdot \cond 0,\cdot \con \eta) , \pote{0}( \cdot \cond 0,\cdot \con \eta) , {\mathfrak{m}}) \big\} = 0 \ ,  
\nonumber
\\
& \text{For } (y,\bx) \in \R \cap \SPoreceta^c, 
&&
\alpha_1^{(\eta)}(y,\bx) = 0 \ .
\end{align}
The other two functions $\beta_1(\bx \con \eta)$ and $\mu(\bx \con \eta)$ are defined by the relationships in Lemma \ref{lemma-beta_and_mu}; i.e.,
\begin{align*}
&
\beta_{1} (\bX \con \eta)
= 
\frac{ \Pr(A=1 \cond \bX \con \eta) / \Pr(A=0 \cond \bX \con \eta) }{\EXP^{(\eta)} \big\{ \alpha_1 (\Yo, \bX \con \eta) \cond A = 0 , \bX \big\}}
\ , \
\mu(\bX \con \eta) 
=
\frac{
\EXP^{(\eta)} \big\{ \mathcal{G}( \Yo ) \alpha_1 ( \Yo, \bX \con \eta) \cond A=0, \bX \big\}
}{
\EXP^{(\eta)} \big\{ \alpha_1( \Yo, \bX \con \eta) \cond A=0, \bX \big\}
}
\ .
\end{align*}
Therefore, the right hand side of \eqref{eq-diffpara} is
\begin{align}              \label{eq-RHS}
& 
\EXP \big\{
s_O(\bO \con \eta^*) \InfFt^*(\bO)		
\big\}
\nonumber
\\
&
=
\EXP 
\left[
\begin{array}{l}
\big\{ s_{0|1}(\Yz \cond \bO_{1} \con \eta^*) +
s_{1}(\bO_{1} \con \eta^*) \big\} \widetilde{\uncInfFt}(\bO_{1} \con \eta^*) 
\\
- 
\big\{ s_{01X|A}(\Yz,\Yo,\bX \cond A \con \eta^*) + s_{A}(A \con \eta^*) \big\}
A \EXP \big\{ \mathcal{G}(\pYo) \cond A=1 \big\}/\Pr(A=1)
\\
+
\big\{ s_{1|0}(\Yo \cond \bO_{0} \con \eta^*) +
s_{0}(\bO_{0} \con \eta^*) \big\} \InfFtAug(\bO_{0} )
\end{array}		
\right]
\nonumber
\\
&
=
\EXP \Big[
s_{1}(\bO_{1} \con \eta^*) \widetilde{\uncInfFt}(\bO_{1} \con \eta^*) 
+ s_{0}(\bO_{0} \con \eta^*) \InfFtAug(\bO_{0} )
\Big]
- s_{A}(1 \con \eta)  \EXP \big\{ \mathcal{G}(\pYo) \cond A=1 \big\}
\ , 
\end{align}
where the score functions are
\begin{align*}
& s_A(a \con \eta) = \frac{\nabla_{\eta} \Pr(A=a \con \eta)}{\Pr(A=a \con \eta)}
\ , \ 
\quad \quad
s_{1} (\bo_{1} \con \eta) = 
\frac{ \nabla_\eta \potf{1AX}(y,a,\bx \con \eta) }{ \potf{1AX}(y,a,\bx \con \eta) }
\ , \
\quad \quad
s_{0} (\bo_{0} \con \eta) = 
\frac{ \nabla_\eta \potf{0AX}(y,a,\bx \con \eta) }{ \potf{0AX}(y,a,\bx \con \eta) }
\ , \\
&
s_{0|1} (\bo \con \eta) = 
\frac{ \nabla_\eta \potf{0|1AX}(y_0 \cond y_1,a,\bx \con \eta) }{ \potf{0|1AX}(y_0 \cond y_1,a,\bx \con \eta) } \ , \
\quad \quad
s_{1|0}(\bo \con \eta)
=
\frac{ \nabla_\eta \potf{1|0AX}(y_1 \cond y_0,a,\bx \con \eta) }{ \potf{1|0AX}(y_1 \cond y_0,a,\bx \con \eta) } \ .
\end{align*}
Of note, the restrictions of the score functions are $\EXP^{(\eta)} \big\{ s_{A} (A \con \eta) \big\} = \EXP^{(\eta)} \big\{ s_{1} (\bO_{1} \con \eta) \big\} = \EXP^{(\eta)} \big\{ s_{0} (\bO_{0} \con \eta) \big\} = \EXP^{(\eta)} \big\{ s_{1|0} (\bO \con \eta) \cond \bO_0 \big\} = \EXP^{(\eta)} \big\{ s_{0|1} (\bO \con \eta) \cond \bO_1 \big\}  = 0$.

Next, we focus on the left hand side of \eqref{eq-diffpara}. From the AIPW representation \eqref{supp:eq-rep-AIPW}, we have $\EXP^{(\eta)} \big\{ \mathcal{G}(\pYo) \cond A=1 \big\} = \EXP^{(\eta)} \big\{ \widetilde{\uncInfFt}(\bO_{1} \con \eta) \big\} $. Therefore, the derivative of $\EXP^{(\eta)} \big\{ \mathcal{G}(\pYo) \cond A=1 \big\}$  is
\begin{align}			\label{eq-pd}
&
\frac{ \partial }{\partial \eta}  \EXP^{(\eta)} \big\{ \mathcal{G}(\pYo) \cond A=1 \big\}
\nonumber
\\
&
=
\frac{\partial }{\partial \eta}
\EXP^{(\eta)} \big\{ \widetilde{\uncInfFt}(\bO_{1} \con \eta) \big\} 
\nonumber
\\
& = 
\EXP^{(\eta)} \Big[ \big\{ s_{0|1}(\bO \con \eta) + s_{1} (\bO_{1} \con \eta) - s_{A}(1 \con \eta) \big\} \widetilde{\uncInfFt}(\bO_{1} \con \eta) \Big]
\nonumber
\\
& 
\hspace*{1cm} +
\EXP^{(\eta)} \bigg[ 
\frac{(1-A) \big\{ s_\alpha(\Yo,\bX \con \eta) + s_\beta(\bX \con \eta) \big\} \beta_1(\bX \con \eta) \alpha_1(\Yo,\bX \con \eta)}{\Pr(A=1 \con \eta)}		
\Big\{ \mathcal{G}(\Yo) - \mu(\bX \con \eta) \Big\}
\bigg]
\nonumber
\\
&  \hspace*{1cm} -
\EXP^{(\eta)} \bigg[ 
\frac{(1-A) \beta_1(\bX \con \eta) \alpha_1(\Yo,\bX \con \eta) - A}{\Pr(A=1 \con \eta)}		
\Big\{
\nabla_\eta \mu (\bX \con \eta)
\Big\}
\bigg] \ ,
\end{align}
where $s_{\alpha} (y, \bx \con \eta) 
= {\nabla_\eta \alpha_1(y, \bx \con \eta)}/{\alpha_1(y, \bx \con \eta)}$ and $s_\beta(\bx \con \eta) = {\nabla_\eta \beta(\bx \con \eta)}/{\beta(\bx \con \eta)}$. 
% \begin{align*}
% 	& 
%  s_{\alpha} (y, \bx \con \eta) 
%            = \frac{\nabla_\eta \alpha_1(y, \bx \con \eta)}{\alpha_1(y, \bx \con \eta)}
%            \ ,
% 	&&
% 	s_\beta(\bx \con \eta) = \frac{\nabla_\eta \beta(\bx \con \eta)}{\beta(\bx \con \eta)} \ .
% \end{align*}
Of note, $s_\alpha(0,\bX \con \eta) = 0$ whereas $s_\beta(\bX \con \eta)$ is unrestricted. We observe that some terms in \eqref{eq-pd}	are simplified as follows. First, we obtain
\begin{align*}
& \EXP^{(\eta)} \Big[  s_{0|1}(\bO \con \eta) \widetilde{\uncInfFt}(\bO_{1} \con \eta) \Big]
=
\EXP^{(\eta)} \Big[  \EXP^{(\eta)} \big[ s_{0|1}(\bO \con \eta) \cond \bO_{1} \big] \widetilde{\uncInfFt}(\bO_{1} \con \eta) \Big]
=
0 \ , 
\\
&
\EXP^{(\eta)} \Big[ s_{A}(1 \con \eta) \widetilde{\uncInfFt}(\bO_1 \con \eta) \Big]
=
s_{A}(1 \con \eta)  \EXP^{(\eta)} \big\{ \mathcal{G}(\pYo) \cond A=1 \big\} \ .
\end{align*}
Second, using \eqref{eq-DRrepresentcorrectmu}, which is established in the proof of Lemma \ref{lemma-DR} in Section \ref{sec:supp:proof-lemma-DR}, we get
\begin{align*}		
& \EXP^{(\eta)} \bigg[ 
\frac{(1-A) s_\beta(\bX \con \eta) \beta_1(\bX \con \eta) \alpha_1(\Yo,\bX \con \eta)}{\Pr(A=1 \con \eta)}		
\Big\{ \mathcal{G}(\Yo) - \mu(\bX \con \eta) \Big\}
\bigg]
=
0 \ .
\end{align*}
Lastly, from the definition of $\beta_1(\bX \con \eta)$, we obtain
\begin{align*}
&
\EXP^{(\eta)} \bigg[ 
\frac{(1-A) \beta_1(\bX \con \eta) \alpha_1(\Yo,\bX \con \eta) - A}{\Pr(A=1 \con \eta)}		
\Big\{
\nabla_\eta \mu (\bX \con \eta)
\Big\}
\bigg]
\\
&
=
\EXP^{(\eta)} \Bigg[ 
\frac{\nabla_\eta \mu (\bX \con \eta)}{\Pr(A=1 \con \eta)}
\bigg[
\beta_1(\bX \con \eta) 
\EXP^{(\eta)} \Big\{ (1-A) \alpha_1(\Yo, \bX \con \eta)
\, \Big| \, \bX
\Big\}
- 
\EXP^{(\eta)} \big( A \cond \bX \big)
\Bigg]
\\
&  =
\EXP^{(\eta)} \Bigg[ 
\frac{\nabla_\eta \mu (\bX \con \eta)}{\Pr(A=1 \con \eta)}
\bigg[
\beta_1(\bX \con \eta) 
\Pr(A=0 \cond \bX \con \eta) 
\EXP^{(\eta)} \Big\{ \alpha_1(\Yo, \bX \con \eta)
\, \Big| \, A=0, \bX
\Big\}
- 
\Pr(A=1 \cond \bX \con \eta) 
\Bigg]
\\
&
=
0 \ .
\end{align*}
Therefore, the pathwise derivative evaluated at $\eta^*$ is 
\begin{align}                              \label{eq-LHS}
& \frac{ \partial }{\partial \eta}  \EXP^{(\eta)} \big\{ \mathcal{G}(\pYo) \cond A=1 \big\} \bigg|_{\eta=\eta^*}
\\
& 
= 
- s_{A}(1 \con \eta)  \EXP \big\{ \mathcal{G}(\pYo) \cond A=1 \big\}
\nonumber
\\
&
\hspace*{1cm}
+
\EXP  \bigg[ s_{1} (\bO_{1} \con \eta^*) \widetilde{\uncInfFt}(\bO_{1} \con \eta^*)
+
\frac{(1-A) s_\alpha (\Yo,\bX \con \eta^*)  \beta_1^*(\bX) \alpha_1^*(\Yo,\bX)}{\Pr(A=1)}		
\Big\{ \mathcal{G}(\Yo) - \mu^*(\bX) \Big\}
\bigg] \ .
\nonumber
\end{align}

Comparing \eqref{eq-RHS} and \eqref{eq-LHS}, we establish \eqref{eq-diffpara} if the following identity holds:
\begin{align}       \label{eq-LHS-RHS}
\hspace*{-0.25cm}
\EXP \Big\{
s_{0}(\bO_{0} \con \eta^*) \InfFtAug(\bO_{0} )
\Big\}
=
\EXP \bigg[ \frac{(1-A) s_\alpha (\Yo,\bX \con \eta^*)  \beta_1^*(\bX) \alpha_1^*(\Yo,\bX)}{\Pr(A=1)}		
\Big\{ \mathcal{G}(\Yo) - \mu^*(\bX) \Big\}
\bigg] \ .
\end{align}

To show \eqref{eq-LHS-RHS}, we re-visit the definition of $\alpha_1(y,\bx \con \eta)$ in \eqref{eqref-alpha1-constraint-eta}. The gradient of the restriction in \eqref{eqref-alpha1-constraint-eta} is always zero, Therefore, the following condition holds:
\begin{align*}
\hspace*{-0.25cm}
0 
& =
\nabla_{\eta}
\overline{\EXP}_{\SPoreceta}^{(\eta)} \big\{ \Psi_{\SPoreceta} ( \bO_{0} \con \alpha_1 (\eta) , \potf{0 | A X}(\eta) , \potf{A | 0 X}(\eta) , {\mathfrak{m}}) \big\}
\\
& =
\overline{\EXP}_{\SPoreceta}^{(\eta)} \Big[
\big\{ s_{1|0} ( \bO \con \eta) +
s_{0}(\bO_{0} \con \eta) \big\} 
\Psi_{\SPoreceta} ( \bO_{0} \con \alpha_1 (\eta) , \potf{0 | A X}(\eta) , \potf{A | 0 X}(\eta) , {\mathfrak{m}}) 
\Big]
\\
& =
\overline{\EXP}_{\SPoreceta}^{(\eta)} \Big[
s_{0}(\bO_{0} \con \eta) 
\Psi_{\SPoreceta} ( \bO_{0} \con \alpha_1 (\eta) , \potf{0 | A X}(\eta) , \potf{A | 0 X}(\eta) , {\mathfrak{m}}) 
\Big]
\\
&
+
\overline{\EXP}_{\SPoreceta}^{(\eta)} 
\left[
\begin{array}{l}
\big[ {\mathfrak{m}}(\Yz,\bX) - \overline{\EXP}_{{\SPoreceta}, 0 | AX}^{(\eta)} \big\{ s_{0 |A X} (\Yz \cond 0 , \bX \con \eta) {\mathfrak{m}}(\Yz, \bX) \cond A=0, \bX \big\} \big] 
\\[0.2cm]	
\quad \times
\big\{ \alpha_1 (\Yz,\bX \con \eta) \big\}^{-A} 
\big\{ A- 
\EXP_{A|0X}^{(\eta)} ( A \cond \Yz=0, \bX) \big\}
\end{array}
\right]
\tag{T1}           \label{eq-EIF-T1}
\\
&
+
\overline{\EXP}_{\SPoreceta}^{(\eta)} 
\left[
\begin{array}{l}
 \big[ {\mathfrak{m}}(\Yz,\bX) - \overline{\EXP}_{{\SPoreceta} , 0 | AX}^{(\eta)} \big\{ {\mathfrak{m}}(\Yz, \bX) \cond A=0, \bX \big\} \big] 
 \\
\big\{
-	\frac{A s_\alpha (\Yz,\bX \con \eta) }{\alpha_1 (\Yz,\bX \con \eta)}
\big\}
\big\{ A- \EXP_{A|0X}^{(\eta)} ( A \cond \Yz=0, \bX) \big\}
\end{array} 
\right]
\\
&
+
\overline{\EXP}_{\SPoreceta}^{(\eta)} \left[
\begin{array}{l}
\big[ {\mathfrak{m}}(\Yz,\bX) - \overline{\EXP}_{{\SPoreceta} , 0 | AX}^{(\eta)} \big\{ {\mathfrak{m}}(\Yz, \bX) \cond A=0, \bX \big\} \big] 
\\[0.2cm]
\quad \times
\big\{ \alpha_1 (\Yz,\bX \con \eta) \big\}^{-A} 
\big[ A- \EXP_{A|0X}^{(\eta)} \big\{ A s_{A|0X}(A \cond \Yz=0,\bX \con \eta)  \cond \Yz=0, \bX \big\}
\big]
\end{array}
\right]
\ ,
\tag{T2}            \label{eq-EIF-T2}
\end{align*}
where $s_{0|AX} (y \cond 0 , \bx \con \eta) 
=
{ \nabla_\eta \potf{0|AX} (y \cond A=0, \bx \con \eta) }/{ \potf{0|AX} (y \cond A=0, \bx \con \eta) }$ and \\
$s_{A|0X} (a \cond 0 , \bx \con \eta)
=
{ \nabla_\eta \potf{A|0X}(a \cond \Yz=0,\bx \con \eta) }/{ \potf{A|0X}(a \cond \Yz=0,\bx \con \eta) }$. 
% \begin{align*}
% 	&
% 	s_{0|AX} (y \cond 0 , \bx \con \eta) 
% 	=
% 	\frac{ \nabla_\eta \potf{0|AX} (y \cond A=0, \bx \con \eta) }{ \potf{0|AX} (y \cond A=0, \bx \con \eta) }
% 	\ , \quad\quad 
% 	s_{A|0X} (a \cond 0 , \bx \con \eta)
% 	=
% 	\frac{ \nabla_\eta \potf{A|0X}(a \cond \Yz=0,\bx \con \eta) }{ \potf{A|0X}(a \cond \Yz=0,\bx \con \eta) } \ .
% \end{align*}
Each score function satisfies $\EXP^{(\eta)} \big\{ 	s_{0|AX} (\Yz \cond 0 , \bX \con \eta)  	\cond A=0, \bX \big\} = 0$ and $\EXP^{(\eta)} \big\{ 	s_{A|0X} (A \cond 0 , \bX \con \eta)  	\cond \Yz=0, \bX \big\} = 0$. 
% \begin{align*}
% 	\EXP^{(\eta)} \big\{ 	s_{0|AX} (\Yz \cond 0 , \bX \con \eta)  	\cond A=0, \bX \big\} = 0 \ , \quad \quad
% 	\EXP^{(\eta)} \big\{ 	s_{A|0X} (A \cond 0 , \bX \con \eta)  	\cond \Yz=0, \bX \big\} = 0  \ .
% \end{align*}
From the AIPW property of $\Psi$ that is shown in Section \ref{sec:supp-OR-DR}, we find \eqref{eq-EIF-T1} and \eqref{eq-EIF-T2} are zero. Therefore, at $\eta^*$, we find the following restriction holds for $\alpha_1(y,\bx \con \eta^*)$: For any function ${\mathfrak{m}}(\Yz,\bX)$, we have
\begin{align}                   \label{eq-restriction}
&
\overline{\EXP}_{\SPorecetastar} \big[
s_{0} (\bO_{0} \con \eta^*) \Psi_{\SPorecetastar} ( \bO_{0} \con \alpha_1(\eta^*) , \potf{0 | A X}^* , \potf{A | 0 X}^* , {\mathfrak{m}}) 
\big]
\nonumber
\\
&
= 
\overline{\EXP}_{\SPorecetastar} \left[
\begin{array}{l} 
\big[ {\mathfrak{m}}(\Yz,\bX) - \overline{\EXP}_{{\SPorecetastar} , 0 | AX}  \big\{ {\mathfrak{m}}(\Yz, \bX) \cond A=0, \bX \big\} \big] 
\\
\displaystyle{ \times \bigg\{ \frac{A s_\alpha(\Yz,\bX \con \eta^*) }{\alpha_1(\Yz,\bX \con \eta^*)} \bigg\}
\big\{ A- \EXP_{A|0X} ( A \cond \Yz=0, \bX) \big\}
}
\end{array} 
\right] \ . 
\end{align}
We choose ${\mathfrak{m}}(\Yz,\bX)$ so that 
\begin{align*}
& 
{\mathfrak{m}}(\Yz,\bX) - \overline{\EXP}_{{\SPorecetastar} , 0 | AX}  \big\{ {\mathfrak{m}}(\Yz, \bX) \cond A=0, \bX \big\} 
= 
\frac{ \potf{1A|X}^*(\Yz , 1 \cond  \bX) }{\potf{0A|X}^*(\Yz , 0 \cond \bX)}
\frac{  \big\{ \mathcal{G}(\Yz) - \mu^*(\bX) \big\} }{\Pr(A=1) \Pr(A=1 \cond \Yz=0, \bX)}
\ .
\end{align*}
Note that we can choose such ${\mathfrak{m}}$ because it satisfies the conditional mean restriction:
\begin{align*}
& 
\overline{\EXP}_{\SPorecetastar} \bigg[
\frac{ \potf{1A|X}^*(\Yz, 1  \cond \bX) }{\potf{0A|X}^*(\Yz , 0 \cond \bX)} \big\{ \mathcal{G}(\Yz) - \mu^*(\bX) \big\} \, \bigg| \, A=0, \bX \Bigg]
\\
& 
=  \EXP \bigg[
\frac{ \potf{1A|X}^*(\Yz, 1  \cond \bX) }{\potf{0A|X}^*(\Yz , 0 \cond \bX)} \big\{ \mathcal{G}(\Yz) - \mu^*(\bX) \big\} \, \bigg| \, A=0, \bX \Bigg]
\\
& =
\int_{\SPz}
\frac{ \potf{1A|X}^*(y , 1 \cond \bX) }{\potf{0A|X}^*(y , 0  \cond \bX)} \big\{ \mathcal{G}(y) - \mu^*(\bX) \big\} \potf{0|AX}^*(y \cond 0 , \bX) \, dy 
\\
& = 
\frac{\Pr(A=1 \cond \bX)}{\Pr(A=0 \cond \bX)}
\int_{\SPorecetastar}
\potf{1|AX}^*(y \cond 1, \bX) 
\big\{ \mathcal{G}(y) - \mu^*(\bX) \big\} \, dy
\\
& = 
\frac{\Pr(A=1 \cond \bX)}{\Pr(A=0 \cond \bX)} 
\Big\{
\mu^*(\bX) - \mu^*(\bX)  \Big\}
= 0 \ .
\end{align*}
The first identity holds because $\potf{1|AX}^*(y \cond 1,\bx)=0$ for $(y,\bx) \in \SPorecetastar^c$. The third identity holds because $\SPot$, the support of $\potf{1|AX}^*$, is equal to $\SPorecetastar$ under Assumptions \ref{assumption:support-1}, \ref{assumption:support-2}, \ref{assumption:consistency}-\ref{assumption:OREC}. The third identity holds from the definition of $\mu^*(\bX) = \EXP \big\{ \mathcal{G} (\pYo) \cond A=1, \bX \big\}$. This choice of ${\mathfrak{m}}$ yields $\Psi_{\SPorecetastar} ( \bO_{0} \con \alpha_1(\eta^*) , \potf{0 | A X}^* , \potf{A | 0 X}^* , {\mathfrak{m}}) $ as follows:
\begin{align*}
&
\Psi_{\SPorecetastar} ( \bO_{0} \con \alpha_1(\eta^*) , \potf{0 | A X}^* , \potf{A | 0 X}^* , {\mathfrak{m}})
\\
& = 
\frac{1}{\Pr(A=1) }
\frac{ \potf{1A|X}^*(\Yz, 1  \cond \bX) }{\potf{0A|X}^*(\Yz , 0 \cond \bX)} \big\{ \mathcal{G}(\Yz) - \mu^*(\bX) \big\}
\big\{ \alpha_1 (\Yz, \bX \con \eta^*) \big\}^{-A}
\frac{ 
\big\{ A - \Pr(A=1 \cond \Yz=0, \bX ) \big\} }{\Pr(A=1 \cond \Yz=0, \bX)
} \ .
\end{align*}
Note that $\alpha_1(y,\bx \con \eta^*) = \alpha_0^*(y,\bx)$ over $(y,\bx) \in \SPorecetastar$. Therefore, at $A=1$, we obtain
\begin{align*}
&
\Psi_{\SPorecetastar} ( \Yz,A=1,\bX \con \alpha_1(\eta^*) , \potf{0 | A X}^* , \potf{A | 0 X}^* , {\mathfrak{m}})
\\
& = 
\frac{1}{\Pr(A=1) }
\frac{ \potf{1A|X}^*(\Yz, 1  \cond \bX) }{\potf{0A|X}^*(\Yz , 0 \cond \bX)} 
\big\{ \mathcal{G}(\Yz) - \mu^*(\bX) \big\}
\frac{ \Pr(A=0 \cond \Yz, \bX) }{ \Pr(A=1 \cond \Yz, \bX) }
\\
& = 
\frac{1}{\Pr(A=1) }
\frac{ \potf{1A|X}^*(\Yz, 1  \cond \bX) }{\potf{0A|X}^*(\Yz , 1 \cond \bX)} 
\big\{ \mathcal{G}(\Yz) - \mu^*(\bX) \big\}
\end{align*}
At $A=0$, we obtain
\begin{align*}
&
\Psi_{\SPorecetastar} ( \Yz,A=0,\bX \con \alpha_1(\eta^*) , \potf{0 | A X}^* , \potf{A | 0 X}^* , {\mathfrak{m}})
= 
- \frac{1}{\Pr(A=1) }
\frac{ \potf{1A|X}^*(\Yz, 1  \cond \bX) }{\potf{0A|X}^*(\Yz , 0 \cond \bX)} 
\big\{ \mathcal{G}(\Yz) - \mu^*(\bX) \big\}
\ .
\end{align*}
Therefore, we find $\Psi_{\SPorecetastar} ( \bO_{0} \con \alpha_1(\eta^*) , \potf{0 | A X}^* , \potf{A | 0 X}^* , {\mathfrak{m}})$ is equivalent to $\InfFtAug(\bO_0)$:
\begin{align*}
\Psi_{\SPorecetastar} ( \bO_{0} \con \alpha_1(\eta^*) , \potf{0 | A X}^* , \potf{A | 0 X}^* , {\mathfrak{m}})
& =
\frac{2A -1 }{\Pr(A=1) }\underbrace{ \frac{\potf{1A|X}^*(\Yz,1 \cond \bX)}{\potf{0A|X}^*(\Yz,A \cond \bX)} }_{=R^*(\Yz,A,\bX)}
\big\{ \mathcal{G}(\Yz) - \mu^*(\bX) \big\} 
= 
\InfFtAug(\bO_0) \ .
\end{align*}  
Therefore, \eqref{eq-restriction} becomes
\begin{align*}
& \overline{\EXP}_{\SPorecetastar} \big[
s_{0} (\bO_{0} \con \eta^*) \Psi_{\SPorecetastar} ( \bO_{0} \con \alpha_1(\eta^*) , \potf{0 | A X}^* , \potf{A | 0 X}^* , {\mathfrak{m}}) 
\big]
\\
& = 
\overline{\EXP}_{\SPorecetastar} \left[
\begin{array}{l} 
\big[ {\mathfrak{m}}(\Yz,\bX) - \overline{\EXP}_{{\SPorecetastar} , 0 | AX}  \big\{ {\mathfrak{m}}(\Yz, \bX) \cond A=0, \bX \big\} \big] 
\\
\displaystyle{ \times \bigg\{ \frac{A s_\alpha(\Yz,\bX \con \eta^*) }{\alpha_1(\Yz,\bX \con \eta^*)} \bigg\}
\big\{ A- \EXP_{A|0X} ( A \cond \Yz=0, \bX) \big\}
}
\end{array} 
\right]
\\
\Rightarrow \quad
& 
\EXP 
\big\{
s_{0} (\bO_{0} \con \eta^*) \InfFtAug(\bO_0)
\big\}
\\
& =
\frac{1}{\Pr(A=1)}
\EXP\Bigg[
\frac{ \potf{1A|X}^*(\Yz , 1 \cond  \bX) }{\potf{0A|X}^*(\Yz , 0 \cond \bX)} 
\frac{\big\{ \mathcal{G}(\Yz) - \mu^*(\bX) \big\}}{\Pr(A=1 \cond \Yz=0, \bX)} \bigg\{ \frac{A s_\alpha(\Yz,\bX \con \eta^*) }{\alpha_1(\Yz,\bX \con \eta^*)} \bigg\}
\big\{ A- \EXP_{A|0X} ( A \cond \Yz=0, \bX) \big\}
\Bigg]
\end{align*}
The last term is represented as follows:
\begin{align}
&
\frac{1}{\Pr(A=1)}
\EXP\Bigg[
\frac{ \potf{1A|X}^*(\Yz , 1 \cond  \bX) }{\potf{0A|X}^*(\Yz , 0 \cond \bX)} 
\frac{\big\{ \mathcal{G}(\Yz) - \mu^*(\bX) \big\}}{\Pr(A=1 \cond \Yz=0, \bX)} \bigg\{ \frac{A s_\alpha(\Yz,\bX \con \eta^*) }{\alpha_1(\Yz,\bX \con \eta^*)} \bigg\}
\big\{ A- \EXP_{A|0X} ( A \cond \Yz=0, \bX) \big\}
\Bigg]
\nonumber
\\
& = 
\EXP\Bigg[
\frac{ \potf{1A|X}^*(\Yz , 1 \cond  \bX) }{\potf{0A|X}^*(\Yz , 0 \cond \bX)} 
\frac{\big\{ \mathcal{G}(\Yz) - \mu^*(\bX) \big\}}{\Pr(A=1 \cond \Yz=0, \bX)} \bigg\{ \frac{ s_\alpha(\Yz,\bX \con \eta^*) }{\alpha_1(\Yz,\bX \con \eta^*)} \bigg\}
\Pr(A=0 \cond \Yz=0, \bX) \, \Bigg| \, A=1
\Bigg]
\nonumber
\\
& = 
\EXP\Bigg[
\frac{ \potf{1A|X}^*(\Yz , 1 \cond  \bX) }{\potf{0A|X}^*(\Yz , 0 \cond \bX)} 
s_\alpha(\Yz,\bX \con \eta^*)
\frac{\Pr(A=0 \cond \Yz,\bX)}{\Pr(A=1 \cond \Yz,\bX)}
\big\{ \mathcal{G}(\Yz) - \mu^*(\bX) \big\}
\, \Bigg| \, A=1
\Bigg]
\nonumber
\\
& = 
\EXP\Bigg[
\frac{ \potf{1|AX}^*(\Yz \cond 1, \bX) }{\potf{0|AX}^*(\Yz \cond 1 , \bX)} 
s_\alpha(\Yz,\bX \con \eta^*)
\big\{ \mathcal{G}(\Yz) - \mu^*(\bX) \big\}
\, \Bigg| \, A=1
\Bigg]
\nonumber
\\
& =
\EXP\Bigg[
\EXP \Big[
s_\alpha(\pYo,\bX \con \eta^*)
\big\{ \mathcal{G}(\pYo) - \mu^*(\bX) \big\}
\, \Big| \, A=1 , \bX \Big]
\, \Bigg| \, A=1
\Bigg]
\nonumber
\\
& =
\iint_{\SPorecetastar} 
s_\alpha(y, \bx \con \eta^*)
\big\{ \mathcal{G}(y) - \mu^*(\bx) \big\} 
\potf{1X}^*(y , \bx \cond 1) \, d(y,\bx) 
\nonumber
\\
& =
\frac{1}{\Pr(A=1)}
\iint_{\SPorecetastar} 
\underbrace{ \frac{ \potf{1AX}^*(y , 1, \bx) }{\potf{1AX}^*(y , 0, \bx)} }_{ = \beta_1(\bx \con \eta^*) \alpha_1(y,\bx \con \eta^*)}
s_\alpha(y, \bx \con \eta^*)
\big\{ \mathcal{G}(y) - \mu^*(\bx) \big\} 
\potf{1X}^*(y , 0, \bx) \, d(y,\bx) 
\nonumber
\\
& = 
\EXP \bigg[ \frac{(1-A) s_\alpha (\Yo,\bX \con \eta^*)  \beta_1^*(\bX) \alpha_1^*(\Yo,\bX)}{\Pr(A=1)}		
\Big\{ \mathcal{G}(\Yo) - \mu^*(\bX) \Big\}
\bigg] \ .
\label{eq-RHS 2}
\end{align}
Therefore, we establish \eqref{eq-LHS-RHS}, and \eqref{eq-diffpara} as well by combining \eqref{eq-RHS}, \eqref{eq-LHS}, and \eqref{eq-LHS-RHS}. This concludes that the conjectured EIF $\InfFt_{0}^*(\bO)$ is the EIF for $\tau_{0}^*(\mathcal{G})$ in model $\mathcal{M}_{\OREC}$. This concludes the proof.

\subsection{Proof of Theorem \ref{thm-AsympNormal}}         \label{sec:supp:thm-AsmptoticNormal}

To reuse some results again in the other proofs, we establish more general results for some quantities. In particular, let $\MM(y,\bx \con \btheta)$ be a uniformly bounded function with finite-dimensional parameter $\btheta$, and let $\mu_\MM^*$ and $\widehat{\mu}_\MM\LSS$ be 
\begin{align*}
& \mu_\MM^*(\bX \con \btheta) = \frac{ \int_{\SPo} \MM(y,\bX \con \btheta) {\alpha}_1^*(y,\bX) \potf{1}^*(y \cond 0,\bX) \, dy }{ \int_{\SPo} \alpha_1^*(y,\bX) \potf{1}^*(y \cond 0,\bX) \, dy } \ ,
\\
& \widehat{\mu}_\MM \LSS (\bX \con \btheta) = \frac{ \int_{\SPo} \MM(y,\bX \con \btheta) \widehat{\alpha}_1\LSS(y,\bX) \hpotf{1}\LSS(y \cond 0,\bX) \, dy }{ \int_{\SPo} \widehat{\alpha}_1\LSS(y,\bX) \hpotf{1}\LSS(y \cond 0,\bX) \, dy } \ .
\end{align*}

To facilitate the proof, we introduce the following propositions.
\begin{proposition}
The convergence rate of $\widehat{\mu}_\MM\LSS$ is 
\begin{align}				\label{eq-rate-mu1}
\big\| \mu_\MM^* (\btheta) -  \widehat{\mu}_\MM \LSS (\btheta) \big\| _{P,2}^2
\precsim \big\| \alpha_1^* - \widehat{\alpha}_1\LSS  \big\|_{P,2}^2 + 
\big\| \potf{1}^* - \hpotf{1}\LSS  \big\|_{P,2}^2 \ .
\end{align}
\end{proposition}

\begin{proof}

$ \widehat{\mu}_\MM\LSS(\bX \con \btheta)  - \mu_\MM^*(\bX \con \btheta) $ is represented as 
\begin{align*}
&
\big\|  \widehat{\mu}_\MM\LSS(\bX \con \btheta)  - \mu_\MM^*(\bX \con \btheta)  \big\| 
\\
&
=
\bigg\|
    \frac{ \int_{\SPo} \MM(y,\bX \con \btheta) \widehat{\alpha}_1\LSS(y,\bX) \hpotf{1}\LSS(y \cond 0,\bX) \, dy }{ \int_{\SPo} \widehat{\alpha}_1\LSS(y,\bX) \hpotf{1}\LSS(y \cond 0,\bX) \, dy }
    -
    \frac{ \int_{\SPo} \MM(y,\bX \con \btheta) \alpha_1^* (y,\bX) \potf{1}^* (y \cond 0,\bX) \, dy }{ \int_{\SPo} \alpha_1^*(y,\bX) \potf{1}^*(y \cond 0,\bX) \, dy }
\bigg\|
\\
&
\begin{array}{ll}
\leq
&
\overbrace{
\textstyle{ \big\{ \int_{\SPo} \widehat{\alpha}_1\LSS(y,\bX) \hpotf{1}\LSS(y \cond 0,\bX) \, dy \big\}^{-1}
\big\{ \int_{\SPo} \alpha_1^*(y,\bX) \potf{1}^*(y \cond 0,\bX) \, dy \big\}^{-1} }
}^{\leq C}
\\
&
\times 
\left\|
\begin{array}{l}
\big\{ \int_{\SPo} \alpha_1^*(y,\bX) \potf{1}^*(y \cond 0,\bX) \, dy  \big\} \big\{ \int_{\SPo}  \MM(y,\bX \con \btheta) \widehat{\alpha}_1\LSS(y,\bX) \hpotf{1}\LSS(y \cond 0,\bX) \, dy \big\}
\\
    - 	\big\{  \int_{\SPo} \widehat{\alpha}_1\LSS(y,\bX) \hpotf{1}\LSS(y \cond 0,\bX) \, dy  \big\} \big\{ \int_{\SPo}  \MM(y,\bX \con \btheta) \alpha_1^*(y,\bX) \potf{1}^*(y \cond 0,\bX) \, dy \big\}
\end{array}
    \right\|
    \end{array}
    \\
& 
\begin{array}{ll}
\precsim
&
\big\| \int_{\SPo} \alpha_1^*(y,\bX) \potf{1}^*(y \cond 0,\bX) - \widehat{\alpha}_1\LSS(y,\bX) \hpotf{1}\LSS(y \cond 0,\bX) \, dy  \big\|
\\
&
\quad \times
\big\| \int_{\SPo}  \MM(y,\bX \con \btheta) \big\{ \widehat{\alpha}_1\LSS(y,\bX) \hpotf{1}\LSS(y \cond 0,\bX) + \alpha_1^*(y,\bX) \potf{1}^*(y \cond 0,\bX) \big\} \, dy \big\|
    \\
    &
    + \big\| \int_{\SPo} \alpha_1^*(y,\bX) \potf{1}^*(y \cond 0,\bX) + \widehat{\alpha}_1\LSS(y,\bX) \hpotf{1}\LSS(y \cond 0,\bX) \, dy  \big\|
\\
&
\quad \times
\big\| \int_{\SPo}  \MM(y,\bX \con \btheta) \big\{ \widehat{\alpha}_1\LSS(y,\bX) \hpotf{1}\LSS(y \cond 0,\bX) - \alpha_1^*(y,\bX) \potf{1}^*(y \cond 0,\bX) \big\} \, dy \big\|
\end{array}
    \\
    &
\begin{array}{ll}
\precsim &
\EXP \LSS \Big\{
\big\| \alpha_1^*(\Yo,\bX) - \widehat{\alpha}_1\LSS(\Yo,\bX) \big\|
+		  \big\|  \potf{1}^*(\Yo \cond 0,\bX) - \hpotf{1}\LSS(\Yo \cond 0,\bX)  \big\|
\, \Big| \, A=0, \bX
\Big\}
\ .
\end{array}
\end{align*} 
To establish the last result, we used Assumption \HL{A3} to bound the following quantities:
\begin{align*}
& 
\overbrace{
\bigg\| \int_{\SPo} \MM(y,\bX \con \btheta) \big\{ \widehat{\alpha}_1\LSS(y,\bX) \hpotf{1}\LSS(y \cond 0,\bX) + \alpha_1^*(y,\bX) \potf{1}^*(y \cond 0,\bX) \big\} \, dy \bigg\|
}^{ \leq C' }
\\
& \hspace*{2cm} \times
\bigg\| \int_{\SPo} \alpha_1^*(y,\bX) \potf{1}^*(y \cond 0,\bX) - \widehat{\alpha}_1\LSS(y,\bX) \hpotf{1}\LSS(y \cond 0,\bX) \, dy  \bigg\|
\\
&
\precsim
\bigg\| \int_{\SPo}  \big\{ \alpha_1^*(y,\bX) - \widehat{\alpha}_1\LSS(y,\bX) \big\} \big\{ \potf{1}^*(y \cond 0,\bX) + \hpotf{1}\LSS(y \cond 0,\bX) \big\}
\, dy  \bigg\|
\\
& \hspace*{2cm}
+
\bigg\|
\int_{\SPo}  \big\{ \alpha_1^*(y,\bX) + \widehat{\alpha}_1\LSS(y,\bX) \big\} \big\{ \potf{1}^*(y \cond 0,\bX) - \hpotf{1}\LSS(y \cond 0,\bX) \big\}
\, dy  \bigg\|
\\
&
\precsim
\int_{\SPo} \Big\{ \big\| \alpha_1^*(y,\bX) - \widehat{\alpha}_1\LSS(y,\bX) \big\| +  \big\| \potf{1}^*(y \cond 0,\bX) - \hpotf{1}\LSS(y \cond 0,\bX) \big\| \Big\}
\potf{1}^*(y \cond 0, \bX) \, dy 
\\
& 
\leq
\EXP \LSS \Big\{
\big\| \alpha_1^*(\Yo,\bX) - \widehat{\alpha}_1\LSS(\Yo,\bX) \big\|
+
\big\|  \potf{1}^*(\Yo \cond 0,\bX) - \hpotf{1}\LSS(\Yo \cond 0,\bX)  \big\|
\, \Big| \, A=0, X
\Big\}
\end{align*}
and
\begin{align*}
& 
\overbrace{
\bigg\| \int_{\SPo}  \big\{ \widehat{\alpha}_1\LSS(y,\bX) \hpotf{1}\LSS(y \cond 0,\bX) + \alpha_1^*(y,\bX) \potf{1}^*(y \cond 0,\bX) \big\} \, dy \bigg\|
}^{\leq C'}
\\
& \hspace*{2cm} \times 
\bigg\| \int_{\SPo} \MM(y,\bX \con \btheta) \big\{ \alpha_1^*(y,\bX) \potf{1}^*(y \cond 0,\bX) - \widehat{\alpha}_1\LSS(y,\bX) \hpotf{1}\LSS(y \cond 0,\bX) \big\} \, dy  \bigg\|
\\
& 
\precsim
\bigg\| \int_{\SPo} \MM(y,\bX \con \btheta) \big\{ \alpha_1^*(y,\bX) - \widehat{\alpha}_1\LSS(y,\bX) \big\} \big\{ \potf{1}^*(y \cond 0,\bX) + \hpotf{1}\LSS(y \cond 0,\bX) \big\}
\, dy  \bigg\|
\\
& \hspace*{2cm}
+
\bigg\| 
\int_{\SPo} \MM(y,\bX \con \btheta) \big\{ \alpha_1^*(y,\bX) + \widehat{\alpha}_1\LSS(y,\bX) \big\} \big\{ \potf{1}^*(y \cond 0,\bX) - \hpotf{1}\LSS(y \cond 0,\bX) \big\}
\, dy  \bigg\|
\\
&
\precsim 
\int_{\SPo} \big\| \MM(y,\bX \con \btheta) \big\| \cdot \Big\{
\big\| \alpha_1^*(y,\bX) - \widehat{\alpha}_1\LSS(y,\bX) \big\|
+
\big\| \potf{1}^*(y \cond 0,\bX) - \hpotf{1}\LSS(y \cond 0,\bX) \big\|
\Big\} \cdot
\potf{1}^*(y \cond 0, \bX) \, dy
\\
& 
\leq 
\EXP \LSS \Big[
\big\| \MM(\Yo,\bX \con \btheta) \big\| 
\cdot \Big\{ 
\big\| \alpha_1^*(\Yo,\bX) - \widehat{\alpha}_1\LSS(\Yo,\bX) \big\| +
\big\|  \potf{1}^*(\Yo \cond 0,\bX) - \hpotf{1}\LSS(\Yo \cond 0,\bX)  \big\|
\Big\}
\, \Big| \, A=0, X
\Big]
\\
& 
\precsim 
\EXP \LSS \Big\{
\big\| \alpha_1^*(\Yo,\bX) - \widehat{\alpha}_1\LSS(\Yo,\bX) \big\|
+
\big\|  \potf{1}^*(\Yo \cond 0,\bX) - \hpotf{1}\LSS(\Yo \cond 0,\bX)  \big\|
\, \Big| \, A=0, X
\Big\} \ .
\end{align*}
Consequently, \eqref{eq-rate-mu1} is established: 
\begin{align*}
&
\big\|  \widehat{\mu}_\MM\LSS(\bX \con \btheta) - \mu_\MM^*(\bX \con \btheta)  \big\|_{P,2}^2
\\
& = 
\EXP \LSS \Big\{ \big\|  \widehat{\mu}_\MM\LSS(\bX \con \btheta)  - \mu_\MM^*(\bX \con \btheta)  \big\|^2 \, \Big| \, A=0 \Big\}
\\
&
=
\EXP \LSS \Bigg[
\bigg\|
    \frac{ \int_{\SPo} \MM(y,\bX \con \btheta) \widehat{\alpha}_1\LSS(y,\bX) \hpotf{1}\LSS(y \cond 0,\bX) \, dy }{ \int_{\SPo} \widehat{\alpha}_1\LSS(y,\bX) \hpotf{1}\LSS(y \cond 0,\bX) \, dy }
    -
    \frac{ \int_{\SPo} \MM(y
    ,\bX \con \btheta) \alpha_1^*(y,\bX) \potf{1}^*(y \cond 0,\bX) \, dy }{ \int_{\SPo} \alpha_1^*(y,\bX) \potf{1}^*(y \cond 0,\bX) \, dy }
\bigg\| ^2
    \, \Bigg| \, A=0 \Bigg]
\\
&
\precsim
\EXP \LSS
\Bigg[
\begin{array}{l}
\EXP \LSS \big\{
\big\| \alpha_1^*(\Yo,\bX) - \widehat{\alpha}_1\LSS(\Yo,\bX) \big\|
\cond A=0, \bX
\big\}^2
\\
+
\EXP \LSS \big\{
\big\|  \potf{1}^*(\Yo \cond 0,\bX) - \hpotf{1}\LSS(\Yo \cond 0,\bX)  \big\|
\cond A=0, \bX
\big\}^2
\end{array}		\Bigg|  \,  A=0 \Bigg]
        \\
&
\precsim
\EXP \LSS
\Bigg[
\begin{array}{l}
\EXP \LSS \big\{
\big\| \alpha_1^*(\Yo,\bX) - \widehat{\alpha}_1\LSS(\Yo,\bX) \big\|^2
\cond A=0, \bX \big\}
\\
+
\EXP \LSS \big\{
\big\|  \potf{1}^*(\Yo \cond 0,\bX) - \hpotf{1}\LSS(\Yo \cond 0,\bX)  \big\|^2
\cond A=0, \bX  \big\}
\end{array} \Bigg| A=0 \Bigg]
\\
&
= 
\EXP \LSS \Big\{
\big\| \alpha_1^*(\Yo,\bX) - \widehat{\alpha}_1\LSS(\Yo,\bX) \big\|^2 \cond A=0 \big\}
+
\big\|  \potf{1}^*(\Yo \cond 0,\bX) - \hpotf{1}\LSS(\Yo \cond 0,\bX)  \big\|^2 \, \Big| \,  A=0 \Big\}
\\
& 
\precsim \big\| \alpha_1^* (\Yo,\bX) - \widehat{\alpha}_1 (\Yo,\bX) \big\|_{P,2}^2 + 
\big\|  \potf{1}^*(\Yo \cond 0,\bX) - \hpotf{1}\LSS(\Yo \cond 0,\bX)  \big\|_{P,2}^2 \ .
\end{align*} 

\end{proof}

\begin{proposition}
The following equalities hold for $t=0,1$ and $(y,\bX) \in \mathcal{S}_t(0)$:
\begin{align}
&
\EXP \big\{ 1-A \cond \potY{0}{t}=y, \bX \big\}
=
\frac{\EXP  \big\{ A \cond \potY{0}{t}=y,\bX \big\}}{\beta_t^*(\bX) \alpha_1^*(y,\bX)}
\ , 
\label{eq-aux01}
\\
&
\beta_t^*(\bX) \alpha_t^*(y,\bX) \potf{t}^* (y,0 \cond \bX)
=
\potf{t}^* (y,1 \cond \bX) \ , 
\label{eq-aux02}
\\
&
\beta_t^*(\bX) \alpha_t^*(y,\bX) \frac{\Pr(A=0 \cond \bX)}{\Pr(A=1 \cond \bX)} 
=
\frac{\potf{t}^*(y \cond 1, \bX)}{\potf{t}^*(y \cond 0, \bX)} \ .
\label{eq-aux03}
\end{align}
\end{proposition}

\begin{proof}
Note that 
\begin{align*}
\beta_t^*(\bX) \alpha_t^*(y,\bX)
=
\frac{ \Pr \big\{ A=0 \cond \potY{0}{t}=y,\bX \big\} }{ \Pr \big\{ A=1 \cond \potY{0}{t}=y,\bX \big\} }
=
\frac{ P \big\{ \potY{0}{t}=y,A=0 \cond \bX \big\} }{ P \big\{ \potY{0}{t}=y, A=1 \cond \bX \big\} } \ .
\end{align*}
Therefore, \eqref{eq-aux01} holds as follows:
\begin{align*} 
\beta_{t}^*(\bX) \alpha_t^*(y,\bX) \EXP \big\{ 1-A \cond \potY{0}{t}=y, \bX \big\}
& =
\frac{ \Pr \big\{ A=1 \cond \potY{0}{t}=y,\bX \big\} }{ \Pr \big\{ A=0 \cond \potY{0}{t}=y,\bX \big\} } \Pr \big\{ A=0 \cond \potY{0}{t}=y,\bX \big\}
\\
&
=\Pr \big\{ A=1 \cond \potY{0}{t}=y,\bX \big\} \ .
\end{align*}
Equation \eqref{eq-aux02} is established as follows:
\begin{align*}
\beta_t^*(\bX) \alpha_t^*(y,\bX) P \big\{ \potY{0}{t}=y, A=1 \cond \bX \big\}
& =
\frac{ P \big\{ \potY{0}{t}=y,A=0 \cond \bX \big\} }{ P \big\{ \potY{0}{t}=y, A=1 \cond \bX \big\} }
P \big\{ \potY{0}{t}=y, A=1 \cond \bX \big\}
\\
&
=
P \big\{ \potY{0}{t}=y, A=0 \cond \bX \big\} \ .
\end{align*}
Equation \eqref{eq-aux03} is trivial from \eqref{eq-aux02}.

\end{proof}

We now return to the proof of Theorem \ref{thm-AsympNormal}. To facilitate the proof, we define $\tau_{n,1}^* = \EXP\{ A \Yo \}$, $\tau_{n,0}^* = \EXP\{ A \pYo\}$, and $\tau_{d}^* = \EXP(A)$. In addition, let $\widehat{\tau}_{n,1} = \AVER(A \Yo)$, $\widehat{\tau}_{n,0} = K^{-1}\sum_{k=1}^{K} \widehat{\tau}_{n,0}^{(k)}$, $ \widehat{\tau}_{n,0}^{(k)} =  \AVER_{\mathcal{I}_k} \big\{ \widehat{\uncInfFt}_0\LSS(\bO)\big\}$, and $\widehat{\tau}_{d} = \AVER(A)$. We will establish that
\begin{align} \label{eq-ssestimator-full}
    \sqrt{N}
    \left\{
        \begin{pmatrix}
            \widehat{\tau}_{n,1}
            \\
            \widehat{\tau}_{n,0}
            \\
            \widehat{\tau}_{d}
        \end{pmatrix}
        -
        \begin{pmatrix}
            \tau_{n,1}^*
            \\
            \tau_{n,0}^*
            \\
            \tau_{d}^*
        \end{pmatrix}
    \right\}
    =
    \frac{1}{N}
    \sum_{i=1}^{N}
    \left\{ 
        \begin{pmatrix}
            A_iY_{1,i} - \tau_{n,1}^*
            \\
            \phi_0^*(\bO_i) - \tau_{n,0}^*
            \\
            A_i - \tau_{d}^*
        \end{pmatrix}
    \right\}
    +
    o_P(1) \ .
\end{align}
Therefore, we have
\begin{align*}
        \sqrt{N}
    \left\{
        \begin{pmatrix}
            \widehat{\tau}_{n,1}
            \\
            \widehat{\tau}_{n,0}
            \\
            \widehat{\tau}_{d}
        \end{pmatrix}
        -
        \begin{pmatrix}
            \tau_{n,1}^*
            \\
            \tau_{n,0}^*
            \\
            \tau_{d}^*
        \end{pmatrix}
    \right\}
    \stackrel{D}{\rightarrow}
    N (0, \Sigma^*) \ , \quad 
    \Sigma^* = 
    \VAR
    \left\{ 
        \begin{pmatrix}
            A_iY_{1,i} - \tau_{n,1}^*
            \\
            \phi_0^*(\bO_i) - \tau_{n,0}^*
            \\
            A_i - \tau_{d}^*
        \end{pmatrix}
    \right\} \ .
\end{align*}
We also show that a consistent estimator for $\Sigma^*$ is $\widehat{\Sigma} = K^{-1} \sum_{k=1}^{K} \widehat{\Sigma}^{(k)}$, where 
\begin{align*}
    \widehat{\Sigma}^{(k)}
    =
    \AVER_{\mathcal{I}_k}
    \left\{ 
\begin{pmatrix}
            A_iY_{1,i} - \widehat{\tau}_{n,1}
            \\
            \phi_0^*(\bO_i) - \widehat{\tau}_{n,0}
            \\
            A_i - \widehat{\tau}_{d}
        \end{pmatrix}^{\otimes 2}
    \right\} \ .
\end{align*}
Since $\tau^* = \{\tau_{n,1}^* - \tau_{n,0}^* \}/\tau_d^*$ and  $\widehat{\tau} = \{ \widehat{\tau}_{n,1} - \widehat{\tau}_{n,0} \} / \widehat{\tau}_d$, we have
\begin{align*}
&
\sqrt{N} 
\Big\{ \widehat{\tau} - \tau^* \Big\}
\\
 &
 =
 \sqrt{N}
 \frac{	\big\{ \widehat{\tau}_{n,1} - \widehat{\tau}_{n,0} \big\} \tau_d^* - \tau^* \widehat{\tau}_d }{\widehat{\tau}_d \tau_d^* } 
 \\
 &
 =
 \sqrt{N}
 \frac{ 1 }{2 \widehat{\tau}_{d} \tau_d^* }
 \left[
    \begin{array}{l} 		
 \big[
 \big\{
    \widehat{\tau}_{n,1} - \widehat{\tau}_{n,0} \big\} - 
    \big\{ \tau_{n,1}^* - \tau_{n,0}^* \big\}
 \big]
 \big\{
    \widehat{\tau}_{d} + \tau_d^*
 \big\}
 \\
 -
 \big[
 \big\{
    \widehat{\tau}_{n,1} - \widehat{\tau}_{n,0} \big\} + 
    \big\{ \tau_{n,1}^* - \tau_{n,0}^* \big\}
 \big]
 \big\{
 \widehat{\tau}_{d} - \tau_d^* 
   \big\}
    \end{array}
 \right]
  \\
 &
 =
 \sqrt{N}
 \frac{ 1 + o_P(1) }{2 \{ \tau_d^*  \}^2 } 
 \left[
    \begin{array}{l}
        \big[
 \big\{
    \widehat{\tau}_{n,1} - \widehat{\tau}_{n,0} \big\} - 
    \big\{ \tau_{n,1}^* - \tau_{n,0}^* \big\}
 \big]
        \big\{ 2 \tau_d^* + o_P(1) \big\}
        \\
        -
        \big[ 2 \big\{ \tau_{n,1}^* - \tau_{n,0}^* \big\} + o_P(1) \big]
        \big\{ \widehat{\tau}_{d} - \tau_d^* \big\}
    \end{array}
 \right] 
  \\
 &
 =
 \frac{1}{\sqrt{N}}
\sum_{i=1}^{N}
 \Bigg[
    \frac{A_i Y_{1,i} - \tau_{n,1}^* - \phi_{0}^*(\bO_i) + \tau_{n,0}^*}{\tau_d^* } 
    -
    \frac{\tau_{n,1}^* - \tau_{n,0}^*}{\{ \tau_d^*   \}^2} \big\{A_i - \tau_d^* \big\}
 \Bigg]
 +
 o_P(1) 
  \\
 &
 =
 \frac{1}{\sqrt{N}}
\sum_{i=1}^{N}
 \Bigg\{
    \frac{ A_i Y_{1,i} - \phi_{0}^*(\bO_i) - A_i \tau^*}{\tau_d^* } 
 \Bigg\}
   \\
 &
 =
 \frac{1}{\sqrt{N}}
\sum_{i=1}^{N}
 \InfFt^* (\bO_i)
 +
 o_P(1) \ ,
\end{align*} 
which is asymptotically normal with the limiting distribution $N \big( 0 ,  \VAR \big\{ \InfFt^*(\bO) \big\} \big)$. In order to establish \eqref{eq-ssestimator-full}, it suffices to show 
\begin{align}						\label{eq-ssestimator}
\big| \mathcal{I}_k \big|^{1/2}
\Big\{
\widehat{\tau}_{n,0}^{(k)} - \tau_{n,0}^*
\Big\}
=
\frac{1}{\big| \mathcal{I}_k \big|^{1/2}}
\sum_{ i \in \mathcal{I}_k } \big\{ \phi_0^*(\bO_i) - \tau_{n,0}^* \big\} + o_P(1) \ .
\end{align} 
In what follows, we establish \eqref{eq-ssestimator}. 

Let $\EMP_{\mathcal{I}_k} (V) = | \mathcal{I}_k |^{-1/2} \sum_{i \in \mathcal{I}_k} \big\{ V_i - \EXP(V_i) \big\} $ be the empirical process of $V_i$ centered by $\EXP(V_i)$. Similarly, let $\EMP_{\mathcal{I}_k}\LSS \big( \widehat{V}\LSS \big) = | \mathcal{I}_k |^{-1/2} \sum_{i \in \mathcal{I}_k} \big\{ \widehat{V}_i\LSS - \EXP\LSS(\widehat{V}\LSS) \big\} $ be the empirical process of $\widehat{V}\LSS$ centered by $\EXP\LSS \{ \widehat{V}\LSS \}$ where $\EXP\LSS (\cdot)$ is the expectation after considering random functions obtained from $\mathcal{I}_k^c$ as fixed functions. The empirical process of $\widehat{\uncInfFt}_0\LSS - \tau_{n,0}^*$ is
\begin{align} 
\big| \mathcal{I}_k \big|^{-1/2}
\sum_{i \in \mathcal{I}_k} \big\{ \widehat{\uncInfFt}_{0}\LSS (\bO_i)  - \tau_{n,0}^* \big\}
= 	&
\  \EMP_{\mathcal{I}_k} \big( \uncInfFt_0^*  - \tau_{n,0}^* \big)
\label{Term1}
\\
& 
+
\big| \mathcal{I}_k \big|^{1/2}
\cdot
\EXP\LSS
\big( \widehat{\uncInfFt}_0\LSS - \uncInfFt_0^* \big) 
\label{Term3}
\\
& +
\EMP_{\mathcal{I}_k}\LSS
\big(
\widehat{\uncInfFt}_0\LSS - \uncInfFt_0^*
\big)
\label{Term2} \ ,
\end{align}
where
\begin{align*}
\EMP_{\mathcal{I}_k} \big( \uncInfFt_0^*  - \tau_{n,0}^* \big)
& =
\EMP_{\mathcal{I}_k} 
\Bigg[
\begin{array}{l}
(1-A)
\beta_{1}^*(\bX) \alpha_1^*(\Yo ,\bX) \big\{ \Yo  - \mu^*(\bX) \big\}
+
A
\mu^*(\bX)
-
\tau_{n,0}^*
\\
+ (2A-1)
R^* (\Yz ,A , \bX)
\big\{
\Yz 
-
\mu^*(\bX)
\big\} 
\end{array}		
\Bigg]
\\
&
=
\big| \mathcal{I}_k \big|^{-1/2}
\sum_{i \in \mathcal{I}_k} \Big\{ \uncInfFt_{0}^*(\bO_i) - \tau_{n,0}^* \Big\}
\\
\EMP_{\mathcal{I}_k}\LSS
\big(
\widehat{\uncInfFt}_0\LSS - \uncInfFt_0^*
\big)
&
=
\EMP_{\mathcal{I}_k}\LSS
\left[
\begin{array}{l}
(1-A)
\widehat{\beta}_{1}\LSS(\bX) \widehat{\alpha}_1\LSS(\Yo ,\bX) \big\{ \Yo  - \widehat{\mu}\LSS(\bX) \big\}
+
A
\widehat{\mu}\LSS(\bX)
\\
-
(1-A)
\beta_1^*(\bX) \alpha_1^* (\Yo ,\bX) \big\{ \Yo  - \mu^*(\bX) \big\}
-
A
\mu^*(\bX)
\\
+ (2A-1)
\widehat{R}\LSS (\Yz ,A ,\bX)
\big\{
\Yz 
-
\widehat{\mu}\LSS (\bX)
\big\} 
\\
- (2A-1)
R^* (\Yz ,A , \bX)
\big\{
\Yz 
-
\mu^*(\bX)
\big\} 
\end{array}		
\right] \ .
\end{align*}
From the derivation below, we find that \eqref{Term3} and \eqref{Term2} are $o_P(1)$, indicating that \eqref{Term1} is asymptotically normal. This implies \eqref{eq-ssestimator} holds.

In the rest of the proof, we show that \eqref{Term3} and \eqref{Term2} are $o_P(1)$ by establish the following more general results:
\begin{align}
& \big| \mathcal{I}_k \big|^{1/2}
\cdot
\EXP\LSS
\big\{ \widehat{\MM}_\EFF\LSS (\bO \con \btheta) - \MM_\EFF^*(\bO \con \btheta) \big\}
= o_P(1)
\label{Term3-General}
\\
& 
\EMP_{\mathcal{I}_k}\LSS
\big\{
\widehat{\MM}_\EFF\LSS (\bO \con \btheta) - \MM_\EFF^*(\bO \con \btheta)
\big\} = o_P(1) \ .
\label{Term2-General}
\end{align}
where $\widehat{\MM}_\EFF\LSS$ and $\MM_\EFF^*$ are given as
\begin{align*}
&
\widehat{\MM}_\EFF \LSS (\bO \con \btheta)
=
\Bigg[
\begin{array}{l}
(1-A)  \widehat{\beta}_1\LSS (\bX) \widehat{\alpha}_1 \LSS(\Yo,\bX)
\big\{ \MM(\Yo,\bX \con \btheta) -  \widehat{\mu}_\MM\LSS(\bX \con \btheta)  \big\}
+
A
\widehat{\mu}_\MM\LSS(\bX \con \btheta) 
\\
+ (2A-1) \widehat{R}\LSS(\Yz,A,\bX) \big\{  \MM(\Yz,\bX \con \btheta) -  \widehat{\mu}_\MM\LSS(\bX \con \btheta)  \big\}
\end{array}	
\Bigg]
\\
&
\MM_\EFF^* (\bO \con \btheta)
=
\Bigg[
\begin{array}{l}
(1-A)
\beta_1^*(\bX) \alpha_1^* (\Yo,\bX)
\big\{ \MM(\Yo,\bX \con \btheta) -  \mu_\MM^*(\bX \con \btheta)  \big\}
+
A
\mu_\MM^*(\bX \con \btheta) 
\\
+ (2A-1) R^*(\Yz,A,\bX) \big\{  \MM(\Yz,\bX \con \btheta) -  \mu_\MM^*(\bX \con \btheta)  \big\}
\end{array}	
\Bigg] \ .
\end{align*}
Note that $\widehat{\uncInfFt}_0\LSS - \uncInfFt_0^*$ is a special case of $ \widehat{\MM}_\EFF\LSS (\bO \con \btheta) - \MM_\EFF^*(\bO \con \btheta)$ with $\Omega(y,\bX \con \btheta) = y - \theta$ where $\btheta$ in the right hand side cancels out.

\begin{itemize}
    \item[(i)] (\textit{Asymptotic Property of \eqref{Term3-General}})
\end{itemize}

Term \eqref{Term3-General} is
\begin{align*}
& \hspace*{-0.5cm} \big| \mathcal{I}_k \big|^{1/2}
\EXP\LSS
\Big\{
\widehat{\MM}_{\EFF}\LSS (\bO \con \btheta)
-
\MM_{\EFF}^* (\bO \con \btheta) 
\Big\}
\\
& \hspace*{-0.5cm} 
= 
\big| \mathcal{I}_k \big|^{1/2}
\EXP\LSS \Big[ 
(1-A)
\widehat{\beta}_1\LSS(\bX) \widehat{\alpha}_1\LSS(\Yo, \bX) \big\{ \MM(\Yo,\bX \con \btheta)  - \widehat{\mu}_\MM \LSS(\bX \con \btheta) \big\}
+
A
\widehat{\mu}_\MM\LSS(\bX\con \btheta)
\tag{T1}	\label{eq-4-1}
\\
& \hspace*{2.5cm} -
\underbrace{
(1-A)
\beta_{1}^*(\bX) \alpha_1(\Yo ,\bX) \big\{  \MM(\Yo,\bX \con \btheta)   - \mu_\MM^*(\bX \con \btheta) \big\}}_{=0}
-
A
\mu_\MM^*(\bX \con \btheta)
\tag{T2}	\label{eq-4-2}
\\
& \hspace*{2.5cm}
+ (2A-1)
\widehat{R}\LSS (\Yz ,A , \bX)
\big\{
\MM(\Yz,\bX \con \btheta)
-
\widehat{\mu}_\MM\LSS(\bX \con \btheta)
\big\} 
\tag{T3}	\label{eq-4-3}
\\
& \hspace*{2.5cm}
-  
\underbrace{
(2A-1)
R^* (\Yz ,A , \bX)
\big\{
\MM(\Yz,\bX \con \btheta)
-
\mu_\MM^*(\bX \con \btheta)
\big\} 
}_{=0}
\Big] \ .
\tag{T4}	\label{eq-4-4}
\end{align*}
Term \eqref{eq-4-1} is 
\begin{align*}
&\eqref{eq-4-1}
\\
& = 
\EXP\LSS
\Big[
(1-A)
\widehat{\beta}_1\LSS(\bX) \widehat{\alpha}_1\LSS(\pYo ,\bX) \big\{ \MM(\pYo,\bX \con \btheta)  - \widehat{\mu}_\MM\LSS(\bX \con \btheta) \big\}
\Big]
+
\EXP\LSS
\big\{ A \widehat{\mu}_\MM\LSS(\bX \con \btheta) \big\}
\\
& = 
\EXP\LSS
\Big[
(1-A)
\big\{
\widehat{\beta}_1\LSS(\bX) \widehat{\alpha}_1\LSS(\pYo ,\bX) 
-
\beta_1^*(\bX) \alpha_1^*(\pYo,\bX)
\big\}
\big\{ \MM(\pYo,\bX \con \btheta)  - \widehat{\mu}_\MM\LSS(\bX \con \btheta) \big\}
\Big]
\\
& \hspace*{2cm}
+
\EXP\LSS \Big[ (1-A) \beta_1^*(\bX) \alpha_1^*(\pYo,\bX) \big\{ \MM(\pYo,\bX \con \btheta)  - \widehat{\mu}_\MM\LSS(\bX \con \btheta) \big\} \Big]
+
\EXP\LSS
\big\{ A  \widehat{\mu}_\MM\LSS(\bX \con \btheta)  \big\} 
\\
& = 
\EXP\LSS
\Big[
(1-A)
\big\{
\widehat{\beta}_1\LSS(\bX) \widehat{\alpha}_1\LSS(\pYo ,\bX) 
-
\beta_1^*(\bX) \alpha_1^*(\pYo,\bX)
\big\}
\big\{ \MM(\pYo,\bX \con \btheta)  - \widehat{\mu}_\MM\LSS(\bX \con \btheta) \big\}
\Big]
\\
& \hspace*{2cm} + \EXP\LSS \big\{ A \MM(\pYo,\bX \con \btheta) \big\}
\end{align*}
Similar to Term \eqref{eq-4-1}, Term \eqref{eq-4-2} is $ - \EXP\LSS \big\{ A \mu_\MM^*(\bX \con \btheta) \big\} = - \EXP\LSS \big\{ A \MM(\pYo,\bX \con \btheta) \big\}$.

Combining the established result, term \eqref{Term3-General} is equivalent to
{
\begin{align}
& \hspace*{-0.5cm}
\eqref{Term3-General}
\nonumber
\\
& \hspace*{-0.5cm}
=
\EXP\LSS
\Big[
(1-A)
\big\{
\widehat{\beta}_1\LSS(\bX) \widehat{\alpha}_1\LSS(\pYo ,\bX) 
-
\beta_1^*(\bX) \alpha_1^*(\pYo,\bX)
\big\}
\big\{ \MM(\pYo,\bX \con \btheta)  - \widehat{\mu}_\MM\LSS(\bX \con \btheta) \big\}
\Big]
\nonumber
\\
& \hspace*{-0.5cm}
+
\EXP \LSS \Big[
A
\widehat{R}\LSS (\Yz , A , \bX)
\big\{
\MM(\Yz,\bX \con {\theta})
-
\widehat{\mu}_\MM\LSS(\bX \con \btheta)
\big\}
\Big]
\nonumber
\\
& \hspace*{-0.5cm} 
-
\EXP \LSS \Big[
(1-A)
\widehat{R}\LSS (\Yz ,A , \bX)
\big\{
\MM(\Yz,\bX \con {\theta})
-
\widehat{\mu}_\MM\LSS(\bX \con \btheta)
\big\}
\Big]	
\nonumber
\\
& \hspace*{-0.5cm}
= 
\EXP\LSS
\Big[
(1-A)
\big\{ \widehat{\beta}_1\LSS(\bX)  - \beta_1^*(\bX) \big\} 
\alpha_1^*(\pYo,\bX)
\big\{ \MM(\pYo,\bX \con \btheta)  - \widehat{\mu}_\MM\LSS(\bX \con \btheta) \big\}
\Big]
\tag{T5} \label{eq-prod1}
\\
& \hspace*{-0.5cm}
+  
\EXP\LSS
\Big[
(1-A)
\widehat{\beta}_1\LSS(\bX) 
\big\{ \widehat{\alpha}_1\LSS(\pYo ,\bX) - \alpha_1^*(\pYo,\bX) 	\big\}
\big\{ \MM(\pYo,\bX \con \btheta)  - \widehat{\mu}_\MM\LSS(\bX \con \btheta) \big\}
\Big]
\tag{T6-1}	\label{eq-prod2-1}
\\
& \hspace*{-0.5cm}
+
\EXP \LSS \Big[
A
\widehat{R}\LSS (\Yz , A , \bX)
\big\{
\MM(\Yz,\bX \con {\theta})
-
\widehat{\mu}_\MM\LSS(\bX \con \btheta)
\big\}
\Big]
\tag{T6-2}	\label{eq-prod2-2}
\\
& \hspace*{-0.5cm}
-
\EXP \LSS \Big[
(1-A)
\widehat{R}\LSS (\Yz ,A , \bX)
\big\{
\MM(\Yz,\bX \con {\theta})
-
\widehat{\mu}_\MM\LSS(\bX \con \btheta)
\big\}
\Big]	
\tag{T6-3}	\label{eq-prod2-3}
\ .
\end{align} }%

Term \eqref{eq-prod1}  is
\begin{align}
& \big\| \eqref{eq-prod1} \big\|
\nonumber
\\
& =
\Big\|
\Pr(A=0)
\EXP \LSS \Big[ \Big\{
\widehat{\beta}_{1}\LSS(\bX) - \beta_{1}^*(\bX)
\Big\}
\alpha_1^*(\pYo,\bX)
\Big\{
\MM(\pYo,\bX \con \btheta)
-
\widehat{\mu}_\MM\LSS(\bX \con \btheta) 
\Big\} \bcond A=0 \Big]
\Big\|
\nonumber
\\
& 
\precsim
\Big\|
\EXP \LSS
\Big[
\Big\{ \widehat{\beta}_{1}\LSS(\bX) - \beta_{1}^*(\bX) \Big\}
\Big\{
\mu_\MM^*(\bX \con \btheta) 
-
\widehat{\mu}_\MM\LSS(\bX \con \btheta) 
\Big\} \bcond A=0 \Big]
\Big\|
\nonumber
\\
&
\leq
\big\| \widehat{\beta}_{1}\LSS(\bX) - \beta_{1}^*(\bX) \big\|_{P,2}
\big\|  \widehat{\mu}_\MM\LSS(\bX \con \btheta)  - \mu_\MM^*(\bX \con \btheta) \big\|_{P,2}
\nonumber
\\
&
\precsim
\big\| \widehat{\beta}_{1}\LSS(\bX) - \beta_{1}^*(\bX) \big\|_{P,2}
\Bigg[
\begin{array}{l}
\big\| \alpha_1^* (\Yo,\bX) - \widehat{\alpha}_1\LSS (\Yo,\bX) \big\|_{P,2}
\\
+
\big\| \potf{1}^* (\Yo \cond 0,\bX) - \hpotf{1}\LSS (\Yo \cond 0,\bX)  \big\|_{P,2} 
\end{array}
\Bigg] \ .
\label{eq-aux04}
\end{align}	
The second inequality is from $\sup_{y \in \SPo} \big\| \alpha_1^*(y,\bX) \big\| < \infty$ and the last line \eqref{eq-aux04} uses \eqref{eq-rate-mu1}.

Next, the conditional expectation of Term \eqref{eq-prod2-1} given $\bX$ is
\begin{align*}
&
\EXP\LSS
\Big[
(1-A)
\widehat{\beta}_1\LSS(\bX) 
\big\{ \widehat{\alpha}_1\LSS(\pYo ,\bX) - \alpha_1^*(\pYo,\bX) 	\big\}
\big\{ \MM(\pYo,\bX \con \btheta)  - \widehat{\mu}_\MM\LSS(\bX \con \btheta) \big\}
\bcond \bX
\Big]
\\
& =
\widehat{\beta}_1\LSS(\bX) 
\Pr(A=0 \cond \bX)
\EXP\LSS
\Big[
\big\{ \widehat{\alpha}_1\LSS(\pYo ,\bX) - \alpha_1^*(\pYo,\bX)	\big\}
\big\{ \MM(\pYo,\bX \con \btheta)  - \widehat{\mu}_\MM\LSS(\bX \con \btheta) \big\}
\bcond A=0 , \bX
\Big]
\\
& =
\widehat{\beta}_1\LSS(\bX) 
\Pr(A=0 \cond \bX) 
\left[
\begin{array}{l}
\EXP\LSS 
\big\{
\widehat{\alpha}_1\LSS (\Yo , \bX) \MM(\Yo,\bX \con \btheta)
\cond A=0, \bX
\big\} 
\\
-
\EXP\LSS 
\big\{
\alpha_1^* (\Yo , \bX) \MM(\Yo,\bX \con \btheta)
\cond A=0, \bX
\big\}		
\\
+
\widehat{\mu}_\MM\LSS(\bX \con \btheta)
\EXP\LSS 
\big\{
\alpha_1^* (\Yo , \bX)
\cond A=0, \bX
\big\} 
\\
-
\widehat{\mu}_\MM\LSS(\bX \con \btheta)
\EXP\LSS 
\big\{
\widehat{\alpha}_1\LSS (\Yo , \bX) 
\cond A=0, \bX
\big\}		
\end{array}
\right]
\\
& =
\widehat{\beta}_1\LSS(\bX) 
\Pr(A=0 \cond \bX) 
\left[
\begin{array}{l}
    \int_{\SPo} \MM(y,\bX \con \btheta) \widehat{\alpha}_1\LSS(y,\bX) \potf{1}^*(y \cond 0,\bX) \, dy  
    \\
    - \int_{\SPo} \MM(y,\bX \con \btheta) \alpha_1^* (y,\bX) \potf{1}^*(y \cond 0,\bX) \, dy
    \\
    +  \widehat{\mu}_\MM\LSS(\bX \con \btheta)  \int_{\SPo} \alpha_1^* (y,\bX) \potf{1}^*(y \cond 0,\bX) \, dy  
    \\
    -  \widehat{\mu}_\MM\LSS(\bX \con \btheta)  \int_{\SPo} \widehat{\alpha}_1\LSS(y,\bX) \potf{1}^*(y \cond 0,\bX) \, dy
\end{array}
\right]
\\
& =
\widehat{\beta}_1\LSS(\bX) 
\Pr(A=0 \cond \bX)
 \bigg\{ \int_{\SPo} \widehat{\alpha}_1\LSS (y, \bX) \hpotf{1}\LSS (y \cond 0,\bX) \, dy \bigg\} ^{-1}
 \\
 & 
 \quad
 \times 
\left[
\begin{array}{l}
    \big\{ \int_{\SPo} \widehat{\alpha}_1\LSS (y, \bX) \hpotf{1}\LSS (y \cond 0,\bX) \, dy \big\}
    \big\{ \int_{\SPo} \MM(y,\bX \con \btheta) \widehat{\alpha}_1\LSS(y,\bX) \potf{1}^*(y \cond 0,\bX) \, dy   \big\}
    \\
    - \big\{ \int_{\SPo} \widehat{\alpha}_1\LSS (y, \bX) \hpotf{1}\LSS (y \cond 0,\bX) \, dy \big\}
    \big\{ \int_{\SPo} \MM(y,\bX \con \btheta) \alpha_1^* (y,\bX) \potf{1}^*(y \cond 0,\bX) \, dy \big\}
    \\
    +  \big\{ \int_{\SPo} \alpha_1^* (y,\bX) \potf{1}^*(y \cond 0,\bX) \, dy   \big\}
    \big\{ \int_{\SPo} \MM(y,\bX \con \btheta) \widehat{\alpha}_1\LSS (y, \bX) \hpotf{1}\LSS (y \cond 0,\bX) \, dy \big\}
    \\
    -   \big\{ \int_{\SPo} \widehat{\alpha}_1\LSS(y,\bX) \potf{1}^*(y \cond 0,\bX) \, dy \big\}
    \big\{ \int_{\SPo} \MM(y,\bX \con \btheta) \widehat{\alpha}_1\LSS (y, \bX) \hpotf{1}\LSS (y \cond 0,\bX) \, dy \big\}
\end{array}
\right]
\end{align*}
Note that $ \widehat{\mu}_\MM\LSS(\bX \con \btheta)   \cdot \big\{ \int_{\SPo} \widehat{\alpha}_1\LSS (y, \bX) \hpotf{1}\LSS (y \cond 0,\bX) \, dy \big\} = \int_{\SPo} \MM(y,\bX \con \btheta) \widehat{\alpha}_1\LSS (y, \bX) \hpotf{1}\LSS (y \cond 0,\bX) \, dy$.

Before we modify Term \eqref{eq-prod2-2} and \eqref{eq-prod2-3}, we remark that
\begin{align*}
\widehat{R}\LSS (y, 1, \bX) \potf{0}^*(y \cond 1, \bX)
& =
\frac{ \widehat{\beta}_1\LSS (\bX) }{ \widehat{\beta}_0\LSS (\bX) }
\frac{  		\hpotf{1}\LSS(y \cond 0, \bX)  }{ \hpotf{0}\LSS(y \cond 0, \bX) }
\potf{0}^*(y \cond 1, \bX)
\\
& =
\beta_0^*(\bX) \alpha_0^* (y,\bX)
\frac{\Pr(A=0 \cond \bX)	}{\Pr(A=1 \cond \bX)}
\frac{ \widehat{\beta}_1\LSS (\bX) }{ \widehat{\beta}_0\LSS (\bX) }
\frac{  		\hpotf{1}\LSS(y \cond 0, \bX)  }{ \hpotf{0}\LSS(y \cond 0, \bX) }
\potf{0}^*(y \cond 0, \bX) \  ,
\\
\widehat{R}\LSS (y, 0, \bX) \potf{0}^*(y \cond 0, \bX)
& =
\widehat{\beta}_1\LSS (\bX) \widehat{\alpha}_1\LSS (y,\bX)
\frac{  		\hpotf{1}\LSS(y \cond 0, \bX)  }{ \hpotf{0}\LSS(y \cond 0, \bX) }
\potf{0}^*(y \cond 0, \bX)
\ .
\end{align*}

The conditional expectation of Term \eqref{eq-prod2-2} given $\bX$ is 
\begin{align*}
& 	\EXP \LSS \Big[
A
\widehat{R}\LSS (\Yz , A , \bX)
\big\{
\MM(\Yz,\bX \con {\theta})
-
\widehat{\mu}_\MM\LSS(\bX \con \btheta)
\big\}
\bcond \bX
\Big]
\\
& =
\Pr(A=1 \cond \bX)
\EXP \LSS \Big[ 
\widehat{R}\LSS (\Yz , 1 , \bX)
\big\{
\MM(\Yz,\bX \con {\theta})
-
\widehat{\mu}_\MM\LSS(\bX \con \btheta)
\big\}
\bcond
A=1, \bX
\Big]
\\
&
=
\Pr(A=1 \cond \bX)
\left[
\begin{array}{l}
\EXP \LSS
\Big\{ 
\MM(\Yz,\bX \con \btheta)
\widehat{R}\LSS(\Yz,1,\bX)
\, \Big| \,
A=1, \bX \Big\}
\\
-
\widehat{\mu}_\MM\LSS(\bX \con \btheta) 
\EXP\LSS
\Big\{ 
\widehat{R}\LSS(\Yz,1,\bX)
\bcond
A=1, \bX \Big\} 
\end{array}
\right]
\\
&
=
\Pr(A=1 \cond \bX) \bigg\{ \int_{\SPo} \widehat{\alpha}_1\LSS (y, \bX) \hpotf{1}\LSS (y \cond 0,\bX) \, dy \bigg\} ^{-1}
\\
& \times
\left[
\begin{array}{l}
\big\{ \int_{\SPo} \widehat{\alpha}_1\LSS (y, \bX) \hpotf{1}\LSS (y \cond 0,\bX) \, dy \big\}
\big\{ \int_{\SPzt} \MM(y,\bX \con \btheta) \frac{ \widehat{\beta}_1\LSS (\bX) }{ \widehat{\beta}_0\LSS (\bX) }
\frac{  		\hpotf{1}\LSS(y | 0, \bX)  }{ \hpotf{0}\LSS(y | 0, \bX) }
\potf{0}^*(y \cond 1, \bX) \, dy \big\}
\\
-
\big\{ \int_{\SPo} \MM(y,\bX \con \btheta) \widehat{\alpha}_1\LSS (y, \bX) \hpotf{1}\LSS (y \cond 0,\bX) \, dy \big\}
\big\{ \int_{\SPzt}  \frac{ \widehat{\beta}_1\LSS (\bX) }{ \widehat{\beta}_0\LSS (\bX) }
\frac{  		\hpotf{1}\LSS(y | 0, \bX)  }{ \hpotf{0}\LSS(y | 0, \bX) }
\potf{0}^*(y \cond 1, \bX) \, dy \big\}
\end{array}
\right]
\\
&
=
\Pr(A=0 \cond \bX) \bigg\{ \int_{\SPo} \widehat{\alpha}_1\LSS (y, \bX) \hpotf{1}\LSS (y \cond 0,\bX) \, dy \bigg\} ^{-1}
\frac{\beta_0^*(\bX) \widehat{\beta}_1\LSS(\bX)}{\widehat{\beta}_0\LSS(\bX)}
\\
& \times
\left[
\begin{array}{l}
\big\{ \int_{\SPo} \widehat{\alpha}_1\LSS (y, \bX) \hpotf{1}\LSS (y \cond 0,\bX) \, dy \big\}
\big\{ \int_{\SPo} \MM(y,\bX \con \btheta)  \alpha_0^*(y,\bX)
\frac{  		\hpotf{1}\LSS(y | 0, \bX)  }{ \hpotf{0}\LSS(y | 0, \bX) }
\potf{0}^*(y \cond 0, \bX) \, dy \big\}
\\
-
\big\{ \int_{\SPo} \MM(y,\bX \con \btheta) \widehat{\alpha}_1\LSS (y, \bX) \hpotf{1}\LSS (y \cond 0,\bX) \, dy \big\}
\big\{ \int_{\SPo}   \alpha_0^*(y,\bX)
\frac{  		\hpotf{1}\LSS(y | 0, \bX)  }{ \hpotf{0}\LSS(y | 0, \bX) }
\potf{0}^*(y \cond 0, \bX) \, dy \big\}
\end{array}
\right]
\end{align*} 
The last line is derived as follows:
\begin{align*}
&
\int_{\SPzt} \MM(y,\bX \con \btheta) \frac{ \widehat{\beta}_1\LSS (\bX) }{ \widehat{\beta}_0\LSS (\bX) }
\frac{ \potf{0}^*(y \cond 1, \bX)  }{ \hpotf{0}\LSS(y | 0, \bX) }
\hpotf{1}\LSS(y | 0, \bX)  \, dy
\\
& = 
\int_{\SPo \cap \SPzt } \MM(y,\bX \con \btheta) \frac{ \widehat{\beta}_1\LSS (\bX) }{ \widehat{\beta}_0\LSS (\bX) }
\frac{ \potf{0}^*(y \cond 1, \bX)  }{ \hpotf{0}\LSS(y | 0, \bX) }
\hpotf{1}\LSS(y | 0, \bX)  \, dy 
\\
& = 
\int_{\SPo } \MM(y,\bX \con \btheta) \frac{ \widehat{\beta}_1\LSS (\bX) }{ \widehat{\beta}_0\LSS (\bX) }
\frac{ \potf{0}^*(y \cond 1, \bX)  }{ \hpotf{0}\LSS(y | 0, \bX) }
\hpotf{1}\LSS(y | 0, \bX)  \, dy 
\\
& =  \int_{\SPo} \MM(y,\bX \con \btheta)  \alpha_0^*(y,\bX)
\frac{  		\hpotf{1}\LSS(y | 0, \bX)  }{ \hpotf{0}\LSS(y | 0, \bX) }
\potf{0}^*(y \cond 0, \bX) \, dy \ .
\end{align*}
The first identity is from $\hpotf{1}\LSS(y | 0, \bX) = 0 \text{ for } (y,\bX) \in \R \cap \SPo^c$. The second identity is from $f_0^*(y \cond 1,\bX)=0$ for $(y,\bX) \in \R \cap \SPzt$. The last identity is from \eqref{eq-aux03}: $f_0^*(y \cond 1,\bX) = f_0^*(y \cond 0, \bX) \beta_0^*(\bX) \alpha_0^*(y,\bX) \Pr(A=0 \cond \bX)/\Pr(A=1 \cond \bX) $ for $(y,\bX) \in \SPo \subseteq \SPz$.

Similarly, the conditional expectation of  Term \eqref{eq-prod2-3} given $\bX$ is
\begin{align*}
&
\EXP \LSS \Big[
(1-A)
\widehat{R}\LSS (\Yz ,A , \bX)
\big\{
\MM(\Yz,\bX \con {\theta})
-
\widehat{\mu}_\MM\LSS(\bX \con \btheta)
\big\}
\bcond \bX
\Big]	
\\
& =
\Pr(A=0 \cond \bX)
\EXP \LSS \Big[ 
\widehat{R}\LSS (\Yz , 0 , \bX)
\big\{
\MM(\Yz,\bX \con {\theta})
-
\widehat{\mu}_\MM\LSS(\bX \con \btheta)
\big\}
\bcond
A=0, \bX 	\Big]
\\
&
=
\Pr(A=0 \cond \bX)
\left[
\begin{array}{l}
\EXP \LSS
\Big\{ 
\MM(\Yz,\bX \con \btheta)
\widehat{R}\LSS(\Yz,0,\bX)
\, \Big| \,
A=0, \bX \Big\}
\\
- 
\widehat{\mu}_\MM\LSS(\bX \con \btheta) 
\EXP\LSS
\Big\{ 
\widehat{R}\LSS(\Yz,0,\bX)
\bcond
A=0, \bX \Big\}
\end{array}
\right]	
\\
&
=
\widehat{\beta}_1\LSS(\bX) 
\Pr(A=0 \cond \bX)  \bigg\{ \int_{\SPo} \widehat{\alpha}_1\LSS (y, \bX) \hpotf{1}\LSS (y \cond 0,\bX) \, dy \bigg\} ^{-1}
\\
& \times
\left[
\begin{array}{l}
\big\{ \int_{\SPo} \widehat{\alpha}_1\LSS (y, \bX) \hpotf{1}\LSS (y \cond 0,\bX) \, dy \big\}
\big\{ \int_{\SPz} \MM(y,\bX \con \btheta)  
\frac{  		\hpotf{1}\LSS(y | 0, \bX)  }{ \hpotf{0}\LSS(y | 0, \bX) }
\potf{0}^*(y \cond 0, \bX) \, dy \big\}
\\
-
\big\{ \int_{\SPo} \MM(y,\bX \con \btheta) \widehat{\alpha}_1\LSS (y, \bX) \hpotf{1}\LSS (y \cond 0,\bX) \, dy \big\}
\big\{ \int_{\SPz}  
\frac{  		\hpotf{1}\LSS(y | 0, \bX)  }{ \hpotf{0}\LSS(y | 0, \bX) }
\potf{0}^*(y \cond 0, \bX) \, dy \big\}
\end{array}
\right]
\\
&
=
\widehat{\beta}_1\LSS(\bX) 
\Pr(A=0 \cond \bX)  \bigg\{ \int_{\SPo} \widehat{\alpha}_1\LSS (y, \bX) \hpotf{1}\LSS (y \cond 0,\bX) \, dy \bigg\} ^{-1}
\\
& \times
\left[
\begin{array}{l}
\big\{ \int_{\SPo} \widehat{\alpha}_1\LSS (y, \bX) \hpotf{1}\LSS (y \cond 0,\bX) \, dy \big\}
\big\{ \int_{\SPo} \MM(y,\bX \con \btheta)  
\frac{  		\hpotf{1}\LSS(y | 0, \bX)  }{ \hpotf{0}\LSS(y | 0, \bX) }
\potf{0}^*(y \cond 0, \bX) \, dy \big\}
\\
-
\big\{ \int_{\SPo} \MM(y,\bX \con \btheta) \widehat{\alpha}_1\LSS (y, \bX) \hpotf{1}\LSS (y \cond 0,\bX) \, dy \big\}
\big\{ \int_{\SPo}  
\frac{  		\hpotf{1}\LSS(y | 0, \bX)  }{ \hpotf{0}\LSS(y | 0, \bX) }
\potf{0}^*(y \cond 0, \bX) \, dy \big\}
\end{array}
\right] \ .
\end{align*}
The last line is from $\hpotf{1}\LSS(y | 0, \bX) = 0 \text{ for } (y,\bX) \in \R \cap \SPo^c$.  

Therefore, the conditional expectation of  \eqref{eq-prod2-1}+\eqref{eq-prod2-2}+\eqref{eq-prod2-3} given $(A=0,\bX)$ is rearranged as follows:
{\small
\begin{align}
& 
\frac{  \int_{\SPo}  \widehat{\alpha}_1\LSS (y, \bX) \hpotf{1}\LSS (y \cond 0,\bX) \, dy }{\Pr(A=0 \cond \bX)  }
\frac{ \widehat{\beta}_0\LSS(\bX)}{ \widehat{\beta}_1\LSS(\bX)}
\EXP \LSS \Big\{  \eqref{eq-prod2-1}+\eqref{eq-prod2-2}+\eqref{eq-prod2-3} \, \Big| \, A=0 , \bX \Big\}
\nonumber \\
& 
=
    \bigg\{  \int_{\SPo} \widehat{\alpha}_1\LSS (y,\bX) \, \hpotf{1} \LSS (y \cond 0, \bX) \, dy \bigg\}
    \bigg[ \int_{\SPo} \MM(y,\bX \con \btheta) \widehat{\beta}_{0}\LSS(\bX) \big\{ \widehat{\alpha}_1\LSS(y,\bX) - \alpha_0^* (y,\bX) \big\} \potf{1}^*(y \cond 0,\bX) \, dy \bigg]
\nonumber \\
& 
    - \bigg\{  \int_{\SPo} \MM(y,\bX \con \btheta) \widehat{\alpha}_1\LSS(y,\bX) \, \hpotf{1} \LSS (y \cond 0, \bX) \, dy \bigg\}
    \bigg[ \int_{\SPo} \widehat{\beta}_{0}\LSS(\bX) \big\{ \widehat{\alpha}_1\LSS(y,\bX) - \alpha_0^* (y,\bX) \big\} \potf{1}^*(y \cond 0,\bX) \, dy \bigg]
\nonumber \\
& 
+
\bigg\{  \int_{\SPo} \widehat{\alpha}_1\LSS(y,\bX) \, \hpotf{1} \LSS (y \cond 0, \bX) \, dy \bigg\}
\nonumber \\
& \times 
\Bigg[
\int_{\SPo} \MM(y,\bX \con \btheta)
\Big\{ 
\alpha_0^*(y, \bX) \beta_0^*(\bX) 
-
\widehat{\alpha}_1\LSS(y,\bX) \widehat{\beta}_{0}\LSS(\bX)
\Big\}			
\hpotf{1} \LSS (y \cond 0, \bX)
\frac{  \potf{0}^*(y \cond 0, \bX) }{ \hpotf{0}\LSS(y \cond 0, \bX) } \, dy			
\Bigg]
\nonumber \\
& 
-
\bigg\{  \int_{\SPo} \MM(y,\bX \con \btheta) \widehat{\alpha}_1\LSS (y,\bX) \, \hpotf{1} \LSS (y \cond 0, \bX) \, dy \bigg\}
\nonumber \\
& \times 
\Bigg[
\int_{\SPo}
\Big\{ 
\alpha_0^*(y, \bX) \beta_0^*(\bX) 
-
\widehat{\alpha}(y,\bX) \widehat{\beta}_{0}\LSS(\bX)
\Big\}
\hpotf{1} \LSS (y \cond 0, \bX)
\frac{  \potf{0}^*(y \cond 0, \bX) }{ \hpotf{0}\LSS(y \cond 0, \bX) }
\, dy			
\Bigg]
\nonumber \\
&  =
\overbrace{
\bigg\{  \int_{\SPo} \widehat{\alpha}_1\LSS (y,\bX) \, \hpotf{1} \LSS (y \cond 0, \bX) \, dy \bigg\}
}^{F_1}
\nonumber \\
& \times 
\overbrace{
\left[
\begin{array}{l}
\int_{\SPo}	\MM(y,\bX \con \btheta)
\big\{ 
\alpha_0^*(y, \bX) \beta_0^*(\bX) 
-
\widehat{\alpha}_1\LSS (y,\bX) \widehat{\beta}_{0}\LSS(\bX)
\big\}
\hpotf{1} \LSS (y \cond 0, \bX)
\frac{  \potf{0}^*(y \, | \, 0, \bX) }{ \hpotf{0}\LSS(y \, | \, 0, \bX) } \, dy			
\nonumber \\
+
\int_{\SPo} \MM(y,\bX \con \btheta) \widehat{\beta}_{0}\LSS(\bX) \big\{ \widehat{\alpha}_1\LSS (y,\bX) - \alpha_0^* (y,\bX) \big\} \potf{1}^*(y \cond 0,\bX) \, dy
\end{array}
\right]
}^{G_2+H_2}
\nonumber \\
& 
-
\underbrace{
\bigg\{  \int_{\SPo} \MM(y,\bX \con {\theta}) \widehat{\alpha}_1\LSS (y,\bX) \, \hpotf{1} \LSS (y \cond 0, \bX) \, dy \bigg\}
}_{F_2}
\nonumber \\
& 
\times 
\underbrace{
\left[
\begin{array}{l}
\int_{\SPo}
\big\{ 
\alpha_0^*(y, \bX) \beta_0^*(\bX) 
-
\widehat{\alpha}_1\LSS (y,\bX) \widehat{\beta}_{0}\LSS(\bX)
\big\}
\hpotf{1} \LSS (y \cond 0, \bX)
\frac{  \potf{0}^*(y \, | \, 0, \bX) }{ \hpotf{0}\LSS(y \, | \, 0, \bX) } \, dy			
\nonumber \\
+
\int_{\SPo} \widehat{\beta}_{0}\LSS(\bX) \big\{ \widehat{\alpha}_1\LSS (y,\bX) - \alpha_0^* (y,\bX) \big\} \potf{1}^*(y \cond 0,\bX) \, dy
\end{array}
\right] }_{G_1+H_1}
\nonumber \\
& = F_1 H_2 - F_2 H_1 + F_1 G_2 - F_2 G_1 \ .
\label{eq-FGH-decomposition}
\end{align}}
In the last equality we defined the following quantities:
\begin{align*}
& F_1 = \int_{\SPo} \widehat{\alpha}_1\LSS (y,\bX) \, \hpotf{1} \LSS (y \cond 0, \bX) \, dy
\\
& F_2 = \int_{\SPo} \MM(y,\bX \con \btheta) \widehat{\alpha}_1\LSS (y,\bX) \, \hpotf{1} \LSS (y \cond 0, \bX) \, dy
\\
& G_1 = \int	_{\SPo}	
\big\{ 
\alpha_0^*(y, \bX) \beta_0^*(\bX) 
-
\widehat{\alpha}_1\LSS (y,\bX) \widehat{\beta}_{0}\LSS(\bX)
\big\}
\hpotf{1} \LSS (y \cond 0, \bX)
\frac{  \potf{0}^*(y \, | \, 0, \bX) }{ \hpotf{0}\LSS(y \, | \, 0, \bX) } \, dy			
\\
& G_2 = \int	_{\SPo}
\MM(y,\bX \con \btheta)
\big\{ 
\alpha_0^*(y, \bX) \beta_0^*(\bX) 
-
\widehat{\alpha}_1\LSS (y,\bX) \widehat{\beta}_{0}\LSS(\bX)
\big\}
\hpotf{1} \LSS (y \cond 0, \bX)
\frac{  \potf{0}^*(y \, | \, 0, \bX) }{ \hpotf{0}\LSS(y \, | \, 0, \bX) } \, dy		
\\
& H_1 = \int_{\SPo} \widehat{\beta}_{0}\LSS(\bX) \big\{ \widehat{\alpha}_1\LSS (y,\bX) - \alpha_0^* (y,\bX) \big\} \potf{1}^*(y \cond 0,\bX) \, dy
\\
& H_2 = \int_{\SPo} \MM(y,\bX \con \btheta) \widehat{\beta}_{0}\LSS(\bX) \big\{ \widehat{\alpha}_1\LSS (y,\bX) - \alpha_0^* (y,\bX) \big\} \potf{1}^*(y \cond 0,\bX) \, dy	
\end{align*}
Additionally, for a function $h$, we denote $\llbracket h \rrbracket = \int_{\SPo} h(y,\bX) \hpotf{1} (y \cond 0,\bX) \, dy$; note that, the ranges of integral are unified to $\SPo$. Additionally, for $y \in \SPo$, we obtain $\alpha_0^*(y,\bX)=\alpha_1^*(y,\bX)$. We find $F_1H_2 - F_2H_1$ is represented as
{\small
\begin{align}	
& \hspace*{-1cm} 
F_1 H_2 - F_2 H_1
\nonumber \\
& \hspace*{-1cm}   = 
\widehat{\beta}_0 \LSS (\bX)
\llbracket \widehat{\alpha}_1\LSS \rrbracket
\bigg[ \int_{\SPo} \MM(y,\bX \con \btheta) \big\{ \widehat{\alpha}_1\LSS(y,\bX) - \alpha_1^*(y,\bX) \big\} \potf{1}^*(y \cond 0,\bX) \, dy \bigg]
\nonumber \\
& \hspace*{-1cm}   \hspace*{0.5cm}
-
\widehat{\beta}_0 \LSS (\bX)
\llbracket \MM(\btheta) \widehat{\alpha}_1\LSS \rrbracket
\bigg[ \int_{\SPo} \big\{ \widehat{\alpha}_1\LSS(y,\bX) - \alpha_1^*(y,\bX) \big\} \potf{1}^*(y \cond 0,\bX) \, dy \bigg]
\nonumber \\
& \hspace*{-1cm}   = 
\widehat{\beta}_0 \LSS (\bX)
\llbracket \widehat{\alpha}_1\LSS \rrbracket
\bigg[ \int_{\SPo} \MM(y,\bX \con \btheta) \big\{ \widehat{\alpha}_1\LSS(y,\bX) - \alpha_1^*(y,\bX) \big\} 
\big\{ \potf{1}^*(y \cond 0,\bX) - \hpotf{1}\LSS(y \cond 0,\bX) \big\} \, dy \bigg]
\nonumber \\
& \hspace*{-1cm}   \hspace*{0.5cm}
+
\widehat{\beta}_0 \LSS (\bX)
\llbracket \widehat{\alpha}_1\LSS \rrbracket
\bigg[ \int_{\SPo} \MM(y,\bX \con \btheta) \big\{ \widehat{\alpha}_1\LSS(y,\bX) - \alpha_1^*(y,\bX) \big\} 
\hpotf{1}\LSS(y \cond 0,\bX)  \, dy \bigg]
\nonumber \\
& \hspace*{-1cm}   \hspace*{0.5cm}
-
\widehat{\beta}_0 \LSS (\bX)
\llbracket \MM(\btheta) \widehat{\alpha}_1\LSS \rrbracket
\bigg[ \int_{\SPo} \big\{ \widehat{\alpha}_1\LSS(y,\bX) - \alpha_1^*(y,\bX) \big\} \big\{ \potf{1}^*(y \cond 0,\bX) - \hpotf{1}\LSS(y \cond 0,\bX) \big\} \, dy \bigg]
\nonumber \\
& \hspace*{-1cm}   \hspace*{0.5cm}
-
\widehat{\beta}_0 \LSS (\bX)
\llbracket \MM(\btheta) \widehat{\alpha}_1\LSS \rrbracket
\bigg[ \int_{\SPo} \big\{ \widehat{\alpha}_1\LSS(y,\bX) - \alpha_1^*(y,\bX) \big\} \hpotf{1}\LSS(y \cond 0,\bX) \, dy \bigg]
\nonumber \\
& \hspace*{-1cm}  
= 
\widehat{\beta}_0 \LSS (\bX)
\llbracket \widehat{\alpha}_1\LSS \rrbracket
\bigg[ \int_{\SPo} \MM(y,\bX \con \btheta) \bigg\{ 
\begin{array}{l}
\widehat{\alpha}_1\LSS(y,\bX) \\
- \alpha_1^*(y,\bX)
\end{array}
\bigg\} 
\bigg\{ 
\begin{array}{l}
\potf{1}^*(y \cond 0,\bX) \\
- \hpotf{1}\LSS(y \cond 0,\bX) 
\end{array}
\bigg\} 
\frac{ \potf{1}^*(y \cond 0,\bX) }{\potf{1}^*(y \cond 0,\bX)} 
\, dy \bigg]
\tag{T7-1}\label{eq-aux101}
\\
& \hspace*{-1cm}  
\hspace*{0.5cm}
- 
\widehat{\beta}_0 \LSS (\bX)
\llbracket \MM(\btheta) \widehat{\alpha}_1\LSS \rrbracket
\bigg[ \int_{\SPo} \bigg\{ 
\begin{array}{l}
\widehat{\alpha}_1\LSS(y,\bX) \\
- \alpha_1^*(y,\bX)
\end{array}
\bigg\} 
\bigg\{ 
\begin{array}{l}
\potf{1}^*(y \cond 0,\bX) \\
- \hpotf{1}\LSS(y \cond 0,\bX) 
\end{array}
\bigg\} 
\frac{ \potf{1}^*(y \cond 0,\bX) }{\potf{1}^*(y \cond 0,\bX)} 
\, dy \bigg]
\tag{T7-2}\label{eq-aux102}
\\
& \hspace*{-1cm}   \hspace*{0.5cm}
- 
\widehat{\beta}_0 \LSS (\bX)
\llbracket  \widehat{\alpha}_1\LSS \rrbracket
\llbracket \MM(\btheta) \alpha_1^* \rrbracket
+ 
\widehat{\beta}_0 \LSS (\bX)
\llbracket \MM(\btheta) \widehat{\alpha}_1\LSS \rrbracket
\llbracket \alpha_1^* \rrbracket
\ .
\tag{T7-3}\label{eq-aux103}
\end{align}}%

Similarly, $F_1 G_2 - F_2 G_1$ is
{\small
\begin{align}
&
F_1G_2 - F_2 G_1
\nonumber \\
& = 
\llbracket \widehat{\alpha}_1\LSS \rrbracket
\bigg[ \int_{\SPz} \MM(y,\bX \con \btheta) \big\{ \alpha_1^*(y,\bX) \beta_{0}^*(\bX) - \widehat{\alpha}_1\LSS(y,\bX) \widehat{\beta}_{0}\LSS(\bX) \big\} 
\hpotf{1}\LSS(y \cond 0,\bX)
\frac{\potf{0}^*(y \cond 0,\bX)}{\hpotf{0}(y \cond 0,\bX)} 
\, dy \bigg]
\nonumber \\
& \hspace*{1cm}
-
\llbracket \MM(\btheta) \widehat{\alpha}_1\LSS \rrbracket
\bigg[ \int_{\SPz} \big\{ \alpha_1^*(y,\bX) \beta_{0}^*(\bX) - \widehat{\alpha}_1\LSS(y,\bX) \widehat{\beta}_{0}\LSS(\bX) \big\} 
\hpotf{1}\LSS(y \cond 0,\bX)
\frac{\potf{0}^*(y \cond 0,\bX)}{\hpotf{0}(y \cond 0,\bX)} 
\, dy \bigg]
\nonumber \\
& = 
\llbracket \widehat{\alpha}_1\LSS \rrbracket
\bigg[ \int_{\SPz} \MM(y,\bX \con \btheta) \bigg\{
\begin{array}{l}
\alpha_1^*(y,\bX) \beta_{0}^*(\bX) \\
- \widehat{\alpha}_1\LSS(y,\bX) \widehat{\beta}_{0}\LSS(\bX)
\end{array}
\bigg\} 
\hpotf{1}\LSS(y \cond 0,\bX)
\frac{\potf{0}^*(y \cond 0,\bX)}{\hpotf{0}(y \cond 0,\bX)} 
\nonumber \\
& \hspace*{2cm}
- \MM(y,\bX \con \btheta) \alpha_1^*(y,\bX) \widehat{\beta}_{0}\LSS(\bX) \hpotf{1}\LSS(y \cond 0, \bX) \, dy 
\bigg]
\tag{T7-4}\label{eq-aux104} \\
& \hspace*{1cm}
-
\llbracket \MM(\btheta) \widehat{\alpha}_1\LSS \rrbracket
\bigg[ \int_{\SPz}
\bigg\{
\begin{array}{l}
\alpha_1^*(y,\bX) \beta_{0}^*(\bX) \\
- \widehat{\alpha}_1\LSS(y,\bX) \widehat{\beta}_{0}\LSS(\bX)
\end{array}
\bigg\} 
\hpotf{1}\LSS(y \cond 0,\bX)
\frac{\potf{0}^*(y \cond 0,\bX)}{\hpotf{0}(y \cond 0,\bX)} 
\nonumber \\
& \hspace*{2cm}
-
\alpha_1^* (y,\bX) \widehat{\beta}_{0}\LSS(\bX) \hpotf{1}\LSS(y \cond 0, \bX) \, dy 
\bigg]
\tag{T7-5}\label{eq-aux105} \\
&
\hspace*{1cm}
+
\widehat{\beta}_{0}\LSS (\bX)
\llbracket \widehat{\alpha}_1\LSS \rrbracket
\llbracket \MM(\btheta) \alpha_1^* \rrbracket
- 
\widehat{\beta}_{0}\LSS (\bX)
\llbracket \MM(\btheta) \widehat{\alpha}_1\LSS \rrbracket
\llbracket \alpha_1^* \rrbracket
\tag{T7-6}\label{eq-aux106} 
\end{align}}%
Note that \eqref{eq-aux103} and \eqref{eq-aux106} cancel out, i.e., $\big\| \eqref{eq-aux103} +\eqref{eq-aux106}  \big\|=0$. Therefore, we focus on \eqref{eq-aux101}+\eqref{eq-aux102} and \eqref{eq-aux104}+\eqref{eq-aux105}. The absolute value of \eqref{eq-aux101}+\eqref{eq-aux102} is upper bounded as follows
\begin{align}
&
\Big\|
\eqref{eq-aux101}+\eqref{eq-aux102} \Big\|
\nonumber
\\
&
\leq 
\EXP_{\SPo} \LSS
\Bigg[
\underbrace{
\Big\| \widehat{\beta}_{0} \LSS  (\bX)
\llbracket \widehat{\alpha}_1\LSS \rrbracket \Big\|
\cdot
\bigg\|
\frac{ \MM(\Yo,\bX \con \btheta) }{ \potf{1}^*(\Yo \cond 0, \bX) }
\bigg\| }_{\leq C \text{ over } \Yo \in \SPo}
\cdot 
\Bigg\| \begin{array}{l}
\widehat{\alpha}_1\LSS(\Yo,\bX) \\
- \alpha_1^*(\Yo,\bX)
\end{array}	 \Bigg\| 
\Bigg\|  \begin{array}{l}
\potf{1}^*(\Yo \cond 0,\bX) \\
- \hpotf{1}\LSS(\Yo \cond 0,\bX)
\end{array}	 \Bigg\| 
\, \Bigg| \, A=0, \bX
\Bigg]
\nonumber
\\
&
\hspace*{0.5cm}
+
\EXP_{\SPo} \LSS
\Bigg[
\underbrace{
\Big\| \widehat{\beta}_{0}\LSS (\bX)
\llbracket \MM(\btheta) \widehat{\alpha}_1\LSS \rrbracket \Big\| 
\cdot
\bigg\|
\frac{1}{ \potf{1}^*(\Yo \cond 0, \bX) }
\bigg\|
}_{\leq C \text{ over } \Yo \in \SPo}
\cdot 
\Bigg\| \begin{array}{l}
\widehat{\alpha}_1\LSS(\Yo,\bX) \\
- \alpha_1^*(\Yo,\bX)
\end{array}	 \Bigg\| 
\Bigg\|  \begin{array}{l}
\potf{1}^*(\Yo \cond 0,\bX) \\
- \hpotf{1}\LSS(\Yo \cond 0,\bX)
\end{array}	 \Bigg\| 
\, \Bigg| \, A=0, \bX
\Bigg]
\nonumber
\\
&
\precsim
\EXP_{\SPo} \LSS
\Big[
\Big\| \widehat{\alpha}_1\LSS(\Yo,\bX) - \alpha_1^*(\Yo,\bX) \Big\| 
\Big\|  \potf{1}^*(\Yo \cond 0,\bX) - \hpotf{1}\LSS(\Yo \cond 0,\bX) \Big\| 
\, \Big| \, A=0, \bX
\Big] \ .
\label{Temporary1} 
\end{align}

To obtain the absolute value of \eqref{eq-aux104}+\eqref{eq-aux105}, we first obtain an alternative representation of it. Let $\widehat{\mathcal{R}}\LSS = \potf{0}^* (y \cond 0,\bX) / \hpotf{0} \LSS(y \cond 0, \bX) - 1$ for $y \in \SPz$. After some algebra, we find \eqref{eq-aux104}+\eqref{eq-aux105} is represented as the summation of the cross-products:
{\small
\begin{align*}
&
\eqref{eq-aux104}+\eqref{eq-aux105}
\\
& = 
\llbracket \widehat{\alpha}_1\LSS\rrbracket \cdot 
\llbracket \MM(\btheta)\alpha_1^* \beta_{0}^* \widehat{\mathcal{R}}\LSS + \MM(\btheta)\alpha_1^* \beta_{0}^* -\MM(\btheta)\widehat{\alpha}_1\LSS \widehat{\beta}_{0} \LSS \widehat{\mathcal{R}}\LSS - \MM(\btheta)\widehat{\alpha}_1\LSS \widehat{\beta}_{0} \LSS - \MM(\btheta)\alpha_1^* \widehat{\beta}_{0} \LSS \rrbracket
\\
& \hspace*{0.5cm}
-
\llbracket \MM(\btheta) \widehat{\alpha}_1\LSS\rrbracket \cdot \llbracket\alpha_1^*\beta_{0}^* \widehat{\mathcal{R}}\LSS +\alpha_1^*\beta_{0}^* -\widehat{\alpha}_1\LSS \widehat{\beta}_{0} \LSS \widehat{\mathcal{R}}\LSS - \widehat{\alpha}_1\LSS \widehat{\beta}_{0} \LSS -\alpha_1^* \widehat{\beta}_{0} \LSS \rrbracket
\\
& =
\beta_{0}^* (\bX) \llbracket\widehat{\alpha}_1\LSS\rrbracket \cdot 
\llbracket \MM(\btheta)\alpha_1^* \widehat{\mathcal{R}}\LSS \rrbracket 
+ \beta_{0}^* (\bX)	\llbracket\widehat{\alpha}_1\LSS\rrbracket \cdot \llbracket \MM(\btheta)\alpha_1^* \rrbracket 
- \widehat{\beta}_{0}\LSS(\bX)
\llbracket\widehat{\alpha}_1\LSS\rrbracket \cdot \llbracket \MM(\btheta)\widehat{\alpha}_1\LSS \widehat{\mathcal{R}}\LSS\rrbracket
\\
&
\hspace*{0.5cm}
- \widehat{\beta}_{0}\LSS(\bX) \llbracket\widehat{\alpha}_1\LSS\rrbracket \cdot \llbracket \MM(\btheta)\widehat{\alpha}_1\LSS \rrbracket 
- \widehat{\beta}_{0}\LSS(\bX) \llbracket\widehat{\alpha}_1\LSS\rrbracket \cdot \llbracket \MM(\btheta)\alpha_1^* \rrbracket
\\
&
\hspace*{0.5cm}
-
\beta_0^*(\bX)
\llbracket \MM(\btheta) \widehat{\alpha}_1\LSS\rrbracket \cdot \llbracket\alpha_1^* \widehat{\mathcal{R}}\LSS\rrbracket
-
\beta_0^*(\bX)
\llbracket \MM(\btheta) \widehat{\alpha}_1\LSS\rrbracket \cdot	\llbracket \alpha_1^* \rrbracket 
+
\widehat{\beta}_{0}\LSS(\bX)
\llbracket \MM(\btheta) \widehat{\alpha}_1\LSS\rrbracket \cdot	\llbracket\widehat{\alpha}_1\LSS \widehat{\mathcal{R}}\LSS\rrbracket 
\\
&
\hspace*{0.5cm}
+ \widehat{\beta}_{0}\LSS(\bX)
\llbracket \MM(\btheta) \widehat{\alpha}_1\LSS\rrbracket \cdot	\llbracket \widehat{\alpha}_1\LSS \rrbracket
+ \widehat{\beta}_{0}\LSS(\bX)
\llbracket \MM(\btheta) \widehat{\alpha}_1\LSS\rrbracket \cdot	\llbracket\alpha_1^* \rrbracket 
\\
& 
=
\llbracket\widehat{\alpha}_1\LSS\rrbracket
\cdot
\Bigg\{
\begin{array}{l}
\beta_{0}^*(\bX) \llbracket \MM(\btheta)\alpha_1^* \widehat{\mathcal{R}}\LSS \rrbracket
-
\beta_{0}^*(\bX) \llbracket \MM(\btheta) \widehat{\alpha}_1\LSS \widehat{\mathcal{R}}\LSS \rrbracket
\\
+
\beta_{0}^*(\bX) \llbracket \MM(\btheta) \widehat{\alpha}_1\LSS \widehat{\mathcal{R}}\LSS \rrbracket
-
\widehat{\beta}_{0} \LSS(\bX) \llbracket \MM(\btheta) \widehat{\alpha}_1\LSS \widehat{\mathcal{R}}\LSS \rrbracket
\end{array}
\Bigg\}
\\
& \hspace*{0.5cm}
- \llbracket \MM(\btheta) \widehat{\alpha}_1\LSS \rrbracket
\cdot 
\Bigg\{
\begin{array}{l}
\beta_{0}^* (\bX) \llbracket\alpha_1^* \widehat{\mathcal{R}}\LSS \rrbracket
-
\beta_{0}^* (\bX) \llbracket \widehat{\alpha}_1\LSS \widehat{\mathcal{R}}\LSS \rrbracket
\\
+
\beta_{0}^* (\bX) \llbracket \widehat{\alpha}_1\LSS \widehat{\mathcal{R}}\LSS \rrbracket
-
\widehat{\beta}_{0} \LSS(\bX) \llbracket \widehat{\alpha}_1\LSS \widehat{\mathcal{R}}\LSS \rrbracket
\end{array} 
\Bigg\}
\\
& \hspace*{0.5cm}
+
\frac{1}{2}
\big\{ \beta_{0}^*(\bX) - \widehat{\beta}_{0} \LSS (\bX)  \big\}
\Big\{ 
\llbracket \widehat{\alpha}_1\LSS -\alpha_1^* \rrbracket \cdot \llbracket \MM(\btheta) \alpha_1^* + \MM(\btheta) \widehat{\alpha}_1\LSS  \rrbracket
+
\llbracket \widehat{\alpha}_1\LSS +\alpha_1^* \rrbracket \cdot \llbracket \MM(\btheta) \alpha_1^* - \MM(\btheta) \widehat{\alpha}_1\LSS  \rrbracket
\Big\}	
\\
& =
\beta_{0}^* (\bX)
\llbracket\widehat{\alpha}_1\LSS\rrbracket
\int_{\SPo} \MM(y,\bX \con \btheta)
\bigg\{
\begin{array}{l}
\widehat{\alpha}_1\LSS(y,\bX) \\
-\alpha_1^*(y,\bX)
\end{array}	
\bigg\} 
\, 
\bigg\{
\begin{array}{l}
\potf{0}^*(y \cond 0,\bX) \\
- \hpotf{0}\LSS(y \cond 0,\bX)
\end{array} 
\bigg\}
\, \frac{\hpotf{1}\LSS(y \cond 0,\bX) \potf{1}^*(y \cond 0, \bX) }{\potf{1}^*(y \cond 0,\bX) \hpotf{0}\LSS(y \cond 0,\bX) }  \, dy
\\
& \hspace*{0.5cm}
+
\llbracket\widehat{\alpha}_1\LSS\rrbracket
\int_{\SPo}
\MM(y,\bX \con \btheta)
\widehat{\alpha}_1\LSS(y,\bX)
\bigg\{
\begin{array}{l}
\beta_{0}^*(\bX) \\
- \widehat{\beta}_{0} \LSS(\bX)
\end{array}
\bigg\}
\bigg\{
\begin{array}{l}
\potf{0}^*(y \cond 0,\bX) \\
- \hpotf{0}\LSS(y \cond 0,\bX)
\end{array}
\bigg\} 
\, \frac{\hpotf{1}\LSS(y \cond 0,\bX) \potf{1}^*(y \cond 0, \bX) }{\potf{1}^*(y \cond 0,\bX) \hpotf{0}\LSS(y \cond 0,\bX) }  \, dy
\\
& \hspace*{0.5cm}
-
\beta_{0}^* (\bX)
\llbracket \MM(\btheta) \widehat{\alpha}_1\LSS \rrbracket
\int_{\SPo}
\bigg\{
\begin{array}{l}
\widehat{\alpha}_1\LSS(y,\bX) \\
-\alpha_1^*(y,\bX)
\end{array}	
\bigg\} 
\, 
\bigg\{
\begin{array}{l}
\potf{0}^*(y \cond 0,\bX) \\
- \hpotf{0}\LSS(y \cond 0,\bX)
\end{array} 
\bigg\}
\, \frac{\hpotf{1}\LSS(y \cond 0,\bX) \potf{1}^*(y \cond 0, \bX) }{\potf{1}^*(y \cond 0,\bX) \hpotf{0}\LSS(y \cond 0,\bX) }  \, dy
\\
& \hspace*{0.5cm}  -
\llbracket \MM(\btheta) \widehat{\alpha}_1\LSS \rrbracket
\int_{\SPo}   \widehat{\alpha}_1\LSS(y,\bX) 
\bigg\{
\begin{array}{l}
\beta_{0}^*(\bX) \\
- \widehat{\beta}_{0} \LSS(\bX)
\end{array}
\bigg\}
\bigg\{
\begin{array}{l}
\potf{0}^*(y \cond 0,\bX) \\
- \hpotf{0}\LSS(y \cond 0,\bX)
\end{array}
\bigg\} 
\, \frac{\hpotf{1}\LSS(y \cond 0,\bX) \potf{1}^*(y \cond 0, \bX) }{\potf{1}^*(y \cond 0,\bX) \hpotf{0}\LSS(y \cond 0,\bX) }  \, dy
\\
& \hspace*{0.5cm}
+ 0.5
\llbracket \MM(\btheta) \alpha_1^* + \MM(\btheta) \widehat{\alpha}_1\LSS \rrbracket
\int_{\SPo}
\bigg\{
\begin{array}{l}
\beta_{0}^*(\bX) \\
- \widehat{\beta}_{0} \LSS(\bX)
\end{array}
\bigg\}
\bigg\{
\begin{array}{l}
\widehat{\alpha}_1\LSS(y,\bX) \\
-\alpha_1^*(y,\bX)
\end{array}	
\bigg\}
\, \frac{\hpotf{1}\LSS(y \cond 0,\bX) \potf{1}^*(y \cond 0, \bX) }{\potf{1}^*(y \cond 0,\bX)}  \, dy
\\
& \hspace*{0.5cm}
+ 0.5
\llbracket\alpha_1^*+\widehat{\alpha}_1\LSS \rrbracket
\int_{\SPo} \MM(y,\bX \con \btheta)
\bigg\{
\begin{array}{l}
\beta_{0}^*(\bX) \\
- \widehat{\beta}_{0} \LSS(\bX)
\end{array}
\bigg\}
\bigg\{
\begin{array}{l}
\widehat{\alpha}_1\LSS(y,\bX) \\
-\alpha_1^*(y,\bX)
\end{array}	
\bigg\}
\, \frac{\hpotf{1}\LSS(y \cond 0,\bX) \potf{1}^*(y \cond 0, \bX) }{\potf{1}^*(y \cond 0,\bX)}  \, dy \ . 
\end{align*}}%
Therefore, the absolute value of \eqref{eq-aux104}+\eqref{eq-aux105} is upper bounded as follows:
{\small
\begin{align*}
&
\Big\| \eqref{eq-aux104}+\eqref{eq-aux105} \Big\|
\\
&
\leq
\EXP_{\SPo} \LSS \Bigg[
{
\bigg\|
\frac{ \beta_0^* (\bX)
\llbracket\widehat{\alpha}_1\LSS\rrbracket 	
\MM(\Yo,\bX \con \btheta)
\hpotf{1} \LSS(\Yo \cond 0,\bX)}{\potf{1}^*(\Yo \cond 0,\bX) \hpotf{0} \LSS(\Yo \cond 0,\bX) }
\bigg\|
}
\cdot 
\bigg\|
\begin{array}{l}
\widehat{\alpha}_1\LSS(\Yo,\bX) \\
- \alpha_1^*(\Yo,\bX)
\end{array} 
\bigg\| \cdot
\bigg\|
\begin{array}{l}
\potf{0}^*(\Yo \cond 0,\bX) \\
- \hpotf{0} \LSS(\Yo \cond 0,\bX)
\end{array} 
\bigg\|
\, \Bigg| \, A=0,\bX \Bigg]
\\
& \hspace*{0.5cm}
+
\EXP_{\SPo} \LSS \Bigg[
{
\bigg\|
\frac{ 
\llbracket \widehat{\alpha}_1\LSS \rrbracket 	
\MM(\Yo,\bX \con \btheta) 
\widehat{\alpha}_1\LSS(\Yo,\bX) 
\hpotf{1} \LSS(\Yo \cond 0,\bX)}{\potf{1}^*(\Yo \cond 0,\bX) \hpotf{0} \LSS(\Yo \cond 0,\bX) }
\bigg\| 
}
\cdot
\bigg\|
\begin{array}{l}
\widehat{\beta}_0\LSS(\bX) \\
- \beta_0^*(\bX)
\end{array} 
\bigg\| \cdot
\bigg\|
\begin{array}{l}
\potf{0}^*(\Yo \cond 0,\bX) \\
- \hpotf{0} \LSS(\Yo \cond 0,\bX)
\end{array} 
\bigg\|
\, \Bigg| \, A=0,\bX \Bigg]
\\
& \hspace*{0.5cm}
+
\EXP_{\SPo} \LSS \Bigg[
{
\bigg\| 
\frac{ 
\beta_0^*(\bX) \llbracket \MM(\btheta) \widehat{\alpha}_1\LSS \rrbracket 	
\hpotf{1} \LSS(\Yo \cond 0,\bX)}{\potf{1}^*(\Yo \cond 0,\bX) \hpotf{0} \LSS(\Yo \cond 0,\bX) }
\bigg\|
}
\cdot
\bigg\|
\begin{array}{l}
\widehat{\alpha}_1\LSS(\Yo,\bX) \\
- \alpha_1^*(\Yo,\bX)
\end{array} 
\bigg\| \cdot
\bigg\|
\begin{array}{l}
\potf{0}^*(\Yo \cond 0,\bX) \\
- \hpotf{0} \LSS(\Yo \cond 0,\bX)
\end{array} 
\bigg\|
\, \Bigg| \, A=0,\bX \Bigg]
\\
& \hspace*{0.5cm}
+
\EXP_{\SPo} \LSS \Bigg[
{
\bigg\|  
\frac{ 
\llbracket \MM(\btheta) \widehat{\alpha}_1\LSS \rrbracket 	
\widehat{\alpha}_1\LSS(\Yo,\bX)
\hpotf{1} \LSS(\Yo \cond 0,\bX)}{\potf{1}^*(\Yo \cond 0,\bX) \hpotf{0} \LSS(\Yo \cond 0,\bX) }
\bigg\|
}
\cdot
\bigg\|
\begin{array}{l}
\widehat{\beta}_0\LSS(\bX) \\
- \beta_0^*(\bX)
\end{array} 
\bigg\| \cdot
\bigg\|
\begin{array}{l}
\potf{0}^*(\Yo \cond 0,\bX) \\
- \hpotf{0} \LSS(\Yo \cond 0,\bX)
\end{array} 
\bigg\|
\, \Bigg| \, A=0,\bX \Bigg]
\\
& \hspace*{0.5cm}
+
0.5
\EXP_{\SPo} \LSS \Bigg[ 
{
\bigg\| 
\frac{ 
\llbracket \MM(\btheta) (\alpha_1^*+\widehat{\alpha}_1\LSS) \rrbracket 	
\hpotf{1} \LSS(\Yo \cond 0,\bX)}{\potf{1}^*(\Yo \cond 0,\bX) }
\bigg\| 
}
\cdot
\bigg\|
\begin{array}{l}
\widehat{\beta}_0\LSS(\bX) \\
- \beta_0^*(\bX)
\end{array} 
\bigg\| \cdot
\bigg\|
\begin{array}{l}
\widehat{\alpha}_1\LSS(\Yo,\bX) \\
- \alpha_1^*(\Yo,\bX)
\end{array} 
\bigg\|
\, \Bigg| \, A=0,\bX \Bigg]
\\
& \hspace*{0.5cm}
+
0.5
\EXP_{\SPo} \LSS \Bigg[ 
{
\bigg\| 
\frac{ 
\llbracket (\alpha_1^*+\widehat{\alpha}_1\LSS) \rrbracket 	
\MM(\Yo,\bX \con \btheta)
\hpotf{1} \LSS(\Yo \cond 0,\bX)}{\potf{1}^*(\Yo \cond 0,\bX) }
\bigg\|
}
\cdot
\bigg\|
\begin{array}{l}
\widehat{\beta}_0\LSS(\bX) \\
- \beta_0^*(\bX)
\end{array} 
\bigg\| \cdot
\bigg\|
\begin{array}{l}
\widehat{\alpha}_1\LSS(\Yo,\bX) \\
- \alpha_1^*(\Yo,\bX)
\end{array} 
\bigg\|
\, \Bigg| \, A=0,\bX \Bigg]
\end{align*}	}%
Note that the first term in each expectation is bounded over $\Yo \in \SPo$ under the assumptions. Therefore, 
\begin{align}
& \Big\| \eqref{eq-aux104}+\eqref{eq-aux105} \Big\|
\nonumber
\\
&
\precsim
\EXP_{\SPo}  \LSS
\Big[
\Big\| \widehat{\alpha}_1\LSS(\Yo,\bX) - \alpha_1^*(\Yo,\bX) \Big\| 
\Big\|  \potf{0}^*(\Yo \cond 0,\bX) - \hpotf{0}\LSS(\Yo \cond 0,\bX) \Big\| 
\, \Big| \, A=0, \bX
\Big]
\nonumber
\\
& \hspace*{1cm}
+
\EXP_{\SPo}  \LSS
\Big[
\Big\| \widehat{\alpha}_1\LSS(\Yo,\bX) - \alpha_1^*(\Yo,\bX) \Big\| 
\Big\|  \beta_0^*(\bX) - \widehat{\beta}_0\LSS(\bX) \Big\| 
\, \Big| \, A=0, \bX
\Big]
\nonumber
\\
& \hspace*{1cm}
+
\EXP_{\SPo}  \LSS
\Big[
\Big\|  \beta_0^*(\bX) - \widehat{\beta}_0\LSS(\bX) \Big\| 
\Big\|  \potf{0}^*(\Yo \cond 0,\bX) - \hpotf{0}\LSS(\Yo \cond 0,\bX) \Big\| 
\, \Big| \, A=0, \bX
\Big] \ .
\label{Temporary2} 
\end{align}
Therefore, from \eqref{eq-FGH-decomposition}, we get $\EXP\LSS \big\{ \eqref{eq-prod2-1}+\eqref{eq-prod2-2}+\eqref{eq-prod2-3} \cond A=0 ,\bX \big\}$ is upper bounded as
\begin{align*}
& 
\Big\|
\EXP \LSS \Big\{  \eqref{eq-prod2-1}+\eqref{eq-prod2-2}+\eqref{eq-prod2-3} \, \Big| \, A=0 , \bX \Big\}
\Big\|
\\
&
\leq
\Bigg\|
\frac{\Pr(A=0 \cond \bX)  }{  \int_{\SPo}  \widehat{\alpha}_1\LSS (y, \bX) \hpotf{1}\LSS (y \cond 0,\bX) \, dy }
\frac{ \widehat{\beta}_1\LSS(\bX)}{ \widehat{\beta}_0\LSS(\bX)}
\Bigg\|
\Big\|
F_1 H_2 - F_2 H_1 + F_1 G_2 - F_2 G_1
\Big\|
\\
& 
\precsim
\big\| \eqref{eq-aux101}+\eqref{eq-aux102} \big\|
+
\big\| \eqref{eq-aux104}+\eqref{eq-aux105} \big\|
+
\big\| \eqref{eq-aux103}+\eqref{eq-aux106} \big\| 
\\
& 
=
\big\| \eqref{eq-aux101}+\eqref{eq-aux102} \big\|
+
\big\| \eqref{eq-aux104}+\eqref{eq-aux105} \big\| 
\ .
\end{align*}
Note that \eqref{eq-aux103} and \eqref{eq-aux106} cancel out, i.e., $\big\| \eqref{eq-aux103} +\eqref{eq-aux106}  \big\|=0$. As a result, $\EXP\LSS \big\{ \eqref{eq-prod2-1}+\eqref{eq-prod2-2}+\eqref{eq-prod2-3} \cond A=0 \big\}$ is upper bounded as follows:
\begin{align}
& \big\| \EXP\LSS \big\{ \eqref{eq-prod2-1}+\eqref{eq-prod2-2}+\eqref{eq-prod2-3} \cond A=0 \big\} \big\|
\nonumber
\\
&
\precsim
\EXP \LSS \big[ 
\big\|  \eqref{eq-aux101}+\eqref{eq-aux102} \big\|
\, \big| \, A=0
\big]
+
\EXP \LSS \big[ 
\big\|  \eqref{eq-aux104}+\eqref{eq-aux105} \big\|
\, \big| \, A=0
\big]		
\nonumber \\
& \precsim
\EXP \LSS
\big[
\big\| \widehat{\alpha}_1\LSS(\Yo,\bX) - \alpha_1^*(\Yo,\bX) \big\| 
\big\|  \potf{1}^*(\Yo \cond 0,\bX) - \hpotf{1}\LSS(\Yo \cond 0,\bX) \big\| 
\, \big| \, A=0 
\big] 
\nonumber \\
& \hspace*{1cm}
+
\EXP \LSS
\big[
\big\| \widehat{\alpha}_1\LSS(\Yo,\bX) - \alpha_1^*(\Yo,\bX) \big\| 
\big\|  \potf{0}^*(\Yo \cond 0,\bX) - \hpotf{0}\LSS(\Yo \cond 0,\bX) \big\| 
\, \big| \, A=0 
\big]
\nonumber \\
& \hspace*{1cm}
+
\EXP \LSS
\big[
\big\| \widehat{\alpha}_1\LSS(\Yo,\bX) - \alpha_1^*(\Yo,\bX) \big\| 
\big\|  \beta_0^*(\bX) - \widehat{\beta}_0\LSS(\bX) \big\| 
\, \big| \, A=0 
\big]
\nonumber \\
& \hspace*{1cm}
+
\EXP \LSS
\big[
\big\|  \beta_0^*(\bX) - \widehat{\beta}_0\LSS(\bX) \big\| 
\big\|  \potf{0}^*(\Yo \cond 0,\bX) - \hpotf{0}\LSS(\Yo \cond 0,\bX) \big\| 
\, \big| \, A=0 
\big] 
\nonumber \\
&
\leq
\big\| \widehat{\alpha}_1\LSS(\Yo,\bX) - \alpha_1^*(\Yo,\bX) \big\|_{P,2} 
\big\|  \potf{1}^*(\Yo \cond 0,\bX) - \hpotf{1}\LSS(\Yo \cond 0,\bX) \big\| _{P,2} 
\nonumber \\
& \hspace*{1cm}
+
\big\| \widehat{\alpha}_1\LSS(\Yo,\bX) - \alpha_1^*(\Yo,\bX) \big\|_{P,2}
\big\|  \potf{0}^*(\Yo \cond 0,\bX) - \hpotf{0}\LSS(\Yo \cond 0,\bX) \big\| _{P,2}
\nonumber \\
& \hspace*{1cm}
+
\big\| \widehat{\alpha}_1\LSS(\Yo,\bX) - \alpha_1^*(\Yo,\bX) \big\| _{P,2}
\big\|  \beta_0^*(\bX) - \widehat{\beta}_0\LSS(\bX) \big\| _{P,2}
\nonumber \\
& \hspace*{1cm}
+
\big\|  \beta_0^*(\bX) - \widehat{\beta}_0\LSS(\bX) \big\| _{P,2}
\big\|  \potf{0}^*(\Yo \cond 0,\bX) - \hpotf{0}\LSS(\Yo \cond 0,\bX) \big\| _{P,2}  \ . 
\label{eq-aux05}
\end{align}
The second inequality holds from \eqref{Temporary1} and \eqref{Temporary2}. The last line holds from the H\"older's inequality.

Therefore, we get the following result by combining \eqref{eq-aux04} and \eqref{eq-aux05} and using $|\mathcal{I}_k| = N/K $:
\begin{align}
\eqref{Term3-General} 
& = 
\big| \mathcal{I}_k \big|^{1/2}
\Pr(A=1) \EXP \big\{  \eqref{eq-prod1} + \eqref{eq-prod2-1} + \eqref{eq-prod2-2} + \eqref{eq-prod2-3} \cond A=0 \big\}
\nonumber
\\
& 
\precsim
N^{1/2}
\left[
\begin{array}{l}
\big\| \widehat{\beta}_{1}\LSS - \beta_{1}^* \big\|_{P,2}
\big\| \widehat{\alpha}_1\LSS - \alpha_1^* \big\|_{P,2}
+
\big\| \widehat{\beta}_{1}\LSS - \beta_{1}^* \big\|_{P,2}
\big\| \hpotf{1}\LSS - \potf{1}^* \big\|_{P,2} 
\\
+
\big\| \widehat{\alpha}_1\LSS - \alpha_1^* \big\|_{P,2}
\big\| \hpotf{1}\LSS - \potf{1}^* \big\|_{P,2} 
+
\big\| \widehat{\beta}_{0}\LSS - \beta_{0}^* \big\|_{P,2}
\big\| \widehat{\alpha}_1\LSS - \alpha_1^* \big\|_{P,2}
\\
+
\big\| \widehat{\beta}_{0}\LSS - \beta_{0}^* \big\|_{P,2}
\big\| \hpotf{0}\LSS - \potf{0}^* \big\|_{P,2} 
+
\big\| \widehat{\alpha}_1\LSS - \alpha_1^* \big\|_{P,2}
\big\| \hpotf{0}\LSS - \potf{0}^* \big\|_{P,2}  
\end{array} 
\right] 
\ .
\label{eq-Bias Final}
\end{align}
Therefore, under the rate conditions on the nuisance function estimates, we establish \eqref{Term3-General} is $o_P(1)$.

\begin{itemize}
    \item[(ii)] (\textit{Asymptotic Property of \eqref{Term2-General}})
\end{itemize}

The expectation of \eqref{Term2-General} conditioning on $\mathcal{I}_k \LSS$ is 0. The variance of \eqref{Term2-General} is
\begin{align*}
\VAR^{(-k)} \big\{ \eqref{Term2-General} \big\}
\leq 
\VAR^{(-k)} \big\{ \widehat{\MM}_\EFF\LSS (\bO \con \btheta )
-
{\MM}_\EFF^* (\bO \con \btheta )   \big\}
\leq
\EXP^{(-k)} \big[ \big\| \widehat{\MM}_\EFF\LSS (\bO \con \btheta )
-
{\MM}_\EFF^* (\bO \con \btheta )  \big\|_2^2 \big] \ .
\end{align*}
Therefore, it suffices to find the rate of $\EXP^{(-k)} \big[ \big\| \widehat{\MM}_\EFF\LSS (\bO \con \btheta )
-
{\MM}_\EFF^* (\bO \con \btheta )  \big\|_2^2 \big]$. Each element $\widehat{\MM}_\EFF\LSS (\bO \con \btheta )
-
{\MM}_\EFF^* (\bO \con \btheta )  $ has the following form:
\begin{align*}
&
(1-A)
\widehat{\beta}_1\LSS(\bX) \widehat{\alpha}_1\LSS(\Yo ,\bX) \big\{ \MM(\Yo,\bX \con \btheta)  -  \widehat{\mu}_\MM\LSS(\bX \con \btheta)  \big\}
+
A
\widehat{\mu}_\MM\LSS(\bX \con \btheta) 
\\
&
\hspace*{0.25cm}
-
(1-A)
\beta_1^*(\bX) \alpha_1^*(\Yo ,\bX) \big\{ \MM(\Yo,\bX \con \btheta)  - \mu_\MM^*(\bX \con \btheta)  \big\}
-
A
\mu_\MM^*(\bX \con \btheta) 
\\
&
\hspace*{0.25cm}
+ (2A-1)
\widehat{R}\LSS (\Yz ,A , \bX)
\big\{
\MM(\Yz,\bX \con \btheta)
-
\widehat{\mu}_\MM\LSS(\bX \con \btheta) 
\big\} 
\\
&
\hspace*{0.25cm}
- (2A-1)
R^* (\Yz ,A , \bX)
\big\{
\MM(\Yz,  \bX \con \btheta)
-
\mu_\MM^*(\bX \con \btheta) 
\big\} 
\\
&
=
(1-A) \MM(\Yo,\bX \con \btheta) \big\{ \widehat{\beta}_1\LSS(\bX) \widehat{\alpha}_1\LSS(\Yo , \bX) - \beta_1^*(\bX) \alpha_1^*(\Yo ,\bX) \big\}
&&
\text{See \HL{S1}}
\\
&
\hspace*{0.25cm} 
-
(1-A) \big\{ \widehat{\beta}_1\LSS(\bX) \widehat{\alpha}_1\LSS(\Yo , \bX)  \widehat{\mu}_\MM\LSS(\bX \con \btheta)  - \beta_1^*(\bX) \alpha_1^*(\Yo ,\bX) \mu_\MM^*(\bX \con \btheta)  \big\}
&&
\text{See \HL{S2}}
\\
& 
\hspace*{0.25cm}
+ A \big\{ \widehat{\mu}_\MM\LSS(\bX \con \btheta)  - \mu_\MM^*(\bX \con \btheta)  \big\}
&&
\text{See \HL{S3}}
\\
&
\hspace*{0.25cm}
+
A \big\{ 
\widehat{R}\LSS (\Yz ,A , \bX)
-
R^* (\Yz ,A , \bX)
\big\} 
\big[
\MM(\Yz,\bX \con \btheta)
- 0.5 \big\{\mu_\MM^*(\bX \con \btheta) +  \widehat{\mu}_\MM\LSS(\bX \con \btheta)  \big\}
\big]
&&
\text{See \HL{S4}}
\\
& \hspace*{0.25cm}
+
0.5
A \big\{ 
\widehat{R}\LSS (\Yz ,A , \bX)
+
R^* (\Yz ,A , \bX)
\big\} \big\{ \mu_\MM^*(\bX \con \btheta)  -  \widehat{\mu}_\MM\LSS(\bX \con \btheta)  \big\}
&&
\text{See \HL{S5}}
\\
&
\hspace*{0.25cm}
-
(1-A) \big\{ 
\widehat{R}\LSS (\Yz ,A , \bX)
-
R^* (\Yz ,A , \bX)
\big\} 
\big[
\MM(\Yz,\bX \con \btheta)
- 0.5 \big\{ \mu_\MM^*(\bX \con \btheta)  +  \widehat{\mu}_\MM\LSS(\bX \con \btheta)  \big\}
\big]
&&
\text{See \HL{S4}}
\\
& \hspace*{0.25cm}
-
0.5
(1-A) \big\{ 
\widehat{R}\LSS (\Yz ,A , \bX)
+
R^* (\Yz ,A , \bX)
\big\} \big\{ \mu_\MM^*(\bX \con \btheta)  -  \widehat{\mu}_\MM\LSS(\bX \con \btheta) \big\} \ .
&&
\text{See \HL{S5}}
\end{align*}
For a finite number of random variables $\{W_1,\ldots, W_K\}$, there exists a constant $C$ satisfying $\EXP\{ (\sum_{j=1}^K W_j )^2 \}  \leq C \cdot \EXP (W_j^2)$. Thus, it suffices to study the rate of the 2-norm of each term, which are given in \HL{S1}-\HL{S5} below:

\begin{itemize}
\item[] \HT{S1}
\begin{align*}
&
\EXP \LSS \big[ (1-A)^2 \big\| \MM(\Yo,\bX \con \btheta) \big\|_2^2 \big\{ \widehat{\beta}_1\LSS(\bX) \widehat{\alpha}_1\LSS(\Yo , \bX) - \beta_1^*(\bX) \alpha_1^*(\Yo ,\bX) \big\}^2 \big]
\\
& 
=
\EXP \LSS \big[ (1-A) \big\| \MM(\Yo,\bX \con \btheta) \big\|_2^2 \big\{ \widehat{\beta}_1\LSS(\bX) \widehat{\alpha}_1\LSS(\Yo , \bX) - \beta_1^*(\bX) \alpha_1^*(\Yo ,\bX) \big\}^2 \big]
\\
& 
=
\Pr(A=0)
\EXP \LSS \big[ \big\| \MM(\Yo,\bX \con \btheta) \big\|_2^2 \big\{ \widehat{\beta}_1\LSS(\bX) \widehat{\alpha}_1\LSS(\Yo , \bX) - \beta_1^*(\bX) \alpha_1^*(\Yo ,\bX) \big\}^2 \, \big| \, A=0 \big]
\\
& 
\precsim
\EXP \LSS \big[   \underbrace{ \big\| \MM(\Yo,\bX \con \btheta) \big\|_2^2 \big\{ \widehat{\alpha}_1\LSS(\Yo , \bX) + \alpha_1^*(\Yo ,\bX) \big\}^2 }_{\leq C \text{ over } \Yo \in \SPo} \big\{ \widehat{\beta}_1\LSS(\bX) - \beta_1^*(\bX) \big\}^2 
\, \big| \, A=0 \big] 
\\
&
\hspace*{0.25cm}
+
\EXP \LSS \big[ \underbrace{ \big\| \MM(\Yo,\bX \con \btheta) \big\|_2^2 \big\{ \widehat{\beta}_1\LSS(\bX) + \beta_1^*(\bX) \big\} ^2 }_{\leq C \text{ over } \Yo \in \SPo}
\big\{ \widehat{\alpha}_1\LSS(\Yo , \bX) - \alpha_1^*(\Yo ,\bX) \big\}^2  \, \big| \, A=0 \big] 
\\
&
\precsim
\big\| \widehat{\beta}_1\LSS(\bX) - \beta_1^*(\bX) \big\|_{P,2}^2
+
\big\| \widehat{\alpha}_1\LSS(\Yo , \bX) - \alpha_1^*(\Yo ,\bX)  \big\|_{P,2}^2 \ .
\end{align*}

\item[] \HT{S2}
\begin{align*}
&
\EXP \LSS \big[ (1-A)^2 \big\| \widehat{\beta}_1\LSS(\bX) \widehat{\alpha}_1\LSS(\Yo , \bX)  \widehat{\mu}_\MM\LSS(\bX \con \btheta)  - \beta_1^*(\bX) \alpha_1^*(\Yo ,\bX) \mu_\MM^*(\bX \con \btheta)  \big\|_2^2 \big]
\\
& 
\precsim
\EXP \LSS \Bigg[
\underbrace{
\Bigg\{ 
\begin{array}{l}
\widehat{\alpha}_1\LSS(\pYo , \bX) \\
+ \alpha_1^*(\pYo,\bX) 
\end{array}
\Bigg\}^2 
}_{\leq C \text{ over } \Yo \in \SPo}
\Bigg\|
\begin{array}{l}
    \big\{ \widehat{\beta}_1\LSS(\bX) - \beta_1^*(\bX)\big\} \big\{  \widehat{\mu}_\MM\LSS(\bX \con \btheta)   +  \mu_\MM^*(\bX \con \btheta)  \big\}
    \\
    +
    \big\{ \widehat{\beta}_1\LSS(\bX) + \beta_1^*(\bX)\big\} \big\{ \widehat{\mu}_\MM\LSS(\bX \con \btheta)   -  \mu_\MM^*(\bX \con \btheta)  \big\}
\end{array}
\Bigg\|_2^2
\, \Bigg| \, A=0
\Bigg]
\\
&
\hspace*{0.25cm}
+
\EXP \LSS \big[ \underbrace{ \big\| \widehat{\beta}_1\LSS(\bX)  \widehat{\mu}_\MM\LSS(\bX \con \btheta)  + \beta_1^*(\bX) \mu_\MM^*(\bX \con \btheta)  \big\|_2^2 }_{\leq C \text{ over } \Yo \in \SPo}
\big\{ \widehat{\alpha}_1\LSS(\pYo , \bX)  - \alpha_1^*(\pYo,\bX) \big\}^2
\, \big| \, A=0
\big]
\\
&
\precsim
\EXP \LSS \big\{ \| \widehat{\beta}_1\LSS(\bX) - \beta_1^*(\bX) \big\|^2 \, \big| \, A=0 \big\}
\\
& \hspace*{0.25cm}
+
\EXP \LSS \big\{ \big\|  \widehat{\mu}_\MM\LSS(\bX \con \btheta)  - \mu_\MM^*(\bX \con \btheta)  \big\|^2 \, \big| \, A=0 \big\}
+
\big\| \widehat{\alpha}_1\LSS(\Yo , \bX) - \alpha_1^*(\Yo ,\bX)  \big\|_{P,2}^2
\\
& 
\precsim
\big\| \widehat{\beta}_1\LSS(\bX) - \beta_1^*(\bX) \big\|_{P,2}^2
+
\big\| \alpha_1 (\Yo,\bX) - \widehat{\alpha}_1\LSS (\Yo,\bX) \big\|_{P,2}^2
+
\big\|  \potf{1}^*(\Yo \cond 0,\bX) - \hpotf{1}\LSS(\Yo \cond 0,\bX)  \big\|_{P,2} ^2 \ .
\end{align*}
The last line is from \eqref{eq-rate-mu1}.

\item[]  \HT{S3}  The result is straightforward from \eqref{eq-rate-mu1}.
\begin{align*}
&
\EXP \big[ A \big\|  \widehat{\mu}_\MM\LSS(\bX \con \btheta)  - \mu_\MM^*(\bX \con \btheta)  \big\|_2^2 \big] 
\\
& \precsim \EXP \big[ \big\|  \widehat{\mu}_\MM\LSS(\bX \con \btheta)  - \mu_\MM^*(\bX \con \btheta)  \big\|_2^2 \, \big| \, A=0 \big] 
\\
&
\precsim 
\big\| \alpha_1 (\Yo,\bX) - \widehat{\alpha}_1\LSS (\Yo,\bX) \big\|_{P,2}^2
+
\big\|  \potf{1}^*(\Yo \cond 0,\bX) - \hpotf{1}\LSS(\Yo \cond 0,\bX)  \big\|_{P,2} ^2 \ .
\end{align*}

\item[]  \HT{S4}
\begin{align*}
&
\EXP\LSS
\big[
A^2 \big\{ 
\widehat{R}\LSS (\Yz , A , \bX)
-
R^* (\Yz ,A , \bX)
\big\}^2
\big\|
\MM(\Yz,\bX \con \btheta) 
- 
0.5 \big\{ \mu_\MM^*(\bX \con \btheta)  +  \widehat{\mu}_\MM\LSS(\bX \con \btheta)  \big\}
\big\|_2^2
\big]
\\
&
\precsim
\EXP\LSS
\big[
A 
\underbrace{
\big\|
\MM(\Yz,\bX \con \btheta)
- 
0.5 \big\{ \mu_\MM^*(\bX \con \btheta)  +  \widehat{\mu}_\MM\LSS(\bX \con \btheta)  \big\}
\big\|_2^2
}_{\leq C \text{ over } \Yz \in \SPzt}
\big\{ 
\widehat{R}\LSS (\Yz , 1 , \bX)
-
R^* (\Yz ,1 , \bX)
\big\}^2 
\big]  
\\
&
\precsim
\EXP\LSS
\big[
\Pr(A=1 \cond \bX)
\EXP\LSS
\big[
\big\{ 
\widehat{R}\LSS (\Yz , 1 , \bX)
-
R^* (\Yz ,1 , \bX)
\Big\}^2 
\, \big| \, A=1, \bX
\big]
\, \big| \, \bX
\big]  
\\
& 
=
\EXP\LSS
\bigg[
\Pr(A=1 \cond \bX)
\int_{\SPzt}
\Big\{ 
\widehat{R}\LSS (y , 1 , \bX)
-
R^* (y ,1 , \bX)
\Big\}^2 
\potf{0}^*(y \cond 1, \bX) \, dy
\, \bigg| \, \bX
\bigg]
\\
&
\precsim
\EXP \LSS
\bigg[
\bigg\{
\frac{ \widehat{\beta}_1\LSS(\bX) \hpotf{1}\LSS(\Yo \cond 0, \bX) }{ \widehat{\beta}_{0} \LSS (\bX) \hpotf{0}\LSS (\Yo \cond 0, \bX) }
-
\frac{  {\beta}_{1}^* (\bX) \potf{1} ^* (\Yo \cond 0, \bX) }{  {\beta}_{0}^* (\bX)  \potf{0}^* (\Yo \cond 0, \bX) }
\bigg\}^2 \, \bigg| \, A=0
\bigg]
\\
&
\precsim \big\| \hpotf{0} \LSS (\Yo \cond 0, \bX) - \potf{0}^* (\Yo \cond 0, \bX) \big\|_{P,2}^2
+
\big\| \widehat{\beta}_{0} \LSS (\bX) - \beta_{0}^* (\bX) \big\|_{P,2}^2
\\
&
\hspace*{1cm}
+
 \big\| \hpotf{1}\LSS(\Yo \cond 0, \bX) - \potf{1}^*(\Yo \cond 0, \bX) \big\|_{P,2}^2
+
\big\| \widehat{\beta}_1\LSS(\bX) - \beta_1^*(\bX) \big\|_{P,2}^2	
\end{align*}
Note that the fifth line holds from the following result:
\begin{align*}
& 
\int_{\SPzt}
\Big\{ 
\widehat{R}\LSS (y , 1 , \bX)
-
R^* (y ,1 , \bX)
\Big\}^2 
\potf{0}^*(y \cond 1, \bX) \, dy
\\
&
=
\int_{\SPzt}
\Bigg\{
\frac{ \widehat{\beta}_1\LSS (\bX) }{ \widehat{\beta}_0\LSS (\bX) }
\frac{  		\hpotf{1}\LSS(y \cond 0, \bX)  }{ \hpotf{0}\LSS(y \cond 0, \bX) }
-
\frac{  {\beta}_{1}^* (\bX) \potf{1} ^* (y \cond 0, \bX) }{  {\beta}_{0}^* (\bX)  \potf{0}^* (y \cond 0, \bX) }
\Bigg\}^2
\potf{0}^*(y \cond 1, \bX) \, dy
\\
&
=
\int_{\SPzt \cap \SPo}
\Bigg\{
\frac{ \widehat{\beta}_1\LSS (\bX) }{ \widehat{\beta}_0\LSS (\bX) }
\frac{  		\hpotf{1}\LSS(y \cond 0, \bX)  }{ \hpotf{0}\LSS(y \cond 0, \bX) }
-
\frac{  {\beta}_{1}^* (\bX) \potf{1} ^* (y \cond 0, \bX) }{  {\beta}_{0}^* (\bX)  \potf{0}^* (y \cond 0, \bX) }
\Bigg\}^2
\potf{0}^*(y \cond 1, \bX) \, dy
\\
&
\leq 
\int_{ \SPo}
\frac{\potf{0}^*(y \cond 1, \bX)}{\potf{1}^*(y \cond 0, \bX)} 
\Bigg\{
\frac{ \widehat{\beta}_1\LSS (\bX) }{ \widehat{\beta}_0\LSS (\bX) }
\frac{  		\hpotf{1}\LSS(y \cond 0, \bX)  }{ \hpotf{0}\LSS(y \cond 0, \bX) }
-
\frac{  {\beta}_{1}^* (\bX) \potf{1} ^* (y \cond 0, \bX) }{  {\beta}_{0}^* (\bX)  \potf{0}^* (y \cond 0, \bX) }
\Bigg\}^2
\potf{1}^*(y \cond 0, \bX) \, dy
\\
&
\precsim	
\int_{ \SPo}
\Bigg\{
\frac{ \widehat{\beta}_1\LSS (\bX) }{ \widehat{\beta}_0\LSS (\bX) }
\frac{  		\hpotf{1}\LSS(y \cond 0, \bX)  }{ \hpotf{0}\LSS(y \cond 0, \bX) }
-
\frac{  {\beta}_{1}^* (\bX) \potf{1} ^* (y \cond 0, \bX) }{  {\beta}_{0}^* (\bX)  \potf{0}^* (y \cond 0, \bX) }
\Bigg\}^2
\potf{1}^*(y \cond 0, \bX) \, dy \ . 
\end{align*} 
The third line holds from $\SPo = \text{supp} \big\{ \hpotf{1}\LSS (y \cond 0, \bX) \big\} = \text{supp} \big\{ \potf{1}^* (y \cond 0, \bX) \big\} $.

Substituting $A R^*(y,1,x)$ with $(1-A) R^*(y,0,x)$, we obtain the similar result:	
\begin{align*}
&
\EXP\LSS
\big[
(1-A)^2 \big\{ 
\widehat{R}\LSS (\Yz , A , \bX)
-
R^* (\Yz ,A , \bX)
\big\}^2
\big\|
\MM(\Yz,\bX \con \btheta) 
- 
0.5 \big\{\mu_\MM^*(\bX \con \btheta)  +  \widehat{\mu}_\MM\LSS(\bX \con \btheta)  \big\}
\big\|_2^2
\big]
\\
&
\precsim
\EXP\LSS
\big[
(1-A) 
\underbrace{ \big\|
\MM(\Yz,\bX \con \btheta)
- 
0.5 \big\{ \mu_\MM^*(\bX \con \btheta)  +  \widehat{\mu}_\MM\LSS(\bX \con \btheta)  \big\}
\big\|_2^2
}_{\leq C \text{ over } \Yz \in \SPz}
\big\{ 
\widehat{R}\LSS (\Yz , 0 , \bX)
-
R^* (\Yz ,0 , \bX)
\big\}^2 
\big]  
\\
&
\precsim
\EXP\LSS
\big[
\Pr(A=0 \cond \bX)
\EXP\LSS
\big[
\big\{ 
\widehat{R}\LSS (\Yz , 0 , \bX)
-
R^* (\Yz ,0 , \bX)
\big\}^2 
\, \big| \, A=0 , \bX
\big]  
\, \big| \, \bX
\big]  
\\
&
\precsim
\EXP\LSS
\bigg[
\Pr(A=0 \cond \bX)
\int_{\SPz}  \Big\{ 
\widehat{R}\LSS (y , 0 , \bX)
-
R^* (y ,0 , \bX)
\Big\}^2 
\potf{0}^*(y \cond 0, \bX) \, dy
\bigg]  
\, \bigg| \, \bX
\bigg]  
\\
&
\precsim
\EXP  \LSS
\bigg[
\bigg\{
\frac{ \widehat{\alpha}_1\LSS(y, \bX) \widehat{\beta}_1\LSS(\bX) \hpotf{1}\LSS(y \cond 0, \bX) }{  \hpotf{0} \LSS (y \cond 0, \bX) }
-
\frac{   {\alpha_1}^* (y, \bX)  {\beta}_{1}^* (\bX)  \potf{1}^*(y \cond 0, \bX) }{   \potf{0}^* (y \cond 0, \bX) }
\bigg\}^2 \, \bigg| \, A=0
\bigg]
\\
&
\precsim
\big\| \hpotf{0} \LSS (\Yo \cond 0, \bX) -  \potf{0}^*(\Yo \cond 0, \bX) \big\|_{P,2}^2
+
\big\| \widehat{\alpha}_1\LSS(\Yo,\bX) - {\alpha_1}^* (\Yo,\bX) \big\|_{P,2}^2
\\
& \hspace*{0.25cm}
+
 \big\| \hpotf{1}\LSS(\Yo \cond 0, \bX) - \potf{1}^*(\Yo \cond 0, \bX) \big\|_{P,2}^2
+
\big\| \widehat{\beta}_1\LSS(\bX) - \beta_1^*(\bX) \big\|_{P,2}^2
\end{align*}
Note that the fifth line holds from the following result: 		
\begin{align*}
& 
\int_{\SPz}
\Big\{ 
\widehat{R}\LSS (y , 0 , \bX)
-
R^* (y ,0 , \bX)
\Big\}^2 
\potf{0}^*(y \cond 0, \bX) \, dy
\\
&
=
\int_{\SPz}
\Bigg\{
\frac{ \widehat{\beta}_1\LSS (\bX) \widehat{\alpha}_1\LSS (y,\bX) \hpotf{1}\LSS(y \cond 0, \bX)  }{ \hpotf{0}\LSS(y \cond 0, \bX) } 
-
\frac{ \beta_1^* (\bX) \alpha_1^* (y,\bX) \potf{1}^*(y \cond 0, \bX)  }{ \potf{0}^*(y \cond 0, \bX) }
\Bigg\}^2
\potf{0}^*(y \cond 0, \bX) \, dy
\\
&
=
\int_{\SPz \cap \SPo}
\Bigg\{
\frac{ \widehat{\beta}_1\LSS (\bX) \widehat{\alpha}_1\LSS (y,\bX) \hpotf{1}\LSS(y \cond 0, \bX)  }{ \hpotf{0}\LSS(y \cond 0, \bX) } 
-
\frac{ \beta_1^* (\bX) \alpha_1^* (y,\bX) \potf{1}^*(y \cond 0, \bX)  }{ \potf{0}^*(y \cond 0, \bX) }
\Bigg\}^2
\potf{0}^*(y \cond 0, \bX) \, dy
\\
&
\leq
\int_{\SPo}
\frac{\potf{0}^*(y \cond 0, \bX)}{\potf{1}^*(y \cond 0, \bX)} 
\Bigg\{
\frac{ \widehat{\beta}_1\LSS (\bX) \widehat{\alpha}_1\LSS (y,\bX) \hpotf{1}\LSS(y \cond 0, \bX)  }{ \hpotf{0}\LSS(y \cond 0, \bX) } 
-
\frac{ \beta_1^* (\bX) \alpha_1^* (y,\bX) \potf{1}^*(y \cond 0, \bX)  }{ \potf{0}^*(y \cond 0, \bX) }
\Bigg\}^2
\potf{1}^*(y \cond 0, \bX) \, dy
\\
&
\precsim
\int_{\SPo}
\Bigg\{
\frac{ \widehat{\beta}_1\LSS (\bX) \widehat{\alpha}_1\LSS (y,\bX) \hpotf{1}\LSS(y \cond 0, \bX)  }{ \hpotf{0}\LSS(y \cond 0, \bX) } 
-
\frac{ \beta_1^* (\bX) \alpha_1^* (y,\bX) \potf{1}^*(y \cond 0, \bX)  }{ \potf{0}^*(y \cond 0, \bX) }
\Bigg\}^2
\potf{1}^*(y \cond 0, \bX) \, dy
\end{align*}

\item[]  \HT{S5} Using \eqref{eq-rate-mu1}, we obtain 
\begin{align*}
&
\EXP \LSS \big[
A^2 \big\{ 
\widehat{R}\LSS (\Yz ,A , \bX)
+
R^* (\Yz ,A , \bX)
\big\}^2 \big\| \mu_\MM^*(\bX \con \btheta)  -  \widehat{\mu}_\MM\LSS(\bX \con \btheta)  \big\|_2^2
\big]
\\
&
\precsim 
\big\| \alpha_1^* (\Yo,\bX) - \widehat{\alpha}_1\LSS (\Yo,\bX) \big\|_{P,2}^2
+
\big\|  \potf{1}^*(\Yo \cond 0,\bX) - \hpotf{1}\LSS(\Yo \cond 0,\bX)  \big\|_{P,2} ^2 \ .
\end{align*}
Similar result holds for $(1-A)^2 \big\{ \widehat{R}\LSS (\Yz ,A , \bX)
+
R^* (\Yz ,A , \bX)
\big\}^2 \big\| \mu_\MM^*(\bX \con \btheta)  -  \widehat{\mu}_\MM\LSS(\bX \con \btheta)  \big\|_2^2$:
\begin{align*}
&
\EXP \LSS \big[ (1-A)^2 \big\{ \widehat{R}\LSS (\Yz ,A , \bX)
+
R^* (\Yz ,A , \bX)
\big\}^2 \big\| \mu_\MM^*(\bX \con \btheta)  -  \widehat{\mu}_\MM\LSS(\bX \con \btheta)  \big\|_2^2 \big]
\\
&
\precsim 
\big\| \alpha_1^* (\Yo,\bX) - \widehat{\alpha}_1\LSS (\Yo,\bX) \big\|_{P,2}^2
+
\big\|  \potf{1}^*(\Yo \cond 0,\bX) - \hpotf{1}\LSS(\Yo \cond 0,\bX)  \big\|_{P,2} ^2  \ .
\end{align*}
\end{itemize}
Combining the results, we find
\begin{align}				\label{eq-variance0}
&
\VAR^{(-k)} \big\{ \eqref{Term2-General} \big\}
\\
&
\leq
\EXP^{(-k)} \big[ \big\| \widehat{\MM}_\EFF\LSS (\bO \con \btheta )
-
{\MM}_\EFF^* (\bO \con \btheta )  \big\|_2^2 \big] 
\nonumber
\\
&
\precsim
\big\| \widehat{\alpha}_1\LSS  - \alpha_1^*  \big\|_{P,2}^2 
+
\big\| \widehat{\beta}_0\LSS  - \beta_0^*  \big\|_{P,2}^2 	
+
\big\| \widehat{\beta}_1\LSS  - \beta_1^*  \big\|_{P,2}^2 	
+
\big\|  \hpotf{0}\LSS  - \potf{0}^* \big\|_{P,2} ^2 
+
\big\|   \hpotf{1}\LSS  - \potf{1}^* \big\|_{P,2} ^2  \ .
\nonumber
\end{align}				
Therefore, under the assumptions, $\VAR^{(-k)} \big\{ \eqref{Term2-General} \big\} = o_P(1)$, indicating \eqref{Term2-General} is $o_P(1)$.

\begin{itemize}
    \item[(iii)] (\textit{Consistent Variance Estimation})
\end{itemize}

The proposed variance estimator is 
\begin{align*}
\widehat{\sigma}^2 
=
\frac{1}{K}
\sum_{k=1}^{K} 
\widehat{\sigma}^{2,(k)}
\ , \
\widehat{\sigma}^{2,(k)}
=
\AVER_{\mathcal{I}_k}
\bigg[
\bigg\{
\frac{AY - \widehat{\phi}_0\LSS(\bO) - A \widehat{\tau} }{ \AVER \big( A \big)}		
\bigg\}^2
\bigg] \ .
\end{align*}
Therefore, it suffices to show that $ 	\widehat{\sigma}^{2,(k)} - \sigma^2 = o_P(1)$, which is represented as follows:
\begin{align}
&
\widehat{\sigma}^{2,(k)} - \sigma^2
\nonumber
\\
&
=
\big\{ \AVER(A) \big\}^{-2}
\AVER_{\mathcal{I}_k}
\Big[
\big\{ AY - \widehat{\phi}_0\LSS(\bO) - A \widehat{\tau}  \big\}^2
\Big] -
\sigma^2
\nonumber
\\
&
=
\big\{ \Pr(A=1) \big\}^{-2}
\AVER_{\mathcal{I}_k}
\Big[
\big\{ AY - \widehat{\phi}_0\LSS(\bO) - A \widehat{\tau}  \big\}^2
\Big] -
\sigma^2 + o_P(1)
\nonumber
\\
&
=
\big\{ \Pr(A=1) \big\}^{-2}
\Big[
\AVER_{\mathcal{I}_k}
\Big[
\big\{ AY - \widehat{\phi}_0\LSS(\bO) - A \widehat{\tau}  \big\}^2
\Big] -
\AVER_{\mathcal{I}_k}
\Big[
\big\{ AY - \phi_0^*(\bO) - A \tau^*  \big\} ^2
\Big]
\Big]
\nonumber
\\
& \hspace*{2cm}
+		 
\Big[
\big\{ \Pr(A=1) \big\}^{-2}
\AVER_{\mathcal{I}_k}
\Big[
\big\{ AY - \phi_0^*(\bO) - A \tau^*  \big\} ^2
\Big]
-
\sigma^2
\Big] + o_P(1)
\nonumber
\\
& = 
\big\{ \Pr(A=1) \big\}^{-2}
\Big[
\AVER_{\mathcal{I}_k}
\Big[
\big\{ AY -  \widehat{\phi}_0\LSS(\bO) - A \widehat{\tau} \big\}^2
-
\Big\{
\big( AY -  \phi_0^*(\bO) - A \tau^*  \big)^2
\Big\}
\Big]
+ o_P(1)
\label{eq-variance1} \ .
\end{align}
The third and fifth lines hold from the law of large numbers.  
Therefore, it is sufficient to show that \eqref{eq-variance1} is also $o_P(1)$. From some algebra, we find the term in  \eqref{eq-variance1} is
\begin{align*}
&
\frac{1}{|\mathcal{I}_k|}
\sum_{i \in \mathcal{I}_k}
\Big[ 
\big\{
A_i Y_{1,i}
-
\widehat{\phi}_{0}\LSS (\bO_i)
-
A_i \widehat{\tau}
\big\}^2
-
\big\{
A_i Y_{1,i}
-
\phi_{0}^* (\bO_i)
-
A_i
\tau^*
\big\}^2
\Big]
\\
& =
\frac{1}{|\mathcal{I}_k|}
\sum_{i \in \mathcal{I}_k}
\left[
\begin{array}{l}
\big[
\big\{
\widehat{\phi}_{0}\LSS (\bO_i)
-
A_i
\widehat{\tau}
\big\}
-
\big\{
\phi_{0}^* (\bO_i)
-
A_i
\tau^*
\big\}
\big]
\\
\times 
\big[
\big\{
A_i Y_{1,i}
-
\widehat{\phi}_{0}\LSS (\bO_i)
-
A_i
\widehat{\tau}
\big\}
+
\big\{
A_i Y_{1,i}
-
\phi_{0}^* (\bO_i)
-
A_i
\tau^*
\big\}
\big]
\end{array} 
\right]
    \\
& = 
\frac{1}{|\mathcal{I}_k|}
\sum_{i \in \mathcal{I}_k}
\left[
\begin{array}{l}
\big[
\big\{
\widehat{\phi}_{0}\LSS (\bO_i)
-
A_i
\widehat{\tau}
\big\}
-
\big\{
\phi_{0}^* (\bO_i)
-
A_i
\tau^*
\big\}
\big] \\
\times \big[
\big\{
\widehat{\phi}_{0}\LSS (\bO_i)
-
A_i
\widehat{\tau}
\big\}
-
\big\{
\phi_{0}^* (\bO_i)
-
A_i
\tau^*
\big\}
+ 
2 \big\{
A_i Y_{1,i}
-
\phi_{0}^* (\bO_i)
-
A_i
\tau^*
\big\}
\big]
\end{array}
\right]
    \\
& = 
\frac{1}{|\mathcal{I}_k|}
\sum_{i \in \mathcal{I}_k}
\Big[
\big\{
\widehat{\phi}_{0}\LSS (\bO_i)
-
A_i
\widehat{\tau}
\big\}
-
\big\{
\phi_{0}^* (\bO_i)
-
A_i
\tau^*
\big\}
\Big]^2
\\
&
\quad 
+
\frac{2 }{|\mathcal{I}_k|}
\sum_{i \in \mathcal{I}_k}
\Big[
\big[
\big\{
\widehat{\phi}_{0}\LSS (\bO_i)
-
A_i
\widehat{\tau}
\big\}
-
\big\{
\phi_{0}^* (\bO_i)
-
A_i
\tau^*
\big\}
\big]
\big\{
A_i Y_{1,i} 
-
\phi_{0}^* (\bO_i)
-
A_i
\tau^*
\big\}
\Big]			\ .
\end{align*}
Let $\widehat{\Delta}_i\LSS = \big\{
\widehat{\phi}_{0}\LSS(\bO_i)
-
A_i
\widehat{\tau}
\big\}
-
\big\{
\phi_{0}^*(\bO_i)
-
A_i
\tau^*
\big\}$. From the H\"older's inequality, we find the absolute value of \eqref{eq-variance1} is upper bounded by
\begin{align*}
    \big\| \eqref{eq-variance1} \big\|
    & \precsim
    \AVER_{\mathcal{I}_k} \Big[  \big\{ \widehat{\Delta} \LSS \big\}^2 \Big]
    +
    2
    \AVER_{\mathcal{I}_k} \Big[  \big\{ \widehat{\Delta} \LSS \big\}^2 \Big]
    \cdot 
    \AVER_{\mathcal{I}_k} \Big\{  \big( A Y_{1} - \phi_0^*(\bO) - A \tau^* \big)^2 \Big\} \ .
\end{align*}
Since $\AVER_{\mathcal{I}_k} \big\{  \big(  A Y_{1} - \phi_0^*(\bO) - A \tau^* \big)^2 \big\} = \Pr(A=1)^2 \sigma^2 + o_P(1) = O_P(1)$, \eqref{eq-variance1} is $o_P(1)$ if $\AVER_{\mathcal{I}_k} \big[  \big\{ \widehat{\Delta} \LSS \big\}^2 \big] = o_P(1)$. From some algebra, we find
\begin{align*}
    \AVER_{\mathcal{I}_k} \Big[  \big\{ \widehat{\Delta} \LSS \big\}^2 \Big]
    & \leq
    \frac{2}{|\mathcal{I}_k|}
\sum_{i \in \mathcal{I}_k} 
\big\{
\widehat{\phi}_{0}\LSS (\bO_i)
-
\phi_{0}^*(\bO_i)
\big\}^2
+
2 \big( \widehat{\tau}  - \tau^* \big)^2
\\
& =
2 \EXP\LSS\Big[
\big\{
\widehat{\phi}_0\LSS(\bO)
-
\phi_0^*(\bO)
\big\}^2 \Big]
+
2 \big( \widehat{\tau}  - \tau^* \big)^2
 + o_P(1)
=
o_P(1) \ . 
\end{align*}
The first line holds from $(\ell_1+A\ell_2)^2 \leq 2\ell_1^2 + 2\ell_2^2$. The second line holds from the law of large numbers applied to $\big\{ \widehat{\phi}_0\LSS
-
\phi_0^*
\big\}^2$. The last line holds from \eqref{eq-variance0} and $\widehat{\tau} = \tau^* + o_P(1)$, which is from the asymptotic normality of the estimator.

\begin{itemize}
    \item[(iv)] (\textit{Consistency of $\widehat{\tau}$ under Assumptions \ref{assumption:support}-\ref{assumption:ConsistentEstimation}})
\end{itemize}

The bias of the estimator $\widehat{\tau}_{n,0}^{(k)}$ can be decomposed as follows:
\begin{align}
\widehat{\tau}_{n,0}^{(k)} - \tau_{n,0}^*
&
=
\frac{1}{\big| \mathcal{I}_k \big|}
\sum_{ i \in \mathcal{I}_k } \big\{ \widehat{\phi}_0\LSS (\bO_i) - \tau_{n,0}^* \big\} 
\nonumber
\\
&
= 
\frac{1}{\big| \mathcal{I}_k \big|^{1/2}} 
 \EMP_{\mathcal{I}_k} \big( \uncInfFt_0^*  - \tau_{n,0}^* \big)
\label{Term1-B}
\\
& 
\ \
+
\EXP\LSS
\big( \widehat{\uncInfFt}_0\LSS - \uncInfFt_0^* \big) 
\label{Term3-B}
\\
& 
\ \
+
\frac{1}{\big| \mathcal{I}_k \big|^{1/2}}
\EMP_{\mathcal{I}_k}\LSS
\big(
\widehat{\uncInfFt}_0\LSS - \uncInfFt_0^*
\big)
\label{Term2-B} \ .
\end{align} 

We analyze these three components individually. First, since $\EMP_{\mathcal{I}_k} \big( \uncInfFt_0^*  - \tau_{n,0}^* \big) \stackrel{D}{\rightarrow} N(0,\sigma^2)$, \eqref{Term1-B} is $o_P(1)$. Second, in the derivation of \eqref{Term2}, we established $\EMP_{\mathcal{I}_k}\LSS
\big(
\widehat{\uncInfFt}_0\LSS - \uncInfFt_0^*
\big) = o_P(1)$, without requiring Assumptions \ref{assumption:PreCross} and \ref{assumption:PostCross}. Therefore, \eqref{Term2-B} is $o_P(1)$ under Assumptions \ref{assumption:support}-\ref{assumption:ConsistentEstimation}. Lastly, in the derivation of \eqref{Term3}, specifically \eqref{eq-Bias Final}, we established
\begin{align*}
\eqref{Term3-B}
=
\frac{\eqref{Term3}}{\big| \mathcal{I}_k \big|^{1/2}}
\lesssim
\left[
\begin{array}{l}
\big\| \widehat{\beta}_{1}\LSS - \beta_{1}^* \big\|_{P,2}
\big\| \widehat{\alpha}_1\LSS - \alpha_1^* \big\|_{P,2}
+
\big\| \widehat{\beta}_{1}\LSS - \beta_{1}^* \big\|_{P,2}
\big\| \hpotf{1}\LSS - \potf{1}^* \big\|_{P,2} 
\\
+
\big\| \widehat{\alpha}_1\LSS - \alpha_1^* \big\|_{P,2}
\big\| \hpotf{1}\LSS - \potf{1}^* \big\|_{P,2} 
+
\big\| \widehat{\beta}_{0}\LSS - \beta_{0}^* \big\|_{P,2}
\big\| \widehat{\alpha}_1\LSS - \alpha_1^* \big\|_{P,2}
\\
+
\big\| \widehat{\beta}_{0}\LSS - \beta_{0}^* \big\|_{P,2}
\big\| \hpotf{0}\LSS - \potf{0}^* \big\|_{P,2} 
+
\big\| \widehat{\alpha}_1\LSS - \alpha_1^* \big\|_{P,2}
\big\| \hpotf{0}\LSS - \potf{0}^* \big\|_{P,2}  
\end{array} 
\right]  \ .
\end{align*}
This term is $o_P(1)$ under Assumptions \ref{assumption:support}-\ref{assumption:ConsistentEstimation}.

 This concludes the proof.

\section{Proof of the Lemmas and Theorems in the Supplementary Material}									\label{sec:supp-proof-supp}

In this section, we use the following shorthand for the conditional distributions for $t=0,1$:
\begin{align*}
& \potg{t}^* (y,a \cond \bx) = \potf{tA|X}^*(y,a \cond \bx)  = P(\potY{0}{t}=y,A=a \cond \bX=\bx)
\ , \\
& \potf{t}^* (y \cond a,\bx) = \potf{t|AX}^* (y \cond a, \bx) = P(\potY{0}{t}=y \cond A=a,\bX=\bx)
\ , \\
& 
\pote{t}^* (a \cond y,\bx) = \potf{A|tX}^*(a \cond y,\bx) = \Pr(A=a \cond \pYz=y, \bX=\bx)
\ .
\end{align*}
Similarly, we denote
\begin{align*}
& \potf{t|X}^* (y \cond \bx) = P(\potY{0}{t}=y \cond \bX=\bx)
\ , 
&& \potf{A|X}^* (a \cond \bx) = \Pr(A=a \cond \bX=\bx) \ .
\end{align*}

\subsection{Proof of Lemma \ref{lemma-PT_OREC}}

The parallel trend assumption implies $\mathcal{T}(\alpha_1,\potf{10} \con \bx) = \mathcal{T}(\alpha_0,\potf{00} \con \bx)$. 
Let $\varphi_{y,\potf{t0}}^{-1}: \mathcal{F}_{t X} \rightarrow \mathcal{F}_X$ be the inverse mapping of $\varphi_{y,\potf{t0}}$, which is well-defined because $\varphi_{t,\potf{t0}}$ is injective.
We find that
\begin{align*}
\varphi_{y,\potf{10}}^{-1} \big( \alpha_1(y,\bx) \big) = \mathcal{T}(\alpha_1,\potf{10} \con \bx) = \mathcal{T}(\alpha_0,\potf{00} \con \bx) = \varphi_{y,\potf{00}}^{-1} \big( \alpha_0(y,\bx) \big) \ . 
\end{align*}
This implies $\alpha_1(y,\bx) = \varphi_{y,\potf{10}} \big( \varphi_{y,\potf{00}}^{-1} \big( \alpha_0(y,\bx) \big) \big)$. 
Therefore, $\alpha_1$ is variationally dependent to $\potf{10}$ unless the mapping $h (y,\bx) \mapsto \phi (h(y,\bx), \potf{00},\potf{10}) := \varphi_{y,\potf{10}} \big(  \varphi_{y,\potf{00}}^{-1} \big(  h(y,\bx)  \big) \big)$ is independent of $\potf{10}$. 
From the same logic, $\alpha_0$ is variationally dependent to $\potf{00}$ unless the mapping $\phi(\cdot , \potf{10},\potf{00})$ is independent of $\potf{00}$.
Therefore, $\phi$ should be a fixed map that does not depend on $\potf{10}$ and $\potf{00}$. 
This implies there exists a one-to-one fixed mapping between $\alpha_1$ and $\alpha_0$. 
Consequently, returning to the \nHL{PT} assumption, we find $\mathcal{T}_{\bx} (\alpha_1,\potf{10}) = \mathcal{T}_{\bx} (\phi(\alpha_0),\potf{10}) = \mathcal{T}_{\bx} (\alpha_0,\potf{00})$.

\subsection{Proof of Lemma \ref{lemma-DR}} \label{sec:supp:proof-lemma-DR}

In the proof, we show a more general result by characterizing the EIF for $ \tau^*(\mathcal{G}) := \EXP \{ \mathcal{G}(\potY{1}{1}) - \mathcal{G}(\potY{0}{1}) \cond A=1 \}$, where $\mathcal{G}(\cdot)$ is a fixed, integrable function. With a slight abuse of notation, we denote  $\mu^*(\bX) = \EXP \big\{ \mathcal{G}(\potY{0}{1}) \cond A=1 , \bX \big\}$.

First, suppose $\beta_1$ is correctly specified whereas $\potf{10}$ is misspecified as $f'$. Then, the outcome regression is misspecified as $\mu' (\bX) = \mu(\bX \con f')$. Additionally, we obtain
\begin{align*} 
&
\EXP \big[
(1-A) 
\beta_1^*(\bX) \alpha_1^*(\Yo , \bX)
\big\{  \mathcal{G}(\Yo) - \mu' (\bX) \big\}
\big]
\\
& = 
\EXP \big[
\Pr(A=0 \cond \bX)
\beta_1^*(\bX) 
\EXP \big[
\alpha_1^*(\Yo , \bX)
\big\{  \mathcal{G}(\Yo) - \mu' (\bX) \big\}  
\cond A=0 , \bX
\big]
\big]
\\
& = 
\EXP \big[\Pr(A=1 \cond \bX)\EXP \big\{
\mathcal{G}(\pYo) - \mu'(\bX)
\cond A=1 , \bX
\big\}
\big]
\\
& = 
\Pr(A=1)
\EXP \big \{ \mathcal{G}(\pYo)- \mu'(\bX) \cond A=1 \big\}
\\
& = 
\EXP \big[  A \big\{  \mathcal{G}(\pYo) - \mu' (\bX) \big\} \big]
\ .
\end{align*}
The third line is from \eqref{eq-IPW-basis}. Combining all, we obtain
\begin{align*}
\EXP \big\{ \InfFt(\bO_{1} \con \beta_1^*,f') \big\}
& =
\frac{1}{\Pr(A=1)}
\Big[ \EXP \big[  A \big\{  \mathcal{G}(\pYo) - \mu' (\bX) \big\} \big] + \EXP \big\{ A \mu'(\bX) \big\} \Big]
\\
&
=
\frac{ \EXP \big\{  A   \mathcal{G}(\pYo) \big\} }{\Pr(A=1)}
=
\EXP \big\{ \mathcal{G}(\pYo) \cond A=1 \big\} \ .
\end{align*}
This concludes the case of correctly specified $\beta_1$ and misspecified $\potf{10}$.

Next, suppose $\beta_1$ is misspecified as $\beta_1'$ and $\potf{10}$ is correctly specified. Then, the outcome regression is correctly specified as $\mu^* (\bX) = \mu(\bX \con \potf{1|AX}^* )$. Moreover, the first term becomes
\begin{align}       \label{eq-DRrepresentcorrectmu}
&
\EXP 
\big[
\beta_1'(\bX) \alpha_1^*(\Yo,\bX)
(1-A) \big\{  \mathcal{G}(\Yo) - \mu^* (\bX) \big\}
\big]
\nonumber
\\
& = 
\EXP \bigg[
\frac{\beta_1'(\bX)}{\beta_1^*(\bX)}
\beta_1^*(\bX)
\alpha_1^*(\Yo, \bX)
(1-A) \big\{  \mathcal{G}(\Yo) - \mu^* (\bX) \big\}
\bigg]
\nonumber
\\
& = 
\EXP \bigg[
\frac{\beta_1'(\bX)}{\beta_1^*(\bX)}
A \big\{  \mathcal{G}(\pYo) - \mu^* (\bX) \big\}
\bigg] 
\nonumber
\\
& = 
\EXP \bigg[
\frac{\beta_1'(\bX)}{\beta_1^*(\bX)}
\Pr(A=1 \cond \bX)
\underbrace{
\Big[ \EXP \big\{  \mathcal{G}(\pYo) \cond A=1, \bX \big\} - \mu^*(\bX) \Big] }_{=0}
\bigg] 
= 0 \ .
\end{align}
The second identity is from the previous result under $(\beta_1^*,f')$ case. The third identity is from the law of iterated expectation. Therefore, 
\begin{align*}
\EXP \big\{ \InfFt(\bO_{1} \con  \beta_1',\potf{1|AX}^*) \big\}
& =
\frac{1}{\Pr(A=1)}
\Big[ 0 + \EXP \big\{ A \mu^*(\bX) \big\} \Big]
=
\EXP \big\{ \mathcal{G}(\pYo) \cond A=1 \big\} \ .
\end{align*}
This concludes the case of correctly specified $\potf{10}$ and misspecified $\beta_1$.

\subsection{Proof of Lemma \ref{lemma-OR-DR}}						\label{sec:supp-OR-DR}

\begin{itemize}
    \item[(i)] Result (i)
\end{itemize}

For simplicity, let $\Psi_{\SPg}$ be
\begin{align*}
& \Psi_{\SPg} ( \bO_{0} \con \alpha, {\mathfrak{m}})  = M_{\SPg}(\Yz,\bX) \alpha(\Yz,\bX)^{-A} \big\{ A - \Pr(A=1 \cond \Yz=0,\bX) \big\}  / \Pr(A=1 \cond \Yz=0,\bX) \ , \\
& 
M_{\SPg}(y,\bX) = \big[ {\mathfrak{m}}(y,\bX) - \overline{\EXP}_{\SPg} \big\{ {\mathfrak{m}}(\Yz, \bX) \cond A=0, \bX \big\} \big] \Pr(A=1 \cond \Yz=0,\bX) \ .
\end{align*}
For any $\alpha$, we find
\begin{align*}
\overline{\EXP}_{\SPg} \big\{ \Psi_{\SPg} ( \bO_{0} \con \alpha, {\mathfrak{m}}) \big\} 
& =
\iint_{\SPg} M_{\SPg}(y,\bx) \alpha(y,\bx)^{-1}  
\frac{ \potf{A|0X}^*(0 \cond 0,\bx) }{\potf{A|0X}^*(1 \cond 0,\bx)}
\potf{A|0X}^*(1 \cond y,\bx) 
\potf{0X}^* (y,\bx) \, d(y,\bx)
\\
& \hspace*{1cm} 
- 
\iint_{\SPg} M_{\SPg}(y,\bx) 
\potf{A|0X}^*(0 \cond y,\bx) 
\potf{0X}^* (y,\bx) \, d(y,\bx)
\\
& =
\iint_{\SPg} M_{\SPg}(y,\bx) \alpha(y,\bx)^{-1}  
\alpha_0^*(y,\bx) \potf{A|0X}^*(0 \cond y,\bx)
\potf{0X}^*(y,\bx) \, d(y,\bx)
\\
& \hspace*{1cm} 
- 
\iint_{\SPg} M_{\SPg}(y,\bx) 
\potf{A|0X}^*(0 \cond y,\bx) \potf{0X}^*(y,\bx) \, d(y,\bx)
\\
&
=
\overline{\EXP}_{\SPg} \Bigg[
(1-A) M_{\SPg} (\Yz,\bX) 
\bigg\{
\frac{\alpha_0^*(\Yz,\bX) }{\alpha(\Yz,\bX)} -1 \bigg\}
\Bigg] \ .
\end{align*}
Therefore, $\overline{\EXP}_{\SPg} \big\{ \Psi_{\SPg} ( \bO_{0} \con \alpha_0^*, {\mathfrak{m}}) = 0 \big\}$ if $\alpha_0^*(y,\bx)/\alpha(y,\bx)-1 = 0$ over $y \in \SPg \cap \SPz$, which indicates $\alpha_0^*(y,\bx) = \alpha(y,\bx)$ over $y \in \SPg \cap \SPz$.

\begin{itemize}
    \item[(ii)] Result (ii)
\end{itemize}

Note that $ \overline{\EXP}_{\SPg} \big\{ k(\bX) (1-A) M_{\SPg} (\Yz, \bX) \big\} = \EXP\big[ k(\bX)  \overline{\EXP}_{\SPg}  \big\{ M_{\SPg} (\Yz, \bX) \cond A=0, \bX \big\} \Pr(A=0 \cond \bX) \big] = 0$ for any function $k$. Therefore, we find the following result holds for any $c(\bx)$:
\begin{align*}
\overline{\EXP}_{\SPg}  \big\{ \Psi_{\SPg} ( \bO_{0} \con \alpha^\dagger, {\mathfrak{m}}) \big\} 
=
\overline{\EXP}_{\SPg}  \Big[ 
(1-A)
M_{\SPg} (\Yz,\bX)
\big[
\big\{ \alpha^\dagger(\Yz,\bX) \big\}^{-1} 
\alpha_0^*(\Yz,\bX)
- c(\bX)
\big]
\Big] \ .
\end{align*}

Now, for any function $c(\bX)$, let $\mathcal{Y}_{\text{diff}}(\bX):= \big\{ y \in \SPg \cap \SPz \cond \alpha_0^*(y , \bX)/\alpha^\dagger(y, \bX) \neq c(\bX) \big\}$ and $d_Y(\bX) := \Pr\big\{ \Yz \in \mathcal{Y}_{\text{diff}}(\bX) \cond A=0, \bX \big\} > 0$. Let $\mathcal{X}_{\text{diff}} := \big\{ \bX \cond d_Y(\bX)  > 0 \big\}$. The goal of the proof is to show that, for some function $c(\bX)$, we have $\Pr(\bX \in \mathcal{X}_{\text{diff}}) = 0$, indicating that $\alpha_0^*(y , \bX)/\alpha^\dagger(y, \bX) = c(\bX)$ almost surely for $y \in \SPg \cap \SPz$.

We take $c(\bx)$ as
\begin{align*}
c(\bx) 
= 
\bigg\{ \int_{\SPg} \potf{0|AX}^*(y \cond 0,\bx) \, dy \bigg\}^{-1}
\bigg\{ \int_{\SPg} \frac{\alpha_0^*(y,\bx)}{\alpha^\dagger(y,\bx)} \potf{0|AX}^*(y \cond 0,\bx) \, dy \bigg\} \ .
\end{align*}
Then, we find $c(\bx)$ is the mean of $\alpha_0^*(y , \bX)/\alpha^\dagger(y, \bX)$:
\begin{align*}
0 = \int_{\SPg} \bigg\{ \frac{\alpha_0^*(y,\bx)}{\alpha^\dagger(y,\bx)} - c(\bx) \bigg\} \potf{0|AX}^*(y \cond 0,\bx) \, dy \ .
\end{align*}
We define two sets $\mathcal{Y}_+(\bx)$ and $\mathcal{Y}_-(\bx)$ as follows:
\begin{align*}
&
\mathcal{Y}_+(\bx) =\big\{ y \in \SPg \cap \SPz \cond \alpha_0^*(y,\bx) \geq c(\bx) \alpha^\dagger(y,\bx) \big\}
\ , 
&&
\mathcal{Y}_-(\bx)=\big\{ y \in \SPg \cap \SPz \cond \alpha_0^*(y,\bx) < c(\bx) \alpha^\dagger(y,\bx) \big\} \ .
\end{align*}
Then, if $d_Y(\bX)>0$, it is trivial that  $ \omega_+(\bX) := \overline{\EXP}_{\SPg} \big[ \ind\{ \Yz \in \mathcal{Y}_+(\bX) \big\} \cond A=0 , \bX \big] >0 $ and  $\omega_-(\bX) := \overline{\EXP}_{\SPg} \big[ \ind\{   \Yz \in \mathcal{Y}_-(\bX) \big\} \cond A=0 , \bX \big] >0$.  
Additionally, if $\omega_+(\bX)>0$ and $\omega_-(\bX)>0$, it means $d_Y(\bX) > 0$. Therefore, we find $\mathcal{X}_{\text{diff}} = \big\{ \bX \cond  \omega_+(\bX)>0 \text{ and } \omega_-(\bX) \big\} =  \big\{ \bX \cond d_Y(\bX)  > 0 \big\}$.

Using $\mathcal{Y}_\pm(\bX)$ and $\omega_\pm(\bX)$, we design a function $M_{\SPg} '(y,\bx)$ as follows
\begin{align*}
M_{\SPg} '(y,\bx) 
=
\omega_- (\bx) \ind \big\{ y \in \mathcal{Y}_+(\bx) \big\} 
- \omega_+ (\bx) \ind \big\{ y \in \mathcal{Y}_-(\bx) \big\}  
\ ,
\end{align*}
which satisfies the condition on $M_{\SPg} $:
\begin{align*}
&
\overline{\EXP}_{\SPg}   \big\{ M_{\SPg} '(\Yz,\bX) \cond A=0, \bX \big\}
\\
& = 
\omega_-(\bX) \overline{\EXP}_{\SPg}  \big[ \ind\{ \Yz \in \mathcal{Y}_+(\bX) \big\} \cond A=0 , \bX \big]
-
\omega_+(\bX) \overline{\EXP}_{\SPg}  \big[ \ind\{ \Yz \in \mathcal{Y}_-(\bX) \big\} \cond A=0 , \bX \big]
\\
& = 
\omega_-(\bX) \omega_+(\bX)
-
\omega_+(\bX) \omega_-(\bX)
=
0 \ .
\end{align*}
Therefore, with this choice of $M_{\SPg} '$, we find 
\begin{align*}
&
\overline{\EXP}_{\SPg}  \big\{ \Psi_{\SPg}  ( \bO_{0} \con \alpha^\dagger, {\mathfrak{m}}) \big\} 
\\
&
=
\overline{\EXP}_{\SPg}  \Big[ 
\overline{\EXP}_{\SPg}  \Big[	
M_{\SPg} '(\Yz,\bX)
\big[
\big\{ \alpha^\dagger(\Yz,\bX) \big\}^{-1} 
\alpha_0^*(\Yz,\bX)
- c(\bX)
\big]
\, \Big| \, A=0, \bX
\Big] 
\Pr(A=0 \cond \bX)
\Big] 
\\
&
=
\overline{\EXP}_{\SPg}  \Bigg[ 
\overline{\EXP}_{\SPg}  \Bigg[	
\underbrace{
\begin{array}{l}
\omega_-(\bX) \ind \big\{ \Yz \in \mathcal{Y}_+(\bX) \big\}
\big[
\big\{ \alpha^\dagger(\Yz,\bX) \big\}^{-1} 
\alpha_0^*(\Yz,\bX)
- c(\bX)
\big]
\\
-
\omega_+(\bX) \ind \big\{ \Yz \in \mathcal{Y}_-(\bX) \big\}
\big[
\big\{ \alpha^\dagger(\Yz,\bX) \big\}^{-1} 
\alpha_0^*(\Yz,\bX)
- c(\bX)
\big]
\end{array} }_{=: (*) }
\, \Bigg| \, A=0, \bX
\Bigg] 
\Pr(A=0 \cond \bX)
\Bigg] \ .
\end{align*}	 
Here, due to the definition of $\mathcal{Y}_+(\bX)$ and $\mathcal{Y}_-(\bX)$, the underbraced term $(*)$ is positive for all $\bX \in \mathcal{X}_{\text{diff}}$, and $\Pr(A=0 \cond \bX)$ is also positive for all $\bX \in \mathcal{X}_{\text{diff}}$. If $\Pr(\bX \in \mathcal{X}_{\text{diff}})>0$, this implies that 
\begin{align*}
& \overline{\EXP}_{\SPg}  \big\{ \Psi_{\SPg}  ( \bO_{0} \con \alpha^\dagger, {\mathfrak{m}}) \big\}
\\
&
=
\underbrace{
\overline{\EXP}_{\SPg}  \big\{ \Psi_{\SPg}  ( \bO_{0} \con \alpha^\dagger, {\mathfrak{m}}) \cond \bX \in \mathcal{X}_{\text{diff}} \big\} }_{>0}
\underbrace{\Pr (\bX \in \mathcal{X}_{\text{diff}} )}_{>0}
+
\underbrace{
\overline{\EXP}_{\SPg}  \big\{ \Psi_{\SPg}  ( \bO_{0} \con \alpha^\dagger, {\mathfrak{m}}) \cond \bX \in \mathcal{X}_{\text{diff}}^c \big\} }_{=0}
\Pr (\bX \in \mathcal{X}_{\text{diff}}^c )
> 0 \ .
\end{align*} 
This result contradicts the definition of $\alpha^\dagger$, a solution to the moment equation $\overline{\EXP}_{\SPg} \big\{ \Psi_{\SPg}  ( \bO_{0} \con \alpha^\dagger, {\mathfrak{m}}) \big\}=0$, indicating that $\Pr (\bX \in \mathcal{X}_{\text{diff}} )$ must be zero. As a result, we have $\alpha_0^* (y,\bx) /\alpha^\dagger  (y,\bx) = c(\bx)$ for some function $c(\bx)$ almost surely for $y \in \SPg \cap \SPz$. Since we have the boundary condition $\alpha_0^*(0,\bx)/\alpha^\dagger(0,\bx)=1$, this means that $\alpha_0^*(y,\bx) = \alpha^\dagger(y,\bx)=1$ almost surely for $y \in  \SPg \cap \SPz$.

\begin{itemize}
    \item[(iii)] Result (iii)
\end{itemize}

We first consider that $\potf{00}(y \cond \bx)$ is correctly specified as $\potf{0 | A X}^*(y \cond 0,\bx)$ over $y \in \SPg \cap \SPz$ whereas $\pote{00}$ may be misspecified. Let $h(\bX) = \EXP_{\pote{00}} \big( A \cond \Yz=0,\bX \big)$. Then, we find the following result for $\Yz \in \SPg \subseteq \SPzt$:
\begin{align*}
&
\EXP \Big[ 
\big\{ \alpha_0^*(\Yz,\bX) \big\}^{-A} 
\big\{ A-\EXP_{\pote{00}} (A \cond \Yz=0, \bX) \big\}
\, \Big| \, \Yz, \bX
\Big]
\\
& =
\Pr(A=1 \cond \Yz, \bX)
\big\{ \alpha_0^*(\Yz,\bX) \big\}^{-1} 
\big\{ 1- h(\bX) \big\}
-
\Pr(A=0 \cond \Yz, \bX)
h(\bX)
\\
& =
\Pr(A=0 \cond \Yz,\bX)
\bigg[ 
\underbrace{
\frac{\Pr(A=1 \cond \Yz=0,\bX)}{\Pr(A=0 \cond \Yz=0,\bX)} 
\big\{ 1 - h(\bX) \big\} - h(\bX)
}_{=H(\bX)}
\bigg]
= \Pr(A=0 \cond \Yz,\bX) H(\bX) \ .
\end{align*}

Let ${\widetilde{\mathfrak{m}}}(\Yz,\bX) = {\mathfrak{m}}(\Yz, \bX)  H(\bX)$. Then, we obtain the zero-mean property of the moment equation:
\begin{align*}
& \overline{\EXP}_{\SPg} \big\{ \Psi_{\SPg} ( \bO_{0} \con \alpha_0^*,  \potf{0 | A X}^*, \pote{00}, {\mathfrak{m}}) \big\} 
\\
& = 
\overline{\EXP}_{\SPg} \bigg[ \Big[ {\widetilde{\mathfrak{m}}}(\Yz,\bX) - \overline{\EXP}_{\SPg} \big\{ {\widetilde{\mathfrak{m}}}(\Yz, \bX) \cond A=0, \bX \big\} \Big] 
\Pr(A=0 \cond \Yz, \bX)
\bigg] 
\\
& = 
\overline{\EXP}_{\SPg} \bigg[ (1-A) \Big[ {\widetilde{\mathfrak{m}}}(\Yz,\bX) - \overline{\EXP}_{\SPg} \big\{ {\widetilde{\mathfrak{m}}}(\Yz, \bX) \cond A=0, \bX \big\} \Big]  \bigg] 
\\
& = 
\EXP \bigg[ \underbrace{ \overline{\EXP}_{\SPg} \Big[ {\widetilde{\mathfrak{m}}}(\Yz,\bX) - \overline{\EXP}_{\SPg} \big\{ {\widetilde{\mathfrak{m}}}(\Yz, \bX) \cond A=0, \bX \big\} \, \Big| \, A=0,\bX \Big]}_{=0} \Pr(A=0 \cond \bX)  \bigg] 
= 0 \ . 
\end{align*}	

Next, we consider that $\pote{00}$ is correctly specified as $\potf{A|0 X}^*$ whereas $\potf{00}$ may be misspecified. From Result (i) with $\alpha=\alpha_0^*$ and $e_{00}=\potf{A|0X}^*$, we get the zero-mean property of the moment equation:
\begin{align*}
& 
\overline{\EXP}_{\SPg} \big\{ \Psi_{\SPg} ( \bO_{0} \con \alpha_0^*,  \potf{00}, \potf{A | 0 X}^*, {\mathfrak{m}}) \big\} 
\\
&
=
\overline{\EXP}_{\SPg} \bigg[ \Big[ {\widetilde{\mathfrak{m}}}(\Yz,\bX) - \overline{\EXP}_ {\SPg,\potf{00}} \big\{ {\widetilde{\mathfrak{m}}}(\Yz, \bX) \cond A=0, \bX \big\} \Big] 
\underbrace{
\Big[
\big\{ \alpha_0^*(\Yz,\bX) \big\}^{-1} 
\alpha_0^*(\Yz,\bX)
-1
\Big]}_{
=0 \text{ over } (\Yz ,\bX) \in \SPg \cap \SPz
} \EXP(1-A \cond \Yz, \bX) \bigg]
= 0 \ .
\end{align*}
This concludes the proof.

\subsection{Proof of Theorem \ref{thm-EIF SB}}	\label{sec:supp-EIF bound}

We prove that $\InfFt^{\dagger}(\bO)$ defined in \eqref{eq-supp-EIF bound} is the EIF for $\tau^{\dagger}$ in the nonparametric model $\mathcal{M}_{\OREC}$ for $\dagger \in \{\text{LB}, \text{UB}\}$. The proof adapts the approach of the proof of Theorem \ref{thm-EIF} in Section \ref{sec:supp-Proof-thm-EIF}. The key additional ingredient is showing that the variation of the optimal weight function $w^{\min}$ or $w^{\max}$ with respect to the data-generating distribution does not contribute to the pathwise derivative. 

We begin by stating Danskin's theorem:
\begin{lemma}[Danskin's Theorem; \citealp{Danskin1966, Bertsekas1999}] \label{lemma-Danskin}
Let $\Theta$ be a compact subset of $\R^d$, let $I$ be an open interval in $\R$, and let $\phi : \Theta \times I \to \R$ be a continuous function such that $\nabla_\eta \phi(\theta, \eta)$ exists and is continuous on $\Theta \times I$. Define $\phi^*(\eta) = \min_{\theta \in \Theta} \phi(\theta, \eta)$. If the minimizer $\theta^*(\eta) = \argmin_{\theta \in \Theta} \phi(\theta, \eta)$ is unique for all $\eta \in I$, then $\phi^*$ is differentiable on $I$ and
\begin{align*}
\nabla_\eta \phi^*(\eta) = \nabla_\eta \phi \big( \theta^*(\eta), \eta \big) \ .
\end{align*}
The analogous result holds for $\phi^*(\eta) = \max_{\theta \in \Theta} \phi(\theta, \eta)$ with unique maximizer.
\end{lemma}

We now provide the details of the proof. Without loss of generality, we focus on $\dagger = \text{UB}$; the case $\dagger = \text{LB}$ is analogous (replacing $w^{\min}$ by $w^{\max}$ and $m^{\min}$ by $m^{\max}$ throughout). Recall from Section \ref{sec:supp-SA} that for continuous outcomes, the weight $w^{\min}(y,\bx)$ minimizing $\mu^*(\bx \con w)$ over $w(y,\bx) \in [\Gamma^{-1}, \Gamma]$ takes the form of a step function with threshold $m^{\min}(\bx) = \mu^{\text{UB}}(\bx)$. Following the proof of Theorem \ref{thm-EIF}, we consider a parametric submodel $\{P(\bO \con \eta)\}$ as in \eqref{eq-Model-OREC-ParaSubmodel} and verify the pathwise differentiability condition:
\begin{align} \label{eq-diffpara-bound}
\frac{\partial}{\partial \eta}
\tau_0^{\text{UB}}(\eta)
\bigg|_{\eta=\eta^*}
=
\EXP \big\{
s_O(\bO \con \eta^*) \InfFt_{0}^{\text{UB}}(\bO)
\big\} \ ,
\end{align}
where $\tau_0^{\text{UB}}(\eta) = \EXP^{(\eta)} \big\{ A \mu^{\text{UB}}(\bX \con \eta) \big\} / \Pr(A=1 \con \eta)$ and $\InfFt_{0}^{\text{UB}}(\bO) = \uncInfFt_{0}^{\text{UB}}(\bO) / \Pr(A=1)$.  The $\InfFt_1^*(\bO)$ component of $\InfFt^{\text{UB}}(\bO)$ satisfies the differentiability condition by the same argument as in the proof of Theorem \ref{thm-EIF}; it remains to verify \eqref{eq-diffpara-bound}.

For the continuous outcome, the proof proceeds in three steps. \\

\noindent \textbf{Step 1: The variation of $w^{\min}$ does not contribute to the pathwise derivative} \\

In the parametric submodel at $\eta$, the conditional counterfactual mean under sensitivity is
\begin{align*}
\mu^{\text{UB}}(\bx \con \eta)
=
\min_{m \in \R} \mu(\bx \con m, \eta)
\ ,
\end{align*}
where the minimization is over the threshold $m := m(x)$ that parameterizes the step function $w^{\min}(y,\bx \con \eta)$. With $g(y, \bx \con \eta) := \alpha_0(y, \bx \con \eta) f_1(y \cond 0, \bx \con \eta)$, the conditional mean as a function of $m$ is $\mu(\bx \con m, \eta) = N(m, \eta) / D(m, \eta)$, where
\begin{align*}
N(m, \eta) &= \Gamma \int_{-\infty}^{m} y \, g(y, \bx \con \eta) \, dy + \Gamma^{-1} \int_{m}^{\infty} y \, g(y, \bx \con \eta) \, dy \ , \\
D(m, \eta) &= \Gamma \int_{-\infty}^{m} g(y, \bx \con \eta) \, dy + \Gamma^{-1} \int_{m}^{\infty} g(y, \bx \con \eta) \, dy \ .
\end{align*}
Since the minimizer $m^{\min}(\eta) := m^{\min}(\bx \con \eta)$ is unique (established in Section \ref{sec:supp-SA} by the strict monotonicity of $\mathcal{E}^{\min}$), Danskin's theorem (Lemma \ref{lemma-Danskin}) gives
\begin{align} \label{eq-ATT bound 1 proof}
\nabla_\eta \mu^{\text{UB}}(\bx \con \eta) = \nabla_\eta \mu(\bx \con m^{\min}(\eta), \eta) \ ,
\end{align}
where the right-hand side is the partial derivative with respect to $\eta$ with $m$ held fixed at $m^{\min}(\eta)	$.

Let $m^{\min,*} := m^{\min}(\eta^*)$. We further establish that $ \nabla_\eta \mu(\bx \con m^{\min}(\eta), \eta) \big|_{\eta = \eta^*} =  \nabla_\eta \mu(\bx \con m^{\min,*}, \eta) \big|_{\eta = \eta^*}$. For any fixed $m(x)$, the partial derivative of $\mu$ with respect to $\eta$ is
\begin{align*}
\frac{\partial \mu}{\partial \eta}(\bx \con m, \eta)
=
\frac{1}{D(m,\eta)}
\bigg[
\int y \, w_{m}(y,\bx) \, \nabla_\eta g(y, \bx \con \eta) \, dy
-
\mu(\bx \con m, \eta) \int w_{m}(y,\bx) \, \nabla_\eta g(y, \bx \con \eta) \, dy
\bigg] \ ,
\end{align*}
where $w_{m}(y,\bx) = \Gamma \ind(y < m(x)) + \Gamma^{-1} \ind(y \geq m(x))$ denotes the step-function weight at threshold $m$. 

On the other hand, by the Leibniz integral rule applied to $N$ and $D$, we have:
\begin{align*}
\nabla_\eta N\big(m^{\min}(\eta), \eta\big)
&=
\Gamma \int_{-\infty}^{m^{\min}(\eta)} y \, \nabla_\eta g(y, \bx \con \eta) \, dy
+
\Gamma^{-1} \int_{m^{\min}(\eta)}^{\infty} y \, \nabla_\eta g(y, \bx \con \eta) \, dy
\\
&
\hspace*{1cm}
+
(\Gamma - \Gamma^{-1}) \, m^{\min}(\eta) \, g\big(m^{\min}(\eta), \bx \con \eta\big) \cdot \nabla_\eta m^{\min}(\eta) \ ,
\\
\nabla_\eta D\big(m^{\min}(\eta), \eta\big)
&=
\Gamma \int_{-\infty}^{m^{\min}(\eta)} \nabla_\eta g(y, \bx \con \eta) \, dy
+
\Gamma^{-1} \int_{m^{\min}(\eta)}^{\infty} \nabla_\eta g(y, \bx \con \eta) \, dy
\\
&
\hspace*{1cm}
+
(\Gamma - \Gamma^{-1}) \, g\big(m^{\min}(\eta), \bx \con \eta\big) \cdot \nabla_\eta m^{\min}(\eta) \ .
\end{align*}
Therefore, we find
\begin{align} \label{eq-ATT bound 2 proof}
&
\nabla_{\eta} \mu(\bx \con m^{\min}(\eta), \eta)
\nonumber
\\
&
=
\frac{ \nabla_\eta N\big(m^{\min}(\eta), \eta\big)
}{
D\big(m^{\min}(\eta), \eta\big)
}
-
\mu(\bx \con m^{\min}(\eta), \eta)
\frac{ \nabla_\eta D\big(m^{\min}(\eta), \eta\big)
}{
 D\big(m^{\min}(\eta), \eta\big)
}
\nonumber
\\
&
=
\frac{\partial \mu}{\partial \eta}\big(\bx \con m, \eta\big) \bigg|_{m=m^{\min}(\eta)}
+
\frac{
(\Gamma - \Gamma^{-1}) \, g\big(m^{\min}(\eta), \bx \con \eta\big) \cdot \nabla_\eta m^{\min}(\eta)
}{
D\big(m^{\min}(\eta),\eta\big)
}
\Big\{ m^{\min}(\eta) - \mu\big(\bx \con m^{\min}(\eta), \eta\big) \Big\}
\nonumber
\\
&
\stackrel{(\star)}{=}
\frac{\partial \mu}{\partial \eta}\big(\bx \con m, \eta\big) \bigg|_{m=m^{\min}(\eta)} \ .
\end{align}
Evaluating at $\eta = \eta^*$ for \eqref{eq-ATT bound 1 proof} and \eqref{eq-ATT bound 2 proof}, we obtain 
\begin{align*}
\nabla_\eta \mu^{\text{UB}}(\bx \con \eta) \big|_{\eta = \eta^*} 
=
 \nabla_\eta \mu(\bx \con m^{\min}(\eta), \eta) \big|_{\eta = \eta^*} =  \frac{\partial \mu}{\partial \eta}(\bx \con m^{\min,*}, \eta) \bigg|_{\eta = \eta^*} \ .
\end{align*}
In other words, the pathwise derivative of $\mu^{\text{UB}}(\bx \con \eta)$ equals the derivative of $\mu(\bx \con m,\eta)$ where $m$ is held fixed at $m^{\min,*}$. 
\\

\noindent \textbf{Step 2: The threshold variation of $w^{\min}$ does not contribute to the $s_\alpha$ term} \\

Along the submodel, $\alpha_1^{\text{UB}}(y, \bx \con \eta) = w^{\min}(y, \bx \con \eta) \cdot \alpha_0(y, \bx \con \eta)$, where $w^{\min}(y,\bx \con \eta)$ depends on $\eta$ through the threshold $m^{\min}(\bx \con \eta)$. By the product rule, the log-derivative of $\alpha_1^{\text{UB}}$ decomposes as
\begin{align} \label{eq-log-deriv-decomp}
s_\alpha^{\text{UB}}(y, \bx \con \eta)
\alpha_1^{\text{UB}}(y, \bx \con \eta)
=
\nabla_\eta \alpha_1^{\text{UB}}(y, \bx \con \eta)
=
\underbrace{
\nabla_\eta w^{\min}(y,\bx \con \eta)
\alpha_0(y, \bx \con \eta)
}_{\text{threshold variation}}
+
\nabla_\eta \alpha_0(y, \bx \con \eta)
w^{\min}(y,\bx \con \eta)
\ . 
\end{align}
We show that the threshold variation component vanishes from the pathwise derivative. In \eqref{eq-pd} (adapted for UB), the $s_\alpha^{\text{UB}}$ term in the pathwise derivative of $\tau_0^{\text{UB}}(\eta)$ is
\begin{align}
&
\EXP^{(\eta)} \bigg[
\frac{(1-A) \, s_\alpha^{\text{UB}}(\Yo,\bX \con \eta) \, \beta_1^{\text{UB}}(\bX \con \eta) \, \alpha_1^{\text{UB}}(\Yo,\bX \con \eta)}{\Pr(A=1 \con \eta)}
\Big\{ \Yo - \mu^{\text{UB}}(\bX \con \eta) \Big\}
\bigg]
\nonumber
\\
&
=
\EXP^{(\eta)} \bigg[
\frac{(1-A) \, \beta_1^{\text{UB}}(\bX \con \eta) \, \nabla \alpha_1^{\text{UB}}(\Yo,\bX \con \eta) }{\Pr(A=1 \con \eta)}
\Big\{ \Yo - \mu^{\text{UB}}(\bX \con \eta) \Big\}
\bigg]
\nonumber
\\
&
=
\EXP^{(\eta)} \bigg[
\frac{(1-A) \, \nabla_\eta w^{\min}(\Yo,\bX \con \eta) \, \beta_1^{\text{UB}}(\bX \con \eta)\, \alpha_0(\Yo,\bX \con \eta)}{\Pr(A=1 \con \eta)}
\Big\{ \Yo - \mu^{\text{UB}}(\bX \con \eta) \Big\}
\bigg] 
\label{eq-threshold-contribution}
\\
&
\qquad \qquad
+
\EXP^{(\eta)} \bigg[
\frac{(1-A) \, w^{\min}(\Yo,\bX \con \eta) \, \beta_1^{\text{UB}}(\bX \con \eta)\, \nabla_\eta  \alpha_0(\Yo,\bX \con \eta)}{\Pr(A=1 \con \eta)}
\Big\{ \Yo - \mu^{\text{UB}}(\bX \con \eta) \Big\}
\bigg] 
 \ . 
 \nonumber
\end{align}
Conditioning on $(A=0, \bX=\bx)$, the inner expectation over $\Yo$ in \eqref{eq-threshold-contribution} is proportional to
\begin{align} 
& 
\EXP^{(\eta)} \bigg[
\frac{ (1-A) \, \nabla_\eta w^{\min}(\Yo,\bX \con \eta) \, \beta_1^{\text{UB}}(\bX \con \eta)\, \alpha_0(\Yo,\bX \con \eta)
\big\{ \Yo - \mu^{\text{UB}}(\bX \con \eta) \big\} }{ \Pr(A=1 \con \eta) }
\, \bigg| \, A=0, \bX=\bx
\bigg] 
\nonumber
\\
&
\propto 
\int \nabla_\eta w^{\min}(y, \bx \con \eta) \cdot g(y, \bx \con \eta) \cdot \big\{ y - \mu^{\text{UB}}(\bx \con \eta) \big\} \, dy \ .
\label{eq-threshold-integral} 
\end{align}
where $g(y, \bx \con \eta) = \alpha_0(y, \bx \con \eta) \, f_1(y \cond 0, \bx \con \eta)$ as in Step 1. 

Note that
\begin{align*} 
\int w^{\min}(y, \bx \con \eta) \cdot 
g(y, \bx \con \eta) \cdot \big\{ y - \mu^{\text{UB}}(\bx \con \eta) \big\}  \, dy
= N\big(m^{\min}(\eta), \eta\big) - \mu^{\text{UB}}(\bx \con \eta) \cdot D\big(m^{\min}(\eta), \eta\big)
= 0 \ , 
\end{align*}
where the last identity holds for all $\eta$ by the definition of $\mu(\bx \con m, \eta) = N(m, \eta) / D(m, \eta)$. Applying the Leibniz integral rule, we have
\begin{align} \label{eq-Leibniz-step2}
0 
&
=
\nabla_{\eta}
\int w^{\min}(y, \bx \con \eta) \cdot 
g(y, \bx \con \eta) \cdot \big\{ y - \mu^{\text{UB}}(\bx \con \eta) \big\}  \, dy
\nonumber
\\
&
=
\int w^{\min}(y, \bx \con \eta) 
\cdot 
\nabla_\eta 
\big[ g(y, \bx \con \eta) \big\{ y - \mu^{\text{UB}}(\bx \con \eta) \big\} \big]
\, 
dy
\nonumber
\\
&
\qquad 
\qquad 
+
(\Gamma - \Gamma^{-1}) \cdot 
g(m^{\min}(\eta), \bx \con \eta) \underbrace{ \cdot \big\{ m^{\min}(\eta) - \mu^{\text{UB}}(\bx \con \eta) \big\} }_{=0 }
 \cdot \nabla_\eta m^{\min}(\eta)
\nonumber
 \\
 &
=
\int w^{\min}(y, \bx \con \eta)
\cdot
\nabla_\eta
\big[ g(y, \bx \con \eta) \big\{ y - \mu^{\text{UB}}(\bx \con \eta) \big\} \big]
\,
dy
  \ ,
\end{align}
where the last equality uses $m^{\min}(\bx \con \eta) = \mu^{\text{UB}}(\bx \con \eta)$. 

Alternatively, we may also write
\begin{align} \label{eq-Leibniz-step3}
0 
&
=
\nabla_{\eta}
\int w^{\min}(y, \bx \con \eta) \cdot 
g(y, \bx \con \eta) \cdot \big\{ y - \mu^{\text{UB}}(\bx \con \eta) \big\}  \, dy
\nonumber
 \\
 &
=
\underbrace{
\int w^{\min}(y, \bx \con \eta) 
\cdot 
\nabla_\eta 
\big[ g(y, \bx \con \eta) \big\{ y - \mu^{\text{UB}}(\bx \con \eta) \big\} \big]
\, 
dy
}_{=0 \text{ from } \eqref{eq-Leibniz-step2}}
+
\int \nabla_\eta w^{\min}(y, \bx \con \eta) \cdot g(y, \bx \con \eta) \cdot \big\{ y - \mu^{\text{UB}}(\bx \con \eta) \big\} \, dy  
\nonumber
\\
&
=
\int \nabla_\eta w^{\min}(y, \bx \con \eta) \cdot g(y, \bx \con \eta) \cdot \big\{ y - \mu^{\text{UB}}(\bx \con \eta) \big\} \, dy \ .
\end{align}
Therefore, \eqref{eq-threshold-contribution} equals zero, meaning that
\begin{align}
&
\EXP^{(\eta)} \bigg[
\frac{(1-A) \, s_\alpha^{\text{UB}}(\Yo,\bX \con \eta) \, \beta_1^{\text{UB}}(\bX \con \eta) \, \alpha_1^{\text{UB}}(\Yo,\bX \con \eta)}{\Pr(A=1 \con \eta)}
\Big\{ \Yo - \mu^{\text{UB}}(\bX \con \eta) \Big\}
\bigg]
\nonumber  
\\
&
=
\EXP^{(\eta)} \bigg[
\frac{(1-A) \, w^{\min}(\Yo,\bX \con \eta) \, \beta_1^{\text{UB}}(\bX \con \eta)\, \nabla_\eta  \alpha_0(\Yo,\bX \con \eta)}{\Pr(A=1 \con \eta)}
\Big\{ \Yo - \mu^{\text{UB}}(\bX \con \eta) \Big\}
\bigg]  
\nonumber
\\
&
=
\EXP^{(\eta)} \bigg[
\frac{(1-A) \, s_{\alpha}(\Yo,\bX \con \eta) \, \beta_1^{\text{UB}}(\bX \con \eta)\, \alpha_1^{\text{UB}} (\Yo,\bX \con \eta)}{\Pr(A=1 \con \eta)}
\Big\{ \Yo - \mu^{\text{UB}}(\bX \con \eta) \Big\}
\bigg]  \ .
\label{eq-log-deriv-identity}
\end{align}

\noindent \textbf{Step 3: The remainder follows from the proof of Theorem \ref{thm-EIF}}  \\

Let 
\begin{align*}
\widetilde{\uncInfFt}^{\text{UB}}(\bO_{1} \con \eta)
&
=
\frac{  \beta_1^{\text{UB}}(\bX \con \eta) \alpha_1^{\text{UB}}(\Yo,\bX \con \eta)
(1-A) 		 \big\{  \Yo - \mu^{\text{UB}}(\bX \con \eta) \big\}
+
A
\mu^{\text{UB}}(\bX  \con \eta) }{\Pr(A=1 \con \eta)} \ ,
\\
\InfFtAug^{\text{UB}} (\bO_{0} \con \eta  ) 
&
=
\frac{(2A-1) R^{\text{UB}}(\Yz,A,\bX \con \eta) \big\{  \Yz - \mu^{\text{UB}}(\bX \con \eta) \big\}}{\Pr(A=1 \con \eta)} 
\\
&
=
\frac{  
(2A-1) \beta_1^{\text{UB}}(\bX \con \eta) \alpha_1^{\text{UB}}(\Yz,\bX \con \eta) 
}{
\Pr(A=1 \con \eta) 
}		
\frac{ \potf{1 A|X}(\Yz,0 \cond \bX \con \eta ) }{ \potf{0 A|X}(\Yz, A \cond \bX \con \eta ) }
\big\{  \Yz - \mu^{\text{UB}} (\bX \con \eta) \big\}
\ .
\end{align*}
Then, the parameter $\tau_0^{\text{UB}}$ admits the AIPW-type representation
\begin{align*}
\tau_0^{\text{UB}}
& =  \EXP \big\{ \widetilde{\uncInfFt}^{\text{UB}}(\bO_{1} \con \eta^*) \big\} \ ,
\end{align*}
which holds by the definitions of $\mu^{\text{UB}}$ and $\beta_1^{\text{UB}}$, via the same algebraic verification as \eqref{supp:eq-rep-AIPW} with $\alpha_1^{\text{UB}}$ in place of $\alpha_1^*$.

The pathwise derivative computation then proceeds identically to equations \eqref{eq-pd}-\eqref{eq-LHS} of the proof of Theorem \ref{thm-EIF}, with all quantities associated with $\text{UB}$. Specifically, the right-hand side of \eqref{eq-diffpara-bound} is equal to
\begin{align} \label{eq-RHS-bound}
&
\EXP \big\{
s_O(\bO \con \eta^*) \InfFt_{0}^{\text{UB}}(\bO)
\big\}
=
\EXP \Big[
s_{1}(\bO_{1} \con \eta^*) \widetilde{\uncInfFt}^{\text{UB}}(\bO_{1} \con \eta^*)
+ s_{0}(\bO_{0} \con \eta^*) \InfFtAug^{\text{UB}}(\bO_{0} )
\Big]
- s_{A}(1 \con \eta^*)  \tau_0^{\text{UB}} \ .
\end{align} 

The left-hand side of \eqref{eq-diffpara-bound} can be derived from the AIPW representation and Step 1, the derivative of $\tau_0^{\text{UB}}(\eta)$ is computed as in \eqref{eq-pd}-\eqref{eq-LHS}. To this end, we obtain 
\begin{align} \label{eq-LHS-bound}
& \frac{ \partial }{\partial \eta}  \tau_0^{\text{UB}}(\eta)
\bigg|_{\eta=\eta^*}
\nonumber
\\
&
=
- s_{A}(1 \con \eta^*)  \tau_0^{\text{UB}}
\nonumber
\\
&
\hspace*{1cm}
+
\EXP  \bigg[ s_{1} (\bO_{1} \con \eta^*) \widetilde{\uncInfFt}^{\text{UB}}(\bO_{1} \con \eta^*)
+
\frac{(1-A) s_\alpha^{\text{UB}} (\Yo,\bX \con \eta^*)  \beta_1^{\text{UB}}(\bX) \alpha_1^{\text{UB}}(\Yo,\bX)}{\Pr(A=1)}
\Big\{ \Yo - \mu^{\text{UB}}(\bX) \Big\}
\bigg] 
\nonumber
\\
&
\stackrel{\eqref{eq-log-deriv-identity}}{=}
- s_{A}(1 \con \eta^*)  \tau_0^{\text{UB}}
\nonumber
\\
&
\hspace*{1cm}
+
\EXP  \bigg[ s_{1} (\bO_{1} \con \eta^*) \widetilde{\uncInfFt}^{\text{UB}}(\bO_{1} \con \eta^*)
+
\frac{(1-A) s_\alpha (\Yo,\bX \con \eta^*)  \beta_1^{\text{UB}}(\bX) \alpha_1^{\text{UB}}(\Yo,\bX)}{\Pr(A=1)}
\Big\{ \Yo - \mu^{\text{UB}}(\bX) \Big\}
\bigg] 
\ .
\end{align} 

Comparing \eqref{eq-RHS-bound} and \eqref{eq-LHS-bound}, we establish \eqref{eq-diffpara-bound} if the following identity holds:
\begin{align}       \label{eq-LHS-RHS-bound}
\EXP \Big\{
s_{0}(\bO_{0} \con \eta^*) \InfFtAug^{\text{UB}}(\bO_{0} )
\Big\}
=
\EXP \bigg[ \frac{(1-A) s_\alpha (\Yo,\bX \con \eta^*)  \beta_1^{\text{UB}}(\bX) \alpha_1^{\text{UB}}(\Yo,\bX)}{\Pr(A=1)}
\Big\{ \Yo - \mu^{\text{UB}}(\bX) \Big\}
\bigg] \ .
\end{align}

We now verify \eqref{eq-LHS-RHS-bound}. From \eqref{eq-restriction} with $\alpha_0(\eta^*)$ in place of $\alpha_1(\eta^*)$, we have the following results for any function $\mathfrak{m}(\Yz,\bX)$:
\begin{align}                   \label{eq-restriction-bound}
&
\overline{\EXP}_{\SPorecetastar} \big[
s_{0} (\bO_{0} \con \eta^*) \Psi_{\SPorecetastar} ( \bO_{0} \con \alpha_0(\eta^*) , \potf{0 | A X}^* , \potf{A | 0 X}^* , \mathfrak{m})
\big]
\nonumber
\\
&
=
\overline{\EXP}_{\SPorecetastar} \left[
\begin{array}{l}
\big[ \mathfrak{m}(\Yz,\bX) - \overline{\EXP}_{{\SPorecetastar} , 0 | AX}  \big\{ \mathfrak{m}(\Yz, \bX) \cond A=0, \bX \big\} \big]
\\
\displaystyle{ \times \bigg\{ \frac{A s_\alpha(\Yz,\bX \con \eta^*) }{\alpha_0(\Yz,\bX \con \eta^*)} \bigg\}
\big\{ A- \EXP_{A|0X} ( A \cond \Yz=0, \bX) \big\}
}
\end{array}
\right] \ .
\end{align}
We now choose $\mathfrak{m}(\Yz,\bX)$ so that
\begin{align*}
&
\mathfrak{m}(\Yz,\bX) - \overline{\EXP}_{{\SPorecetastar} , 0 | AX}  \big\{ \mathfrak{m}(\Yz, \bX) \cond A=0, \bX \big\}
\\
&
=
\frac{ w^{\min,*}(\Yz,\bX) \potf{1A|X}^*(\Yz , 1 \cond  \bX) }{\potf{0A|X}^*(\Yz , 0 \cond \bX)}
\frac{  \big\{ \Yz - \mu^{\text{UB}}(\bX) \big\} }{\Pr(A=1) \Pr(A=1 \cond \Yz=0, \bX)}
\ .
\end{align*}
This choice is valid because the conditional mean restriction holds:
\begin{align*}
&
\overline{\EXP}_{\SPorecetastar} \bigg[
w^{\min,*}(\Yz,\bX) \frac{ \potf{1A|X}^*(\Yz, 1  \cond \bX) }{\potf{0A|X}^*(\Yz , 0 \cond \bX)} \big\{ \Yz - \mu^{\text{UB}}(\bX) \big\} \, \bigg| \, A=0, \bX \Bigg]
\\
&
=
\frac{\Pr(A=1 \cond \bX)}{\Pr(A=0 \cond \bX)}
\int_{\SPorecetastar}
w^{\min,*}(y,\bX) \alpha_0^*(y,\bX) \potf{1|AX}^*(y \cond 0, \bX)
\big\{ y - \mu^{\text{UB}}(\bX) \big\} \, dy
\\
& =
\frac{\Pr(A=1 \cond \bX)}{\Pr(A=0 \cond \bX)}
\Big\{
N\big(m^{\min,*},\eta^*\big) - \mu^{\text{UB}}(\bX)  D\big(m^{\min,*},\eta^*\big) \Big\}
= 0 \ ,
\end{align*}
where the last identity holds by the definition $\mu^{\text{UB}}(\bX) = N/D$ from Step 1.

With this choice of $\mathfrak{m}$, we evaluate $\Psi_{\SPorecetastar}$ at $A=1$ and $A=0$. Specifically, same calculation as in equations \eqref{eq-LHS-RHS}-\eqref{eq-restriction} with the factor $w^{\min,*}$ gives:
\begin{align*}
&
\Psi_{\SPorecetastar} ( \bO_{0} \con \alpha_0(\eta^*) , \potf{0 | A X}^* , \potf{A | 0 X}^* , \mathfrak{m})
=
\frac{(2A -1)}{\Pr(A=1) }\underbrace{ w^{\min,*}(\Yz,\bX) \frac{\potf{1A|X}^*(\Yz,1 \cond \bX)}{\potf{0A|X}^*(\Yz,A \cond \bX)} }_{= R^{\text{UB}}(\Yz,A,\bX)}
\big\{ \Yz - \mu^{\text{UB}}(\bX) \big\}
=
\InfFtAug^{\text{UB}}(\bO_0) \ .
\end{align*} 
Therefore, the left-hand side of \eqref{eq-restriction-bound} becomes $\EXP \big\{ s_{0}(\bO_{0} \con \eta^*) \InfFtAug^{\text{UB}}(\bO_{0}) \big\}$, as required for \eqref{eq-LHS-RHS-bound}.

For the right-hand side of \eqref{eq-restriction-bound}, we follow the same chain of results as in equations \eqref{eq-LHS-RHS}-\eqref{eq-restriction}. The factor $w^{\min,*}(\Yz,\bX)$ is carried through the conditioning steps, so that the final identity, corresponding to \eqref{eq-RHS 2}, becomes:
\begin{align*}
&
\EXP \bigg[ \frac{(1-A) s_\alpha (\Yo,\bX \con \eta^*)  w^{\min,*}(\Yo,\bX) \alpha_0^*(\Yo,\bX)}{\Pr(A=1)}
\Big\{ \Yo - \mu^{\text{UB}}(\bX) \Big\}
\bigg]
\\
& =
\EXP \bigg[ \frac{(1-A) s_\alpha (\Yo,\bX \con \eta^*)  \beta_1^{\text{UB}}(\bX) \alpha_1^{\text{UB}}(\Yo,\bX)}{\Pr(A=1)}
\Big\{ \Yo - \mu^{\text{UB}}(\bX) \Big\}
\bigg] \ ,
\end{align*}
where the last step uses $w^{\min,*} \alpha_0^* = \alpha_1^{\text{UB}}$ and absorbs $\beta_1^{\text{UB}}$ by the same density ratio argument as in \eqref{eq-LHS-RHS}-\eqref{eq-restriction}. This establishes \eqref{eq-LHS-RHS-bound}.

Combining \eqref{eq-RHS-bound}, \eqref{eq-LHS-bound}, and \eqref{eq-LHS-RHS-bound} establishes \eqref{eq-diffpara-bound}, proving that $\InfFt^{\text{UB}}(\bO)$ is the EIF for $\tau^{\text{UB}}$ in model $\mathcal{M}_{\OREC}$. The proof for $\tau^{\text{LB}}$ is analogous.

For the binary outcome, $w^{\min}(y,\bx) = \ind(y = y_R) + \Gamma^{-1} \ind(y \neq y_R)$ does not depend on a threshold. Thus, the weight $w^{\min}$ is a fixed function of $y$ and $\bx$ that does not vary with the data-generating distribution, so $\alpha_1^{\text{UB}}(y,\bx \con \eta) = w^{\min}(y,\bx) \cdot \alpha_0(y,\bx \con \eta)$ directly gives $s_\alpha^{\text{UB}} = s_\alpha$ (Step 2), and the remainder of the proof (Step 3) proceeds identically.

This concludes the proof.

\subsection{Proof of Theorem \ref{thm-EIF-General}}

The proof is similar to the proof of Theorem \ref{thm-EIF} in Section \ref{sec:supp-Proof-thm-EIF}. In the parametric submodel at $\eta$, we obtain
\begin{align*}
&
\MM_{\EFF}(\bO \con \btheta , \eta)
\nonumber
\\
&
=
\underbrace{
(1-A) \beta_1(\bX \con \eta) \alpha_1(\Yo,\bX \con \eta)
\big\{
\MM ( \Yo,\bX \con \btheta )
-
\mu_{\MM}(\bX \con \btheta, \eta)
\big\}
+
A \mu_{\MM}(\bX \con \btheta, \eta)
}_{=: \MM_{\text{DR}}(\bO_1 \con \btheta , \eta)}
\nonumber
\\
& \hspace*{1cm}
+
\underbrace{ (2A-1) \beta_1(\bX \con \eta) \alpha_1(\Yo , \bX \con \eta) 
\frac{\potf{1A|X}(\Yz,0 \cond \bX \con \eta)}{\potf{0A|X}(\Yz,A \cond \bX \con \eta)}
\big\{
\MM ( \Yo,\bX \con \btheta )
-
\mu_{\MM}(\bX \con \btheta, \eta)
\big\} }_{=: \MM_{\text{Aug}}(\bO_0 \con \btheta , \eta)} \ .
\end{align*}
Let the Jacobian matrices of $\mu_{\MM}(\bX \con \btheta, \eta)$ be given as
\begin{align*}
&
\nabla_\theta \mu_{\MM} (\bX \con \btheta,\eta)
=
\nabla_\theta \EXP^{(\eta)} \big\{  \MM ( \pYo,\bX \con \btheta) \cond A=1, \bX \big\}
=
\bigg[ \frac{\partial \mu_{\MM,i} (\btheta)}{\partial \theta_j} \bigg]_{i,j}
\in \R^{p \times p} \ ,
\\
&
\nabla_\eta \mu_{\MM} (\bX \con \btheta,\eta)
=
\nabla_\eta
\EXP^{(\eta)} \big\{ \MM ( \pYo,\bX \con \btheta) \cond A=1, \bX \big\}
=
\bigg[
\frac{\partial \mu_{\MM,1} (\btheta)}{\partial \eta} \ , \ \ldots \ , \ \ \frac{\partial \mu_{\MM,p} (\btheta)}{\partial \eta} 
\bigg] \T
\in \R^{p \times 1} \ .
\end{align*}
Let $\btheta(\eta)$ be the solution to the moment equation:
\begin{align}\label{eq-MomentEquation-General}
& 0 =
\EXP^{(\eta)} \big\{ 
\MM_{\EFF}(\bO \con \btheta(\eta) , \eta)
\big\}  \ .
\end{align}	
We take the derivative of the moment equation \eqref{eq-MomentEquation-General} at $\eta$, which yields
\begin{align*}
0
& = 
\frac{\partial}{\partial \eta}
\EXP^{(\eta)} 
\Big\{ \MM_{\EFF}(\bO_1 \con \btheta(\eta) , \eta) \Big\}
\\
& =
\EXP^{(\eta)}
\Big\{
s_1(\bO_1 \con \eta) \MM_{\text{DR}} (\bO_1 \con \btheta(\eta), \eta) 
\Big\}
\\
& \hspace*{1cm}
+
\EXP^{(\eta)}
\Big[
(1-A) s_{\alpha} (\Yo, \bX \con \eta)
\beta_1(\bX \con \eta) \alpha_1(\Yo, \bX \con \eta)
\big\{
\MM ( \Yo,\bX \con \btheta (\eta) )
-
\mu_{\MM} (\bX \con \btheta (\eta), \eta)
\big\}
\Big]
\\
& \hspace*{1cm}
+
\underbrace{ \EXP^{(\eta)}
\Big[
(1-A) s_\beta (\bX \con \eta)
\beta_1(\bX \con \eta) \alpha_1(\Yo, \bX \con \eta)
\big\{
\MM ( \Yo,\bX \con \btheta (\eta) )
-
\mu_{\MM} (\bX \con \btheta (\eta), \eta)
\big\}
\Big] }_{=0}
\\	
& \hspace*{1cm}
+
\underbrace{
\nabla_{\eta} 
\EXP^{(\eta)}
\Big[
\big\{
(1-A) \beta_1(\bX \con \eta) \alpha_1(\Yo, \bX \con \eta) - A \big\}
\mu_{\MM} (\bX \con \btheta (\eta), \eta)
\Big]
}_{=0} 
\\
&
\hspace*{1cm}
+
\nabla_{\theta}\T
\EXP^{(\eta)}
\Big[
(1-A) 
\beta_1(\bX \con \eta) \alpha_1(\Yo, \bX \con \eta)
\MM ( \Yo,\bX \con \btheta (\eta) )		
\Big]
\frac{\partial \btheta(\eta)}{\partial \eta}
\\	
& \hspace*{1cm}
+
\underbrace{ 
\nabla_{\theta}\T
\EXP^{(\eta)}
\Big[
\big\{
(1-A) \beta_1(\bX \con \eta) \alpha_1(\Yo, \bX \con \eta) - A \big\}
\mu_{\MM} (\bX \con \btheta (\eta), \eta)
\Big]}_{=0}
\frac{\partial \btheta(\eta)}{\partial \eta}
\\
&
=
\EXP^{(\eta)}
\Big\{
s_1(\bO_1 \con \eta) \MM_{\text{DR}} (\bO_1 \con \btheta(\eta), \eta) 
\Big\}
\\
& \hspace*{1cm}
+
\EXP^{(\eta)}
\Big[
(1-A) s_{\alpha} (\Yo, \bX \con \eta)
\beta_1(\bX \con \eta) \alpha_1(\Yo, \bX \con \eta)
\big\{
\MM ( \Yo,\bX \con \btheta (\eta) )
-
\mu_{\MM} (\bX \con \btheta (\eta), \eta)
\big\}
\Big]
\\
& \hspace*{1cm}
+
\nabla_{\theta}\T
\EXP^{(\eta)}
\Big[
(1-A) 
\beta_1(\bX \con \eta) \alpha_1(\Yo, \bX \con \eta)		
\MM ( \Yo,\bX \con \btheta (\eta) )		
\Big]
\frac{\partial \btheta(\eta)}{\partial \eta} \ .
\end{align*}
The underbraced terms are zero, which are shown in \eqref{eq-pd}. Therefore, we obtain
\begin{align*}
\frac{\partial \btheta(\eta)}{\partial \eta}
= 
& 
-
\Big[ 
\underbrace{
\nabla_{\theta} \T
\EXP^{(\eta)}
\Big[
(1-A) 
\beta_1(\bX \con \eta) \alpha_1(\Yo, \bX \con \eta)
\MM ( \Yo,\bX \con \btheta (\eta) )		
\Big]}_{=:V_\EFF(\btheta(\eta), \eta)} \Big]^{-1}
\\
& 
\hspace*{1cm}
\times
\left[
\begin{array}{l}
\EXP^{(\eta)}
\Big\{
s_1(\bO_1 \con \eta) \MM_{\text{DR}} (\bO_1 \con \btheta(\eta), \eta) 
\Big\}
+
\EXP^{(\eta)}
\Bigg[
\begin{array}{l}
(1-A) s_{\alpha} (\Yo, \bX \con \eta)
\beta_1(\bX \con \eta) \alpha_1(\Yo, \bX \con \eta)
\\
\times
\big\{
\MM ( \Yo,\bX \con \btheta (\eta) )
-
\mu_{\MM} (\bX \con \btheta (\eta), \eta)
\big\}
\end{array}
\Bigg]
\end{array}
\right]	.
\end{align*}

Recall that the conjectured EIF is $\InfFt^*(\bO \con \btheta^*)
= 
-
\Big\{
V_\EFF^* (\btheta^*) \Big\}^{-1}
\MM_{\EFF}^* (\bO \con \btheta^*)$ where 
\begin{align*}  
V_\EFF^* (\btheta^*)
& =
\nabla_{\theta}\T 
\EXP \big\{ \MM_{\EFF}(\bO \con \btheta) \big\} \big|_{\btheta=\btheta^*}
=
\nabla_{\theta}\T 
\EXP \big\{ A \MM ( \Yo,\bX \con \btheta ) \big\} \big|_{\btheta=\btheta^*}
\\
& =
\nabla_{\theta} \T
\EXP
\big\{
(1-A) 
\beta_1^*(\bX) \alpha_1^*(\Yo, \bX)
\MM ( \Yo,\bX \con \btheta^* )		 \big\} \big|_{\btheta=\btheta^*}
= V_\EFF(\btheta(\eta^*), \eta^*) \ .
\end{align*}

Recall that the tangent space of the model $\mathcal{M}_{\OREC}$, defined in \eqref{eq-tangentspace}, is the entire Hilbert space of mean-zero, square-integrable functions of $\bO$. Therefore, to show that $\InfFt^*$ is the EIF, it suffices to show that $\btheta$ is a differentiable parameter; i.e., 
\begin{align*}
\EXP \big\{ s_O (\bO \con \eta^*) \InfFt^*(\bO \con \btheta^*)  \big\}
=
\frac{\partial \btheta(\eta)}{\partial \eta} \bigg|_{\eta=\eta^*}
\ .
\end{align*}
Since $V_\EFF^* (\btheta^*)$ is included in both sides, it is sufficient to show
\begin{align*}
\EXP \big\{ s_O(\bO \con \eta^*) \MM_{\EFF}^* (\bO \con \btheta^*) \big\}
& =
\EXP
\Big\{
s_1(\bO_1 \con \eta^*) \MM_{\text{DR}}^* (\bO_1 \con \btheta^*) 
\Big\}
\\
& 
\hspace*{0.25cm}
+
\EXP
\Big[
(1-A) s_{\alpha} (\Yo, \bX \con \eta^*)
\beta_1^*(\bX) \alpha_1^*(\Yo, \bX)
\big\{
\MM ( \Yo,\bX \con \btheta^* )
-
\mu_{\MM}^* (\bX \con \btheta^*)
\big\}
\Big] \ .
\end{align*}

The left hand side is
\begin{align*}
\EXP \big\{ s_O(\bO \con \eta^*) \MM_{\EFF}^* (\bO \con \btheta^*) \big\}
&
=
\underbrace{
\EXP \big\{ s_{0|1}(\Yz \cond \bO_1 \con \eta^*) \MM_{\text{DR}}^*(\bO_1 \con \btheta^*) \big\}
}_{=0}
+
\EXP \big\{ s_{1}( \bO_1 \con \eta^*) \MM_{\text{DR}}^*(\bO_1 \con \btheta^*) \big\}
\\
& \hspace*{1cm}
+
\underbrace{
\EXP \big\{ s_{1|0}(\Yo \cond \bO_0 \con \eta^*) \MM_{\text{Aug}}^*(\bO_0 \con \btheta^*) \big\}
}_{=0}
+
\EXP \big\{ s_{0}(\bO_0 \con \eta^*) \MM_{\text{Aug}}^*(\bO_0 \con \btheta^*) \big\}
\\
& 
=
\EXP \big\{ s_{1}( \bO_1 \con \eta^*) \MM_{\text{DR}}^*(\bO_1 \con \btheta^*) \big\}
+
\EXP \big\{ s_{0}(\bO_0 \con \eta^*) \MM_{\text{Aug}}^*(\bO_0 \con \btheta^*) \big\} \ .
\end{align*}
Therefore, it suffices to show 
\begin{align*}
&
\EXP
\Big[
(1-A) s_{\alpha} (\Yo, \bX \con \eta^*)
\beta_1^*(\bX) \alpha_1^*(\Yo, \bX)
\big\{
\MM ( \Yo,\bX \con \btheta^* )
-
\mu_{\MM}^* (\bX \con \btheta^*)
\big\}
\Big]
=
\EXP \big\{ s_{0}(\bO_0 \con \eta^*) \MM_{\text{Aug}}^*(\bO_0 \con \btheta^*) \big\} \ ,
\end{align*}
which can be established in the same way as \eqref{eq-LHS-RHS}.
This concludes that $\btheta(\eta)$ is a differentiable parameter, i.e.,
\begin{align*}
\EXP \big\{ s_O (\bO \con \eta^*) \InfFt^*(\bO \con \btheta^*)  \big\}
=
\frac{\partial \btheta(\eta)}{\partial \eta} \bigg|_{\eta=\eta^*}
\ .
\end{align*}
Consequently, $\InfFt^*(\bO \con \btheta^*)$ is the EIF of $\btheta^*$ in the model $\mathcal{M}_{\OREC}$.

\subsection{Proof of Theorem \ref{thm-AsympNormal-General}}

\begin{itemize}
    \item[(i)] (\textit{Consistency of $\widehat{\btheta}$})
\end{itemize}

Since $\widehat{\btheta} = K^{-1} \sum_{k=1}^{K} \widehat{\btheta}^{(k)}$, it suffices to show that $\widehat{\btheta}^{(k)} = \btheta^* + o_P(1)$. We will apply Theorem 5.9 of \citet{Vaart1998} to $\widehat{\btheta}^{(k)}$, which is given below:
\begin{theorem}[Theorem 5.9 of \citet{Vaart1998}] \label{supp-thm-59}
Let $\Psi_n$ be random vector-valued functions and let $\Psi$ be a fixed vector-valued function of $\theta$ such that 
\begin{itemize}
\item[(C1)] $\sup_{\theta \in \Theta} \big\| \Psi_n(\theta) - \Psi(\theta) \big\| = o_P(1)$.
\item[(C2)] For every $\epsilon>0$, $\inf_{\theta: d(\theta,\theta_0) \geq \epsilon} \big\| \Psi(\theta) \| > 0 = \big\| \Psi(\theta_0) \big\|$.
\item[(C3)] An estimator $\widehat{\theta}_n$ satisfies $\Psi_n(\widehat{\theta}_n) = o_P(1)$.
\end{itemize}
Then, $\widehat{\theta}_n = \theta_0+o_P(1)$.
\end{theorem}

We establish the assumptions of Theorem \ref{supp-thm-59}. From the law of large numbers, we find $\AVER_{\mathcal{I}_k} \big\{ \widehat{\MM}_{\EFF} \LSS (\bO \con \btheta) \big\}
-
\EXP \LSS \big\{ \widehat{\MM}_{\EFF}\LSS (\bO \con \btheta) \big\} = O_P\big( \big|\mathcal{I}_k \big|^{-1/2} \big)$ holds for any $\btheta$. Additionally, from \eqref{Term3-General}, we find $\EXP \LSS \big\{ \widehat{\MM}_{\EFF}\LSS (\bO \con \btheta) \big\} 
-
\EXP \LSS \big\{ \MM_{\EFF}^* (\bO \con \btheta) \big\} 
=
o_P\big( \big|\mathcal{I}_k \big|^{-1/2} \big)$.  Combining these two results, we find (C1) of Theorem \ref{supp-thm-59} is satisfied as $\AVER_{\mathcal{I}_k} \big\{ \widehat{\MM}_{\EFF} \LSS (\bO \con \btheta) \big\}
-
\EXP \LSS \big\{ \MM_{\EFF}^* (\bO \con \btheta) \big\} 
=
o_P (1) \text{ for all }\btheta \in \Theta$. Next, (C2) of Theorem \ref{supp-thm-59} is already satisfied because it is the same as Regularity condition \HL{R3}. Lastly, note that $\widehat{\btheta}^{(k)}$ is the solution satisfying (C3) of Theorem \ref{supp-thm-59} with $o_P(N^{-1/2}) = \AVER_{\mathcal{I}_k} \big\{ \widehat{\MM}_\EFF \LSS (\bO \con \widehat{\btheta}^{(k)} ) \big\}$. Therefore, we have $\widehat{\btheta}^{(k)} = \btheta^* + o_P(1)$ from Theorem \ref{supp-thm-59}.

\begin{itemize}
    \item[(ii)] (\textit{Asymptotic Normality of $\widehat{\btheta}$})
\end{itemize}

If we show that $\widehat{\btheta}^{(k)}$ has the asymptotic representation as
\begin{align}                       \label{eq-ssestimator-general}
\big| \mathcal{I}_k \big|^{1/2}
\Big\{
\widehat{\btheta}^{(k)} - \btheta^*
\Big\}
=
\frac{1}{\big| \mathcal{I}_k \big|^{1/2}}
\sum_{ i \in \mathcal{I}_k } \InfFt(\bO_i \con \btheta^*) + o_P(1) \ , 
\end{align}
then we establish $\widehat{\btheta} = K^{-1} \sum_{k=1}^{K} \widehat{\btheta}^{(k)}$ has the asymptotic representation as 
\begin{align*}
\sqrt{N} \Big( \widehat{\btheta} - \btheta^* \Big)
=
\frac{1}{\sqrt{N}}
\sum_{i=1}^{N}  \InfFt^*(\bO_i \con \btheta^*) + o_P(1) \ .
\end{align*}
Therefore, the asymptotic normality result holds from the central limit theorem. Thus, we focus on showing that  \eqref{eq-ssestimator-general} holds.

To show the asymptotic normality, we use Theorem 5.21 of \citet{Vaart1998}:
\begin{theorem}[Theorem 5.21 of \citet{Vaart1998}] \label{supp-thm-521}
For each $\theta$ in an open subset of Euclidean space, let $x \mapsto \psi_\theta(x)$ be a measurable vector-valued function such that
\begin{itemize}
\item[(C1)] $\EMP_n(\psi_{\widehat{\theta}_n}) - \EMP_n (\psi_{\theta_0}) = o_P(1)$.        
% For every $\theta_1$ and $\theta_2$ in a neighborhood of $\theta_0$ and a measurable function $\dot{\psi}$ with $\EXP(\dot{\psi})<\infty$, $\big\| \psi_{\theta_1}(x) - \psi_{\theta_2}(x) \big\| \leq \dot{\psi}(x) \big\| \theta_1 - \theta_2 \big\|$.
\item[(C2)] $\EXP \big\{ \| \psi_{\theta_0} \|^2 \big\} < \infty$ and that the map $\theta \mapsto \EXP ( \psi_\theta)$ is differentiable at zero $\theta_0$ with nonsingular derivative matrix $V_{\theta_0}$.
\item[(C3)] $\AVER_n \psi_{\widehat{\theta}_n} = o_P(N^{-1/2})$.
\item[(C4)] $\widehat{\theta}_n = \theta_0+o_P(1)$.
\end{itemize}
Then, $\displaystyle{\sqrt{n} \big( \widehat{\theta}_n - \theta_0 \big) = - V_{\theta_0}^{-1} \frac{1}{\sqrt{n}} \sum_{i=1}^{n} \psi_{\theta_0}(X_i) + o_P(1)}$.
% \begin{align*}
%      \ .
% \end{align*}    
\end{theorem}

We first show that $\EXP \big\{ \big\| \MM_\EFF^*(\bO \con \widehat{\btheta}^{(k)}) - \MM_\EFF^*(\bO \con \btheta^*)  \big\|_2^2 \big\} \precsim \big\| \widehat{\btheta}^{(k)} - \btheta^* \big\|_2^2 = o_P(1)$. For given nuisance functions ${\eta}$, we find
\begin{align}							\label{eq-gen-EE-1}
\hspace*{-1cm}
\MM_\EFF(\bO \con \btheta_1, {\eta}) - \MM_\EFF(\bO \con \btheta_2, {\eta})
=
\left[
\begin{array}{l}
(1-A)  {\beta}_1 (\bX) {\alpha}_1 (\Yo,\bX)
\big\{ \MM(\Yo,\bX \con \btheta_1) -  \MM(\Yo,\bX \con \btheta_2) \big\}
\\
-(1-A)  {\beta}_1 (\bX) {\alpha}_1 (\Yo,\bX)
\big\{ \mu_\MM( \bX \con \btheta_1, {\eta}) -  \mu_\MM( \bX \con \btheta_2, {\eta}) \big\}
\\
+ A \big\{ \mu_\MM(\bX \con \btheta_1, {\eta}) -  \mu_\MM(\bX \con \btheta_2, {\eta}) \big\}
\\
+ (2A-1) R(\Yz,A,\bX) \big\{ \MM(\Yz,\bX \con \btheta_1) -  \MM(\Yz,\bX \con \btheta_2) \big\}
\\
- (2A-1) R(\Yz,A,\bX) \big\{ \mu_\MM(\bX \con \btheta_1, {\eta}) -  \mu_\MM(\bX \con \btheta_2, {\eta}) \big\}
\end{array}		
\right].
\end{align}
If the nuisance functions are uniformly bounded, $\big\| \MM_\EFF(\bO \con \btheta_1, {\eta}) - \MM_\EFF(\bO \con \btheta_2, {\eta})  \big\|_2^2$ is represented as 
\begin{align}						\label{eq-gen-EE-2}
\hspace*{-1cm}
\big\| \MM_\EFF(\bO \con \btheta_1, {\eta}) - \MM_\EFF(\bO \con \btheta_2, {\eta})  \big\|_2^2 
& 
=
\left[
\begin{array}{l}
(1-A)  {\beta}_1 (\bX) {\alpha}_1 (\Yo,\bX)
\big\| \MM(\Yo,\bX \con \btheta_1) -  \MM(\Yo,\bX \con \btheta_2) \big\|_2^2
\\
-(1-A)  {\beta}_1 (\bX) {\alpha}_1 (\Yo,\bX)
\big\| \mu_{\MM}(\bX \con \btheta_1, {\eta}) -  \mu_{\MM}(\bX \con \btheta_2, {\eta}) \big\|_2^2
\\
+ A \big\| \mu_{\MM}(\bX \con \btheta_1, {\eta}) -  \mu_{\MM}(\bX \con \btheta_2, {\eta}) \big\|_2^2
\\
+ (2A-1) R(\Yz,A,\bX) \big\| \MM(\Yz,\bX \con \btheta_1) -  \MM(\Yz,\bX \con \btheta_2) \big\|_2^2
\\
- (2A-1) R(\Yz,A,\bX) \big\| \mu_{\MM}(\bX \con \btheta_1, {\eta}) -  \mu_{\MM}(\bX \con \btheta_2, {\eta}) \big\|_2^2
\end{array}		
\right]
\nonumber
\\
& \hspace*{-1cm} 
\precsim
\left[
\begin{array}{l}
    (1-A) \beta_1^*(\bX) \alpha_1^*(\Yo,\bX) 
    \big\| \MM(\Yo,\bX \con \btheta_1) -  \MM(\Yo,\bX \con \btheta_2) \big\|_2^2
    \\
    + \big\| \mu_{\MM}(\bX \con \btheta_1, {\eta}) -  \mu_{\MM}(\bX \con \btheta_2, {\eta}) \big\|_2^2
    \\
+ (1-A) R^*(\Yz,A,\bX) \big\| \MM(\Yz,\bX \con \btheta_1) -  \MM(\Yz,\bX \con \btheta_2) \big\|_2^2
\\
+ A R^*(\Yz,A,\bX) \big\| \MM(\Yz,\bX \con \btheta_1) -  \MM(\Yz,\bX \con \btheta_2) \big\|_2^2
\end{array}
\right].
\end{align}
From the Taylor expansion, we find
\begin{align*}
\mu_{\MM}(\bX \con \btheta_1, {\eta}) -  \mu_{\MM}(\bX \con \btheta_2, {\eta}) 
= \mathcal{J}(\bx \con \btheta_2, {\eta}) \big( \btheta_1 - \btheta_2 \big) + r_\MM(\bx \con \btheta_1,\btheta_2, {\eta}) (\btheta_1-\btheta_2)
\end{align*}
where $\mathcal{J}(\bx \con \btheta, {\eta})$ is the Jacobian matrix $\nabla_\theta\T \mu_\MM(\bX \con \btheta \con {\eta})$, and the remainder $r_\MM(\bx \con \btheta_1,\btheta_2, {\eta})$ is uniformly bounded and satisfies $\lim_{\btheta_1 \rightarrow \btheta_2} r_\MM(\bx \con \btheta_1,\btheta_2, {\eta}) = 0$. This indicates $\big\| \mu_{\MM}(\bX \con \btheta_1, {\eta}) -  \mu_{\MM}(\bX \con \btheta_2, {\eta}) \big\|_2^2 \leq \omega_1(\bX, {\eta}) \big\| \btheta_1 - \btheta_2 \big\|_2^2$ for some bounded function $\omega_1$. Then, the expectation of \eqref{eq-gen-EE-2} is
\begin{align}                      \label{eq-gen-EE-3}
& \EXP \big\{ \big\| \MM_\EFF(\bO \con \btheta_1, {\eta}) - \MM_\EFF(\bO \con \btheta_2, {\eta})  \big\|_2^2 \big\}
\nonumber
\\
& \precsim
\EXP 
\left[
\int_{\SPo} \alpha_1^*(y,\bX) 
    \big\| \MM(y,\bX \con \btheta_1, {\eta}) -  \MM(y,\bX \con \btheta_2, {\eta}) \big\|_2^2
    \potf{1|AX}^*(y \cond 0,\bX) \, dy
+ \big\| \mu_{\MM}(\bX \con \btheta_1, {\eta}) -  \mu_{\MM}(\bX \con \btheta_2, {\eta}) \big\|_2^2
\right]
\nonumber
\\
& \precsim
\EXP 
\left[
\int_{\SPo} \alpha_1^*(y,\bX) 
    \big\{ \MM(y,\bX \con \btheta_1) -  \MM(y,\bX \con \btheta_2) \big\}^2
    \potf{1|AX}^*(y \cond 0,\bX) \, dy
    + \omega_1(\bX , {\eta} ) \big\| \btheta_1 - \btheta_2 \big\|_2^2
\right]
\nonumber
\\
& \precsim
\EXP 
\big\{ \omega(\bX, {\eta}^* ) + \omega_1(\bX, {\eta}) \big\}
\big\| \btheta_1 - \btheta_2 \big\|_2^2
\leq C({\eta}) \cdot  \big\| \btheta_1 - \btheta_2 \big\|_2^2 \ ,
\end{align}
The first inequality holds from \eqref{eq-gen-EE-2}, and the second inequality is from the established result above. The third inequality is from Regularity condition \HL{R6}. The last line is from the boundedness of $\omega$ and $\omega_1$.

We first show that Condition (C1) of Theorem \ref{supp-thm-521} is satisfied. From Condition \HL{R5}, we find $\big\{ \MM_\EFF^*(\bO \con \btheta) \cond \btheta \in \Theta \}$ is $P$-Donsker. Additionally, from \eqref{eq-gen-EE-3}, $\EXP \big\{ \big\| \MM_\EFF^*(\bO \con \widehat{\btheta}^{(k)}) - \MM_\EFF^*(\bO \con \btheta^*)  \big\|_2^2 \big\} \precsim \big\| \widehat{\btheta}^{(k)} - \btheta^* \big\|_2^2 = o_P(1)$ by taking ${\eta}$ as the true nuisance components. Therefore, we obtain $\EMP_{\mathcal{I}_k} \big\{ \MM_\EFF^* (\bO \con \widehat{\btheta}^{(k)} ) \big\} - \EMP_{\mathcal{I}_k} \big\{ \MM_\EFF^* (\bO \con \btheta^* ) \big\} = o_P(1)$ from Lemma 19.24 of \citet{Vaart1998}:

\begin{lemma}[Lemma 19.24 of \citet{Vaart1998}]
Suppose that $\mathcal{F}$ is a $P$-Donsker class of measurable functions and $\widehat{f}_n$ is a sequence of random functions that take their values in $\mathcal{F}$ such that $\int (\widehat{f}_n(x) - f_0(x))^2 \, dP(x)$ converges in probability to 0 for some $f_0 \in L_2(P)$. Then, $\EMP_n(\widehat{f}_n -f_0) =o_P(1)$ and hence $\EMP_n(\widehat{f}_n) \stackrel{D}{\rightarrow} \EMP_P f_0$.
\end{lemma}

Condition (C2)  of Theorem \ref{supp-thm-521} is implied by Regularity condition \HL{R2} and \HL{R4}. 

To show (C3), we first find 
\begin{align}
0 
& = \AVER_{\mathcal{I}_k} \big\{ \widehat{\MM}_\EFF\LSS (\bO \con \widehat{\btheta}^{(k)} ) \big\}
\nonumber
\\
&
= \AVER_{\mathcal{I}_k} \big\{ \MM_\EFF^* (\bO \con \widehat{\btheta}^{(k)} ) \big\}
+
\big[ \AVER_{\mathcal{I}_k} \big\{ \widehat{\MM}_\EFF\LSS (\bO \con \widehat{\btheta}^{(k)} ) \big\} 
- 
\AVER_{\mathcal{I}_k} \big\{ \MM_\EFF^* (\bO \con \widehat{\btheta}^{(k)} ) \big\} \big] \ .
\label{eq-supp-general-C3-1}
\end{align}
Consider a class $\Delta_{\MM} := \big\{ \widehat{\MM}_{\EFF}(\bO \con \btheta) - \Omega_{\MM}(\bO \con \btheta) \cond \theta \in \Theta \}$. Note that $\Delta_{\Omega}$ is $P$-Donsker because $\big\{ \widehat{\MM}_\EFF \LSS (\bO \con \btheta) \cond \btheta \in \Theta \big\}$ and $ \big\{{\MM}_{\EFF}^* (\bO \con \btheta) \cond \btheta \in \Theta \big\}$ are $P$-Donsker from Regularity condition \HL{R5}, and a pairwise sum of Donsker classes is also Donsker \citep{VW1996}[Example 2.10.7]. Therefore, applying Lemma 19.24 of \citet{Vaart1998}, we obtain  
\begin{align}
&
\big| \mathcal{I}_k \big|^{1/2}
\Big[
\AVER_{\mathcal{I}_k} \big\{ \widehat{\MM}_\EFF\LSS (\bO \con \widehat{\btheta}^{(k)} ) \big\}
-
\AVER_{\mathcal{I}_k} \big\{ \MM_\EFF^* (\bO \con \widehat{\btheta}^{(k)} ) \big\}
\Big] 
= o_P(1) \ .
\label{eq-supp-general-C3-2}
\end{align}
Combining \eqref{eq-supp-general-C3-1} and \eqref{eq-supp-general-C3-2}, we establish $ \AVER_{\mathcal{I}_k} \big\{ \MM_\EFF^* (\bO \con \widehat{\btheta}^{(k)} ) \big\} = o_P\big( | \mathcal{I}_k |^{-1/2} \big)$, satisfying (C3) of Theorem \ref{supp-thm-521}. 

Condition (C4) is established from Theorem \ref{supp-thm-59}. 

Since all conditions are met, the estimator $\widehat{\btheta}^{(k)}$ has the asymptotic representation
\begin{align*}
\big| \mathcal{I}_k \big|^{1/2} 
\big\{ \widehat{\btheta}^{(k)} - \btheta^* \big\}
& =
\big| \mathcal{I}_k \big|^{-1/2}
\sum_{i \in \mathcal{I}_k} 
- \big\{ V_{\EFF}^*(\btheta^*) \big\}^{-1}
\MM_\EFF^*(\bO_i \con \btheta^*) 
+
o_P(1)
=
\big| \mathcal{I}_k \big|^{-1/2}
\sum_{i \in \mathcal{I}_k} 
\InfFt^*(\bO_i \con \btheta^*)
+
o_P(1) \ . 
\end{align*}
Here, we find the Jacobian of $\MM_\EFF^*(\bO \con \btheta^*)$ is $V_\EFF^*(\btheta^*)$ as follows:
\begin{align*}
& \nabla_\theta \T \EXP \big\{ \MM_\EFF^*(\bO \con \btheta) \big\} \big|_{\btheta = \btheta^*}
= \nabla_\theta \T \EXP \left[ 
\begin{array}{l}
(1-A) \beta_1^*(\bX) \alpha_1^*(\Yo, \bX) \big\{ \MM(\Yo,\bX \con \btheta) - \mu_\MM^*(\bX \con \btheta) \big\} + A \mu_\MM^*(\bX \con \btheta) 
\\
+ (2A-1) R^*(\Yz,A,\bX) \big\{ \MM(\Yz,\bX \con \btheta) - \mu_\MM^*(\bX \con \btheta) \big\}
\end{array}
\right] \Bigg|_{\btheta=\btheta^*}
\\
&
= \nabla_\theta \T \EXP \Big\{ A \mu_\MM^*(\bX \con \btheta)  \Big\}  \Big|_{\btheta=\btheta^*}
= \nabla_\theta \T \EXP \Big\{ (1-A) \beta_1^*(\bX) \alpha_1^*(\Yo,\bX) \MM(\Yo,\bX \con \btheta) \Big\}  \Big|_{\btheta=\btheta^*}
= V_\EFF^*(\btheta^*)
\end{align*}

\begin{itemize}
    \item[(iii)] (\textit{Consistency of Variance Matrix})
\end{itemize}

For notational brevity, let ${v}^{\otimes 2} = {v} {v}\T$. Let $\Sigma_M$ and $\Sigma_B$ be the ``meat'' and ``bread'' of the sandwich variance matrix, i.e., $\Sigma_M := \EXP \big\{ \MM_\EFF(\bO \con \btheta^*)^{\otimes 2} \big\} $ and $\Sigma_B :=  V_\EFF^* (\btheta) = \nabla_\theta \T \EXP \Big\{ A \mu_\MM^*(\bX \con \btheta)  \Big\}  \Big|_{\btheta=\btheta^*}$. 
% \begin{align*}
%     & 
%     \ , \
%     &&
%     \Sigma_B :=  V_\EFF^* (\btheta) = \nabla_\theta \T \EXP \Big\{ A \mu_\MM^*(\bX \con \btheta)  \Big\}  \Big|_{\btheta=\btheta^*}
%      \ . 
% \end{align*}
Recall that the variance estimator can be written as  $\widehat{\Sigma}
=
\widehat{\Sigma}_B^{-1} \widehat{\Sigma}_M \widehat{\Sigma}_B^{-\intercal}$ where
\begin{align*}
& 
\widehat{\Sigma}_B
=
K^{-1} \sum_{k=1}^{K}  \widehat{\Sigma}_B \LSS
\ , \
&&
\widehat{\Sigma}_B^{(k)} 
= 
\AVER_{\mathcal{I}_k} \Big\{ A \widehat{\mathcal{J}} \LSS (\bX \con \widehat{\btheta}) \Big\}
\ , \
&&
\widehat{\Sigma}_M
=
K^{-1} \sum_{k=1}^{K} \widehat{\Sigma}_M^{(k)}
\ , \
&&
\widehat{\Sigma}_M^{(k)} = \AVER_{\mathcal{I}_k}
\Big\{ \widehat{\MM}_\EFF\LSS (\bO \con \widehat{\btheta} ) ^{\otimes 2} \Big\} \ .
\end{align*}
To show the consistency of the variance estimator, we first consider the convergence of the numerator $\widehat{\Sigma}_M$; note that it suffices to show that $\widehat{\Sigma}_M^{(k)}$ is consistent for $\Sigma_M$. We find $\widehat{\Sigma}_M^{(k)} - \Sigma_M$ is represented as
\begin{align*}
& \AVER_{\mathcal{I}_k}
\Big[ \widehat{\MM}_\EFF\LSS (\bO \con \widehat{\btheta} ) ^{\otimes 2} \Big]
- 
\EXP \Big[ {\MM}_\EFF^* (\bO \con \btheta^* ) ^{\otimes 2} \Big]
\\
& = 
\AVER_{\mathcal{I}_k}
\Big[ \widehat{\MM}_\EFF\LSS (\bO \con \widehat{\btheta} ) ^{\otimes 2} \Big]
-
\AVER_{\mathcal{I}_k}
\Big[ {\MM}_\EFF^* (\bO \con \btheta^* ) ^{\otimes 2} \Big]
+
\underbrace{
\AVER_{\mathcal{I}_k}
\Big[ {\MM}_\EFF^* (\bO \con \btheta^* ) ^{\otimes 2} \Big] 
- 
\EXP \Big[ {\MM}_\EFF^* (\bO \con \btheta^* ) ^{\otimes 2} \Big] }_{=o_P(1)} \ ,
\end{align*}
where the latter term is $o_P(1)$ from the law of large numbers. Therefore, it suffices to show the first term is $o_P(1)$, which is further decomposed as follows.
\begin{align*}
\AVER_{\mathcal{I}_k}
\Big[
\widehat{\MM}_\EFF\LSS (\bO \con \widehat{\btheta}  ) ^{\otimes 2} - {\MM}_\EFF^* (\bO \con \btheta^* ) ^{\otimes 2}
\Big]
& =
\AVER_{\mathcal{I}_k} \Big[ 
\big\{ \widehat{\MM}_\EFF\LSS (\bO \con \widehat{\btheta}  ) -  {\MM}_\EFF^* (\bO \con \btheta^* ) \big\} ^{\otimes 2}
\Big]
\\
& \hspace*{2cm}
+
\AVER_{\mathcal{I}_k} \Big[ 
\big\{ {\MM}_\EFF^* (\bO \con \btheta^* ) \big\}
\big\{ \widehat{\MM}_\EFF\LSS (\bO \con \widehat{\btheta} ) -  {\MM}_\EFF^* (\bO \con \btheta^* ) \big\} \T
\Big]
\\
& \hspace*{2cm}
+
\AVER_{\mathcal{I}_k} \Big[ 
\big\{ \widehat{\MM}_\EFF\LSS (\bO \con \widehat{\btheta}  ) -  {\MM}_\EFF^* (\bO \con \btheta^* ) \big\}
\big\{ {\MM}_\EFF^* (\bO \con \btheta^* )  \big\}\T
\Big] \ .
\end{align*}
Therefore, the 2-norm of the above term is upper bounded as follows:
\begin{align*}
& \Big\| \AVER_{\mathcal{I}_k}
\Big[
\widehat{\MM}_\EFF\LSS (\bO \con \widehat{\btheta}  ) ^{\otimes 2} - {\MM}_\EFF^* (\bO \con \btheta^* ) ^{\otimes 2}
\Big] \Big\|_2
\\
& \leq 
\Big\|
\AVER_{\mathcal{I}_k} \Big[ 
\big\{
\widehat{\MM}_\EFF\LSS (\bO \con \widehat{\btheta} ) -  {\MM}_\EFF^* (\bO \con \btheta^* ) 
\big\}^{\otimes 2}
\Big] \Big\|_2
+
2 \AVER_{\mathcal{I}_k} \Big[ 
\big\| {\MM}_\EFF^* (\bO \con \btheta^* ) \big\|_2
\big\| \widehat{\MM}_\EFF\LSS (\bO \con \widehat{\btheta}^{(k)} ) -  {\MM}_\EFF^* (\bO \con \btheta^* ) \big\|_2
\Big]
\\
& \leq 
\AVER_{\mathcal{I}_k} \Big[ 
\big\| \widehat{\MM}_\EFF\LSS (\bO \con \widehat{\btheta}  ) -  {\MM}_\EFF^* (\bO \con \btheta^* ) \big\|_2^2
\Big] 
+
2 \Big[ \AVER_{\mathcal{I}_k} \big[ 
\big\| {\MM}_\EFF^* (\bO \con \btheta^* ) \big\|_2^2 \big] \Big]^{1/2}
\Big[
\AVER_{\mathcal{I}_k} \big[ 
\big\| \widehat{\MM}_\EFF\LSS (\bO \con \widehat{\btheta} ) -  {\MM}_\EFF^* (\bO \con \btheta^* ) \big\|_2^2
\big]
\Big]^{1/2} \ .
\end{align*}
From the law of large numbers, we have $\AVER_{\mathcal{I}_k} \big[ 
\big\| {\MM}_\EFF^* (\bO \con \btheta^* ) \big\|_2^2 \big] = \EXP \big[ 
\big\| {\MM}_\EFF^* (\bO \con \btheta^* ) \big\|_2^2 \big] + o_P(1) = O_P(1)$. Therefore, to show the consistency of the numerator,  it suffices to show that $ \AVER_{\mathcal{I}_k} \Big[ 
\big\| \widehat{\MM}_\EFF\LSS (\bO \con \widehat{\btheta}  ) -  {\MM}_\EFF^* (\bO \con \btheta^* ) \big\|_2^2
\Big]  = o_P(1)$. We further obtain
\begin{align}	
\AVER_{\mathcal{I}_k} \Big[ 
\big\| \widehat{\MM}_\EFF\LSS (\bO \con \widehat{\btheta}  ) -  {\MM}_\EFF^* (\bO \con \btheta^* ) \big\|_2^2
\Big]
& \leq 
\AVER_{\mathcal{I}_k} \Big[ 
\big\| \widehat{\MM}_\EFF\LSS (\bO \con \widehat{\btheta}^{(k)} ) -  \widehat{\MM}_\EFF\LSS (\bO \con \btheta^*) \big\|_2^2
\Big]
                \label{eq-sandwich-numer-1}
\\
&
+
\AVER_{\mathcal{I}_k} \Big[ \big\| \widehat{\MM}_\EFF\LSS (\bO \con \btheta^*) - {\MM}_\EFF^* (\bO \con \btheta^*) \big\|_2^2 \Big] \ . 
                \label{eq-sandwich-numer-2}
\end{align}

To study the first term \eqref{eq-sandwich-numer-1}, we first establish that $\widehat{\MM}_\EFF\LSS(\bO \con \btheta)$ is uniformly bounded.  From Regularity condition \HL{R2} and Assumption \HL{A7}, we find $ \widehat{\mu}_\MM\LSS (\bX \con \btheta)$ is uniformly bounded:
\begin{align*}
\big\| \widehat{\mu}_\MM\LSS (\bX \con \btheta) \big\|_2
=  \frac{  \int_{\SPo} \big\| \MM(y,\bX \con \btheta) \big\|_2 \widehat{\alpha}_1\LSS(y,\bX) \hpotf{1}\LSS(y \cond 0,\bX) \, dy  }{  \int_{\SPo}  \widehat{\alpha}_1\LSS(y,\bX) \hpotf{1}\LSS(y \cond 0,\bX) \, dy  }
\leq C \ .
\end{align*}
Therefore, we find $\widehat{\MM}_\EFF\LSS(\bO \con \btheta)$ is also uniformly bounded:
\begin{align*}
&
\big\|
\widehat{\MM}_\EFF\LSS(\bO \con \btheta)
\big\|_2
\\
& 
\leq 
\big\| (1-A) \widehat{\beta}_1\LSS (\bX) \widehat{\alpha}_1\LSS(\Yo,\bX) 
\big\{ \MM(\Yo,\bX \con \btheta) -  \widehat{\mu}_\MM\LSS (\bX \con \btheta) \big\} \big\|_2
+ \big\| A   \widehat{\mu}_\MM\LSS (\bX \con \btheta) \big\|_2
\\
& \hspace{1cm}
+ \big\| (2A-1) \widehat{R}\LSS (\Yz,A,\bX) \big\{ \MM(\Yz,\bX \con \btheta) -  \widehat{\mu}_\MM\LSS (\bX \con \btheta) \big\} \big\|_2
\\
& 
\leq 
\big\| \widehat{\beta}_1\LSS (\bX) \widehat{\alpha}_1\LSS(\Yo,\bX) \big\| _2
\big\{  \big\| \MM(\Yo,\bX \con \btheta) \big\|_2
+ \big\| \widehat{\mu}_\MM\LSS (\bX \con \btheta) \big\| _2 \big\}
+ \big\| \widehat{\mu}_\MM\LSS (\bX \con \btheta) \big\|_2
\\
& \hspace{1cm}
+ \big\| \widehat{R}\LSS (\Yz,A,\bX) \big\|_2 \big\{ \big\|\MM(\Yz,\bX \con \btheta) \big\|_2 +  \big\|\widehat{\mu}_\MM\LSS (\bX \con \btheta) \big\|_2 \}
\\
& \leq C \ .
\end{align*}
Let us consider a class of functions $\Xi_\MM := \big\{ \widehat{\MM}_\EFF \LSS (\bO \con \btheta) \cond \btheta \in \Theta \big\}$; from Regularity condition \HL{R5}, we find $\Xi_\MM$ is $P$-Donsker, indicating that $\Xi_\MM$ is $P$-Glivenko-Cantelli \citep[page 82]{VW1996}. Let $\big\{ \Xi_\MM(\btheta^*)  \big\} = \big\{ \widehat{\MM}_\EFF\LSS (\bO \con \btheta^*)  \big\}$, which is $P$-Glivenko-Cantelli because it is a singleton set and integrable \citep[page 270]{Vaart1998}. Next, we consider a class of functions 	$ \Xi_M := \big\{ \big\| \widehat{\MM}_\EFF\LSS (\bO \con \btheta ) -  \widehat{\MM}_\EFF\LSS (\bO \con \btheta^*) \big\|_2^2 \cond \btheta \in \Theta \big\} = \lambda( \Xi_\MM, \big\{\Xi_\MM(\btheta^*)\big\} )$ where $\lambda(x_1,x_2) = \big\| x_1 - x_2 \big\|_2^2$ is continuous. Then, since $\big\| \widehat{\MM}_\EFF\LSS (\bO \con \btheta ) -  \widehat{\MM}_\EFF\LSS (\bO \con \btheta^*) \big\|_2^2 \leq C $  for a constant $C$, we can take $C$ as an envelope function. Therefore, from Theorem 3 of \citet{VW2000}, we show that $\Xi_M$ is $P$-Glivenko-Cantelli. Therefore, we find the empirical mean in \eqref{eq-sandwich-numer-1} converges to its expectation in probability, i.e.,
\begin{align*}
\AVER_{\mathcal{I}_k} \Big[ 
\big\| \widehat{\MM}_\EFF\LSS (\bO \con \widehat{\btheta} ) -  \widehat{\MM}_\EFF\LSS (\bO \con \btheta^*) \big\|_2^2
\Big]
& =
\int \big\| \widehat{\MM}_\EFF\LSS (\bO \con \widehat{\btheta} ) -  \widehat{\MM}_\EFF\LSS (\bO \con \btheta^*) \big\|_2^2
\, dP(\bO) + o_P(1)
\\
& 
\precsim C(\widehat{{\eta}}) \cdot \big\| \widehat{\btheta} - \btheta^* \big\|_2^2  + o_P(1)
= o_P(1) \ .
\end{align*}
The second line holds from \eqref{eq-gen-EE-3}. The last line holds from the consistency of $\widehat{\btheta}$.

Next, we study the second term \eqref{eq-sandwich-numer-2}:
\begin{align*}
& 
\AVER_{\mathcal{I}_k} \Big[ \big\| \widehat{\MM}_\EFF\LSS (\bO \con \btheta^*) 
- {\MM}_\EFF^* (\bO \con \btheta^*) \big\|_2^2 \Big] 
\\
& =
\EXP \LSS \Big[ \big\| \widehat{\MM}_\EFF\LSS (\bO \con \btheta^*) 
- {\MM}_\EFF^* (\bO \con \btheta^*) \big\|_2^2 \Big] 
+
o_P(1) 
\\
&
\precsim
\big\| \widehat{\alpha}_1\LSS  - \alpha_1^*  \big\|_{P,2}^2 
+
\big\| \widehat{\beta}_0\LSS  - \beta_0^*  \big\|_{P,2}^2 	
+
\big\| \widehat{\beta}_1\LSS  - \beta_1^*  \big\|_{P,2}^2 	
+
\big\|  \hpotf{0}\LSS  - \potf{0}^* \big\|_{P,2} ^2 
+
\big\|   \hpotf{1}\LSS  - \potf{1}^* \big\|_{P,2} ^2
+
o_P(1) 
\\
& = o_P(1) \ .
\end{align*}
The second line holds from the law of large numbers, and the third line holds from \eqref{eq-variance0}. Combining the results, we find
\begin{align*}
& \AVER_{\mathcal{I}_k} \Big[ 
\big\| \widehat{\MM}_\EFF\LSS (\bO \con \widehat{\btheta}  ) -  {\MM}_\EFF^* (\bO \con \btheta^* ) \big\|_2^2
\Big]
\\
& \leq 
\AVER_{\mathcal{I}_k} \Big[ 
\big\| \widehat{\MM}_\EFF\LSS (\bO \con \widehat{\btheta}^{(k)} ) -  \widehat{\MM}_\EFF\LSS (\bO \con \btheta^*) \big\|_2^2
\Big]
+
\AVER_{\mathcal{I}_k} \Big[ \big\| \widehat{\MM}_\EFF\LSS (\bO \con \btheta^*) - {\MM}_\EFF^* (\bO \con \btheta^*) \big\|_2^2 \Big] 
= o_P(1) \ .
\end{align*}
This concludes that $\widehat{\Sigma}_M = \Sigma_M + o_P(1)$.

Next, we show the consistency of the ``bread'' part $\widehat{\Sigma}_B$; note that it suffices to show that $\widehat{\Sigma}_B^{(k)}$ is consistent for $\Sigma_B = \EXP \big\{ A \mathcal{J} (\bX \con \btheta^*)  \big\}$ where $\mathcal{J}(\bx \con \btheta, {\eta})$ is the Jacobian matrix $\nabla_\theta\T \mu_\MM(\bX \con \btheta \con {\eta})$. Therefore, we find $\widehat{\Sigma}_B^{(k)} - \Sigma_B$ is
\begin{align} 		\nonumber
\widehat{\Sigma}_B^{(k)} 
-
\Sigma_B
& = 
\AVER_{\mathcal{I}_k} \Big\{ A \widehat{\mathcal{J}} \LSS (\bX \con \widehat{\btheta}) \Big\}
-
\EXP \Big\{ A \mathcal{J}^*(\bX \con \btheta^*) \Big\}
\nonumber
\\
&
= 
\AVER_{\mathcal{I}_k} \Big\{ A \widehat{\mathcal{J}} \LSS (\bX \con \widehat{\btheta}) \Big\}
-
\AVER_{\mathcal{I}_k} \Big\{ A \mathcal{J}^* (\bX \con \btheta^*) \Big\}
\label{eq-sandwich-denom-1}
\\
&
\quad
+
\AVER_{\mathcal{I}_k} \Big\{ A \mathcal{J}^* (\bX \con \btheta^*) \Big\}
-
\EXP \Big\{ A \mathcal{J}^* (\bX \con \btheta^*) \Big\}
\label{eq-sandwich-denom-2}
\ .
\end{align}

Let us consider a class of functions $\Xi_{\mathcal{J},ij} := \big\{ A [\widehat{\mathcal{J}}\LSS (\bX \con \btheta)]_{ij}  \cond \btheta \in \Theta \big\}$ where $[B]_{ij}$ is the $(i,j)$th element of matrix $B$. Note that 
(i) $\Theta$ is compact; 
(ii) $A  \widehat{\mathcal{J}}\LSS (\bX \con \btheta)$ is continuous with respect to $\btheta$ for each $\bX$ from Regularity condition \HL{R4}; and 
(iii) the functions in $\Xi_{\mathcal{J},ij}$ are uniformly bounded, indicating that there exists a constant that is an integrable envelope function of $\Xi_{\mathcal{J},ij}$. 
Therefore, by Example 19.8 of \citet{Vaart1998}, we find $\Xi_{\mathcal{J},ij}$ is $P$-Glivenko-Cantelli. Additionally, $\big\{ \Xi_{\mathcal{J},ij} (\btheta^*) \big\} =  \big\{ A  [\widehat{\mathcal{J}}\LSS (\bX \con \btheta^*)]_{ij} \big\}$ is also $P$-Glivenko-Cantelli because it is a singleton set and integrable \citep[page 270]{Vaart1998}.  Next,  let us consider a class of functions $\Xi_{D,ij} := \big\{ 
A [ \widehat{\mathcal{J}}\LSS(\bX \con \btheta) - \widehat{\mathcal{J}}\LSS(\bX \con \btheta^*) ]_{ij}
\cond \btheta \in \Theta \big\} = \lambda ( \Xi_{\mathcal{J},ij}, \big\{ \Xi_{\mathcal{J},ij} (\btheta^*) \big\} )$ where $\lambda(x_1, x_2) =  x_1 - x_2 $. Then, since $\big| A [ \widehat{\mathcal{J}}\LSS(\bX \con \btheta)  -  \widehat{\mathcal{J}}\LSS(\bX \con \btheta^*) ]_{ij} \big| \leq C $  for a constant $C$, we can take $C$ as an envelope function. Therefore, from Theorem 3 of \citet{VW2000}, we show that $\Xi_{D,ij}$ is $P$-Glivenko-Cantelli. Therefore, we find that each element of the empirical mean in \eqref{eq-sandwich-denom-1} converges to its expectation in probability, i.e.,
\begin{align*}
\AVER_{\mathcal{I}_k} \Big[ 
A \big[
\widehat{\mathcal{J}}\LSS (\bX \con \widehat{\btheta} ) 
-  \widehat{\mathcal{J}}\LSS (\bX \con \btheta^*) 
\big]_{ij}
\Big]
& =
\int A \big[ \widehat{\mathcal{J}}\LSS (\bX \con \widehat{\btheta} ) 
-  \widehat{\mathcal{J}}\LSS (\bX \con \btheta^*) \big]_{ij}
\, dP(\bX) + o_P(1)
\\
& 
\precsim 
\int \big\{ \omega(\bx \con {\eta}) \big\}^{1/2} \, dP(\bX)
\cdot \big\| \widehat{\btheta} - \btheta^* \big\|_2  + o_P(1)
= o_P(1) \ .
\end{align*}
The second line holds from Regularity condition \HL{R4}. The last line holds from the consistency of $\widehat{\btheta}$. This implies that \eqref{eq-sandwich-denom-1} is $o_P(1)$. 

Lastly, \eqref{eq-sandwich-denom-2} is $o_P(1)$ from the law of large numbers. This shows that $\widehat{\Sigma}_B^{(k)} = {\Sigma}_B + o_P(1)$, and $\widehat{\Sigma}_B = \Sigma_B+o_P(1)$. Thus, we obtain $\widehat{\Sigma}_B^{-1} = \Sigma_B^{-1} + o_P(1)$. 

Combining all, we obtain 
\begin{align*}
\widehat{\Sigma}
=
\widehat{\Sigma}_B^{-1} \widehat{\Sigma}_M \widehat{\Sigma}_B^{-\intercal}
=
\big\{ \Sigma_B^{-1} + o_P(1) \big\} \big\{ \Sigma_M + o_P(1) \big\} \big\{ \Sigma_B^{-1} + o_P(1) \big\} \T =
\Sigma_B^{-1} \Sigma_M \Sigma_B^{-\intercal}
+ o_P(1)
=
\Sigma + o_P(1) \ .
\end{align*}
This concludes the proof.

\newpage

\bibliographystyle{apa}
\bibliography{UDID.bib}

\end{document}